\documentclass[twocolumn,showpacs,rmp]{revtex4}

\usepackage{psfrag,graphicx}

\usepackage{dcolumn}
\usepackage{amsmath,amssymb}
\usepackage{bm}
\usepackage[dvips]{color}

\definecolor{mygrey}{gray}{0.35}
\definecolor{mygreen}{rgb}{0.85,1,0.9}
\definecolor{myzard}{cmyk}{0,0,0.05,0}
\definecolor{mywhite}{rgb}{1,1,1}
\definecolor{myred}{rgb}{1,0,0}

\def\C{{\mathbb{C}}} \def\F{{\mathbb{F}}}
\def\N{{\mathbb{N}}} 
\def\R{{\mathbb{R}}} \def\Z{{\mathbb{Z}}}

\def\cH{{\mathcal H}}

\def\bfsigma{\boldsymbol{\sigma}}
\def\bfmu{\boldsymbol{\mu}}

\def\dd{\mathord{\rm d}} 
\def\dist{\mathop{\rm dist}} \def\ee{\mathord{\rm e}}
 
\def\id{\mathord{\rm id}} \def\ii{\mathord{\rm i}}
\def\min{\mathord{\rm min}} \def\mod{\mathord{\rm mod}}
\def\prob{\mathord{\rm prob}} \def\tr{\mathop{\rm Tr}}
\def\half{\textstyle\frac{1}{2}} 
\def\fourth{\textstyle\frac{1}{4}}

\def\vec#1{{\bf{#1}}} \def\vect#1{\vec{#1}}

\def\bra#1{\langle#1|} \def\ket#1{|#1\rangle}
\def\braket#1#2{\langle#1|#2\rangle}
 \def\ve#1{\langle#1\rangle}

\begin{document}

\preprint{HEP/123-qed}

\title[Short Title]{ \null\vskip-11.5mm\hskip0.63\textwidth
{\normalsize\rm {\it Reviews of Modern Physics} to apperar}\\[1.5mm]
Information and Computation: Classical and Quantum Aspects}

\author{A. Galindo} \email{galindo@eucmos.sim.ucm.es}

\author{M.A. Mart\'{\i}n-Delgado}

\email{mardel@miranda.fis.ucm.es}

\affiliation{Departamento de F\'{\i}sica T\'eorica I.  Facultad de
Ciencias F\'{\i}sicas.  \newline Universidad Complutense.  28040
Madrid.  Spain.}

\begin{abstract}
Quantum theory has found a new field of applications in
the realm of information and computation during the recent years.
This paper reviews how quantum physics allows information coding in
classically unexpected and subtle nonlocal ways, as well as
information processing with an efficiency largely surpassing that of
the present and foreseeable classical computers.  Some outstanding
aspects of classical and quantum information theory will be addressed
here.  Quantum teleportation, dense coding, and quantum cryptography
are discussed as a few samples of the impact of quanta in the
transmission of information.  Quantum logic gates and quantum
algorithms are also discussed as instances of the improvement in
information processing by a quantum computer.  We provide finally some
examples of current experimental realizations for quantum computers
and future prospects.
\end{abstract}


\maketitle

\tableofcontents

\section{Introduction}
\label{sec1:level1}

The twentieth century we have just left behind opened with the
discovery of quanta by Planck (1900) and followed with the formulation
of the quantum theory during the first decades.  As the century went
by, we have witnessed a continuous and growing increase in the number
of applications of quantum mechanics, which began with atomic physics
and then the number kept growing (nuclear and particle physics,
optics, condensed matter, \ldots) and became countless.  As the
century was closing we have come across an unexpected new field of
applications that have given quantum physics a refreshing twist,
keeping the pace even with the newest trends of discoveries, such as
the field of new technologies of information and computation.  In a
sense and having in mind the times we live, those of the information
era and the new technologies, it seems inevitable that physics gets
affected by the presence of computers all over around, which are more
and more powerful and have revolutionized many areas of science.  What
is more surprising is the fact that quantum physics may influence the
field of information and computation in a new and profound way,
getting at the very root of their foundations.  For instance,
fundamental aspects of quantum mechanics such as those entering the
EPR (Einstein, Podolsky and Rosen, 1935) states have found unexpected
applications in information transmission and cryptography.

But, why has this happened?  It all begun by realizing that
information  has physical nature (Landauer, 1991; 1996; 1961).  It is
printed on a physical support (the rocky wall of a cave, a clay
tablet, a parchment, a sheet of paper, a magneto-optic disk, etc.), it
cannot be transmitted faster than light in vacuum, and it abides by
the natural laws.  The statement that information is physical does not
simply mean that a computer is a physical object, but in addition that
information itself is a physical entity.  In turn, this implies that
the laws of information are restricted or governed by the laws of
physics.  In particular, those of quantum physics.  In fact these
ones, through their linearity, entanglement of states, nonlocality and
indetermination principle make possible new and powerful transmission
tools and information treatments, as well as a really prodigious
efficiency of computation.

A typical computation is implemented through an algorithm in a
computer.  This algorithm is now regarded as a set of physical
operations and the registers of the quantum computer are considered to
be states of a quantum system.  Moreover, the familiar operation of
initializing the data for a program to run is replaced by the
preparation of an initial quantum state, and the usual tasks of
writing programs and running them correspond, in the new formulation,
to finding appropriate Hamiltonians for their time evolution operators
to lead to the desired output.  This output is retrieved by a quantum
measurement of the register, and this fact has deep implications on
the way quantum information must be handled.

We shall see that information and computation blend well with quantum
mechanics.  Their combination brings unexpected results on the way
information can be transmitted and processed, extending the
capabilities known so far in the field of classical information to
unsuspected limits, sometimes entering the realm of science-fiction,
sometimes surpassing it.

The advance has been remarkable mainly in the field of cryptography,
where it has provided systems absolutely secure for the quantum
distribution of keys.  Quantum computation is also one of the hot
research fields in current physics; the same applies to the challenge
posed by the experimental realization of a computer complex enough to
implement the new algorithms that exploit the fantastic possibilities
of the massive parallelism characterizing those quantum computers, and
that would amount to a dramatic improvement for solving hard or
classically untractable problems.

We first review the essentials of quantum information theory and then
discuss several of their consequences and applications, some of them
specifically quantum such as quantum teleportation, dense coding; some
of them with a classical echo such as quantum cryptography.  Next we
review the fundamentals of quantum computation describing the notion
of a quantum Turing machine and its practical implementation with
quantum circuits.  We describe the notion of elementary quantum gates
for universal computation and how this extends the classical
counterpart.  We also provide a discussion of the basic quantum
algorithms and finally we give a general overview of some of the
possible physical realizations of quantum computers.

Both in the information and computation parts we make special emphasis
in presenting first an introduction to the classical aspects of these
disciplines in order to better clarify what quantum theory adds to
them in the new formulations of these theories.  Actually, this is
also what we do in physics.

\section{Classical Information}
\label{sec2:level1}

Information is discretized: it comes in irreducible packages.  The
elementary unit of classical information is the {\em bit} (or {\em
cbit}, for classic bit), a classical system with only two states 0 and
1 (False and True, No and Yes, \ldots).  Any text can be coded into a
string of bits: for instance, it is enough to assign to each symbol
its ASCII code number in binary form, appended with a parity check
bit.  Example: {\tt quanta} can be coded as

{\footnotesize
\begin{verbatim}
11100010 11101011 11000011 11011101 11101000 11000011
\end{verbatim}
}

Each bit can be stored physically; in classical computers, each bit is
registered as a charge state of a capacitor ($0 = \text{discharged},
1=\text{charged}$).  They are distinguishable macroscopic states, and
robust enough or stable.  They are not spoiled when they are read in
(if carefully done) and they can be cloned or replicated without any
problem.

Information is not only stored; it is usually transmitted
(communication), and sometimes processed (computation).

\subsection{The Theorems of Shannon}
\label{sec2A:level2}

The classical theory of information is due to Shannon (1948,1949), who
in two seminal works definitively laid down its principles in 1948.
With his celebrated {\em noiseless coding theorem} he showed how much
{\em compressible} a message can be, or equivalently, how much
redundancy it has.  Likewise with his {\em coding theorem in a noisy
channel} he also found what is the minimum redundancy that must be
present into a message in order to be {\em comprehensible} when
reaching the receiver, despite of the noise.

Let $A:=\{a_1,...,a_{|A|}\}$ be a finite alphabet, endowed with a
probability distribution $p_A: a_i\mapsto p_A(a_i)$, with $\sum_{1\leq
i\leq |A|}p_A(a_i)=1$.  Sometimes we shall be write this as
$A:=\{a_i,p_A(a_i)\}_{i=1}^{|A|}$.  Let us consider messages or
character strings $x_1x_2...x_n\in A^n$, originating from a memoryless
source, i.e., a symbol $a$ appears in a given place with probability
$p_A(a)$, independently of the symbols entering the remaining sites in
the chain.\footnote{The natural languages are not like these (for
instance, in the usual Spanish there exists no digram like {\sc
q\~n}).  Nevertheless, they can be considered, to a good
approximation, as limit of ergodic Markovian languages to which the
Shannon theorem can be extended (Welsh, 1995).}  The first Shannon's
theorem asserts that, if $n\gg 1$, the information supplied by a
generic message of $n$ characters (and thus ($n\log_2|A|$)-bits long)
essentially coincides with that transmitted by another shorter
message, of bit length $nH(A)$, where $H$ is the so called Shannon's
entropy
\begin{equation}
H(A) = -\sum_{1\leq i\leq |A|}p_A(a_i)\log_2p_A(a_i) \in[0,\log_2|A|].
\label{qi1}
\end{equation}

\noindent In other words, each character is compressible up to $H(A)$
bits on the average; moreover, this result is optimal (Welsh, 1995;
Roman 1992; Schumacher, 1995; Preskill, 1998).

The basic idea underlying the proof is simple: it amounts to take
notice only of the {\em typical} messages.  Let us assume for clarity
a binary alphabet ($A=\{0,1\}$).  Let $p,1-p$ be the probabilities of
0,1, respectively.  In a long message of $n$ bits ($n \gg 1$), there
will be approximately $np$ 0s.  Let us call typical messages those
with a number of 0s of the order of $np$.  Asymptotically
($n\to\infty$), there are $2^{nH(A)}$ many of them, among a total of
$2^n$ messages.  The probability $P:(x_1,...,x_n)\mapsto
p(x_1)...p(x_n)$ of the messages ($n \gg 1$)-bits long tends to get
concentrated on this reduced ensemble consisting of the typical
strings, which explains Shannon's result.  The atypical messages are
ignorable in probability.  It suffices to transmit through the
communication channel (assumed perfect, noiseless) the binary number
of length $nH(A)$ assigned to each typical message upon common
agreement between the sender and the recipient, so that the emitted
message can be identified on reception.\footnote{There exist very
practical methods for classical coding with an efficiency close to the
optimal value, such as the Huffman code (Roman, 1992), with multiple
applications (facsimile, digital TV, etc.).  The essence of this code
is to assign shorter binary strings to the most frequent symbols.} The
optimality of Shannon's first theorem is easily arguable: all
$2^{nH(A)}$ typical sequences are asymptotically equiprobable and thus
they cannot be represented faithfully with less than $nH(A)$ bits.

If the transmission channel is noisy (the common case), the
information fidelity gets lost, since some bits may get corrupted
along the way.  To fight the noise of a given channel one resorts to
redundancy, by cleverly coding each symbol with more bits than
strictly necessary so that the erroneous bits might be easily detected
and restored.  A price is payed however, since the transmission of
essential information gets clearly slower.  Shannon's wonderful second
theorem quantifies this issue.

Let $X$ be the alphabet of the transmitter station (of a memoryless
source), and $Y$ be the one of the receiver station.  Let
$(p_{Y|X}(y_j|x_i))$ be the stochastic matrix for that channel, with
entries given by the probabilities that the input symbol $x_i\in X$
appears as $y_i \in Y$ on output.  The marginal probability
distribution for $Y$ is given by $p_Y(y_j)=\sum_i(p_{Y,X}(y_j,x_i):=
\sum_i p_{Y|X}(y_j|x_i)p_X(x_i))$.  The channel ability to transmit
information is measured by its {\em capacity}
$C:=\sup_{p_X}I(X:Y)=\max_{p_X}I(X:Y)$, where $I(X:Y)=I(Y:X)$ is the
{\em mutual information}
\begin{equation}
I(X:Y):=\sum_j\sum_i p_{Y,X}(y_j,x_i)\log_2\frac{p_{Y,X}(y_j,x_i)}
{p_Y(y_j)p_X(x_i)}
\label{qi2}
\end{equation}

\noindent or the information about $X$ ($Y$) conveyed by $Y$ ($X$).
The convexity of the log makes $I(X:Y)\geq 0$ (knowing $Y$ can never
lower the information about $X$).

The capacity $C$ may be viewed as the number of output bits per input
symbol which are correctly transmitted.  Its computation is usually
very difficult.

Many channels are binary symmetric: each transmitted bit has the same
probability $p$ of being reversed, i.e., of being erroneous upon
arrival.  These are the channels considered here.  For them we have
$C=1-H_2(p)=:C(p)$, with $H_2(p):=-p\log_2p-(1-p)\log_2(1-p)$.  Note
that $C(\half)=0$, being such a channel totally useless for
transmission since it transforms any input binary word into a random
ouput sequence. Thus we will assume that $p<\half$.

In the transmission of a word $w\in \{0,1\}^n$, an error $e\in
\{0,1\}^n$ may be produced such that the received word is
$w^\prime=w+e$ (addition mod 2).  A subset of words ${\cal
C}_n\subset\{0,1\}^n$ encoding (i.e. in bijective correspondence with)
a collection of messages is said to be an {\em error-correcting
classical code} (ECCC) for $e\in {\cal E}_n\subset \{0,1\}^n$ if
$(w+{\cal E}_n)\cap(w^\prime+{\cal E}_n)=\emptyset$ for any $w\neq
w^\prime\in{\cal C}_n$.  That is, no matter the distortion produced by
the errors on a codeword $w\in{\cal C}_n$, there is no overlapping
between the different sets $w+{\cal E}_n$, and the decoding is
possible without ambiguities.  If upon previous agreement, it is known
which specific message corresponds to each codeword, it will be enough
to send this one instead of the message; the latter will be capable of
being recovered at the other side of the channel after ``cleaning-up"
the received word from the possible errors which can affect it.  In
this way the transmitted codeword can be identified and its decoding
done afterwards.  In the practical use of a code ${\cal C}_n$,
mistakes can occur in the restoration of the messages, caused by
errors outside ${\cal E}_n$, that is, out of the security framework of
the code.  But as long as the frequency of failures remains very low,
the risk will be bearable.  It is apparent that for this to happen it
will be convenient to put very distant apart (in the Hamming sense,
that is, in the number of bits in which they differ) the different
words of the code, for the possibility that the errors will cause
collisions between two distinct words of code will diminish in this
fashion.

One defines the {\em rate} of the code ${\cal C}_n$ as
$R:=\log_2|{\cal C}_n|/n$.  It measures the number of informative bits
per transmitted bit.  It is easy to argue that in order for the code
to be reliable, its rate must not overcome the capacity of the
channel: $R\leq C$.  In fact, when transmitting a codeword $w$ with
length $n$, there will be produced a number of $np$ reversed bits on
average, and hence an error $e$ which will be likely one of the
$2^{nH_2(p)}$ typical sequences.  For the decoding to be reliable,
there should be no overlapping between the error spheres with  centers
at the codewords, and thus $2^{nH_2 (p)}|{\cal C}_n|\leq 2^n$, thereby
$R\leq C$.  This result suggests that the capacity $C$ is an upper
bound to all faithful transmission rates.

The second Shannon's theorem closes this issue in the asymptotic
limit.  Suppose given a binary symmetric channel, a transmission rate
$R$ not exceeding the capacity of the channel ($0<R<C$), an
$\epsilon>0$ arbitrarily small and any sequence $\{N_n\}_1^\infty$ of
integers such that $1\leq N_n\leq 2^{nR}$.  Then, the theorem asserts
that there exist codes $\{{\cal C}_n\subset \Z_2^n\}_1^\infty$ with
$N_n$ elements (codewords), appropriate decision schemes for decoding,
and an integer $n(\epsilon)$, such that the {\em fidelity} $F({\cal
C}_n)$ or probability that a given decoded message coincides with the
original is $\geq 1-\epsilon$ (that is, the maximum probability of
error in the identification of the codeword on reception is
$\leq\epsilon$) for all $n\geq n(\epsilon)$ (Roman, 1992; Welsh,
1995).  Moreover, it is possible to make the error probabilities to
tend to 0, exponentially in $n$.

The theorem is optimal: the capacity $C$ should not be exceeded if the
transmission is to be faithful.  As a matter of fact, it is known that
for each sequence of codes $\{{\cal C}_n\}_1^\infty$ with $|{\cal
C}_n|=\lceil 2^{nR}\rceil$, whose rate exceeds the capacity of the
channel ($R>C$), the average error probability tends asymptotically to
1.

The proof of this Shannon's theorem relies on codes chosen at random
and decoding schemes based on the maximum likelihood principle;
unfortunately, it is not constructive, but existential, leaving open
the practical problem of finding out codes which cleverly combine a
good efficiency in correcting errors, a simple decoding and a high
rate.

\subsection{Classical Error Correction}
\label{sec2B:level2}

Errors in the storage and processing of the information are
unavoidable.  A classical way of correcting them is resorting to {\em
redundancy} (repetition codes): each bit is substituted by a string of
$n\geq 3$ bits equal to it,
\begin{equation}
0\mapsto\underset{n\;0{\rm s}}{\underbrace{00...00}}, \quad
1\mapsto\underset{n\;1{\rm s}}{\underbrace{11...11}},
\label{qi3}
\end{equation}

\noindent and, if by any chance, an error occurs in such a way that
one of the bits in one of those strings gets reversed (for instance
$00000\mapsto 01000$), to correct the error it is enough to invoke the
majority vote.  Let $p$ be probability for any bit to get spoiled.  In
general, several bits of the $n$-tuple may be reversed.  When
$p<\half$, the probability for the majority rule to fail can be made
as smaller as desired, taking $n$ sufficiently large.  It is apparent
that if the $n$-tuples of bits are systematically and frequently
examined, so that it is very unlikely that errors occur at two or more
bits, then the application of this simple method will clean-up the
$n$-tuples from errors and their error-free state will be restored.
However, the price to pay might be too high since with codes of length
$n$ sufficiently large so as to insure a small error during the
detection, the transmission rate can turn up prohibitively small (in
our case it is $1/n$ source bits per channel bit).

So far, we have been describing correction codes ${\cal
C}\subset\{0,1\}^n$ for errors in ${\cal E}\subset\{0,1\}^n$.  More
generally, we can consider $q$-ary alphabets (whose symbols we shall
assume to be the elements of the finite field $\F_{q}$ with $q=p^f$
elements, $p$ being a prime).  Given two words
$x,y\in\{0,1,\ldots,q-1\}^n$, let $d_{\rm H}(x,y)$ be its Hamming
distance (number of locations in which $x,y$ differ).  Let $d:=d_{\rm
H}({\cal C}):=\inf_{x\neq y\in {\cal C}}d_{\rm H}(x,y)$ be the minimum
distance of the code.  Then, the code ${\cal C}$ allows the correction
of errors that affect to a maximum number
$t:=\lfloor\half(d-1)\rfloor$ of positions:\footnote{Notation:
$\lfloor x\rfloor$ ($\lceil x\rceil$) is the largest (smallest)
integer $\leq x$ ($\geq x$).} it is enough to replace each received
word by the closest codeword in the Hamming metric.\footnote{For
instance, for the repetition code ${\cal C} =
\{0\ldots0,1\ldots1,\ldots,$$(q-1)\ldots(q-1)\}$, with $q$ codewords
of length $n$, we have $d=n$, and thus it exactly corrects
$\lfloor(n-1)/2\rfloor$ errors.}  Therefore, the most convenient codes
are those with a high $d$, but this is at the expense of decreasing
$|{\cal C}|$.  If $M$ is the number of codewords, we shall call it a
$(n,M,d)_q$ code.  Its rate is defined as $R:=n^{-1}\log_q M$.

When ${\cal C}$ is a linear subspace of $\F_{q}^n$, the code is called
{\em linear}.  Therefore the linear codes are of the form
$(n,q^k,d)_q$, where $k$ is the dimension of the linear subspace
${\cal C}$; for them $d$ coincides with the minimal Hamming length of
a non-vanishing codeword, and the searching of the codeword nearest to
each received word is greatly simplified.  It is customary to
represent them as $[n,k,d]_q$, or simply as $[n,k]_{q}$ when $d$ is
irrelevant.  Their rate is $k/n$.  Given a code $\cal C$ of type
$[n,k]_q$, the matrix $G$, $k\times n$, with rows given by the
components of the vectors in a basis of $\cal C$ is called a {\em
generator matrix} for $\cal C$.  Defining now in $\F_{q}^n$ a scalar
product in the canonical way, we can introduce the {\em dual} code
${\cal C}^\perp$ of $\cal C$.  A generator matrix $H$ for ${\cal
C}^\perp$ is known as a {\em parity-check matrix} for $\cal C$; notice
that ${\cal C}=\{u\in\F_{q}^n:Hu=0\}$, what justifies in part the name
given to $H$, for it allows us to easily ``check" whether a vector in
$\F_{q}^n$ belongs or not to the subspace $\cal C$.

The coding applies bijectively and linearly $\F_{q}^k$ onto a code
${\cal C}\subset\F_{q}^n$ of type $(n,q^k,d)_q$, and it is implemented
as follows.  Let $\{e_1,\ldots,e_k\}\subset \F_{q}^n$ be a basis of
${\cal C}$.  Given a source word $w^{\rm
t}=(w_1,\ldots,w_k)\in\F_{q}^k$, it gets assigned a codeword
$c(w):=\sum_i w_ie_i$.  In terms of the generator matrix, $w^{\rm
t}\mapsto w^{\rm t}G$.  Let us call $\pi:w\mapsto c(w)$ this
injection.  During the transmission, $c(w)$ could get corrupted,
becoming $u:=c(w)+e$, where $e\in {\cal E}$ is a possible error
vector.  It is evident that $e\in u+{\cal C}$.  In order to decode it,
the criterion of minimal Hamming distance is applied, replacing $u$ by
$\pi^{-1}(u-u_0)$, where $u_0$ is an element of the coset $u+{\cal C}$
which minimizes the distance to the origin (such $u_0$ is known as a
{\em leader} of $u+{\cal C}$).  The linearity of the code allows us to
economize in this last step.  We make a look-up table containing for
each coset $v+{\cal C}\in \F_{q}^n/{\cal C}$ its {\em syndrome} $Hv$
(which uniquely characterizes the coset) and a leader $v_0$.  Upon
receiving $u$ as a message, the syndrome $Hu$ is computed and its
corresponding leader $u_0$ is searched in the table; next, decoding
proceeds as stated before (Macwilliams and Sloane, 1977; Roman, 1992;
Welsh, 1995).  The original message is faithfully retrieved iff the
error coincides with one of the leaders in the table.

Some of the most relevant linear codes are (Macwilliams and Sloane,
1977; Roman, 1992; Welsh, 1995):

1. The repetition code ${\cal
C}=\{0\ldots0,1\ldots1,\ldots,(q-1)\ldots(q-1)\}$ is of type
$[n,1,n]_q$, and although for it the minimum distance is optimal, its
rate is dreadful.

2.  The Hamming codes H$_q(r)$ are arguably the most famous of them
all.  They are codes of the type $[n=1+q+...+q^{r-1},k=n-r,d=3]_q$,
and they are {\em perfect}, in the sense that the set of Hamming
spheres with radius $\lfloor(d-1)/2\rfloor$ and center at each
codeword fill $\F_{q}^n$. These  codes have rates $R=1-r/n$ which tend
to 1 as $n\to\infty$, but they only correct one error.

For instance, H$_2(3)$ is of type $[7,4,3]_2$ and rate 4/7.  A
parity-check matrix for this code is
\begin{equation}
H=\begin{pmatrix}0&0&0&1&1&1&1\\ 0&1&1&0&0&1&1\\ 1&0&1&0&1&0&1
\end{pmatrix}.
\end{equation}
Its decoding is particularly simple.  Let $u$ be the word received
instead of the codeword $w$, and assume that $u$ has only one
corrupted bit.  The syndrome $s(u):=Hu$ coincides in this case with
the binary expression of the position occupied by the erroneous bit.
Negating this single bit will thus suffice to clean the word to get
the correct codeword.  For example, if $u=0110001$, then $s(u)=110$,
so that the incorrect bit is the sixth one, and hence $w=0110011$.

3.  The Golay codes G$_{24}$ and G$_{23}$ are binary, of type
$[24,12,8]_2$ and $[23,12,8]_2$, respectively.  They are probably the
most important codes.

The code G$_{24}$ is {\em self-dual}, i.e. ${\cal C}={\cal C}^\perp$,
what simplifies decoding.  Its rate is $R=1/2$, and allows the
correction of up to 3 errors; it was used by NASA in 1972-82 for the
transmission of color images of Jupiter and Saturn from the Voyagers.

The code G$_{23}$ is {\em perfect}, and it gives rise to G$_{24}$ when
augmented with a parity bit.

The Golay codes G$_{12}$ and G$_{11}$ are ternary, of type
$[12,6,6]_3$ and $[11,6,5]_3$, respectively.  As before, G$_{12}$ is
self-dual, while G$_{11}$ is perfect and originates G$_{12}$ when
appended with a parity bit.

The codes G$_{24}$ and G$_{12}$ have very peculiar combinatorial
properties; their groups of automorphisms are M$_{24}$ and 2.M$_{12}$,
where M$_{24}$ y M$_{12}$ are the famous sporadic groups of Mathieu.
This latter group is the subgroup of $S_{12}$ generated by two special
permutations of 12 cards labeled from 0 to 11: $0,1,2,...,11\mapsto
11,10,9,...,0$ and $0,1,2,...,11\mapsto 0,2,4,6,8,10,11,9,7,5,3,1$.
It is also the group of motions of the form $\tau_i\tau_j^{-1}$ on a
``Rubick'' icosahedron, where $\tau_i$ indicates a rotation of angle
$2\pi/5$ degrees around the $i$-th axis of the icosahedron (Conway and
Sloane, 1999).  As a matter of fact, it was the discovery of the Golay
codes what drove further the study of the sporadic groups which
resulted into the complete classification of the finite simple groups,
with the discovery by Griess in 1983 of the ``monster'' o ``friendly
giant'' group, finite and simple, an enormous subgroup of SO$(47\times
59\times 71)$ with about $10^{54}$ elements.

4.  The Reed-Muller binary codes RM$(r,m)$, with $0\leq r\leq m$, are
of type $[n=2^m,k=\sum_{k\leq r}\binom{m}{k},d=2^{m-r}]_2$.  Their
rates, for fixed $r$, tend to 0 when increasing $m$.  They rank among
the oldest codes known.  The code RM$(1,5)$, of type $(32,64,16)_{2}$,
is able to correct up to 7 errors with a rate of $R=3/16$. It was used
in 1969-72 to transmit from the  Mariners the white-and-black photos
of Mars.

5.  The Reed-Solomon codes generalize the Hamming codes.  They have
been heavily employed by NASA in the transmission of information
during the Galileo, Ulysses and Magellan missions to the deep outer
space, and currently they are used all over, from CD-ROMs to the
hard-disks of computers.

6.  The algebraic-geometric Goppa codes G$_q(D,G)$ are in turn
interesting generalizations of the Reed-Solomon codes.  They have
allowed to obtain families of codes {\em asymptotically good}, that
is, families containing infinite sequences $\{[n_i,k_i,d_i]_q\}$ of
codes, with $n_i\to\infty$, such that the sequences
$\{k_i/n_i,d_i/n_i\}$ of rates and minimum relative distances are
bounded from below by certain positive numbers (Macwilliams and
Sloane, 1977; Roman, 1992; Stichtenoth, 1993; Blake et al., 1998).

\subsubsection{Some asymptotic bounds for linear codes}

To obtain good encodings it is advisable to use long codes which
permit not only sending many different messages but also present a
large minimum distance which allows for correcting sufficiently many
many errors.  Given a code ${\cal C}=[n,k,d]_{q}$, let $R({\cal
C}):=k/n$ be its rate and $\delta({\cal C}):=d/n$ its minimum relative
distance.  A theorem of Manin asserts that the set of limit points of
$\{(\delta({\cal C}),R({\cal C}))\in[0,1]^2: {\cal C} \text{ is a code
on }\F_q\}$ is of the form $\{(\delta,R)\in[0,1]^2: \delta\in[0,1],
0\leq R\leq \alpha_{q}(\delta)\}$, where $\alpha_{q}(\delta)$ is a
continuous function of $\delta\in[0,1]$, decreasing in $[0,1-q^{-1}]$,
and such that $\alpha_{q}(0)=1$, $\alpha_{q}(\delta)=0$ if
$1-q^{-1}\leq\delta\leq 1$ \cite{stichtenoth}.

\begin{figure}[tbp]
\includegraphics[width=6 cm]{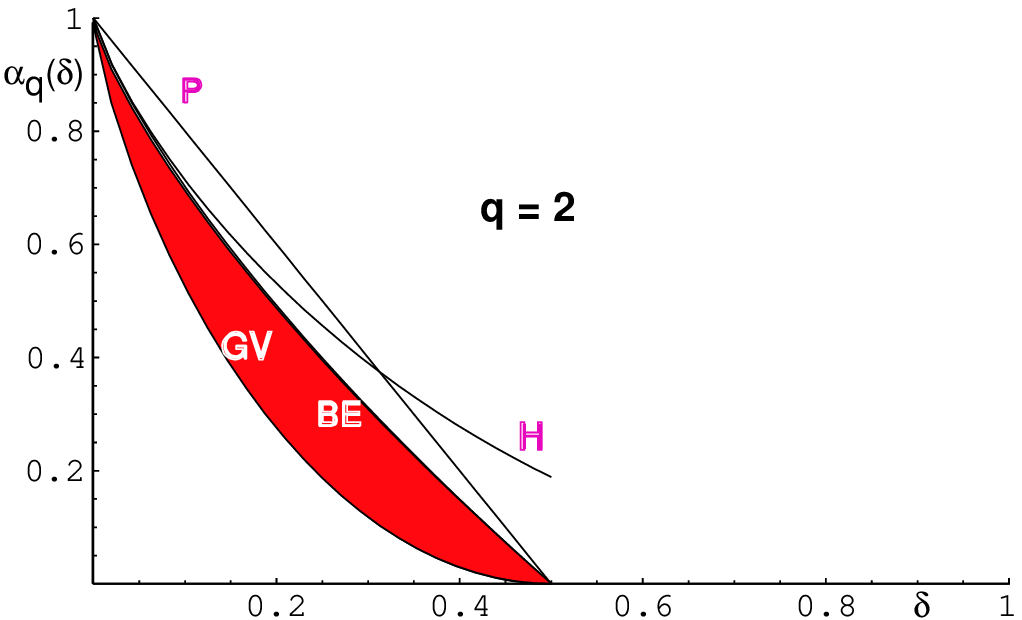}\\ 
\includegraphics[width=6 cm]{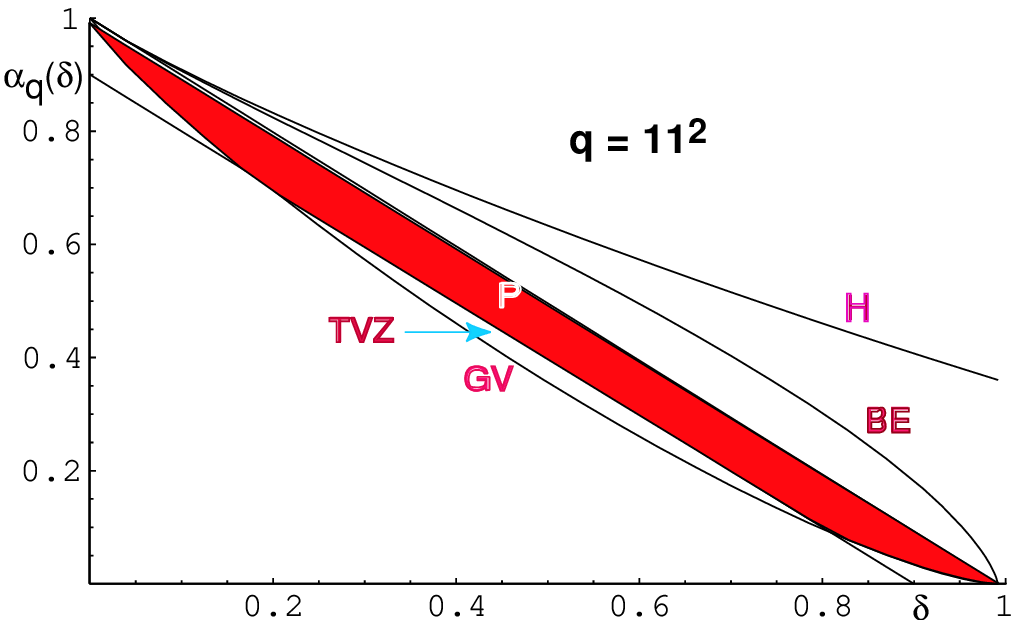}
\caption{Asymptotic bounds.  The dark zone is limited by the lower and
upper bounds mentioned in the text.}
\label{fig:cotas}
\end{figure}

Let $H_{q}$ be the $q$-ary entropy function
$H_{q}(x\in[0,1-q^{-1}]):=x\log_q(q-1)-x\log_q x-(1-x)\log_q(1-x)$.
The following bounds for the function $\alpha_{q}(\delta)$ in the
relevant interval $\delta\in[0,1-q^{-1}]$ are known
\cite{roman,stichtenoth,bhhw}:

\begin{itemize}
\item Plotkin's upper bound:
\begin{equation}
\alpha_{q}(\delta)\leq 1-(1-q^{-1})^{-1}\delta
\end{equation}
\item Hamming's or sphere-packing upper bound:
\begin{equation}
\alpha_{q}(\delta)\leq 1-H_q(\delta/2)
\end{equation}
\item Bassaligo-Elias' upper bound:
\begin{equation}
\alpha_{q}(\delta)\leq
1-H_q(\theta-\sqrt{\theta(\theta-\delta)}),\;{\rm con}\;
\theta:=(1-q^{-1})
\end{equation}
\item Gilbert-Varshamov' lower bound:
\begin{equation}
\alpha_{q}(\delta)\geq 1-H_q(\delta)
\end{equation}
This last one is very important, since it ensures the existence of
codes as long as desired with minimum relative distance $\delta$ and
rate $R$ both asymptotically positive.
\item
Tsfasman-Vl\u{a}du\c{t}-Zink' lower bound: if $q$ is a square, then on
$[0,1-({\sqrt{q}-1})^{-1}]$ one has
\begin{equation}
\alpha_{q}(\delta)\geq \Biggl(1-\frac{1}{\sqrt{q}-1}\Biggr)-\delta
\end{equation}
which is stronger than Gilbert-Varshamov' bound in some places from
$q=7^2$ on.
\end{itemize}
For an illustration see Fig.~\ref{fig:cotas}.

\section{Quantum Information}
\label{sec3:level1}

The quantum information theory, being an extension of the classical
theory, is essentially a product of the past decade (Bouwmeester, Ekert and 
Zeilinger, 2000; Nielsen and Chuang, 2001).

In quantum information, the analogue of the classical bit is called
{\em qubit} or {\em quantum bit} (Schumacher, 1995).  It is a
two-dimensional quantum system (for instance, a spin $\half$, a photon
polarization, an atomic system with two relevant states, etc.), with
Hilbert space isomorphic to ${\C}^2$.  Besides the two basis states
$\ket{0},\ket{1}$, the system can have infinitely many other (pure)
states given by a coherent linear superposition
$\alpha\ket{0}+\beta\ket{1}$.  The Hilbert space of $n$ qubits is the
tensor product ${\C}^2\otimes...\otimes{\C}^2 ={\C}^{2^n}$, and its
natural basis vectors are
$\ket{0}\otimes...\otimes\ket{0}=:\ket{0...0}$,
$\ket{0}\otimes...\otimes\ket{1}=:\ket{0...1}$,...,
$\ket{1}\otimes...\otimes\ket{1}=:\ket{1...1}$.  For this basis, also
known as the computational basis, we shall assume the lexicographic
ordering.  When appropriate, we shall briefly write $\ket{x}$ to
denote $\ket{x_{n-1}...x_0}$, with $x:=x_0+2x_1+...+2^{n-1}x_{n-1}$.
Thus, $\ket{5}=\ket{0...0101}$.

\begin{figure}[ht]
\psfrag{1}{{\color{mywhite}$\ket{0}$}} \psfrag{2}{$\ket{1}$}
\psfrag{3}{{\color{mywhite}$\ket{\Psi}$}} 
\includegraphics[width=5 cm]{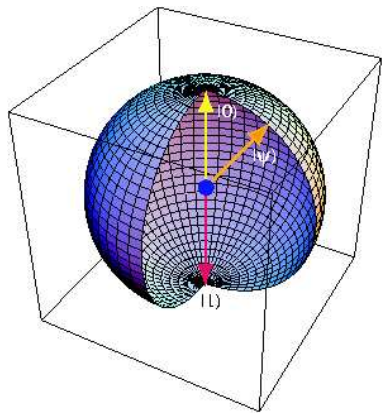}
\caption{Parameterization of the states of one qubit: the Bloch
sphere.}
\label{sphere2}
\end{figure}

There exists the possibility of extending the two-level qubits to {\em
qudits} or $d$-dimensional systems ($d\geq 2$)  (Rungta et al., 2000).
This leads to an extension of the binary quantum logic.  Using $d$
computational levels we can reduce the number $n_2$ of qubits needed
for a computation by a factor of $\lfloor\log_{2}d \rfloor$, since the
Hilbert space of $n_d$ qudits contains the space of $n_2$ qubits
provided that $d^{n_d} \geq 2^{n_{2}}$.

Given an arbitrary state vector $\ket{\Psi}=c_0 \ket{0}+c_1 \ket{1}$
of a qubit, the complex coefficients $c_0,c_1\in \C$ amount to 4 real
parameters.  However, if we parameterize them as $c_i=r_i
\ee^{\ii\phi_i},i=0,1$ and factor out a global irrelevant phase, we
find $\ket{\Psi}=r_0 \ket{0}+r_1 \ee^{\ii (\phi_1-\phi_0)} \ket{1}$.
Imposing $\ket{\Psi}$ to be of unit norm, we can write it as
\begin{equation}
\ket{\psi}=(\cos\half\theta)\ket{0}+\ee^{\ii\phi}
({\sin}\half\theta)\ket{1}
\label{qc1-1}
\end{equation}

\noindent where $r_0,r_1$ are now parameterized by the angles
$\theta$, $\phi:=\phi_1-\phi_0$.

These two angles represent a point in a $S^2$ sphere, called the Bloch
sphere, as shown in Fig.  \ref{sphere2}.  Thus, the (projective)
Hilbert space of pure states of a single qubit can be parameterized by
the points on this sphere.  As a byproduct, this construction provides
a nice representation of the ``classical'' bits as particular points
on the sphere.  The classical bit 0 (better the qubit state $\ket{0}$)
marks the north pole and the 1 sits on the south pole.  Any other
point on the sphere amounts to a non-trivial linear superposition of
the basis states.  The angle $\theta$ is related to the proportion of
$\ket{1}$ to $\ket{0}$ in the composition of that state, while the
angle $\phi$ is their relative quantum phase.

It leaps to the eye from Fig.~\ref{sphere2} that the information
contained in a qubit is infinite as compared to the information in a
classical bit.  In other words, at a given time, a bit can take on
only one of the two values, either 0 or 1, while a qubit can be in any
of the infinitely many possible quantum states in (\ref{qc1-1}).  As
we shall see later in detail, this fact is basic to what is known as
``quantum parallelism'', a source of the unprecedented capabilities
exhibited by a quantum computer.

A {\em quantum logic gate}\footnote{A more extended study of quantum
logic gates and their classical counterparts is presented in
Sec.~\ref{sec9B:level2} and Sec.~\ref{sec8D:level2}.} acting on a
collection or {\em quantum register} of $k$ qubits is just any unitary
operator in the associated Hilbert space ${\C}^{2^k}$ (Deutsch, 89).
For instance, besides the identity, we have for 1 qubit the 1-ary
gates $X$ (or $U_{\rm NOT}$), $Y$, $Z$, given by the Pauli matrices
$\sigma_a$ (in the natural basis $\{\ket{0},\ket{1}\}$):
\begin{equation}
U_{\rm NOT}:=X:=\sigma_x,\quad Y:=-\ii\sigma_y, \quad Z:=\sigma_z.
\label{xyz}
\end{equation}
The particular linear combination $U_{\rm H}:=2^{-1/2}(X+Z)$ is  the
important {\em Hadamard gate}.

The unary gates are easy to implement (for instance, on polarized
photons, with $\half\lambda, \frac{1}{4} \lambda$ plates).

On 2 qubits, the most important gate is {\em controlled NOT} ($U_{\rm
CNOT}$), or {\em exclusive OR} ($U_{\rm XOR}$), defined by $U_{\rm
CNOT},U_{\rm XOR}:\ket{x}\ket{y}\mapsto \ket{x}\ket{x\oplus y}$, where
$x,y$ are either 0,1, and $\oplus$ means  addition mod 2. This gate
can be represented by the matrix
\begin{equation}
\begin{split}
U_{\rm CNOT}:&=U_{\rm XOR}:=\ket{0}\bra{0}\otimes 1+
\ket{1}\bra{1}\otimes U_{\rm NOT}\\ &=\half(1+\sigma_z)\otimes
1+\half(1-\sigma_z)\otimes\sigma_x.
\end{split}
\label{qicnot}
\end{equation}
The physical implementation of this gate is central to the
applications of quantum information and will be addressed later in
Sec.~\ref{sec11:level1}.

The quantum partner of the Shannon entropy is the Von Neumann entropy

\begin{equation}
S(\rho):=-\tr(\rho\log_2\rho),
\label{qivn}
\end{equation}
where $\rho$ is the density operator describing a normal quantum
state.  Given a convex decomposition $\rho=\sum_{i\in I}
p_i\ket{\phi_i}\bra{\phi_i}$ in pure states, it can be shown that
$S(\rho)\leq H(I):=-\sum_i p_i\log_2 p_i$, equality holding if and
only if the state vectors $\phi_i$ are pairwise orthogonal.  The Von
Neumann entropy has the well-known properties of concavity, strong
subadditivity and triangularity (Thirring, 1983; Galindo and Pascual,
1990a; Galindo and Pascual, 1989):
\begin{equation}
\begin{split}
&\lambda_1S(\rho_1)+\lambda_2S(\rho_2) \leq
S(\lambda_1\rho_1+\lambda_2\rho_2),\\ &S(\rho_{ABC})+S(\rho_{B})\leq
S(\rho_{AB})+S(\rho_{BC}),\\ &|S(\rho_{A})-S(\rho_{B})|\leq
S(\rho_{AB})\leq S(\rho_{A})+S(\rho_{B}),
\end{split}
\label{qi4}
\end{equation}

\noindent with $\lambda_{1,2}\geq 0, \lambda_{1}+\lambda_{2}=1$.  The
subscripts $A,B,C$ denote subsystems.

The first two relations also hold in the classical theory of
information.  But the third property (whose second part is just the
property of simple subadditivity) is peculiar. While in Shannon's
theory the entropy of a composite system can never lower the entropy
of any of its parts, quantumly this is not the case.  The EPR states
of the form
$2^{-1/2}(\ket{aa^\prime}+\ket{bb^\prime})$,\footnote{Actually, they
are EPR states {\em \`a la} Bohm, that is, EPRB states (Bohm, 1951).}
where $a,b$ and $a^\prime,b^\prime$ are given orthonormal pairs,
provide us with an explicit counterexample.

\subsubsection{No-cloning theorem}

A basic difference between classical and quantum information is that
while classical information can be copied perfectly, quantum cannot.
This is relevant to quantum communication protocols for should a
quantum copier exist, then safe eavesdropping of quantum channels
would be possible.  In particular, we cannot create a duplicate of a
quantum bit in an unknown state without uncontrollably perturbing the
original.  This follows from the no-cloning theorem of Wootters and
Zurek (1982).  The statement is the following: let $\cH:=\cH_{\rm
orig}\otimes\cH_{\rm copy}$ be the joint Hilbert space of the original
and of the copy, and let $U_{{\rm QCM}}$ be the linear (unitary)
operator in $\cH$ representing the action of an alleged {\em quantum
copier machine}:
\begin{equation}
U_{{\rm QCM}}: \ket{\Psi}_{{\rm orig}}\ket{\phi_0} \mapsto
\ket{\Psi}_{{\rm orig}} \ket{\Psi}_{{\rm copy}}, \forall \ket{\Psi}
\in \cH_{{\rm orig}},
\end{equation}

\noindent where $\ket{\phi_0}$ is the ``blank'' state of the copy.

We claim that such a machine cannot exist.  This is a remarkably
simple application of the linearity of quantum mechanics.  For a
contradiction, suppose it does exist.  Assume for simplicity that the
object to copy is just a single qubit, and let $\ket{\Psi}_{{\rm
orig}}= \alpha_0\ket{0}+\alpha_1 \ket{1}$.  Then, linearity implies
\begin{equation}
U_{{\rm QCM}} \ket{\Psi}\ket{\phi_0} = \alpha_0 \ket{0}\ket{0} +
\alpha_1 \ket{1}\ket{1}
\label{qcm1}
\end{equation}

\noindent whereas the definition of a quantum copier yields
\begin{equation}
\begin{split}
& U_{{\rm QCM}} \ket{\Psi}\ket{\phi_0}=\ket{\Psi} \ket{\Psi} \\ & =
\alpha_0^2 \ket{0}\ket{0} + \alpha_0\alpha_1 \ket{0}\ket{1} +
\alpha_1\alpha_0 \ket{1}\ket{0} + \alpha_1^2 \ket{1}\ket{1}
\label{qcm2}
\end{split}
\end{equation}

\noindent The results (\ref{qcm1}), (\ref{qcm2}) are in general
incompatible, what proves the assertion.

A more general proof of the no-cloning theorem takes into account the
environment and makes use of the unitarity of $U_{{\rm QCM}}$: now
$\cH:=\cH_{\rm orig}\otimes\cH_{\rm copy}\otimes\cH_{\rm env}$, and
\begin{equation}
U_{{\rm QCM}} \ket{\Psi}_{{\rm orig}}\ket{\phi_0}\ket{E_0}
=\ket{\Psi}_{{\rm orig}} \ket{\Psi}_{{\rm copy}} \ket{E_{{\Psi}}},
\forall \ket{\Psi} \in \cH_{{\rm orig}},
\end{equation}
where $\ket{E_0}$ is the ``rest''  state of the ``remaning world''
(environment) before copying, and $\ket{E_{{\Psi}}}$ its state after
copying.  Let us consider two actions of the QCM,
\begin{equation}
\begin{split}
& U_{{\rm QCM}} \ket{\Psi_1}\ket{\phi_0}\ket{E_0}=\ket{\Psi_1}
\ket{\Psi_1} \ket{E_{{\Psi_1}}} \\ & U_{{\rm QCM}}
\ket{\Psi_2}\ket{\phi_0}\ket{E_0}=\ket{\Psi_2} \ket{\Psi_2}
\ket{E_{{\Psi_2}}}.
\label{qcm3}
\end{split}
\end{equation}
Taking the scalar product of these two actions and using unitarity
yields
$\braket{\Psi_1}{\Psi_2}=\braket{\Psi_1}{\Psi_2}^2\braket{E_{\Psi_1}}
{E_{\Psi_2}}$. Therefore, since all these probability amplitudes have
modulus $\leq 1$, then either $\braket{\Psi_1}{\Psi_2}=0$ or 1, and
hence copying two different and non-orthogonal states
${\Psi_1},{\Psi_2}$ is impossible.

However, a known  quantum state can be copied at will.  Moreover,
dropping the requirement that copies be perfect, approximate quantum
copying machines may exist (Buzek and Hillery, 1996).  Should it be
possible to make close to perfect copies then quantum  cryptographic
schemes might still be at risk.  Quantum copying can also become
essential in storage and retrieval of information in quantum computers.

\subsection{Entanglement and Information}
\label{sec3O:level2}

A quantum pure state $\ket{\Psi}$ in a Hilbert space ${\cal
H}=\bigotimes_{i=1}^n {\cal H}_i$ of $n$ qubits is said to be {\em
separable} (with respect to the factor spaces ${\{\cal H}_1,{\cal
H}_2,\ldots,{\cal H}_n$) when it can be factorized as follows:
\begin{equation}
\ket{\Psi} = \otimes_{i=1}^n \ket{\psi_i}, \, \ket{\psi_i}\in {\cal
H}_i.
\label{qi3b}
\end{equation}
Otherwise the state $\ket{\Psi}$ is called {\em entangled}.  Famous
examples of entangled states are the EPR pairs (Einstein, Podolsky and
Rosen, 1935) or Bell states like
\begin{equation}
\begin{split}
\ket{\Psi^{\pm}}:={1\over \sqrt{2}}[\ket{01}\pm \ket{10}] \\
\ket{\Phi^{\pm}}:={1\over \sqrt{2}}[\ket{00}\pm \ket{11}]
\end{split}
\label{qiEPR}
\end{equation}
which physically may be represented by a spin-$\half$ singlet and
triplet  or by entangled polarized (vertical and horizontal) photons
(Kwiat et al., 1995), and the GHZ state (Greenberger, Horne and
Zeilinger, 1989)
\begin{equation}
\ket{{\rm GHZ}}:={1\over \sqrt{2}}[\ket{000}+\ket{111}],
\label{ghz}
\end{equation}
which has been observed experimentally in polarization entanglement of
three spatially separated photons (Bouwmeester et al., 1999).

The concept of entanglement is the distinctive and responsible feature
that allows quantum information to overcome some of the limitations
posed by classical information, as exemplified by the new phenomena of
teleportation, dense coding, etc., to be explained in the following
sections. Although it is simple  to state mathematically, entanglement
leads however to profound experimental consequences like non-local
correlations: when two distant apart parties A (Alice) and B (Bob)
share say an EPR pair,\footnote{It is usual in information theory to
introduce a set of characters named as Alice (the sender), Bob (the
recipient), and Eve (the eavesdropper).}  the measurement by A of her
state univocally determines the state on the B side. Apparently, this
implies instant information transmission, in sharp constract with
Einstein's relativity. However, to reconcile both facts we must notice
that the only way the B side has to know about his state (without
measuring it) is by receiving a classical communication from the A
side, which does propagate no faster than the speed of light.

For these basic reasons, entanglement is considered as a resource in
quantum information (Bennett, 1998), something that we must have
available if we want to take advantage of the new communication
possibilities exhibited by quantum protocols.

When the system has two parts, namely ${\cal H}:={\cal H}_{\rm
A}\otimes {\cal H}_{\rm B}$, it is called {\em bipartite}. In general,
a {\em multipartite} system is of the form ${\cal
H}:=\bigotimes_{i=1}^n {\cal H}_i$.  We may think of entanglement as a
manifestation of the superposition principle when applied to bipartite
or multipartite systems.  Thus, genuine multiparticle or many-body
states exhibit entanglement properties, which in the theory of
strongly correlated systems are known as quantum correlations (Fulde,
1993).\footnote{These type of correlations are responsible for novel
quantum phase transitions (Sachdev, 1999) where the transition is
driven by quantum fluctuations instead of standard thermal
fluctuations.}  We may state that entanglement and quantum
correlations are closely linked.

Being a non-local concept, entanglement must be independent of local
manipulations performed on each of the A and B parties. These
operations are represented by unitary operators $U_{\rm A}\otimes
U_{\rm B}$, in a factorized form, acting on the states of ${\cal
H}={\cal H}_{\rm A}\otimes {\cal H}_{\rm B}$, or they may be local
measurements on either side.  Moreover, classical communication is
also permitted by the two parties.  Entanglement cannot be created by
these local operations. However,  factorized states can be obtained by
local operations, like measurements.  Altogether, these type of local
operations plus classical communications are known as LOCC
transformations. The set LOCC is not a group, but a semigroup for the
inverse of a given transformation is not guaranteed to exist, due to
possible irreversible measurements by each party.

The characterization of entanglement for general quantum states (pure
or mixed, bipartite or multipartite) is very difficult, in part due to
the type of transformations allowed in the set LOCC.  For entangled
pure states of 2 qubits or general bipartite systems A and B with
dimensions $d_{\rm A},  d_{\rm B}$ respectively, entanglement is well
understood in terms of their Schmidt (1906) decomposition: given an
arbitrary state
\begin{equation}
\ket{\Psi}_{{\rm AB}}:=\sum_{i=1}^{d_{\rm A}} \sum_{j=1}^{d_{\rm B}}
C_{ij} \ket{a_i}_{\rm A} \ket{b_j}_{\rm B} \in {\cal H}= {\cal H}_{\rm
A}\otimes {\cal H}_{\rm B}
\label{qi3AB}
\end{equation}
with $\{\ket{a_i}_{\rm A}\}_1^{d_{\rm A}}$,  $\{\ket{b_i}_{\rm
B}\}_1^{d_{\rm B}}$ orthonormal bases of  ${\cal H}_{\rm A}, {\cal
H}_{\rm B}$, then it admits a biorthonormal decomposition of the form
\begin{equation}
\ket{\Psi}_{\rm AB}=\sum_{k=1}^r \sqrt{w_k} \ket{u_k}_{\rm A}
\ket{v_k}_{\rm B}, \ w_k>0, \  \sum_{k=1}^r w_k=1,
\label{qi3dmrg}
\end{equation}
where $\{\ket{u_k}_{\rm A}\}_1^r$  and $\{\ket{v_k}_{\rm B}\}_1^r$ are
sets of orthonormal vectors for subsystems A and B, and $r\leq d:={\rm
min}\{ d_{\rm A}, d_{\rm B} \}$ is the so called {\em Schmidt rank} of
$\ket{\Psi}_{{\rm AB}}$  (Schmidt, 1906;  Hughston, Jozsa and
Wootters, 1993; Ekert and Knight, 1995).\footnote{The Schmidt
decomposition  is  equivalent to the {\em Singular Value
Decomposition} (SVD) of the $d_{\rm A}\times d_{\rm B}$ matrix $C:=
(C_{ij})$ in linear algebra (Press et al., 1992). Let $d_{\rm A}\leq
d_{\rm B}$. Then $C=UDV^{\rm t}$, where $U$ is an orthogonal $d_{\rm
A}\times d_{\rm A}$ matrix ($U^{\rm t}U=1_{d_{\rm A}}$), $V$ is a
$d_{\rm A}\times d_{\rm B}$ matrix representing a Euclidean isometry
from $\C^{d_{\rm A}}$ to $\C^{d_{\rm B}}$ (i.e. $VV^{\rm t}=1_{d_{\rm
A}}$), and $D$ is the $d_{\rm A}\times d_{\rm A}$ diagonal matrix
${\rm diag}(\sqrt{w_1},...,\sqrt{w_r},0,...,0)$.  Using the SVD
$C_{ij}=\sum_{k=1}^{d_{\rm A}} U_{ik} \sqrt{w_k} V_{jk}$  in
(\ref{qi3AB}) we inmediately arrive at the Schmidt decomposition
(\ref{qi3dmrg}).} The coefficients $w_k$ are called {\em Schmidt
weights}.

The Schmidt decomposition is essentially unique in the following
sense: the weights (multiplicities included) are unique (up to order),
and hence the rank; given a nondegenerate weight $w_k$, the state
vectors    $\ket{u_k}_{\rm A}, \ket{v_k}_{\rm B}$, are unique up to
reciprocal phase factors; when the weight $w_k$ is degenerate, the
corresponding states in Alice's side are unique up to an arbitrary
unitary transformation $U_{\rm A}$ to be compensated by a simultaneous
unitary transformation $U_{\rm B}=U_{\rm A}^\ast$ on the associated
vectors in Bob's side.

From the Schmidt decomposition it inmediately follows that  a
bipartite pure state $\ket{\Psi}_{{\rm AB}}$  is entangled if and only
if its Schmidt rank $r>1$.

From the point of view of the subsystem A, the description  of its
quantum properites is realized by means of the {\em reduced density
matrix} $\rho_{\rm A}$  (and likewise for subsystem B with $\rho_{\rm
B}$):
\begin{equation}
\begin{split}
\rho_{\rm A}:={\rm Tr}_{\rm B} \ket{\Psi}_{\rm AB}\bra{\Psi} \\
\rho_{\rm B}:={\rm Tr}_{\rm A} \ket{\Psi}_{\rm AB}\bra{\Psi}
\end{split}
\label{qitrace1}
\end{equation}
where ${\rm Tr}_{\rm B}$ denotes the {\em partial trace} over the B
subsystem (similarly for ${\rm Tr}_{\rm A}$ and subsystem B).  The
Schmidt decomposition (\ref{qi3dmrg}) implies that
\begin{equation}
\begin{split}
\rho_{\rm A}= \sum_{k=1}^r w_k \ket{u_k}_{\rm A}\bra{u_k}\\ \rho_{\rm
B}= \sum_{k=1}^r w_k \ket{v_k}_{\rm B}\bra{v_k}
\end{split}
\label{qitrace2}
\end{equation}
Another important implication of (\ref{qi3dmrg}) is that as  $r\leq
d$,  when a qubit state $d_{\rm A}=2$ is entangled to a qudit state
$d_{\rm B}\geq 2$ then the Schmidt decomposition has at most two
terms, no matter how large $d_{\rm B}$ is.

Interestingly enough, the Schmidt decomposition has appeared
independently again in the field of strongly correlated systems
through the density matrix renormalization group method DMRG (White,
1992; 1993).\footnote{The Schmidt weights govern  the truncation
process inherent to the DMRG method: the highest weights are retained
while the smallest (beyond a certain desired value) are eliminated.
This truncation makes the exponentially large problem much more
amenable.}

Once we know whether a given bipartite pure state is entangled or not,
next question is to get entanglement ordered: given two states
$\ket{\Psi_1}_{\rm AB}, \ket{\Psi_2}_{\rm AB}$, which one is more
entangled?  No sufficiently general answer is known to this question.
A tentative simple choice would be to measure entanglement through the
partial Von Neumann entropies (Bennett et al., 1996a):
\begin{equation}
E(\ket{\Psi_{\rm AB}}):=S(\rho_{\rm A})=S(\rho_{\rm B})
\label{qidistil01}
\end{equation}
Such entropies do not increase under LOCC, but having
$E(\ket{\Phi_{\rm AB}})<E(\ket{\Psi_{\rm AB}})$ does not guarantee
that an LOCC action may bring $\ket{\Psi_{\rm AB}}$ to $\ket{\Phi_{\rm
AB}}$.

The {\em theory of majorization} provides us with a criterium to
ascertain when any two entangled states can be LOCC connected
(Nielsen, 1999). Given two vectors $x=(x_1,x_2,\ldots,x_d)$,
$y=(y_1,y_2,\ldots,y_d)$ in ${\R}^d$, decreasingly ordered $x_1\geq
x_2\geq \ldots x_d$, $y_1\geq y_2\geq \ldots y_d$, we say that $x$ is
{\em majorized} by $y$, denoted $x\prec y$, (equivalently, $y$ {\em
majorizes} $x$) if the following series of relations hold true:
\begin{equation}
\begin{aligned}
 x_1 &\leq  y_1 \\ x_1+x_2 &\leq  y_1+y_2 \\ \vdots & \\ x_1+x_2\ldots
 x_{d-1} &\leq   y_1+y_2\ldots y_{d-1} \\ x_1+x_2\ldots x_{d} & =
 y_1+y_2\ldots y_{d} \\
\end{aligned}
\label{qimaj1}
\end{equation}
The majorization relation is a partial order in ${\R}^d$: 1/ $x\prec
x$, $\forall x$;  2/ $x\prec y$ and $y\prec x$ iff $x=y$;  3/ if
$x\prec y$ and $y\prec z$ then $x\prec z$. When the components of the
vector $x$ are positive $x_k\geq 0$ and normalized  $\sum_k x_k=1$,
they may be thought of as probabilitiy distributions as is
Sec.~\ref{sec2:level1}.  The central result  is the following: a
bipartite state $\ket{\Psi}_{AB}$ can be transformed via LOCC
operations into another state $\ket{\Phi}_{AB}$ iff $w(\ket{\Psi})$ is
majorized by  $w(\ket{\Phi})$,
\begin{equation}
\ket{\Psi}_{AB} \longrightarrow \ket{\Phi}_{AB} \Longleftrightarrow
w(\ket{\Psi}) \prec w(\ket{\Phi})
\label{qimaj2}
\end{equation}
where $w(\ket{\Psi})$ is the ordered vector of eigenvalues or weights
(multiplicities included) of the reduced density matrix $\rho_{\rm A}$
(\ref{qitrace1}),(\ref{qitrace2}) associated with $\ket{\Psi}_{AB}$
(similarly for $w(\ket{\Phi})$).

For example, let us consider the parties A and B sharing this couple
of qutrit states in the basis $\{ \ket{0},\ket{1},\ket{2}\}$:
\begin{equation}
\begin{split}
\ket{\Psi}_{AB}=&\ \frac{2}{3} \ket{00}+ \frac{2}{3} \ket{11}+
\frac{1}{3} \ket{22}\\ \ket{\Phi}_{AB}=&\ \sqrt{\frac{2}{3}} \ket{00}+
\sqrt{\frac{1}{6}} \ket{11}+ \sqrt{\frac{1}{6}} \ket{22}
\end{split}
\end{equation}
Both states are entangled, but $\ket{\Psi}_{AB}$ cannot be transformed
into $\ket{\Phi}_{AB}$ or viceversa: they possess different types of
entanglement. They are said to be {\em incomparable} or {\em
incommensurate} (Nielsen, 1999; Vidal, 1999).

However, for general multipartite systems the issue of how to relate
the LOCC action with entanglement  in a given pure state is an open
question (Lewenstein et al., 2000).

A definition of entanglement  for finite dimensional systems with
mixed states characterized by a density matrix $\rho$ goes as follows
(Werner, 1989):   $\rho$ is called separable when it can be written as
a convex combination of product states
\begin{equation}
\rho = \sum_{k=1}^r \lambda_k \otimes_{j=1}^n \rho_k^{(j)}, \,
\lambda_k\geq 0, \,\sum_{k} \lambda_k =1.
\label{qi3c}
\end{equation}
When $\rho$ is not separable, one calls it an entangled mixed
state. The situation about quantifying and qualifying entanglement  is
even worse for mixed quantum states (Horodecki et al., 1996a; Peres,
1996; D\"{u}r, Cirac and Tarrach, 1999; Giedke et al., 2001).  There
are partial characterizations of entanglement like the  Peres
criterion (1996): a necessary condition for separability of  $\rho$ is
that the matrices $\rho^{{\rm t},j}$,  $j=1,...,r$, obtained by {\em
partial transposition}\footnote{Note that $\rho^{{\rm t},j}:=
\sum_{k=1}^r \lambda_k \rho_k^{(1)}\otimes...\otimes \rho_k^{(j),\rm
t} \otimes...\otimes \rho_k^{(n)}\geq 0$, since the coefficients and
each factor matrix are non-negative, no matter which basis is chosen
in ${\cal H}_j$ to define the transpose.}  of $\rho$ with respect to
an arbitrary orthonormal basis of the factor space ${\cal H}_j$ of the
$j$-component, is non-negative ($\rho^{{\rm t},j}\geq 0$).   The
converse is true in the special cases $\C^2\otimes \C^2$,  and
$\C^2\otimes \C^3$ (Horodecki et al., 1996b).

There are also complete characterizations of entanglement in terms of
{\em entanglement witness operators} and {\em positive maps}
(Horodecki et al., 1996a), but their classifications turns out to be
as complicate as the original problem of entangled mixed states.

\subsection{Quantum Coding and Schumacher's Theorem}
\label{sec3A:level2}

Let now $A:=\{\ket{\phi_i},p_i\}_{i=1}^{|A|}$ be a ``quantum
alphabet'' consisting of a set of distinct pure states (not
necessarily orthogonal) and their corresponding probabilities
($\sum_ip_i=1$).  We assign to it the following density operator
$\rho(A):=\sum_i p_i \ket{\phi_i}\bra{\phi_i}$.  A message emitted by
a source of quantum signals is now a sequence $\phi_{i_1...i_n}:=
\ket{\phi_{i_1}}\ket{\phi_{i_2}}...\ket{\phi_{i_n}}$ of ``quantum
characters" or ``quantum symbols", each produced with probability
$p_{i_j}$ independently of the others.  The collection of messages
with $n$ symbols is representable by the density operator
$\rho^{\otimes n}$, which lives in a Hilbert space of maximum
dimension $|A|^n=2^{n\log_2|A|}$.  The question naturally arises again
as to whether it is possible to compress the information contained in
$\rho^{\otimes n}$.  And the answer, found by Schumacher (Schumacher,
1995), is similar to Shannon's first theorem: asymptotically ($n\gg
1$) the state $\rho^{\otimes n}$ is compressible to a state in a
Hilbert space of dimension $2^{n S(\rho)}$, with a {\em fidelity} $F$
(probability that the decoded state coincides with the state prior to
coding) arbitrarily close to 1.  In other words, it is compressible to
$n S(\rho)$ qubits.  Then $S(\rho)$ can be thought of as the average
number of qubits of essential quantum information, per character of
the alphabet.

The idea of the proof follows the same guideline as for the classical
theorem (Schumacher, 1995; Jozsa and Schumacher, 1994; Preskill,
1998).  Let us diagonalize $\rho=\sum_r\lambda_r\ket{r}\bra{r}$.  The
Von Neumann entropy $S(\rho)$ clearly coincides with the Shannon
entropy $H(D)$ of the classical alphabet
$D:=\{r,\lambda_r\}_{r=1}^{|D|}$.  Introducing the typical messages as
those strings or tensor-product vectors
$\psi_{i_1...i_n}:=\ket{\psi_{i_1}}...\ket{\psi_{i_n}}$ in the
orthonormal basis that diagonalizes $\rho$, such that its probability
$\lambda_{i_1...i_n}:=\prod_j\lambda_{i_j}$ satisfies
$\lambda_{i_1...i_n}\sim 2^{-nH(D)}$ for $n\gg 1$, it is shown that
$\rho^{\otimes n}$ is asymptotically concentrated on the {\em typical
subspace} $T$ spanned by them: $\tr(P_T\rho^{\otimes n})\sim 1$.  Here
$P_T$ is the orthogonal projection onto $T$.  The strategy of
compression amounts to make a measurement that projects the original
message $\phi_{i_1...i_n}$ either onto $T$, or onto $T^\perp$.  If the
former is the case, the projected state $P_T\phi_{i_1...i_n}$ is
faithfully sent, upon coding it into $nH(D)$ qubits.  What one does in
the remaining case is irrelevant, for the probability that the result
be $(1-P_T)\phi_{i_1...i_n}$ is asymptotically negligible.

The average fidelity in this procedure is perfect in the limit
$n\to\infty$, and as in the classical theory, the quantum compression
thus obtained is optimal.

If the alphabet $A:=\{\rho_i,p_i\}_{i=1}^{|A|}$ is made up of mixed
states, the issue of the message compressibility gets more involved.
To properly measure it, the Shannon entropy $S(\rho:=\sum_ip_i\rho_i)$
must yield to another more general concept, the so called {\em Holevo
information} of the alphabet or ensemble
$A:=\{\rho_i,p_i\}_{i=1}^{|A|}$ (Levitin 1969; Holevo, 1973; Preskill,
1998):
\begin{equation}
\chi(A):=S(\rho)-\sum_ip_iS(\rho_i). \label{qi5}
\end{equation}

The Holevo information is similar to the classical mutual information.
As $I(X:Y)$ measures how the entropy of $X$ gets reduced when $Y$ is
known, $\chi(A)$ represents the reduction of the entropy $S(\rho)$ of
$\rho$, when the actual preparation of this state as a convex
combination $\rho=\sum_ip_i\rho_i$ is known.

Assuming the states $\rho_i$ of the alphabet to be mutually
orthogonal, that is, $\tr(\rho_i\rho_j)=0$ for $i\neq j$, it is not
difficult to see that the state $\rho^{\otimes n}$ is asymptotically
$(n\gg 1)$ compressible to a state of $n\chi(A)$ qubits, with fidelity
tending to 1.  Moreover, this result is optimal.

When the states are not orthogonal, the results are only partial: it
is known that there does not exist an asymptotically faithful
compression below $\chi(A)$ per letter of the alphabet, but it is
still open the problem of whether a compression of $\chi(A)$
qubits/character is or not accessible in the limit $n\to\infty$.

\subsection{Capacities of a Quantum Channel}
\label{sec3B:level2}

For a quantum transmission channel we can consider its capacity $C$
for transmitting classical data, its capacity $Q$ for transmitting
quantum states exactly, and its mixed capacities $Q_{1,2}$ for
transmitting quantum states, also exactly, but with the assistance of
a classical side-channel between sender and receiver.

Given a quantum channel ${\cal N}$, usually noisy, Shannon's second
theorem suggests to define the classical capacity $C({\cal N})$ as the
supremum of the transmission rates $R:=k/n$ of classical words
$k$-cbits long such that: 1/ Transmission is carried out after an
appropriate word coding as $n$-bits words that are sent by $n$ forward
uses of the channel ${\cal N}$, followed by an associated decoding
upon arrival (yielding words of $k$ bits).  2/ The fidelity of the
transmission is asymptotically 1.  The quantum capacity $Q({\cal N})$
is defined similarly by replacing the classical input/output words of
$k$ cbits by pure/mixed states of $k$ qubits (Bennett and Shor, 1998).

The assisted quantum capacities $Q_{1,2}({\cal N})$ are defined in a
similar fashion as $Q({\cal N})$, but now the coding-decoding protocol
may include arbitrary local operations on input and output states, and
may resort to a classical communication channel in the input-to-output
direction (subscript 1), or in both directions (subscript 2).

It is possible to show that $Q=Q_{1}$ (Bennett et al. 1996; Bennett
and Shor, 1998); that is, sending classical messages from origin to
destination does not increase the channel capacity. On the other hand,
it is evident that $Q\leq Q_{2}$, and using orthogonal states to
transmit cbits leads to $Q\leq C$. But it is not known whether
$C<Q_{2}$ holds or not.  Channels are known for which $Q < Q_{2}$, and
others for which $Q_{2} < C$.

As asymptotically defined, it is not surprising that the computation
of these capacities is usually difficult.  In some instances they are
known, as in the case of the so called {\em quantum erasure channel},
in which there is a probability $p$ that the channel replaces the
qubit by an erasure symbol orthogonal to the states
$\{\ket{0},\ket{1}\}$, and the complementary probability $1-p$ that
the qubit goes through exactly.  For this type of channel
$C=Q_{2}=1-p$, and $Q=\max\{0,1-2p\}$ (Bennett, DiVincenzo and Smolin,
1997; Bennett and Shor 1998).

Unlike the classical case, where the capacity can be computed
maximizing the mutual information between input and output in a single
use of the channel, the capacities (whether classical or quantum) of
the quantum channels do not usually allow for a similar computation.
This is because in this quantum case it is allowed to code by
entangling several successive states on input, and to decode by means
of joint measurements on several states on output.  However, for the
case $C_{\rm cq}$ (classical capacity with classical encoding and
quantum decoding), it is known that $C_{\rm cq}({\cal N}) =
\sup_{\rho}\chi({\cal N}(\rho))$ (Bennett and Shor, 1998).

Finally, prior entanglement between sender and receiver improves the
transmission capacity.  Let $C_{\rm E}, Q_{\rm E}$ be the classical
and quantum entanglement-assisted capacities of a quantum channel.  A
direct consequence of the dense coding and quantum teleportation, to
be described later, is the relation $C_{\rm E}=2C$ for noiseless
quantum channels, and the relation $Q\leq Q_{\rm E}=\half C_{\rm E}$
for any quantum channel (Bennett et al., 1999).

\subsection{Quantum Error Correction}
\label{sec3C:level2}

It is not possible in the quantum case just to plainly imitate the
classical methods of error corrections, for merely trying to check
which qubits have been affected by errors irremediably damages the
information content.  Neither can we make strings of equal quantum
states, for the unitarity of quantum mechanics forbids the cloning of
arbitrary unknown quantum states.  This explains the initial pessimism
about the possible functioning of a quantum computer (Landauer 1994;
Unruh, 1995).  Then, what to do?  Fortunately enough, in 1995 Shor
provided us with a first solution showing an encoding system (of 9:1
bits) capable of detecting and correcting one erroneous
qubit.\footnote{Actually, the very first idea of quantum error
correction, at the time called ``recoherence'', was proposed by
Deutsch during his talk at the Rank Prize Funds Symposium on Quantum
Communication and Cryptography (1993, Broadway, UK).  This idea was
later on developed further  (Berthiaume, Deutsch and Jozsa, 1994;
Barenco et al., 1997).  Even the idea of decoherence free subspaces
(Palma, Suominen and Ekert, 1996) preceded Shor's 9-qubit code.}  Soon
after, new and more economical codes were discovered, such as the 7:1
code of Steane (1996a; 1996b), Calderbank and Shor (1996), and the 5:1
code of Bennett et al.  (1996).\footnote{An $n:1$ code embeds 1 qubit
into the space of $n$ qubits.} It is not possible to present here a
full account of the many remarkable contributions in this field during
the last six years.  It is currently a developing field which, as it
happened with the classical error correction codes, it has also been
found unexpected connections with pure mathematics (Shor and Sloane,
1998).

The underlying idea of quantum error correction is to hide the
information into subspaces of $\C^{2^n}$ in order to protect it
against decoherence and errors that only affect to a few qubits.  To
this end, if our system has $k$ qubits (called ``logical qubits''), a
quantum error correction code (QECC) encodes their states by means of
a linear isometric embedding $\pi:\C^{2^k}\hookrightarrow\C^{2^n}$,
with $n>k$.  We shall denote by ${\cal Q}$ the image subspace of
$\pi$, and its states will be called code states (or codewords).  The
additional $n-k$ qubits help us in protecting the information.  The
map $\pi$ should disguise the information by delocalizing it, with the
aim that errors (which often affect locally just one or a few qubits)
may alter it nothing or the least possible (Preskill, 1998; Steane,
1997; Aharonov, 1998).

A system of $n$ qubits in an initial pure state $\psi$ is not
absolutely isolated.  Upon interaction with the environment in a state
$a_{\rm in}$, it suffers a transformation of the form $\psi\otimes
a_{\rm in}\mapsto \sum_r (E_r\psi)\otimes a_r$, where the operators
$E_r, 0\leq r\leq 2^{2n}-1,$ are Pauli operators (elements of the set
${\cal P}^{(n)}:=\{1,X,Y,Z\}^{\otimes n}$) and the environment states
$a_r$ are not necessarily orthogonal neither normalized.  Let us call
the {\em weight} of an element in ${\cal P}^{(n)}$ to the number of
its nontrivial (i.e. $X,Y,Z$) tensor factors.  If $\psi$ is a code
state, then each term $(E_r\psi)\otimes a_r$ represents a component
with a number of errors equal to the weight of $E_r$.

Given a collection of errors ${\cal E}\subset {\cal P}^{(n)}$ formed
by all the Pauli operators of weight $\leq t$, a QECC is said to amend
up to $t$ errors when it is capable of correcting every error in
${\cal E}$.  For that to happen it is necessary and sufficient that
$\bra{\bar j}E_s^\dagger E_r\ket{\bar i}= m_{sr}\delta_{ji}$ be
fulfilled, for any arbitrary orthonormal basis $\{\ket{\bar i}\}$ of
the code subspace ${\cal Q}$ and all $E_{r,s}\in{\cal E}$, $m$ being a
selfadjoint matrix.  This condition means something quite natural:
first, that given any two orthogonal codewords $\ket{\bar i},\ket{\bar
j}$, the sets $E_r\ket{\bar i}$, $E_r\ket{\bar j}$ of corrupted
codewords must be mutually orthogonal, otherwise the perfect
distinguishability of those words might get lost, and second, should
$\bra{\bar i}E_s^\dagger E_r\ket{\bar i}$ depend on $\ket{\bar i}$,
the detection of the error would yield information about the code
state, thereby perturbing it.  If $m=\id$, the code is called {\em
nondegenerate}, and the error subspaces $E_r{\cal Q}, 1\neq
E_{r}\in{\cal E}$ are orthogonal to the code subspace ${\cal Q}$ and
perpendicular one another.  In this case it suffices to make a
measurement, which is possible because of the orthogonality, that
determines in which subspace the ($n$-qubits
system)$\otimes$environment lies.  If the result of that measurement
is $(E_r\psi)\otimes a_r$, by applying to the resulting state of the
system the unitary operator $E_r^\dagger$ we shall retrieve the
original state $\psi$ free of error.  In the degenerate case, an error
syndrome does not singularize the error, and the retrieval strategy
gets more involved.

The {\em distance} $d$ of a QECC is defined as the lowest weight of a
Pauli operator $E$ such that $\bra{\bar j}E\ket{\bar i}\neq
c_E\delta_{ji}$.  In analogy with the notation for CECCs, we shall
write $[[n,k,d]]_2$ to denote a binary QECC (i.e., with qubits) of
parameters $n,k,d$.  It is easy to see that a code $[[n,k,d]]_2$
allows the correction of $t:=\lfloor(d-1)/2\rfloor$ errors.

There are also asymptotic bounds for the QECCs $[[n,k,d]]_2$ similar
to those presented for CCCEs (Ekert and Macchiavello, 1996; Preskill,
1998).
\begin{itemize}
\item Hamming's quantum upper bound:
\begin{equation}
R:=k/n\leq 1-H_2(t/n)-(t/n)\log_2 3,\quad n\gg 1.
\end{equation}

\item
Gilbert-Varshamov' quantum lower bound:
\begin{equation}
R \geq 1-H_2(2t/n)-(2t/n)\log_2 3,\quad n\gg 1.
\end{equation}
\end{itemize}

As in the classical case, there exist QECCs which are asymptotically
good.  A different question (still open) is their explicit
construction.

\noindent{\em Example of QECC: CSS codes.} Let ${\cal C}_1$ be a
linear and binary CECC of type $[n,k_1,d_1]_2$, and ${\cal
C}_2\subset{\cal C}_1$ a subcode $[n,k_2,d_2]_2$ of ${\cal C}_1$, with
$k_2<k_1$.  Let ${\cal C}:={\cal C}_1/{\cal C}_2$ be the quotient
space, of dimension $2^{k_1-k_2}$.

Let us introduce a QECC ${\cal Q}\subset\C^{2^n}$ of dimension $2^k$,
with $k=k_1-k_2$, spanned by the vectors
\begin{equation}
\ket{\bar w}:=2^{-k_2/2}\sum_{v\in{\cal C}_2}\ket{w+v},\quad w\in{\cal
C}
\label{qi6}
\end{equation}

\noindent Note that this definition does not depend on the element $w$
chosen to represent the class $w+{\cal C}$, and that the vectors
$\ket{\bar w}$ thus constructed form an orthonormal system.

It can be shown that this quantum code recognizes and corrects (up to)
$t_{\rm b}:=\lfloor(d_1-1)/2\rfloor$ bit-flip errors $X$, and $t_{\rm
ph}:=\lfloor(d_2^\perp-1)/2\rfloor$ phase-flip errors $Z$, where
$d_2^\perp$ is the distance of the code ${\cal C}_2^\perp$ dual to
${\cal C}_2$.  Likewise, the distance $d$ of this quantum code
satisfies $d\geq\min(d_1,d_2^\perp)$.

The QECCs $[[n,k,d]]_2$ thus constructed are called CSS
(Calderbank-Shor-Steane) codes (Steane, 1996a; Steane, 1996b;
Calderbank and Shor, 1996; Preskill, 1998).

The simplest and most illustrative example of a CSS code is the
$[[7,1,3]]_2$ code of Steane, or quantum code of 7 qubits.  It is
obtained taking as ${\cal C}_1$ the Hamming code H$_2(1)$ of type
$[7,4,3]_2$, and as ${\cal C}_2$ its {\em dual} (${\cal C}_2={\cal
C}_1^\perp$), which is of type $[7,3,4]_2$, and coincides with the
even subcode (that is, the code formed by the codewords of even
weight)\footnote{The weight of a binary word is defined as the number
of its nonzero coordinates.} of ${\cal C}_1$.  It corrects one
bit-flip error $X$, and one phase-flip error $Z$.  Thus, it also
corrects a mixed error $Y$, but not a double bit-flip (or phase-flip)
error.

A generator matrix for  H$_2(1)$ is
\begin{equation}
G:=\begin{pmatrix} 1&0&1&0&1&0&1\\ 0&1&1&0&0&1&1\\ 0&0&0&1&1&1&1\\
1&1&1&0&0&0&0
\end{pmatrix}
\label{qi7}
\end{equation}

\noindent and an associated parity matrix (generator for the dual) is
\begin{equation}
H:=\begin{pmatrix} 1&0&1&0&1&0&1\\ 0&1&1&0&0&1&1\\ 0&0&0&1&1&1&1
\end{pmatrix}
\label{qi8}
\end{equation}

\noindent Thus, a basis of code states is given by
\begin{equation}
\begin{split}
&\ket{\bar 0}:=8^{-1/2}(\ket{1010101}+\ket{0110011}+ \\
&\phantom{\ket{\bar 0}:=8^{-1/2}(}\ket{0001111}+\ket{0000000}+
\ket{1100110}+ \\ &\phantom{\ket{\bar
0}:=8^{-1/2}(}\ket{1011010}+\ket{0111100}+ \ket{1101001}) \\
&\ket{\bar 1}:=8^{-1/2}(\ket{0100101}+\ket{1000011}+ \\
&\phantom{\ket{\bar 0}:=8^{-1/2}(}\ket{1111111}+\ket{1110000}+
\ket{0010110}+ \\ &\phantom{\ket{\bar
0}:=8^{-1/2}(}\ket{0101010}+\ket{1001100}+ \ket{0011001})
\end{split}
\label{qi9}
\end{equation}

Let us assume that we have a qubit with a state coded as
$\ket{\bar\phi}:=\alpha\ket{\bar 0}+ \beta \ket{\bar 1}$, in which a
bit flip has occurred at the third place ($X_3$ error).  How can we
detect and correct it?  With the help of an auxiliary system or {\em
ancilla} $A$ of ($n-k_1=3$)-qubits long we form the state
$(X_3\ket{\bar\phi})\otimes\ket{000}_A$, which we transform by the
unitary map defined on $\C^{2^n}\otimes\C^{2^3}$ by
$\ket{v}\otimes\ket{000}_A \mapsto\ket{v}\otimes\ket{Hv}_A$, with the
result $(X_3\ket{\bar\phi})\otimes\ket{He}_A$, where $e:=0010000$ is
the binary word that signals the place number 3 at which the bit-flip
error occurred.  But $He=110$, which is also number 3 in (reversed)
binary form.  That is, we have marked in the ancilla the syndrome of
the error made.  It is essential that the ancilla remains in a state
depending only on the error, and not on the particular state of the
system.  Now, it is enough to measure the state of the ancilla in
order to find out that the error made has been $X_3$, to apply the
operator $X_3^{-1}$ to the system in order to retrieve the state free
of error $\ket{\bar\phi}$, and to bring back the ancilla  to its
neutral state $\ket{000}_A$.  Finally suppose instead that the error
to detect and correct is a phase flip at the fifth place ($Z_5$
error).  Since $Z_5=U_{\rm H}^{\otimes 7}X_5U_{\rm H}^{\otimes 7}$,
with $U_{\rm H}$ being the unary Hadamard application, it is enough
for the system to go through the operation $U_{\rm H}^{\otimes 7}$, to
apply then the previous strategy, and finally to act with $U_{\rm
H}^{\otimes 7}$ once more.

\subsection{Entanglement Distillation}
\label{sec3D:level2}

In addition to quantum error-correction codes (QECC) there is another
method to beat decoherence which is specially suitable when
communicating over noisy channels.  It is based on the notion of {\em
entanglement distillation} or {\em purification}: given two spatially
separated parties A and B sharing a collection of entangled pairs,
they are allowed to perform quantum local operations and classical
communication (LOCC) (\ref{sec3O:level2}) to extract a reduced sample
of pairs with a higher purity of entanglement.  Entanglement
distillation serves as a useful tool for quantum communication
providing us with more powerful protocols for dealing with errors
(decoherence) than quantum error correction (Bennett et al., 1996a).

We need an {\em entanglement measure} (Vedral and Plenio, 1998).   In
distillation an apropriate entanglement measure for a pure bipartite
state $\ket{\Psi_{\rm AB}}$ is $E(\ket{\Psi_{\rm AB}})$
(\ref{qidistil01}).  The reason comes from the fact that given $n$
pure bipartite states $\ket{\Psi_{\rm AB}}$,  local actions and
classical communications are enough to prepare $m$ perfect singlet
states with a yield $\frac{m}{n}$ approaching $E(\ket{\Psi_{\rm AB}})$
as $n\rightarrow \infty$ (Bennett et al., 1996a; Bouwmeester, Ekert and 
Zeilinger, 2000).

Finding optimal purification procedures in full generality is open.
However, explicit  examples of {\em entanglement distillation
protocols} EDP are known to work at least with particular types of
mixed states, like the initial EDP introduced by Bennett et
al. (1996a), which shall be referred as the {\em BBPSSW96
protocol}. It is neither  optimal nor fully general, but it is the
basic protocol known from which other generalizations are derived.

\paragraph*{BBPSSW96 Protocol.}

There are two parties A and B, Alice and Bob, which communicate over a
noisy channel. They share entangled pairs of states and they aim at
obtaining singlets (\ref{qiEPR}) from them.  Their basic strategy is
to coordinate their actions through classical messages sacrifying some
of the entangled pairs to increase the purity of the remaining ones.

Alice and Bob want to distill some pure entanglement, say in the form
of singlet states $\ket{\Psi^-}$ (\ref{qiEPR}), from a given
collection of shared entangled pairs in an arbitrary bipartite mixed
state $\rho$.  The purity of $\rho$ is measured through the fidelity
\begin{equation}
F:=\bra{\Psi^-}\rho\ket{\Psi^-}
\label{qiF}
\end{equation}
relative to a perfect singlet.

To be specific, in this protocol Alice and Bob share two entangled
pairs, each one in the state
\begin{equation}
\begin{split}
&W_{F}:=F\ket{\Psi^-}\bra{\Psi^-} + \\
&\frac{1}{3}(1-F)\left[\ket{\Psi^+}\bra{\Psi^+} +
\ket{\Phi^+}\bra{\Phi^+} + \ket{\Phi^-}\bra{\Phi^-}\right]
\end{split}
\label{qiw1}
\end{equation}
These are called Werner states (1989). Note that they are depolarized
in  the space orthogonal to the singlet. The initial state in $({\cal
H}_{{\rm A}_1}\otimes{\cal H}_{{\rm B}_1})\otimes ({\cal H}_{{\rm
A}_2}\otimes{\cal H}_{{\rm B}_2})$ is therefore
\begin{equation}
\rho_0:=W_{F}\otimes W_{F}.
\label{qidistil1}
\end{equation}
We assume that the Werner pairs have fidelity $F>1/2$.

\noindent {\em Step 1.} Unilaterally, Alice (or Bob) applies the gate
$Y$ on each of her (his) two pairs of qubits. This brings $\rho_0$ to
\begin{equation}
\rho_1:= (Y\otimes 1)\otimes (Y\otimes 1)\rho_0 (Y\otimes 1)\otimes
 (Y\otimes 1)
\label{qidistil2}
\end{equation}

\noindent The Pauli operators map the Bell states (\ref{qiEPR}) onto
one another in a 1:1 pairwise fashion, leaving no state unchanged (up
to irrelevant phase factors which we will ignore); in particular
$Y\otimes 1: \ket{\Psi^{\pm}} \leftrightarrow \ket{\Phi^{\mp}}$. Then
\begin{equation}
\rho_1= W'_{F}\otimes W'_{F}
\label{qidistil3}
\end{equation}

\noindent with

\begin{equation}
\begin{split}
&W'_{F}:=F\ket{\Phi^+}\bra{\Phi^+} +\\
&\frac{1}{3}(1-F)\left[\ket{\Phi^-}\bra{\Phi^-} +
\ket{\Psi^-}\bra{\Psi^-} + \ket{\Psi^+}\bra{\Psi^+}\right]
\end{split}
\label{qidistil4}
\end{equation}
The outcome is a new bipartite state with a large component $F>1/2$ of
$\ket{\Phi^+}$ and equal components of the other three Bell states.

\noindent {\em Step 2.} Bilaterally, Alice and Bob apply a CNOT
operation (\ref{qicnot}) to each of their pairs of qubits.  Let us
denote this joint operation as $U_{\rm BCNOT}$. Thus
\begin{equation}
\rho_1\mapsto\rho_2:=U_{\rm BCNOT}\rho_1 U_{\rm BCNOT}.
\label{qidistil5}
\end{equation}
This composite operation acts conditionally on qubits 3 and 4 (target
qubits) depending on the states of qubits 1 and 2  (source qubits),
namely
\begin{equation}
\begin{split}
&U_{\rm BCNOT}:=\\ &(\ket{0}\bra{0}\otimes 1\otimes 1 \otimes 1 +
\ket{1}\bra{1}\otimes 1\otimes U_{\rm NOT} \otimes 1).\\ & (1\otimes
\ket{0}\bra{0}\otimes 1\otimes 1 +1\otimes \ket{1}\bra{1}\otimes
1\otimes U_{\rm NOT})
\end{split}
\label{qidistil6}
\end{equation}
The possible results of acting with BCNOT on the Bell states as source
and target states are summarized in Table~\ref{tableBCNOT}.

\begin{table}
\begin{ruledtabular}
\begin{tabular}{cccc}
  \multicolumn{2}{c}{Before}&  \multicolumn{2}{c}{After} \\ \colrule
  \multicolumn{1}{c}{source} &\multicolumn{1}{c}{target}
  &\multicolumn{1}{c}{source} &\multicolumn{1}{c}{target} \\ \colrule
  $\ket{\Phi^{\pm}}$ & $\ket{\Phi^{+}}$ & {\rm n.c.} & {\rm n.c.}  \\
  $\ket{\Phi^{\pm}}$ & $\ket{\Psi^{+}}$ & {\rm n.c.} & {\rm n.c.}  \\
  $\ket{\Psi^{\pm}}$ & $\ket{\Phi^{+}}$ & {\rm n.c.} & $\ket{\Psi^+}$
  \\ $\ket{\Psi^{\pm}}$ & $\ket{\Psi^{+}}$ & {\rm n.c.} &
  $\ket{\Phi^+}$  \\ $\ket{\Phi^{\pm}}$ & $\ket{\Phi^{-}}$
  &$\ket{\Phi^{\mp}}$ & {\rm n.c.}  \\ $\ket{\Phi^{\pm}}$ &
  $\ket{\Psi^{-}}$ & $\ket{\Phi^{\mp}}$ & {\rm n.c.}  \\
  $\ket{\Psi^{\pm}}$ & $\ket{\Phi^{-}}$ & $\ket{\Psi^{\mp}}$ &
  $\ket{\Psi^-}$  \\ $\ket{\Psi^{\pm}}$ & $\ket{\Psi^{-}}$ &
  $\ket{\Psi^{\mp}}$ & $\ket{\Phi^-}$  \\
\end{tabular}
\end{ruledtabular}
\caption{The two columns on the right list the states after the action
of BCNOT (\ref{qidistil6}) starting from the states on the left two
columns. The notation is n.c.=no change.}
\label{tableBCNOT}
\end{table}

\noindent {\em Step 3.} Alice and Bob measure (with respect to the
computational basis) their target qubits, i.e., Alice measures qubit 3
and Bob qubit 4. Then, they share their results by classical
communication. If their results agree, they both keep their unmeasured
source qubits, otherwise they discard them.

The source state $\rho_{\rm s}'$ thereby obtained is a convex
combination  of the Bell projections, with a weight of
$\ket{\Phi^{+}}\bra{\Phi^{+}}$  given by
\begin{equation}
F':=\frac{F^2 + \frac{1}{9} (1-F)^2}{F^2 + \frac{2}{3} F(1-F) +
\frac{5}{9} (1-F)^2}.
\label{qidistil7}
\end{equation}
The rest $1-F'$ is not equally distributed among the other three Bell
states.

\noindent {\em Step 4.} Unilaterally, Alice (or Bob) applies $Y$ on
her (his) source qubit in order to convert $\rho_{\rm s}'$  into a
state $\rho_{\rm s}$ of fidelity $F'$ (relative to $\ket{\Psi^-}$).

\noindent {\em Step 5.} The state $\rho_{\rm s}$ is not a Werner
state. But there is a depolarizing procedure, called bilateral random
operation, which  mutates it back into a such one while preserving its
fidelity  (Bennett et al., 1996b).

The net result of this protocol is that with probability greater than
$\fourth$, one Werner pair of fidelity $F'>F>\half$ (\ref{qidistil7})
is distilled out of two Werner pairs of fidelity $F>\half$.

An initial supply of $N$ Werner states of fidelity $F$ is halved by  a
single run of the above protocol to a sample of  Werner states of
fidelity $F'>F$.  Iterating the procedure as much as necessary, Werner
states  of purity $F_{\rm out}$ arbitrarily close to 1 can be
distilled from a supply of input mixed states $\rho$ of any purity
$F_{\rm in}>\half$.\footnote{The map $F\mapsto F'$ is strictly
increasing in the interval $[\half, 1]$, and has an atractive fixed
point at $F=1$.}

The overall result of the BBPSSW96 protocol is to simulate a noiseless
quantum channel by a noisy one assisted with local actions and
classical communication (LOCC). It assumes tacitly that the quantum
channel is shorter than its coherence length; otherwise one may resort
to the assistance of {\em quantum repeaters} (D\"ur et al. 1999).

There exist also EDP protocols using one single pair of qubits (Gisin,
1996; Kwiat et al., 2001).

Finding the optimal distillation protocols for a general state and any
number of copies is the unsolved {\em distillability problem}.
Despite this lack of knowledge, a surprising result is the existence
of entangled states that cannot be distilled and are called  {\em
bound entangled} (Horodecki et al., 1998). Explicit examples of
entangled mixed states of two qutrits that cannot be distilled were
found by Horodecki et al. (1999).  These states are useless for
quantum communication protocols and it is important to distinguish
them form distillable states that are also called {\em free
entangled}.   In some general instances, it is  possible to conclude
that a mixed state is bound entangled: if $\rho$ is entangled and
satisfies the Peres criterion $\rho^{\rm t,j}\geq 0$
(Sec.~\ref{sec3O:level2}) then $\rho$ is a bound entangled state
(Horodecki et al., 1998).

In summary, entanglement is a new resource for computation processing
and communication that can change information theory both
qualitatively and quantitatively. The concept of entanglement is an
genuinely quantum phenomenon that allows us to extend the theory of
information beyond its classical limitations.  We have already seen
error-correction codes as one essential application of entanglement
and more genuine examples like teleportation, dense coding, quantum
key distribution, quantum computations, etc.  are addressed in the
forthcoming sections.

\section{Quantum Teleportation}
\label{sec4:level1}

Copying classical states (be it an Etruscan fibula, a Goya painting,
or a banknote) has never posed unsurmountable difficulties to experts.
It suffices to thoroughfully observe the original as much as it may be
required, taking care of not damaging it, to retrieve the information
needed to make a copy of it.  This careful observation does not alter
in a noticeable way its state.  But if the original to be reproduced
is a quantum system in an unknown state $\phi$, then any measurement
(incompatible with $P_\phi$) made on the system to get information on
$\phi$ will perturb uncontrollably the state destroying the original
(Sec.~\ref{sec3:level1}).  Moreover, even having an unlimited number
of copies of that state, infinitely many measurements will be
necessary to determine that unknown state.

For example, let us assume that Alice has a qubit (say one spin
$\half$) in a pure state. Bob needs it, but Alice does not have any
quantum channel to transmit it to him.  If Alice knows the precise
state of her qubit (for example, if she knows that her spin $\half$ is
oriented in the direction $\vect{n}$), it is enough for her to give
Bob in a letter (classical channel) that information (the components
of $\vect{n}$) to enable him preparing a qubit exactly equal to
Alice's.  But if she happens not to know the state, she may choose to
confess it to Bob, who would then be inevitably driven to prepare his
qubit in a random way, obtaining a 50\% fidelity on average.  But
Alice can also try to be more cooperative, making for example a
measurement on her qubit of $\vect{n}\cdot\bfsigma$, with $\vec{n}$
arbitrarily chosen, and then transmitting to Bob both the components
of $\vec{n}$ and the result $\epsilon=\pm 1$ thus obtained.  Armed
with this information, Bob can prepare his qubit in the state
$\half(1+\epsilon\vect{n}\cdot\bfsigma)$.  The average fidelity so
obtained is larger than before: 2/3.  However, it is not enough.

\begin{figure}[ht]
\psfrag{a}[Bc][Bc][0.6][0]{Alice} \psfrag{s}[Bc][Bc][0.6][0]{Bob}
\psfrag{x}[Bc][Bc][0.6][0]{cbit} \psfrag{q}[Bc][Bc][0.6][0]{qubit}
\psfrag{d}[Bc][Bc][0.6][0]{decoder} \psfrag{c}[Bc][Bc][0.6][0]{coder}
\psfrag{b}[Bc][Bc][0.6][0]{cbit} \psfrag{E}[Bc][Bc][0.6][0]{EPR
Source} \psfrag{p}[Bc][Bc][0.6][0]{$\psi$}
\psfrag{f}[Bc][Bc][0.6][0]{$\Phi$} 
\includegraphics[width=6 cm]{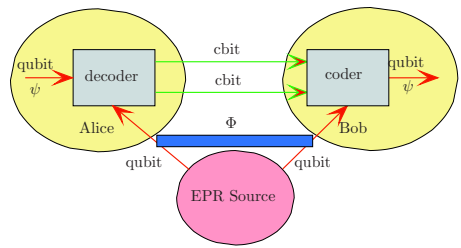}
\caption{Scheme for quantum teleportation.}
\label{teleporte}
\end{figure}

If Alice and Bob share an EPR pair, there exists a protocol, devised
by Bennett et al.  (1993), known as {\em quantum teleportation}, which
resorting to the quantum entanglement of states and the non-locality
of quantum mechanics, it allows Bob to reproduce Alice's unknown
quantum state with the assistance of only 2 cbits of information sent
by Alice to Bob through a classical channel.  This procedure
necessarily destroys Alice's state (otherwise it would violate the
quantum no-cloning theorem, Sec.~\ref{sec3:level1}).  Let us have a
closer look at the aforementioned protocol (see Fig.\ref{teleporte})
(Rieffel and Polack, 1998).

Let $\ket{\psi}=\alpha\ket{0}+\beta\ket{1}$ be Alice's qubit, with
$\alpha=\cos\half\theta$, $\beta=\ee^{\ii\phi}{\sin}\half\theta$ .
And let $\ket{\Phi}:=2^{-1/2}(\ket{00}+\ket{11})$ be the EPR state
shared by Alice and Bob, with Alice having the first of its qubits,
and Bob the second.  The initial state is thus
$\ket{\psi}\otimes\ket{\Phi}$, of which Alice can locally manipulate
its two first bits and Bob the third one.

\noindent {\em Step 1.} Alice applies to the initial state the unitary
operator $U:=((U_{\rm H}\otimes 1)U_{\rm CNOT})\otimes 1$, acting with
the CNOT gate on the first two qubits and next with the Hadamard gate
H on the first one.  The resulting state is
\begin{equation}
\half(\ket{00}\otimes\ket{\psi}+\ket{01}\otimes X\ket{\psi}+
\ket{10}\otimes Z\ket{\psi}+\ket{11}\otimes Y\ket{\psi}).
\label{qi10}
\end{equation}

\noindent {\em Step 2.} Alice then measures the first two qubits,
obtaining $\ket{00}, \ket{01}, \ket{10}$,  or $\ket{11}$
equiprobably.\footnote{Steps 1+2 amount to performing a Bell
measurement on the initial state, thus correlating the Bell states
$00\pm 11,01\pm 10$ of Alice's two qubits with the states of Bob's
qubit.  It suffices to note that
\begin{equation*}
\begin{split}
&\ket{\psi}\ket{\Phi}=\frac{1}{\sqrt{2}}
\ket{\psi}(\ket{00}+\ket{11})=\frac{1}{2\sqrt{2}}
((\ket{00}+\ket{11})\ket{\psi}+\\ &(\ket{01}+\ket{10})X\ket{\psi}+
(\ket{00}-\ket{11})Z\ket{\psi}+ (\ket{01}-\ket{10})Y\ket{\psi}).
\end{split}
\end{equation*}
} Alice lets Bob know the result thus obtained, sending him two cbits:
the pair of binary digits 00, 01, 10, 11 that characterizes it.  As a
byproduct of Alice's measurement, the first bit ceases to be in the
original state $\ket{\psi}$, while the third qubit gets projected onto
$\ket{\psi},X\ket{\psi},Z\ket{\psi},Y\ket{\psi}$, respectively.

\noindent {\em And step 3.} Once Bob receives the classical
information sent by Alice, he just needs to apply on his qubit the
corresponding gate 1, $X, Z, Y$, in order to drive it to the desired
state $\ket{\psi}$.

Notice that this teleportation sends an unknown quantum state from one
place (whence its vanishes) to another place (where it shows up)
without really traversing the intermediate space.  It does not
violates causality, though.  In the first part of the process, quantum
correlations get established between the Bell states obtained by Alice
and the associated states of Bob's qubit.  In the remaining part to
conclude the teleportation, information is transmitted by classical
means, in the standard non-superluminal fashion.  Notice also that in
this ``noncorporeal'' process, it is the information about the quantum
state, the qubit, and not the physical state itself, what gets passed
from Alice to Bob.  There has been no transportation whatsoever of
matter, energy or information at a speed larger than the speed of
light.

It is nevertheless surprising in the quantum teleportation that all
the information needed to reproduce the state
$\ket{\psi}=(\cos\half\theta)\ket{0}+\ee^{\ii\phi}
({\sin}\half\theta)\ket{1}$ (information that is infinite for it
requires to fix a point $(\theta,\phi)$ on the Bloch sphere with
infinite precision, thus requiring infinitely many qubits), can be
accomplished with only 2 cbits, provided an EPR state is shared.  This
state, by itself, only generates potentially an infinite number of
random and correlated bit pairs.

An {\em ebit} is the amount of entanglement in a two-qubit state
maximally entangled (usually, in a bipartite pure state with
entanglement entropy 1) (Bennett et al., 1996).  As an ``exchange
currency'', one ebit is a computing resource made up of a shared EPR
pair.  Writing $a\triangleleft b$ to indicate that a resource $a$ is
implementable upon spending the resource $b$, the following relations
are quite apparent: $\text{1 cbit}\triangleleft\text{1 qubit}$ (to
transmit 1 cbit it is enough to send 1 qubit in one out of two
orthogonal states), $\text{1 ebit}\triangleleft\text{1 qubit}$ (to
have 1 ebit it is enough to produce an EPR pair and to send one half
of it to the other partner).  With this formulation, the quantum
teleportation allows us to write: $\text{1 qbit}\triangleleft\text{1
ebit}+\text{2 cbits}$ (Bennett, 1995a).

Quantum teleportation was realized experimentally with photons for the
first time in two laboratories (Bouwmeester et al., 1997; Boschi et
al., 1998).  This is at least what these authors claim, although
several critiques have been raised (Braunstein and Kimble, 1998;
Vaidman, 1998; Braunstein, Fuchs and Kimble, 1999) (see however
Bouwmeester et al.  (1998; 1999)).  In the experiment by the Roma
group (Boschi et al., 1998), the initial state to be teleported from
Alice to Bob was a photon polarization, but not an arbitrary one, for
it coincided with that of the Alice's photon in the shared EPR photon
pair.  In the experiments by the Innsbruck group (Bouwmeester et al.,
1997), however, the teleported state was arbitrary.  Teleportation was
reached with a high fidelity of $0.80\pm 0.05$,\footnote{This fidelity
overcomes the value $\frac{2}{3}$ corresponding to the case in which
Alice measures her qubit and communicates the result to Bob
classically.} but with a reduced efficiency (a 25\% of cases).

It does not seem to be easy to implement the theoretical protocol with
a 100\% effectiveness.  The Bell operator (which distinguishes among
the four Bell states of 2 qubits) cannot be measured unless both
qubits interact appreciably one each other (as it occurs with the CNOT
gate used in the protocol explained above), something which is very
hard to achieve with photons.  However, with atoms in EM cavities the
hopes are high.

Teleportation has also been realized of states which are parts of
entangled states (Pan et al., 1998).

It is also worthwhile mentioning quantum teleportation of states of
infinite dimensional systems (Furuzawa et al., 1998), namely, the
teleportation of coherent optical states leaning on pairs of EPR
squeezed states.  In this experiment, whose fidelity is $0.58\pm 0.02$
(higher than the maximum $\half$ expected without resorting to
entanglement), a third party, the {\em verifier} Victor, supplies
Alice with one state that is known to him, but not to her.  After
teleporting that state from Alice to Bob, Victor verifies on output if
Bob's state is similar to the one he provided to Alice.  In this
sense, this experiment is different from all the others, and led the
authors to claim priority in the realization of teleporting.

Quantum teleportation, which doubtlessly will be extended to entangled
states from different kinds of systems (photons and atoms, ions and
phonons, etc.), might have in the future remarkable applications for
quantum computers and in computer networks (for example, combined with
prior distillation of good EPR pairs), as well as in the production of
quantum memory records by means of teleportation of information on
systems such as photons to other systems as trapped, well-isolated
ions in cavities (Bennett, 1995a; Bouwmeester et al, 1997).

\section{Dense Coding}
\label{sec5:level1}

Classical information can also be sent through quantum channels: to
transmit the word 10011, it is enough that Alice prepares 5 qubits in
the states $\ket{1},\ket{0}, \ket{0},\ket{1},\ket{1}$, sends them to
Bob through the quantum channel, and Bob measures each of them in the
basis $\{\ket{0},\ket{1}\}$.  Each qubit carries a cbit, and this is
the most it can do in isolation.  But if Alice and Bob share
beforehand an entangled state, then 2 cbits of information can be sent
from Alice to Bob with a single qubit.  This is cast in the formula:
$\text{2 cbits}\triangleleft\text{1 ebit}+\text{1 qubit}$.

As a matter of fact, entanglement is a computing resource that allows
more efficient ways of coding information (Bennett and Wiesner, 1992).
One of them goes under the name of {\em quantum dense coding} (or
superdense coding).  Assume, for instance, an entangled state of two
photons.  One of the photons goes to Alice, the other one to Bob.  She
performs one of the following operations on the polarization of her
arriving photon: identity, flipping (that is,
$\leftrightarrow\rightleftarrows\updownarrow$, or
$\circlearrowright\rightleftarrows\circlearrowleft$), change of $\pi$
in the relative phase, and the product of the last two.  Once this is
done, she sends back the photon to Bob, who measures in which of the
four Bell states the photon pair is.  Then, in this fashion we have
been able to send 2 bits of information over one single particle with
only 2 states, that is, by means of a qubit.  It doubles what can be
accomplished classically.  Thereby the name of dense coding.
Moreover, if Eve intercepts the qubit, she cannot get from it alone
any information whatsoever for its state is $\half I$.  All the
information lies in the entangled state, and Bob possesses half of the
pair.  Actually, Alice has sent Bob 2 qubits, but the first one  long
ago, as part of the initial entangled state.  This fact has allowed
them to communicate more efficiently, resorting to the entangled state
they shared.

Dense coding is kind of the inverse process to teleportation.  In the
latter the communication of two cbits allows us to reproduce a qubit
state, while in the former the communication of a qubit carries along
two cbits of information.

\begin{figure}[ht]
\psfrag{0}[Bc][Bc][0.6][0]{$0,1,2,3$} \psfrag{c}[Bc][Bc][0.6][0]{cbit}
\psfrag{q}[Bc][Bc][0.6][0]{qubit} \psfrag{A}[Bc][Bc][0.6][0]{Alice}
\psfrag{B}[Bc][Bc][0.6][0]{Bob} \psfrag{k}[Bc][Bc][0.6][0]{coder}
\psfrag{d}[Bc][Bc][0.6][0]{decoder} \psfrag{o}[Bc][Bc][0.6][0]{$\Phi$}
\psfrag{E}[Bc][Bc][0.6][0]{EPR Source}
\psfrag{f}[Bc][Bc][0.6][0]{$\Phi,Z_1 \Phi,X_1 \Phi, Y_1 \Phi$}
\includegraphics[width=6 cm]{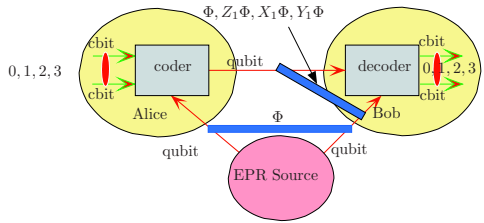}
\caption{Scheme for dense quantum coding.}
\label{dense}
\end{figure}

The following is a protocol that thoroughfully implements what we have
just explained (Rieffel and Polack, 1998): an EPR source supplies
Alice and Bob with EPR two-particle states like
$\ket{\Phi}:=2^{-1/2}(\ket{00}+\ket{11})$, one of whose particles goes
to Alice and the other one to Bob, who keep them.  Alice is supplied
with 2 cbits, which represent the numbers 0, 1, 2, 3 as 00, 01, 10, 11
(see figure \ref{dense}).

\noindent {\em Step 1.  Coding.} According to the value of that
number, Alice effects on her EPR half the unitary operation $1, Z, X,
Y$, which brings the EPR state to 00+11, 00-11, 10+01, 10-01.  Once
this is done, she sends her half to Bob.

\noindent {\em Step 2.  Decoding.} Upon reception, Bob effects on the
EPR pair first a CNOT operation, such that the state becomes 00+10,
00-10, 11+01, 11-01.  He then measures the second qubit; if the finds
0, he already knows that the message was 0 or 1, and if he finds 1,
the message was 2 or 3.  That is, he has gotten the first bit of the
two-bit message.  In order to know the second one, Bob next applies a
Hadamard transformation on the first qubit, thereby the state becomes
00, 10, 01, -11, and measuring the first bit, if he finds 0, he knows
that the message was 0 or 2, and if he finds 1, the message was 1 or
3, that is, he has just gotten the second bit of the message.

An experiment of this nature has been performed in Innsbruck (Mattle
et al., 1996), using as a source of entangled photons the parametric
down conversion that a non-linear crystal of $\beta$-barium borate
produces: UV photons get disintegrated (though with low probability)
in a pair of softer photons, with polarizations which in a certain
geometric configuration they are entangled.  In that experiment it was
achieved to send 1 qutrit/qubit, that is, $\log_{2}3=1.58$ cbits per
qubit.

In a recent experiment, in which the qubits are the spins of $^1$H y
$^{13}$C in a clorophorm molecule $^{13}$CHCl$_3$ marked with
$^{13}$C, and RMN techniques are employed to initialize, manipulate
and read out the spins, the authors claim to have reached the 2 cbits
per qubit (Fang et al., 1999).

The initial preparation of the entangled pair and the posterior
transmission of the information qubit may have opposite senses; for
example, Bob sends to Alice one half of the entangled state, keeping
the other half for himself, and then Alice uses her qubit to send to
Bob the desired information.  This may be of interest if the cost in
the transmission in one way is higher than in the reverse way.  Being
the distribution of the entangled state prior to the communication,
transmission hours at lower charges can be profited from.

On the other hand, intercepting the message from Alice to Bob does not
provide a trifle of information to an eavesdropper, for the message is
entangled with the part of the EPR system possessed by Bob.  Therefore
it is automatically an encrypted emission (except if Eve intercepts
both the original pair and the message and she replaces them).

\section{Cryptography}
\label{sec6:level1}

\subsection{Classical Cryptography}
\label{sec6A:level2}

Cryptography is a very important part of information theory since
1949, with the pioneering works by Shannon at Bell Labs.  He proved
that there exist unbreakable codes or {\em perfectly secret} systems
(Shannon, 1949).  As a matter of fact, one was known since 1918 (but
not that it were unbreakable): the {\em one-time pad} system ({\sc
onetimepad}).  It is also named {\sc vernam} code (Vernam, 1926), for
it was devised by the young engineer Vernam at AT\&T in December 1917
and proposed to the company in 1918 (Kahn, 1967); with Vernam's system
both ciphering and deciphering of messages became automatic tasks for
the first time.

\subsubsection{One-time pad}

To encode with the one-time pad one starts from the {\em plain}  or
{\em source} text to be ciphered, written as a series $\{p_1, p_2,
..., p_N\}$ of integers $p_j\in\Z_B$; then a {\em key} $\{k_1, k_2,
..., k_M\}\in\Z_B^M, M \geq N$, randomly chosen, is used to produce a
{\em ciphered} text or {\em cryptogram}  $\{c_1, c_2, ..., c_N\}$ by
combining the key with the plain text in modular arithmetic $c_j:=p_j
+ k_j\; \mod\,B, 1\leq j\leq N$.  The module $B$ is the maximum number
of distinct symbols (2 for binary, 10 for digits, 27 for letters
(English text and blank space symbol), etc.).

Both the sender (Alice) and the receiver (Bob) need to have the same
key of random numbers, so that upon reception of the cryptogram, Bob
undoes the algorithm with that key recovering thereby the original
text.

Possible repetitions in the source text (to which codebreakers resort
for decoding) are washed out by the random key.  The length of the
random sequence must be greater than or equal to that of the source
text, and must not be employed more than once.\footnote{If two binary
cryptograms encoded with the same key are intercepted, their sum
modulo 2 eliminates the key and makes it possible to decrypt messages
with certain ease (Collings, 1992).}  Shannon showed that if the key
length is smaller than the text length and one reuses cyclically the
key to encrypt the message, then it is possible to extract information
from the encoded text (Shannon, 1949).  These requirements make this
procedure very burdensome when there are lots of information to
encrypt.  Moreover, it is not easy to have long series of really
random numbers at our disposal.

This cipher system was used by German and Russian diplomats during the
Second World War, and by the soviet espionage during the cold war
(Hughes et al., 1995).  It is popularly known as ``one-time pad"
because the keys were written on a notebook or pad, and each time one
was used, the corresponding sheet with the key was torn off and
destroyed.  It is said that the continued use of the same key allowed
to unmask the Rosenberg spy ring and the atom-spy Fuchs (Hughes et
al., 1995).  It was also used by Che Guevara to communicate secretly
with Fidel Castro from Bolivia (Bennett, Brassard and Ekert,1992).
And it is routinely used for White Hose and Kremlin communications
through the ``hot line''.

Although invulnerable, the {\sc vernam} cryptosystem has the
shorthcoming of demanding keys so long at least as the text to be
ciphered.  This is why it is only used to cipher highly valuable
information.  For less delicate or sensitive business it is replaced
by shorter though breakable encryptation keys.

It was precisely the spur for breaking secret messages what fostered
the development of computers.

\subsubsection{{\sc pkc} System}

The {\sc pkc} system ({\em Public Key Cryptographic System}) is of
great interest since it avoids some of the shorthcomings of the {\sc
vernam} system.  It was devised in the middle of the 70s by Diffie and
Hellman at Stanford (Diffie and Hellman, 1976; Diffie, 1992; Hellman,
1979) and later implemented at MIT by Rivest, Shamir and Adleman
(1978).\footnote{Apparently, some years before Diffie and Hellman, the
British Secret Service knew about this system, but as classified
record (military secret) (Ellis, 1970; Ekert, Hayden and Inamori,
2000).}  This system is nowadays used worldwide, for instance in
Internet.

Two keys are involved: one person $X$ gives away a public key, which
anybody can use, and he/she keeps secret a private key, which is the
inverse of the former.  The public key is used by any sender $S$ to
send coded messages to $X$; on receipt, $X$ decodes them with the
private key.  It is pretty clear that this is of interest only if $X$
alone, but nobody else, knows how to undo the coding at a reasonable
cost.  How can we get this done?  In a subtle and cunning way: to
encrypt messages, the {\sc pkc} system uses trapdoor one-way
functions.  These are injective maps of complexity {\bf P}, i.e.,
(computationally) {\em tractable} functions, the inverses of which are
{\em untractable} in practice, that is, high costly to evaluate unless
additional information is supplied ({\bf NP} problem). See Appendix
for details.  Integer factorization stands out among this type of
inverse functions, as well as discrete logarithms in finite fields and
on elliptic curves (Koblitz, 1994; Welsh, 1995).

The {\sc pkc} system affords to leave wide open both the encryptation
algorithm and ``half'' of the total key, namely the public key,
without suffering from any extra insecurity; this contrasts sharply
with the controversial {\sc des} system ({\em Data Encryption
Standard}), which discloses only the algorithm, but whose
vulnerability has been shown up (Electronic Frontier Foundation, 1998).

\subsubsection{{\sc rsa} System}

One of the most interesting ways of implementing the {\sc pkc} system
is the {\sc rsa} method of Rivest, Shamir, and Adleman, 1978, based on
the extreme difficulty of factoring large integer numbers.  In
particular, it is used to protect the electronic bank accounts (for
instance, against bank transfers electronically xxrequested).  The
public key of $X$ consists of a pair of integers $(N(X),c(X))$, the
first one very big, say of 200-300 digits, and the other one in the
interval $(1,\varphi(N(X)))$ and coprime to $\varphi(N(X))$, where
$\varphi$ is Euler's totient function ($\varphi(n)$ is the number of
coprimes to $n$ in the interval $[0,n)$).  Upon transforming the
sender $S$ his/her message $M$ into an integer following some public
bijective prescription which both sender and receiver have agreed
upon, he/she partitions it into blocks $B_j<N(X)$ as lengthy as
possible, encodes each block $B$ as
\begin{equation}
B \mapsto C(B) \equiv B^{c(X)}\; \mod\,N(X),
\label{qi12}
\end{equation}

\noindent and sends the sequence of {\em cryptograms} $\{C(B_j)\}$ to
$X$.  Let us denote this coding operation as $M\mapsto P_X(M)$, with
the symbol $P_X$ meaning that it was done with the {\em public key}
$c(X)$ of $X$.  The receiver $X$ decodes each $C(B)$ as
\begin{equation}
C(B) \mapsto B \equiv C(B)^{d(X)}\; \mod\,N(X),
\label{qi13}
\end{equation}

\noindent where the exponent $d(X)$ for decoding is the {\em private
key}, which is nothing but a solution to
\begin{equation}
c(X)d(X) \equiv 1\; \mod\,\varphi(N(X)).
\label{qi14}
\end{equation}

\noindent That solution is (Koblitz, 1994)
\begin{equation}
d(X)\equiv c(X)^{\varphi(\varphi(N(X)))-1}\; \mod\,\varphi(N(X)).
\label{qi15}
\end{equation}

\noindent We shall indicate the decoding as $P_X(M)\mapsto
S_X(P_X(M))=M$, where the symbol $S_X$ refers to the secret key of $X$.

In principle, since $c(X)$ and $N(X)$ are known, anybody can compute
$d(X)$, and hence break up the secret.  But it is here where the
shrewdness of $X$ enters the stage.  In order to make it extremely
difficult to Eve (spy character that intercepts messages, and listens
to them without permission before delivering them again), it is better
that $X$ abides by certain rules (Salomaa, 1996), among which we
highlight the following:

\begin{enumerate}
\item He/she must choose $N(X)$ as the product $p_{1},p_{2}$ of two
large and random prime numbers (with at least one hundred digits
each), not very close one another (for this it is enough that the
lengths of their expressions differ in a few bits), and avoiding also
that they be tabulated or have some special form.  Algorithms for
testing primality like the probabilistic algorithm of Miller-Rabin
(Miller, 1980; Rabin, 1976), or the deterministic APRCL, discovered by
Adleman, Pomerance, and Rumely (1983), and later simplified and
improved by Lenstra and Cohen (Cohen and Lenstra 1984; Cohen, 1993)
facilitate enormously the election of $p_{1},p_{2}$.

\item As $X$ knows $p_{1},p_{2}$, he/she knows how to compute
$\varphi(N(X))$, namely, $\varphi(N(X))=(p_1-1)(p_2-1)$.  Now $X$ has
to choose an integer $d(X)$ (the private key) randomly in the interval
$(1,\varphi(N(X)))$, coprime to $\varphi(N(X))$, and then compute the
public key $c(X)$ by means of
\begin{equation}
c(X)\equiv d(X)^{\varphi(\varphi(N(X)))-1}\; \mod\,\varphi(N(X)),
\label{qi16}
\end{equation}

\noindent or, much better, by solving $c(X)d(X) \equiv 1$ $ \mod$
$\varphi(N(X))$ with the classical Euclid's algorithm.

One should discard small private keys $d(X)$, in order to avoid their
disclosure by plain trial and error.  That is why it is convenient to
start by fixing $d(X)$.  It is not advisable to have $c(X)$ very small
either, for then the interception of the same message sent to several
addressees sharing the same public key could lead to its break-up
without much effort.

\end{enumerate}

Anybody knowing only $N(X)$ but not its factors, should ``apparently''
factorize first $N(X)$ to compute $\varphi(N(X))$, and hence to find
out the exponent for decoding;\footnote{``Apparently'', for it is
unknown so far whether there exist alternative procedures to decode
$C(B)$ which do not go through getting the inverse exponent, nor
whether the computation of this one necessarily requires to know the
prime factors of $N$.} but factorization of a number 250 digits long
would take about 10 million years on a 200 MIPS\footnote{Million of
instructions per second; it gives a general idea of a computer's
speed, but only refers to CPU speed (real speed depends also on other
factors like input/output speed).} workstation with the best algorithm
known nowadays (Hughes, 1997).

The {\sc rsa} system also allows digital {\em authentication} of
messages, as well as appending to them an {\em electronic} or {\em
digital signature} (van der Lubbe, 1998; Koblitz, 1994; Stinson, 1995;
Welsh, 1995).

\paragraph{The RSA numbers.} In 1977 Martin Gardner
published an encoded message in his {\sc Mathematical Games} of
Scientific American using the {\sc rsa} method, with the promise of a
\$\,100 reward (payable by the Rivest {\em et al.} group at MIT) for
the first person who would decode it (Gardner, 1977):

\begin{verbatim}
96869613754622061477140922254355882905759991124
57431987469512093081629822514570835693147662288
3989628013391990551829945157815154
\end{verbatim}

This cryptomessage had been obtained using the {\sc rsa} method
starting from an English sentence and the dictionary
$\sqcup\,(\text{blank space})\mapsto 00, {\tt a}\mapsto 01,\ldots,{\tt
z}\mapsto 26$), and using as  public key (RSA-129,9007), where RSA-129
was the following number 129 digits long:

\begin{verbatim}
RSA-129 =  114381625757888867669235779976146612
01021829672124236256256184293570693524573389783
0597123563958705058989075147599290026879543541
\end{verbatim}

Decoding this message required to factorize RSA-129 into two prime
factors of 64 and 65 digits each.  It was estimated by then that the
time to reach that goal would be about $4\times 10^{16}$ years, at
least.  In 1994 new factorization algorithms\footnote{There exist
efficient methods, like those based on the quadratic sieve (QS)
(Pomerance, 1982; Gerber, 1983; Pomerance, 1996), elliptic curves (EC)
(Lenstra, 1987), and the general number field sieve (GNFS) (Lenstra,
1993; Pomerance, 1996).  Their complexities are subexponential, but
superpolynomial:
\begin{align*}
&\text{QS: } {\mathcal O}({\rm e}^{(1+o(1))\sqrt{\log N\log\log N}})
\\ &\text{EC: } {\mathcal O}({\rm e}^{(1+o(1))\sqrt{\log p\log\log
p}}) \\ &\text{GNFS: } {\mathcal O}({\rm e}^{(1.923+o(1))(\log
N)^{1/3}(\log\log N)^{2/3}})
\end{align*}
where $p$ is the smallest prime factor of $N$.  From 120-130 digits
on, the number field sieve seems to overcome the other methods.}  and
the combined effort in idle time of a cluster of about a thousand
workstations on the Internet did factorize it in about 8 months, after
a CPU time of 5000 MIPS years, using the quadratic sieve algorithm
(QS).  These factors are

\begin{verbatim}
34905295108476509491478496199038981334177646384  
93387843990820577 x
32769132993266709549961988190834461413177642967  
992942539798288533
\end{verbatim}

With this knowledge, it is straightforward to recover the original
message: {\tt the magic words are squeamish ossifrage} (Atkins, 1995).

Two years later, RSA-130 was broken with the most powerful
factorization algorithm till date (the general number field sieve
(GNFS)), and after a computation time almost one order of magnitude
lower than that employed for RSA-129.  In February 1999, the
factorization of the next number in the RSA list was over: the
RSA-140, after about 2000 MPIS-years and the same GNFS method.  And in
August 1999 the factorization of RSA-155 was achieved, also using GNFS
and after about 8000 MIPS-years.\footnote{We thank A.K.  Lenstra and
H.te.Riele for sharing with us their information about the latest
RSA's factorizations.} It has 512 bits and is the product of two prime
numbers 78 digits long.  Just to figure out the magnitude of this
problem, in its solution 35.7 CPU years have been employed to do the
sieve, distributed in about three hundred workstations and PC's, and
224 CPU hours of CRAY C916 and 2 Gbytes of central memory in order to
find the relations between the rows of a giant sparse matrix of 6.7
million rows and as many columns, with an average of 62.27
non-vanishing elements per row.

\begin{figure}
\psfrag{w}[Bc][Bc][0.8][1]{\color{red} $\tau_{{\rm a}}(n)$}
\psfrag{x}[Bc][Bc][0.8][1]{\color{red}computer fabrication year}
\includegraphics[width=7 cm]{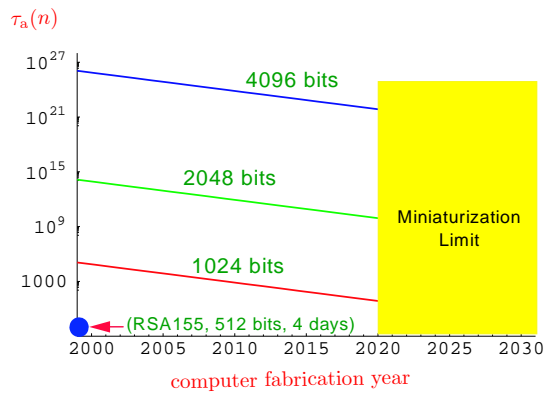}
\caption{Factorization with 1000 workstations with increasing power
according to Moore's law starting from 800 MIPS in 2000.  The vertical
axis shows the factorization time $\tau_{{\rm a}}(n)$, in years, for
an integer number of $n$ bits.  The horizontal axis shows the calendar
year.}
\label{factime1}
\end{figure}

A few years ago, it was considered as very safe the usage of 512-bits
modules.\footnote{The number of bits in the integer $N$ is
$\lfloor\log_2 N\rfloor+1$.} The preceeding example shows that the
GNFS factorization algorithm renders this bit length insufficient.
Nowdays, the use of (768, 1024, 2048)-bits modules is recommended for
(personal, corporative, highly security)-use.  In Fig.~\ref{factime1},
the estimated factorization times under the joint use of 1000
workstations is represented, assuming that the processing power
follows the so called {\em Moore's law} (doubling every 18 months)
(Hughes, 1997).  See Sec.~\ref{sec7:level1} for more details.  We take
the RSA-155 time as reference.\footnote{Miniaturization of classical
devices has the atomic/molecular scale as a limit, which at Moore
law's pace will be reached within a couple of decades.}

Even though the factorization problem remains as a hard problem in
computer science, nobody knows for sure whether one day a
mathematician may come up with a radically new faster algorithm such
that the ordinary classical computers can cope with the task of
factorizing large integer numbers in polynomial time.  As a matter of
fact, quantum computation has raised high expectations in this regard,
with Shor's algorithm (Shor, 1994) to be discussed in
Sec.~\ref{sec10D:level2}.  That is why security agencies closely
follow the new advances in number theory and computation to see what
they are up to!

\subsection{Quantum Cryptography}
\label{sec6B:level2}

Quantum physics provide us with a secure method for coding, guaranteed
by the very laws of physics.  The pioneering idea dates back to
Stephen Wiesner, who already by 1969\footnote{His work was finally
published in 1983, but after being rejected from the journal to which
it was first submitted.  An unpublished version appeared in 1970.}
suggested this possibility, as well as the fabrication of
forgery-proof banknotes, {\em quantum} banknotes (Wiesner, 1983).  In
the middle '80s Bennett and Brassard (1984) devised a quantum
cryptosystem based on the Heisenberg principle, which soon afterwards
was implemented experimentally by sending secret information with
polarized photons to a distance 30 cm apart (Bennett et al., 1992).
This system employs quantum states, not all mutually orthogonal, in
order to keep them from being cloned by a possible interceptor; as it
uses 4 distinct states, it is coined the {\em four-state scheme}.
Using non-local quantum correlations in pairs of entangled photons
(produced, for example, by parametric down conversion)  was
subsequently proposed by Ekert (1991). Within this E91 system the Bell
inequalities (Bell, 1964; 1966; 1987) are in charge of keeping the
security; hence this system is also labeled {\em EPR scheme}. For a
detailed recent review see Gisin et al., 2001.

\subsubsection{Counterfeit-safe ``quantum'' banknotes}

A possible forger-proof banknote could be a banknote provided with a
printed number and a small collection of (say twenty) photons trapped
indefinitely in individual cells of perfectly reflecting walls, and
with secret polarizations
$\circlearrowright,\circlearrowleft,\updownarrow, \leftrightarrow$
randomly distributed, that the issuing bank would keep in secret
correspondence with the identification number.  The bank therefore
could at any moment check the legitimacy of the note, without ruining
it, because it would know beforehand how to place the polarizers to
check each photon polarization without destroying it.  Any forger that
attempts to copy a note, however, ignorant of the directions in which
the photons were polarized, would perturb the initial polarization
projecting it onto some of two corresponding orientations of the
polarizer chosen to measure with (Wiesner,1983; Bennett, 1992b).

\begin{figure}[ht]
\centering \includegraphics[width=4 cm]{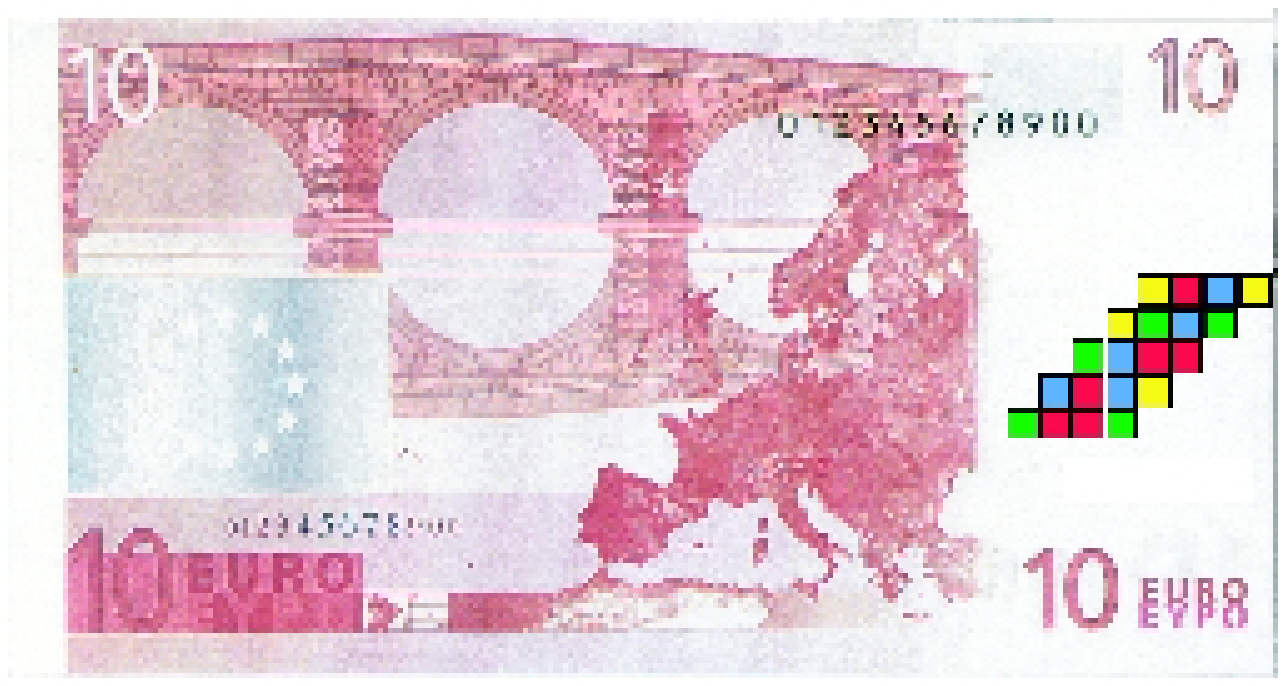} \hfill
\includegraphics[width=4 cm]{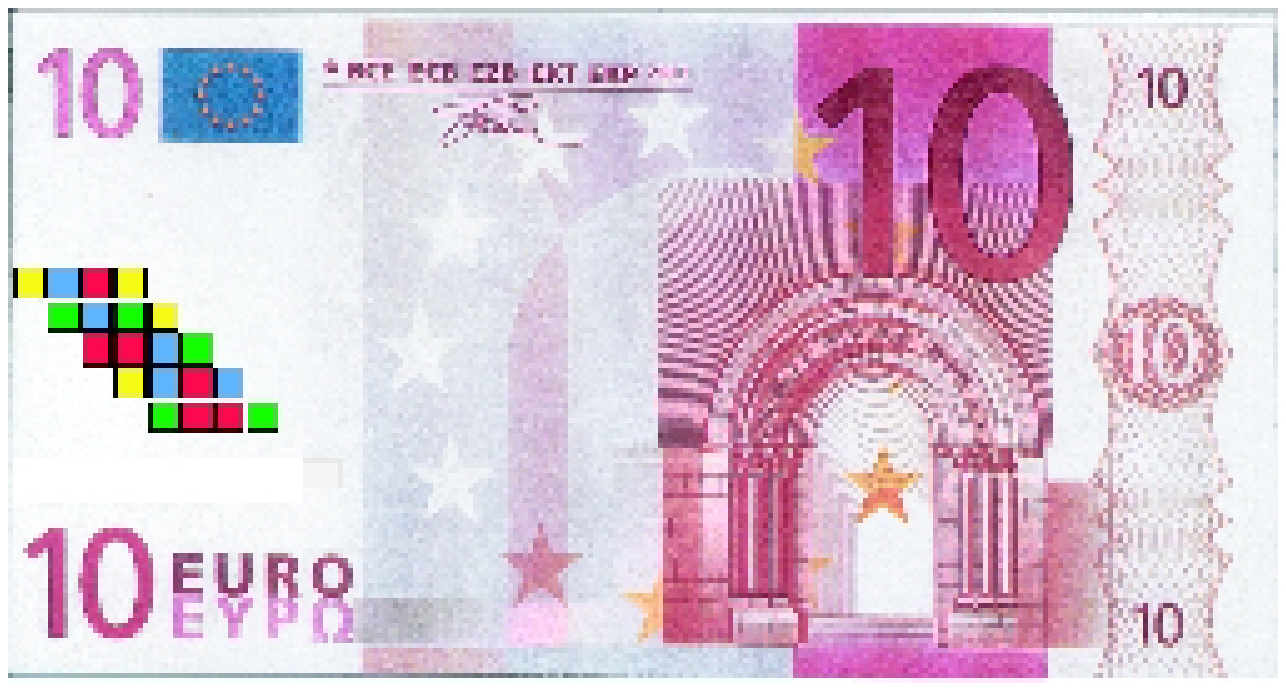}
\caption{Counterfeit-safe banknotes: the identification number is
correlated with the secret polarizations of photons trapped in
individual cells.}
\label{euro}
\end{figure}

\subsubsection{QKD: quantum key distribution}

Although the quantum notes business may look a seer fantasy, this is
not the case for systems of quantum key distribution.  Among the
communication protocols, we may highlight the BB84 of Bennett and
Brassard (1984), E91 of Ekert (1991), B92 of Bennett (1992a), and EPR
without Bell's inequalities, due to Bennett, Brassard and Mermin
(1992).  These protocols provide a way for two parties to share keys
absolutely secret in principle,  and thus they are an ideal complement
to the Vernam code.

Alice and Bob want to exchange secret information, without recourse to
middlemen who bring key pads from one to the other, and without fear
that someone breaks their code.  To this end, they must share a key,
known only to them.  They proceed according to a given communication
{\em protocol}, or set of instructions either to detect any
non-authorized eavesdropper, or else to settle down the secret key
that only they will share for coding and decoding.

\paragraph{BB84 Protocol, or four-state scheme.}

This is the first protocol devised in quantum cryptography.  Alice and
Bob are connected by two channels, one quantum and another public and
classic.  If photons are the vehicle carrying the key, the quantum
channel is usually an optical fiber.  The public channel can also be
so, but with one difference: in the quantum channel, there is in
principle only one photon per bit to be transported, while in the
public channel, in which eavesdropping by any non-authorized person
does not matter, the intensity is hundreds of times bigger.

\noindent {\em Step 1.} Alice prepares photons with linear
polarizations randomly chosen among the angles $0^\circ$, $45^\circ$,
$90^\circ$ and $135^\circ$, which she sends ``in a row'' through the
quantum channel, while keeping a record of the sequence of the
prepared states, as well as of the associated sequence of 0s and 1s
obtained representing by 0 the choices of 0 and 45 degrees, and by 1
otherwise.  This sequence of bits is clearly random.  For instance,
denoting by H, V, D and A the horizontal, vertical, 45$^\circ$ and
135$^\circ$ polarizations, respectively, and by $+$, $\times$ the
polarization basis \{H,V\}, \{D,A\}, possible Alice's sequences are:
{\small
\begin{verbatim}
++++x+xx+x++++xx+xx++xxx++x+++x+xxx+xxx++x+++++x...
VVVHAVAAVAHVHHDDVDDHHAAAVHDHVVDVDADVDAAHVDVHHHVA...
111011111101000010000111100011010101011010100011...
\end{verbatim}
}

\noindent {\em Step 2.} Bob has two analyzers, one ``rectangular" (+
type), the other ``diagonal" ($\times$ type).  Upon receiving each
Alice's photon, he decides at random what analyzer to use, and writes
down the aleatory sequence of analyzers used, as well as the result of
each measurement.  He also produces a bit sequence associating 0 to
the cases when the measurement produces a 0$^\circ$- or
45$^\circ$-photon, and 1 in cases 90$^\circ$ and 135$^\circ$.  With
the following analyzers chosen at random by Bob, a possible result of
Bob's action on the previous Alice's sequence is {\small
\begin{verbatim}
x+x+xxxx+++x++x+x+xxxx+++++++xxxx+++x+xxxxxx++x+...
DVAHADAAVVHDHHDHAVDADAHHVHVHVDDADHVVDVAAADADHHDH...
011010111100000011010100101010010011011110100000...
\end{verbatim}
}

\noindent {\em Step 3.} Next they communicate each other through the
public channel the sequences of polarization basis and analyzers
employed, as well as Bob's failures in detection, but never the
specific states prepared by Alice in each basis nor the resulting
states obtained by Bob upon measuring.  {\small
\begin{verbatim}
Alice to Bob: ++++x+xx+x++++xx+xx++xxx++x+++x+xxx...   
Bob to Alice: x+x+xxxx+++x++x+x+xxxx+++++++xxxx++...
\end{verbatim}
}

\noindent {\em Step 4.} They discard those cases in which Bob detects
no photons, and also those cases in which the preparation basis used
by Alice and the analyzer type used by Bob differ.  After this
distillation, both are left out with the same random subsequence of
bits 0, 1, which they will adopt as the shared secret key:{\small
\begin{verbatim}
Alice 111011111101000010000111100011010101011010...
      ++++x+xx+x++++xx+xx++xxx++x+++x+xxx+xxx++x...  
 Bob  x+x+xxxx+++x++x+x+xxxx+++++++xxxx+++x+xxxx...
      011010111100000011010100101010010011011110...
\end{verbatim}
\begin{verbatim}
Alice -1-01-111-0-000---0--1--10-01-0-0--10-1--0...   
Bob   -1-01-111-0-000---0--1--10-01-0-0--10-1--0...
\end{verbatim}
}

Therefore the distilled key is {\small 1011110000011001001010...}, and
its length is, on average, and assuming no detection failures, one
half of the length of each initial sequence.

\paragraph{Eavesdropping effects.}
All this holds in the ideal case that there are not eavesdroppers,
neither noises in the transmission nor defects in the production,
reception and analysis: the distilled keys of Alice and Bob coincide.
But let us assume that Eve ``taps" the quantum channel, and that,
having the same equipment as Bob's, analyzes the polarization state of
each photon, forwarding them next to Bob.  Ignoring Eve, much like
Bob, the state of each photon sent by Alice, she will use the wrong
analyzer with probability 1/2, and will replace Alice's photon by
another one, so that upon measurement Bob will get Alice's state only
with probability 3/8, instead of the probability 1/2 in absence of
eavesdropping.  Therefore this intervention of Eve induces on each
photon a probability of error 1/4.  Returning to the previous example,
let us assume that Eve's measurements on Alice's photons produce the
following results:{\small
\begin{verbatim}
Eve x++x++++x++xxx++++++x+xxxx++xx+x+++x+xxx+x...
    DVVAVVVVDVHADAVHVHHHAVAAADHHADHDVVVDHAADVD...
\end{verbatim}
}

These Eve's states are now those reaching Bob, who with his sequence
of analyzers will obtain, for instance {\small
\begin{verbatim}
x+x+xxxx+++x++x+x+xxxx+++++++xxxx+++x+xxxxxx++x+...
DVDVADADHVHAHHDHAHAAAAHHHHHHHDDDAVVVAVADDDAAHHAH...
010110100101000010111100000000001111111000110010...
\end{verbatim}
} Proceeding as in step 4: {\small
\begin{verbatim}
Alice 111011111101000010000111100011010101011010...
      ++++x+xx+x++++xx+xx++xxx++x+++x+xxx+xxx++x...   
Bob   x+x+xxxx+++x++x+x+xxxx+++++++xxxx+++x+xxxx...
      010110100101000010111100000000001111111000...
\end{verbatim}
\begin{verbatim}
Alice -1-01-111-0-000---0--1--10-01-0-0--10-1--0...   
Bob   -1-11-100-0-000---1--1--00-00-0-1--11-1--0...
\end{verbatim}
} We see that the coincidences in the distilled lists get disrupted:
in 1 out of 4 cases, the coincidence disappears.  Sacrificing  for
verification a piece of the lists  taken at random from the final
sequences, Alice and Bob can publicly compare them, and their
differences will detect the intervention of Eve.  If the length of
that checking partial sequence is $N$, the probability that Eve's
listening has not produced discrepancies is (3/4)$^N$, and thus
negligible for $N$ large enough.  Therefore, should they not find any
discordance, they can feel safe about the absence of eavesdroppers.
But that binary string they have made public, they must clearly
disregard it and not use it for coding.  However, in practice both the
emitting source, as well as the receiving equipment and the
transmission channel display noise, which necessarily spoils, even
with no snooping Eve, the perfect fit of the bit sequences distilled
by Alice and Bob.  It is necessary then to coexist with error,
whenever this stays under a tolerable limit.  In these circumstances,
Eve will try to behave herself taking care that the effects of her
listening stay below a certain threshold and do not shoot the alarm.

Cryptanalysts like Eve usually are quite more subtle in their
perversity than what the previous simple analysis might suggest.
Aware as they are of the quantum subtleties, they are not satisfied to
incoherently tapping the quantum channel qubit to qubit; they are
quite well knowledgeable that the coherent attack to strands of
qubits, with probes analyzed after the public exchange of information
between Alice and Bob, can be much more rewarding.  To prove the
safeness of a protocol such as this BB84 under any type of imaginable
attack by the malicious and cunning Eve is neither a trivial nor
uninteresting issue, specially having in mind that other protocols
resorting to quantum laws and considered as unconditionally secure
have fallen down, as for example the {\em bit commitment} quantum
protocol: Alice sends something to Bob under the firm commitment of
having chosen a bit $b$ that Bob completely ignores, but such that
Alice can later show it to him when he claims it.  Resorting to
entangled EPR states makes it possible that any party of the couple
behave dishonestly (that a cheating Alice change her commitment at the
end without Bob being aware, or that a villain Bob gets some
information on $b$ without any request to Alice) (Mayers, 1996; 1997;
Brassard et al., 1997).

There exits a proof of unconditional security of QKD through noisy
channels and up to any distance, by means of a protocol based upon the
sharing of EPR pairs and their purification, and under the hypothesis
that both parties (Alice and Bob) have fault-tolerant quantum
computers (Lo and Chau, 1999).  Likewise, it is also claimed the
unconditional security of the BB84 protocol (Mayers, 1998).

\paragraph{B92 Protocol.}

Unlike the previous protocol, that uses a system in four states,
pairwise orthogonal, in this somewhat simpler protocol B92 systems in
only two non-orthogonal states are involved.  Its analysis is similar
to the previous one and shall be skipped.

\subsubsection{EPR Protocols}

In 1991 Ekert, relying on previous ideas of Deutsch, proposed an
elegant method for secret key distribution, where the generalized
Bell's inequality is on the watch to safeguard the confidentiality in
the transmission of pairs of spin $\half$ particles entangled {\em \`a
la} EPRB (Deutsch, 1985; Ekert, 1991).

Six months after appearing Ekert's work, Bennett, Brassard and Mermin
(1992) presented a very simple scheme for key distribution that keeps
using EPRB states in the singlet state
($2^{-1/2}(\ket{01}-\ket{10})$), but does not need to invoke Bell's
theorem to detect Eve's listening.  Alice and Bob measure the spin of
their respective subsystems (halves of EPRB pairs) randomly along $Ox$
or $Oz$.  Through a public channel, they inform each other about their
sequences of selected observables, but not of the results $\pm\half$
obtained.  They discard the cases in which their selections differ.
They keep the remainder; the results of the latter are evidently
anticorrelated.  Bob reverses now all his outcomes
($\pm\half\mapsto\mp\half$), and then both Alice and Bob add $\half$
to their results, thereby obtaining the secret key to be shared.
Sacrificing as before a piece of the key for its public comparison,
they can detect Eve's listening.

Although it can be shown that this protocol is essentially equivalent
to the BB84 (Bennett, Brassard and Mermin, 1992), it presents a
potential bonus (Collins, 1992): the users (Alice and Bob) could wait
for the key to show up just when they were about to use it (should
they know how to keep the EPR states expectant for a while between
their production and use), removing this way the possibility of
robbery by Eve of the shared key.

\subsection{Practical Implementation of QKD}
\label{sec6C:level2}

The BB84 protocol was implemented by the first time in the IBM
T.J. Watson Research Center (1989-1992) with polarized photons over 32
cm in air (Brassard, 1989; Bennett et al., 1992).  In 1995 the B92
protocol was realized experimentally, also with polarized photons,
transmitted this time through optical fibre  22.8 km long in the
Swisscom cable connecting the cities of Geneva and Nyon under the
Leman lake (Muller, Breguet and Gisin, 1993; Muller, Zbinden and
Gisin, 1996).

The use of photon polarization states for long distances has a
disadvantage: birefringency in the nonstraight parts of the fiber
transforms linearly polarized states into states of elliptic
polarization, with accompanying losses in transmission, and further
produces dispersion of the orthogonal polarization modes.  Thereby the
interest in other ways to codify the states, like for example by means
of phases instead of polarizations.  A group from the British Telecom
from UK accomplished it (1994) with optical fiber over 30 km distance,
using interferometry with phase-encoded photons (Marand and Townsend,
1995).  There are no major difficulties in reaching around 50 km.  In
1999 a group from Los Alamos has reached 48 km using this procedure
(Hughes et al., 1996; 1999a; 1999b).  For that reason it can be used
to safely connect diverse agencies of the Government in Washington.
To cover distances larger than 100 km would require the use of safe
repeaters where key material for re-broadcasting might be generated.

\begin{figure}[ht]
\centering \includegraphics[height=4 cm, width=9 cm]{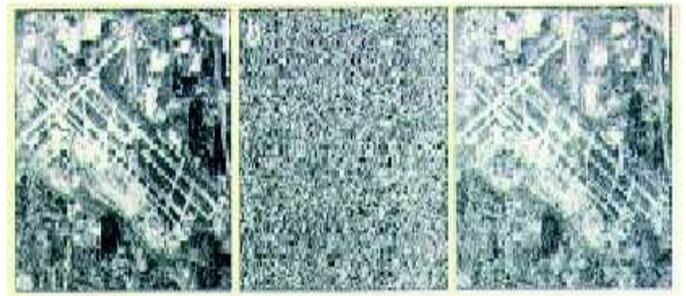}
\caption{Air view of St.  Louis airport (left), encrypted image with a
quantically generated key (center), and decrypted image (right).}
\label{aeropuerto}
\end{figure}

With the protocol B92 again, it was possible in 1998 to quantumly
transmit the secret key, at a rate of 5 kHz and over 0.5 km in broad
daylight and free space, with polarized photons (Hughes et al., 1999a;
1999c).  With this key Alice encrypted a photograph (with 8 bits per
pixel), which Bob decrypted to reconstruct the primitive image, with
the results shown in Fig.~\ref{aeropuerto}.

In the near future this procedure can be used to generate secret keys,
shared by earth-satellite or satellite-satellite, that allow to
protect the confidentiality of the transmissions.

More recently, QKD over 360 m has been achieved using variants of E91
and BB84 (Jennewein et al. 1999). They used pairs of entangled photons
to generate keys at a rate 0.4-0.8 kHz with an error  bit rate  of
about 3\%.

\section{Quantum Computation}
\label{sec7:level1}

A simple and intuitive way to arrive at the notion of quantum
computation is through the {\em miniaturization}.\footnote{The famous
Feynmann's speech addressing the American Physical Society (1959),
with his provocative bets on building microengines and writing on pin
heads, signals the birth of {\em nanotechnology.}} This has been the
driving force in the modern upgrade of ordinary computers.  As a
matter of fact, the electronic industry of computers grows at the same
time as the integrated circuits decrease in size.  This rapid growth
in the industry will continue as long as it is possible to include
more and more circuits in a single chip.  However, this pace cannot
last forever and at some point it will reach the limits of the
integrated circuits technology.  Even if we can overcome these
technological barriers, this trend will head us to the quantum realm
where the quantum laws of physics will impose fundamental limitations
on the size of the circuit components and on their performance.  Thus,
if the computer industry is to keep growing at the same rate, it will
require another technological revolution.

Although this may look quite well ahead, it is estimated that about
the year $2020$ we shall reach the atomic size for storing one bit.
Instead of just waiting for this situation to come, some theoretical
physicists decided to move ahead and started to wonder about the
radical changes and possible advantages that a computer may have if
based upon the principles of the quantum mechanics.

\begin{figure}[ht]
\psfrag{0}[Bc][Bc][0.75][0]{$1$}  \psfrag{1}[Bc][Bc][0.75][0]{$10$}
\psfrag{2}[Bc][Bc][0.75][0]{$10^{2}$}
\psfrag{3}[Bc][Bc][0.75][0]{$10^{3}$}
\psfrag{4}[Bc][Bc][0.75][0]{$10^{4}$}
\psfrag{5}[Bc][Bc][0.75][0]{$10^{5}$}
\psfrag{a}[Bc][Bc][0.75][0]{1975}  \psfrag{b}[Bc][Bc][0.75][0]{1980}
\psfrag{c}[Bc][Bc][0.75][0]{1985}  \psfrag{d}[Bc][Bc][0.75][0]{1990}
\psfrag{e}[Bc][Bc][0.75][0]{1995}  \psfrag{f}[Bc][Bc][0.75][0]{2000}
\psfrag{h}[Bc][Bc][0.75][0]{4004}  \psfrag{i}[Bc][Bc][0.75][0]{8086}
\psfrag{j}[Bc][Bc][0.75][0]{80286}  \psfrag{k}[Bc][Bc][0.75][0]{80386}
\psfrag{l}[Bc][Bc][0.75][0]{80486}  \psfrag{m}[Bc][Bc][0.75][0]{P5
(Pentium)}  \psfrag{n}[Bc][Bc][0.75][0]{P6 (P.Pro)}
\psfrag{o}[Bc][Bc][0.75][0]{P7}  \psfrag{x}[Bc][Bc][0.75][0]{Calendar
Year}  \psfrag{y}[Bc][Bc][0.75][0]{Thousands of Transistors}
\psfrag{r}[Bc][Bc][0.75][0]{2.0 years}
\psfrag{s}[Bc][Bc][0.75][0]{1.5 years}
\psfrag{t}[Bc][Bc][0.75][0]{Intel CPUs} 
\includegraphics[width=6 cm]{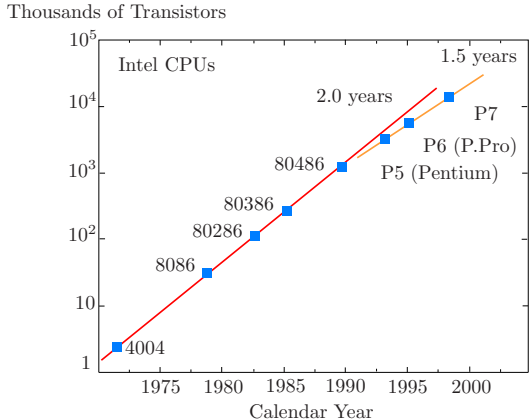}
\caption{Moore's law for processors capacity (number of transistors
per square inch).}
\label{moore3}
\end{figure}

The estimations for reaching the atomic scale are based in a
remarkable observation made by Gordon Moore (1965), later known as
{\em Moore's law}, that the number of transistors per square inch on
integrated circuits had doubled every year since the integrated
circuit was invented.  Explicitely, the original curve for the density
of silicon integrated circuits (transistors per square inch) was
$\propto 2^{(t - 1962)}$ where $t$ is the calendar year.  In
subsequent years, the trend slowed down a bit, but chip capacity has
doubled approximately every 18-24 months, and this is the current
definition of Moore's law (see Fig.~\ref{moore3}).

\section{Classical Computers}
\label{sec8:level1}

To pave the way to the concept of quantum computers it proves
convenient to discuss a classical concept, namely, the notion of {\em
classical parallel computation}.  To properly understand this let us
recall first the basic principles operating most of the ordinary
computers we work with as they were introduced first by Turing in 1936
and subsequently developed by Von Neumann in 1945 (Von Neumann, 1945;
1946), among others.

\subsection{The Turing Machine}
\label{sec8A:level2}

The concept of a {\em Turing Machine} (TM) has become the foundation
of the modern theory of computation and computability: the study of
what computers can and cannot do.  Turing arrived at this concept in
1936 (Turing, 1936) in his quest to answer one of the questions posed
by Hilbert.  This was the {\em problem of decidability}
(Entscheidungsproblem): Does it exist, at least in principle, a
definite method or process by which all mathematical questions can be
decided? (Hodges, 1992).

Turing realized that addressing this problem would require a precise
and compelling definition of what a {\em definite method} is, as it
appears in the statement of Hilbert's problem.  This is what Turing
achieved by analyzing what a person does during a methodically process
of reasoning.  His guiding idea was how to translate the human process
of thought into something purely ``mechanical'', and then he went on
to map that process into a ``theoretical machine'' which would operate
on symbols on a paper tape according to precisely defined elementary
rules.  Turing also provided convincing arguments that the
capabilities of such a machine would be enough to encompass everything
that would amount to a {\em definite method}, which in modern language
is what we call an {\em algorithm}.

We shall see later how Turing answered the question of decidability in
the negative using his concept of a TM, which we should first
introduce.

\smallskip A {\em Turing Machine} is a type of Finite State Machine
(FSM) which has a finite set of states ${\cal S}=\{ s_1, s_2,\ldots,
s_S; s_{S+1}=s_{\rm halt} \}$, a finite alphabet of symbols ${\cal
A}=\{ a_1, a_2, \ldots, a_A; a_{A+1}={\rm blank}\}$ and a finite set
of instructions ${\cal I}=\{ i_1, i_2,\ldots, i_I \}$.  In addition,
it has an external infinitely long memory tape.  This is called a
($S$-state,$A$-symbol) TM.

The states $s_i$ correspond to the functioning modes of the machine
and the TM is exactly in one of these states at any given time.  The
symbols in the alphabet serve to encode the information processed by
the machine: they are used to code input/output data and to store the
intermediate operations.  The instructions are associated to the
states in ${\cal S}$ and they tell the machine what action to perform
if it is currently scanning a certain symbol, and what state to go
into after performing this action.  There is a single {\em halt} state
$s_{\rm halt}$ (or halt, for short)  from which no instructions
emerge, and this halt state is not counted in the total number of
states.  There is also a {\em blank} symbol which serves to separate
strings of data coded with the rest of the alphabet symbols.

All these elements $({\cal S,A,I})$ are physically arranged as
follows.  A TM consists of three components:

\smallskip
\noindent The {\em tape}, which is a doubly-infinite tape divided into
distinct sections or cells. Each cell can hold only one symbol $a_i
\in {\cal A}$.

\noindent A {\em Read/Write (R/W) head} or {\em cursor}, which can
read or write the symbol $a_i \in {\cal A}$ in each tape cell.

\noindent A {\em control unit}, which is a device (or box) that
controls the movements of the R/W head based on the current state of
the TM and the content of the cell currently scanned by the R/W Head,
i.e., based on a pair $(s_i,a_i)$.

The R/W head is capable of only three actions:

\noindent {\em Write} on the tape (or erase from tape), only the cell
being scanned.

\noindent {\em Change} the internal state.

\noindent {\em Move} the head one cell to the left or right.  Let us
denote this variable as $\gamma\in\{ L,R\}$.

The behaviour of a TM is governed by the set of instructions ${\cal
I}$.  These are rules which describe the transition from an initial
pair (state, symbol) to a final pair plus the movement of the R/W
head.   Thus, each instruction $j \in {\cal I}$ is a 5-tuple
$[(s_i,a_i),(s_f,a_f;\gamma)]$ representing the following transition
\begin{equation}
{\cal I} \ni j: (s_i,a_i) \longmapsto (s_f,a_f;\gamma).
\label{qc1}
\end{equation}

\noindent A consistency condition is demanded: no two instructions
$j_1,j_2 \in {\cal I}$ have the same initial pair $(s_i,a_i)$.

\begin{figure}[h]
\includegraphics[width=8 cm]{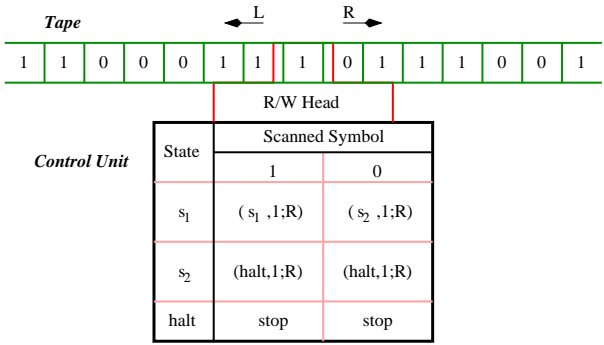}
\caption{A picture showing the components of a  Turing Machine.  The
alphabet \{1;0\} is unary, with 0 denoting blank. Stop means that
$(s_{\rm halt},.)$ has no assigned instruction.}
\label{turing1}
\end{figure}

In Fig.~\ref{turing1}  we plot a schematic picture of a TM.

\begin{figure}[h]
\includegraphics[width=8 cm]{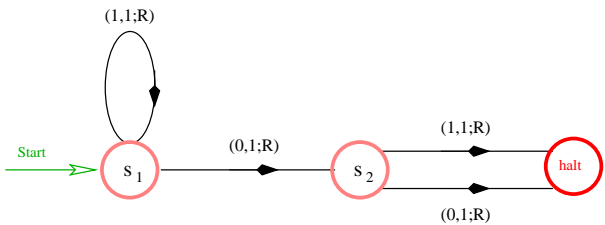}
\caption{An example of flow diagram for a (2-state,1-symbol) Turing
Machine as shown in Fig.~\ref{turing1}.}
\label{turing2}
\end{figure}

An alternative and efficient way to describe a TM is by means of a
{\em flow} or {\em state diagram} (see Fig.~\ref{turing2}).  Here each
state $s_i \in {\cal S}$ is enclosed in a circle, and the instructions
associated to a couple of states are represented by arrows showing
also the change of symbols on the tape and the head  movement.

In Fig.~\ref{turing2} we show a (2-state,1-symbol) TM. It is customary
in this case to use a $1$ for the symbol and $0$ for the blank, i.e.,
${\cal A}=\{1;0\}$.  When $A=1$ and $S=2$ we talk of a 2-state TM for
brevity.  Then, this is a unary machine, which should not be confused
with a binary system, since each number $n$ is represented as a string
of $n$ $1$s on the tape, and not by its binary representation.  The
state set is ${\cal S}=\{s_1,s_2; {\rm halt}\}$.  In this simple
example of TM, when it is in state $s_1$ scanning a $1$, the machine
will move Right one cell and stay in state $s_1$ (this is the loop in
Fig.~\ref{turing2}).  When it is in state $s_1$ scanning a blank
symbol, it will change this symbol to a $1$ and go to state $s_2$.
When it is in state $s_2$, it will just move Right and stop.

In summary, unless it is in the halt state, this simple TM will march
rightward as long as it scans $1$s, and  when it meets its first blank
symbol, it will change this into a $1$ and then it will move Right
twice and stop.

\begin{figure*}
\includegraphics[width=16 cm]{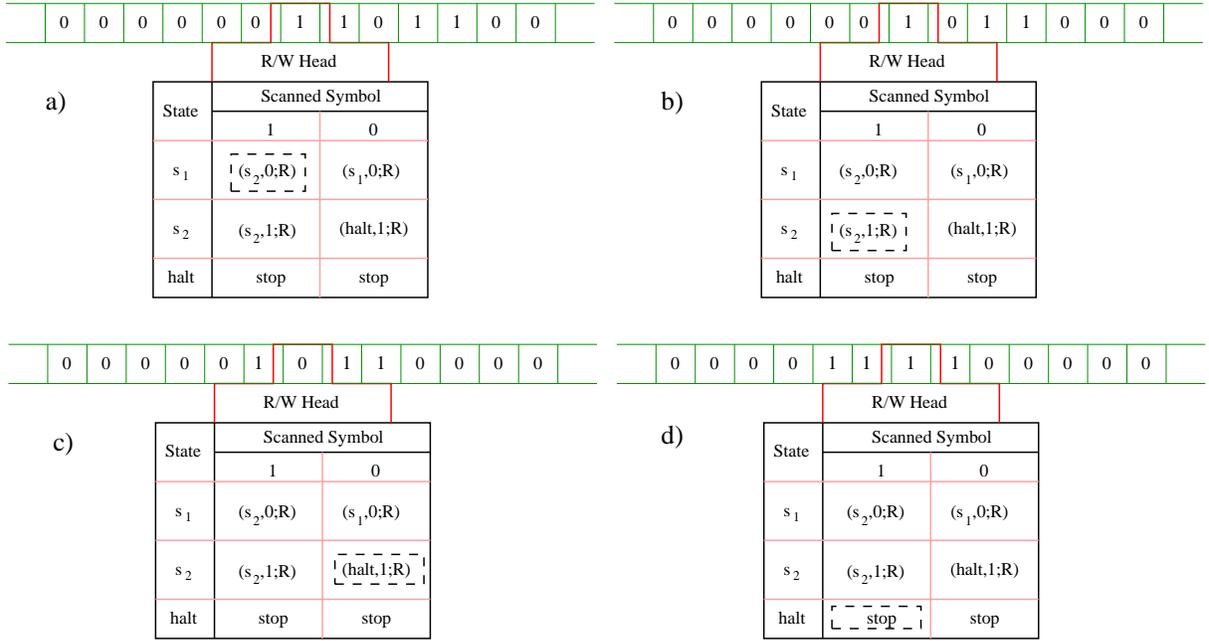}
\caption{An example of Adding Turing Machine: following the sequence
of instructions in the Control Unit the machine performs $2+2=4$.}
\label{turing3}
\end{figure*}

Let us now describe a TM performing a more interesting task like
adding two numbers.  This is a Adding TM. Suppose we want to sum $n_1
+ n_2$.  The input data in the tape is a string of $n_1$ $1$s
separated by a $0$ from another string of $n_2$ $1$s.  The output data
in the tape must be a string of $n_1+n_2$ $1$s.  To achieve this
output, we need to remove the leftmost $1$ in $n_1$ and convert the
$0$ into a $1$.  Then we can use a $2$-state TM defined as follows
(see Fig.~\ref{turing3}).  When it is in state $s_1$ and the R/W Head
scans a $1$, there is a transition to state $s_2$, the $1$ is replaced
by $0$ and the head moves to the right.  Similarly, there are other 3
instructions which we plot in Fig.~\ref{turing3}  in the form of a
chart table of instructions.  In this Fig.~\ref{turing3}  the input is
$2+2$ and the output $4$.

\subsubsection{Computability}

Despite their simplicity, Turing machines can be devised to compute
remarkably complicated functions.  In fact, a TM can compute anything
that the most powerful ordinary classical computer can compute.  Until
the formulation of Quantum Computing, none had yet proposed a model of
computation more powerful than the TM. Thus, if we stick to classical
machines and we had to solve problems which a TM cannot solve, it
seems that we would have to resort to ``supermachines'' performing
infinitely many steps in a finite time or to guess the answer out of
the blue or something similar.  The formalization of this idea into a
proposition was done independently by A. Church and A.  Turing and
goes by the name of {\em Church-Turing hypothesis} (Church, 1936;
Turing, 1936; 1950; Hodges, 1992).  Following Turing, it is stated as:
Every function that would naturally be regarded as computable can be
computed by some Turing Machine.

This is a hypothesis because it cannot be proved unless we provide a
formal definition of what {\em naturally} means.  This hypothesis has
not been refuted within the realm of classical physics, but we shall
see that the notion of a Quantum Turing Machine requires to
reformulate the Church-Turing thesis.

As a consequence of the Church-Turing hypothesis, a function is called
{\em computable} when it can be computed by a TM, while it is declared
a {\em noncomputable} function otherwise.

\subsubsection{The Universal Turing Machine}

A further crucial concept introduced by Turing is that of the
Universal Turing Machine (UTM) (Turing, 1936).  So far we have
considered TMs built for a specific purpose and for that purpose only.
The Universal TM allows us to run all TMs on a general machine.  Thus,
a UTM is defined as a single machine which comprises all Turing
Machines and is therefore capable of computing any algorithm.

Just as an ordinary TM is defined by a set (${\cal S}, {\cal A}, {\cal
I}$) with the instructions in ${\cal I}$ being described by a 5-tuple
$[(s_i,a_i),(s_f,a_f;\gamma)]$, a UTM is constructed likewise by
providing a set (${\cal S}_{\rm U}, {\cal A}_{\rm U}, {\cal I}_{\rm
U}$) and a description of its instructions
$[(S_i,A_i),(S_f,A_f;\Gamma)]$.  These instructions of a UTM must be
general enough to accommodate any possible TM. This is accomplished by
supplying it with the information of a TM and the data of its tape.

There are several ways to construct explicitly a UTM (Herken, 1995;
Feynman, 1996; Minsky, 1967).  For simplicity, let us assume that the
alphabet ${\cal A}_{\rm U}=\{a_1=0,a_2=1;{\cal A}'_{\rm U}\}$ has a
binary part corresponding to ${\cal A}$.  This is not a restriction
since any alphabet ${\cal A}$ can be mapped onto a binary alphabet.
At any given step of the functioning of a UTM, the initial pair
$(S_i,A_i)$ will know about the current description of the TM's tape,
and as it also knows about the set of instructions ${\cal I}$, then
the UTM will output exactly the same data as the TM it is simulating.
In order to implement this, we need to accommodate quite a lot of, but
finite, information in the UTM's tape.  Namely, the input data for the
UTM's tape is precisely all we need to know about the TM it
reproduces: ($\tau; (\cal{S,A,I})$), where $\tau$ denotes the TM's
tape.  These elements are disposed on the UTM's tape consecutively and
separated by marks belonging to ${\cal A}'_{\rm U}$.  The R/W head of
the UTM is positioned at the initial cell of the string encoding the
data pair $(s_0,a_0)$ of the TM.  Then the UTM starts working,
resorting to its set of instructions ${\cal I}_{\rm U}$.  Without
going into further details, this set contains rules specifying how to
bring the R/W head to read a pair $(s_i,a_i)$, change it to a new pair
$(s_f,a_f)$ and find the movement $\gamma$ of the tape $\tau$.  This
is repeated all over until the given TM is fully imitated.

The number of states $S_{\rm U}$ and symbols $A_{\rm U}$ is variable
in a UTM. Minsky has constructed one with $S_{\rm U}=7,A_{\rm U}=4$
(Minsky, 1967).  In fact, one can in principle construct always a UTM
with only $S_{\rm U}=2$ and finitely many symbols, or only $A_{\rm
U}=2$ and finitely many states.

The importance of the universal machine is clear.  We do not need to
have an infinity of different machines doing different jobs.  A single
one will suffice.  The engineering problem of producing various
machines for various jobs is replaced by the office work of {\em
programming} the universal machine to do these jobs (Turing, 1948).
In summary, a TM is comparable to an algorithm much like the UTM is to
a programmable computer.

\subsubsection{Undecidability. The Halting Problem}

With the aid of a TM, Turing was able to answer the problem of
decidability.  This can be rephrased in terms of TMs: is it possible
to compute {\em any} function by designing an appropriate TM? Turing
showed that this is not possible because the set of possible functions
is much larger that the set of possible TMs.  In fact, the set of TMs
is denumerable (and so is the set of inputs).  This is because any TM
can be encoded into a finite binary string.  However, it is possible
to find sets of functions which are uncountable.  Turing provided one
such example due to Cantor: the set $\cal F$ of all functions $f: \N
\to \N$.  Cantor had shown fifty years earlier, with his dilemma of
diagonalization,  that this set $\cal F$ was not countable.  The proof
is simple, by reductio ad absurdum: assume $\cal F$ is denumerable,
then label each function $f\in\cal F$ with an integer: ${\cal F} =
\{f_0,f_1,\ldots,f_n\ldots\}$.  Next construct a function $g: \N \to
\N$ by defining $g(k):=f_k(k)+1$, $\forall k$.  This function $g$ is
new, it is not contained in the initial set $\cal F$ since it differs
for at least one value of the argument from each function in $\cal F$.
Thus, the set $\cal F$ is not complete.  Contradiction.

This analysis implies that there must be noncomputable functions.
Turing provided the first explicit example known as {\em the halting
problem}: is it possible to design a TM $H$ which tells us whether
{\em any} TM will halt or not, when executing its procedure for {\em
any} input?  Turing showed that there does not exist such a TM $H$
(Turing, 1936), in other words, the halting decision problem is
undecidable, or equivalently, the predicate ($\{0,1\}$-valued
function) $h:\N\times\N\ni (i,j)\mapsto 1$ if the $i$-th TM $T_i$ will
halt for input $j$, $h:(i,j)\mapsto 0$ otherwise, is
noncomputable.\footnote{Any form of input/output can be encoded into
nonnegative integers (Salomaa, 1989).}  In fact, suppose that the
contrary holds, i.e. that there exists $H$ which computes $h$, and
define a function $\bar h:x\mapsto 1$ if $h(x,x)=0$, $\bar h(x)$ being
undefined otherwise.\footnote{Note that the same integer $x$ singles
out here both a TM and an input.} The function $\bar h$ is computable
by a TM $\bar H$ obtained from $H$ just by replacing 0 by 1 when $H$
halts and outputs 0, and by entering an endless loop when $H$ is ready
to halt with output 1.  Let $\bar H= T_{i(\bar H)}$; if $\bar h(i(\bar
H),i(\bar H))=1$, then $h(i(\bar H),i(\bar H))=0$ and thus $\bar H$
should not halt for input $i(\bar H)$.  Contradiction.  Similarly, if
$\bar h(i(\bar H),i(\bar H))$ is not defined, then $h(i(\bar H),i(\bar
H))=1$ and thus $\bar H$ should halt for input $i(\bar H)$.
Contradiction again.  Therefore $H$ cannot exist.

Another example was provided by T. Rado (1962) with the so called
Rado's $\Sigma$-function: assume that the TM has $S$ states, $A=1$
symbols and the input data is a tape completely blank.  Then,
$\Sigma(S)$ is defined as the maximum number of $1$s left on the tape
after this $S$-state TM halts.  This type of TM is now known as {\em
the busy-beaver problem}.  Busy beavers TMs are difficult to find for
two reasons (Shallit, 1998): firstly, the search space is extremely
large -- there are $[4(S+1)]^{2S}$ TMs with $S$ states (for each
non-halting state there are two transitions out, so the total of
transitions is $2S$, and each transition has 2 possibilities for the
symbol being written, $2$ possibilities for the direction to move
$\gamma=L,R$, and $S+1$ possibilities for what state to go to --
including the halting state).  Secondly, due to the halting problem,
it is in general not possible to determine whether a particular TM
will halt.  We have to content ourselves with finding busy beavers for
small $S$ by a brute-force approach.  In Table~\ref{tablebusy} we show
the current status of this search.  Another Rado's function
$\Sigma'(S)$ appears which is the maximum number of moves performed by
the TM before halting.  Clearly, $\Sigma'(S) \geq \Sigma (S)$.

\begin{table}[ht]
\begin{ruledtabular}
\begin{tabular}{ccc}
 $S$ & $\Sigma(S)$ & $\Sigma' (S)$ \\ \colrule 1 &1 &1$^{\rm a}$\\ 2
&4 &6$^{\rm a}$\\ 3 &6 &21$^{\rm a}$\\ 4 &13 &107$^{\rm b}$\\ 5 &
$\geq $ 4098 & $\geq$ 47\,176\,870$^{\rm c}$
\end{tabular}
\end{ruledtabular}
$^{\rm a}$ (Lin and Rado, 1965).  $^{\rm b}$ (Brady, 1983).  $^{\rm
c}$ (Marxen and Buntrock, 1990).
\caption{This is a table of busy-beaver TMs for small $S$ number of
states.  For $S=6$, $\Sigma(6) \geq95\,524\,079$, $\Sigma'(6) \geq
8\,690\,333\,381\,690\,951$ (Marxen, 1997).}
\label{tablebusy}
\end{table}

\begin{figure}[ht]
\includegraphics[width=8 cm]{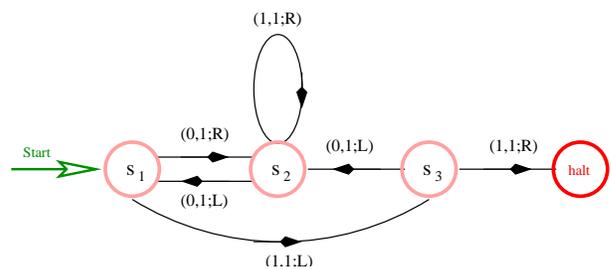}
\caption{A $3$-state busy-beaver Turing Machine.}
\label{turing4}
\end{figure}

In Fig.~\ref{turing4} we plot an explicit flow diagram of a 3-state
busy beaver (Shallit, 1998).  When this TM starts with input data a
completely blank tape, it executes 13 moves and writes six $1$s.
Thus, $\Sigma(3)\geq 6$ and $\Sigma'(3)\geq 13$.  Lin and Rado showed
(1965) that for $S=3$ the $\Sigma(3)$ lower bound yields in fact the
correct solution.  From $S=5$ on, only lower bounds are known. For
example, $\Sigma(8)>10^{44}$ (Rozenberg and Salomaa, 1994).

The proof that $\Sigma (S)$ is a noncomputable function goes by
reductio ad absurdum.  One shows that $\Sigma (S)$ grows with $S$
faster than any computable function, i.e. if $F(S)$ is an arbitrary
computable function, then there exists $S_0$ such that $\Sigma
(S)>F(S)$ for $S\geq S_0$ (Shallit, 1998).  As a byproduct,
$\Sigma'(S)$ is not computable either.

\subsubsection{Other Types of Turing Machines}

The TMs considered so far are {\em deterministic}: the instructions
$i\in {\cal I}$ follow the transition rules in (\ref{qc1}).  It is
possible to design other TMs called {\em nondeterministic Turing
machine} (NDTM) for which, given an initial pair $(s_i,a_i)$, there
exists a bunch of possible final triplets (Yan, 2000).  This means
that the transition mapping (\ref{qc1}) in no longer a function, but a
{\em relation} given by
\begin{equation}
({\cal S}, {\cal A})  \longrightarrow {\rm Subsets}({\cal S}, {\cal
A};\gamma)
\label{qc1b}
\end{equation}

\noindent where ${\rm Subsets}({\cal S}, {\cal A};\gamma)$ denote all
possible subsets of the Cartesian product ${\cal S}\times{\cal
A}\times\gamma$.  A {\em probabilistic Turing Machine} (PTM) is a type
of nondeterministic Turing machine with some distinguished states
called {\em coin-tossing states}.  When the machine goes into one of
these coin-tossing states, the control unit chooses between two
possible legal next triplets in ${\cal S}\times{\cal
A}\times\gamma$. The computation of a probabilistic TM is
deterministic except that in coin-tossing states the machine tosses an
unbiased coin to decide between two possible legal next moves.  The
class of NDTMs is more powerful than the class of deterministic Turing
machines in the sense that anything computable with a TM is also
computable with a NDTM and usually faster.  A nondeterministic TM is
closer to the idea of a Quantum Computer, but still it is far from one
of them as we shall see in Sec.~\ref{sec9:level1}.

The Turing Machines introduced so far are {\em irreversible}: given
the output of a computation we cannot generally reconstruct the input
data.  A {\em reversible} TM is one for which the input determines the
output and conversely, the output determines the input.  More
explicitely, to each Turing machine $M$ we can associate a directed
configuration graph $\Gamma(M)$: each node of the graph is a possible
configuration $C\in{\cal S}\times{\cal A}$, and two nodes $C,C'$ are
arc-connected when there is some instruccion $i\in{\cal I}$ of $M$
bringing $C$ to $C'$ in a single computation step.

{\em Reversible Turing Machine}:  A Turing machine $M$ is reversible
iff its graph of configurations $\Gamma(M)$ has only nodes with
indegree and outdegree\footnote{The indegree (outdegree) of a node is
the number of incoming (outgoing) lines.} $\leq 1$.

We know that a non-reversible Turing machine has outdegrees $\leq 1$.
It is apparent that demanding indegrees $\leq 1$ implies that $M$ can
be executed in reverse deterministically, since every configuration
has only one possible predecessor.

Lecerf (1963) and independently Bennett showed (1973) that an
irreversible Turing machine can be simulated with a reversible Turing
machine, at the expense of extra computer space and time.  This is a
remarkable fact for quantum computing since a quantum Turing machine
must be reversible (see Sec.~\ref{sec9:level1}).

Not only Turing devised a theoretical computer, but he also pursued
the practical construction of one of them.  At the end of the war
Turing was invited by the National Physical Laboratory (NPL) in London
to design a computer.  His report proposing the Automatic Computing
Engine (ACE) was submitted in March 1946.  Turing's design was at that
point an original detailed design and prospectus for a computer in the
modern sense.  The size of storage he planned for the ACE was regarded
by most who considered the report as hopelessly over-ambitious and
there were delays in the project being approved.  In the long run, the
NPL design made no advance and other computer plans at Cambridge and
Manchester took the lead.  One year earlier von Neumann had pushed
forward another project for constructing a computer machine.

\subsection{The von Neumann Machine}
\label{sec8B:level2}

The foundations of von Neumann's work on computers were laid down in
the ``First Draft of a Report on the EDVAC," written in the spring of
1945 and distributed to the staff of the Moore School of Engineering
at the University of Pennsylvania (where the EDVAC was originally
developed) in late June (Aspray, 1990).  It presented the first
written description of the {\em stored-program} concept and explained
how a stored-program computer does process information.  Von Neumann
collaborated with Mauchly and Eckert on the design for EDVAC.

\begin{figure}[ht]
\psfrag{p}[Bc][Bc][0.8][0]{Processor}  \psfrag{c}[Bc][Bc][1][0]{CPU}
\psfrag{m}[Bc][Bc][0.8][0]{Memory}
\psfrag{l}[Bc][Bc][0.7][0]{location $n$}
\psfrag{a}[Bc][Bc][0.8][0]{address $n$}
\psfrag{d}[Bc][Bc][0.8][0]{data}
\psfrag{i}[Bc][Bc][0.7][0]{instructions}  
\includegraphics[width=5 cm]{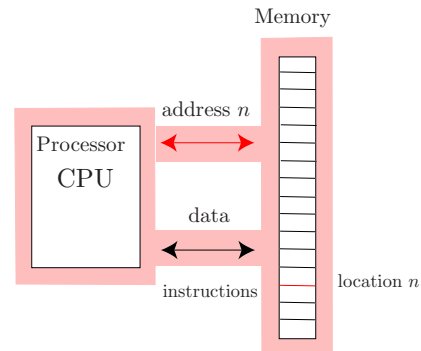}
\caption{Von Neumann Machine.}
\label{VNM}
\end{figure}

We can summarize the functioning of an ordinary computer by saying
{\em one single thing at a time}.  Von Neumann was the first to
formalize the principles of a ``program-registered calculator" based
in the {\em sequential} execution of the programs registered in the
memory of the computer.  This is called a {\em von Neumann machine}
(VNM).  A VNM has the following parts which are depicted in
Fig.~\ref{VNM}:

{\em Processor}: The active part of the computer where the information
contained in the programs is processed step by step.  It is in turn
divided into three main parts:

i) {\em Control Unit}: The unit which controls all the parts of the
computer in order to carry out all the operations requested by other
parts, such as extracting data from the memory, executing and
interpreting instructions, etc.

ii) {\em Registers}: A very fast memory unit inside the processor
which contains that part of the data which is currently being
processed.

iii) {\em ALU}: The Arithmetic and Logic Unit which is devoted to the
real computations such as sums, multiplications, logic operations,
etc., executed on the data supplied by the registers or memory upon
demand by the control unit.

{\em Memory}: The part of the computer devoted to the storage of the
data and instructions to be processed.  It is divided into individual
cells which are accesible by means of a number called {\em address}.

The functioning of a VNM is cyclic.  One of these cycles contains the
following operations: the control unit reads one program instruction
from the memory, which is executed after being decoded.  Depending on
the type of instruction, a piece of data can either be read from or
written in the memory, or an instruction be executed.  In the next
cycle to be performed, the control unit reads another program
instruction which is precisely next in the memory to the one processed
in the previous cycle.

It is the simplicity of this sequentially operating model which makes
it rather advantageous for many purposes because it facilitates the
design of machines and programs.

\subsection{Classical Parallelism}
\label{sec8C:level2}

There are complex problems which demand a very large number of
operations to be performed as well as a large amount of computer
resources.  These problems include image processing such as satellite
images, meteorological predictions, scientific calculations arising in
strongly correlated many-body systems, computation of the hadronic
spectrum in QCD (Quantum Chromodynamics) on the lattice,  real-time
calculations in plasma physics, turbulence in fluids, and many more.
It was noticed soon that an ordinary computer based on the VNM
architecture would have a very long way to cope with such a type of
problems where a massive number of operations is needed to be done in
a very short period of time.

A classical parallel computer is the natural way to address these
problems.  The idea of parallelism is also simply summarized as {\em
many things at a time.} We shall see that a quantum computer would
realize this goal at the highest possible degree of parallelism.

Although the idea of parallelism is very simple to state, its
practical implementation has faced many obstacles for several reasons
we shall briefly describe.  This will be quite illustrative later when
we refer to the principles of quantum computation.

The way to extend the sequential VNM into a parallel computer is not
unique.  The components entering a parallel machine (PM) are already
present in the VNM, but its number and organization differs.  One way
to understand the various possibilities is by recalling the
organization of a program in any computer.  A program is divided into
{\em instructions} and {\em data.} These are its building blocks.
This distinction means that we may have several degrees of parallelism
depending on how many instructions and/or data the PM handles at a
time.  This leads to a first classification of PM's known as {\em
Flynn's classification} (1966; 1972) which describes in four
categories how a computer functions without entering the details of
its architecture:

i) {\em SISD}: Single Instruction stream, Single Data stream.
Executes one instruction at a time (single instruction stream) and
fetches/stores one data value at a time (single data stream).  It has
only one CPU. Example: the von Neumann machine (specifically,
processors like Motorola, Intel and AMD, etc.).

ii) {\em MISD}: Multiple Instruction stream, Single Data stream.  This
corresponds to multiple programs operating on the same data
(performing different computations) Example: none is available.  This
category does not seem to be useful.

iii) {\em SIMD}: Single Instruction stream, Multiple Data stream.
Executes one instruction at a time (single instruction stream) and the
same operation is performed on many data values at the same time
(multiple data stream).  Example: The {\em vector machines} like
Thinking Machine's Connection Machine CM-2.  A vector operation with
$n$ elements can be executed by one instruction cycle on a SIMD
parallel machine.

iv) {\em MIMD}: Multiple Instructions stream, Multiple Data stream.
These are multiprocessor systems, each processor executing a different
program on its own data.  Thus, there are multiple instruction streams
(programs) and multiple data streams.  Example: most distributed
memory parallel processors, like Thinking Machine's Connection Machine
CM-5, Cray T3D, IBM SP-2, workstation clusters, fit in this category.

\begin{figure}[ht]
\psfrag{P}[Bc][Bc][0.5][0]{P} \psfrag{M}[Bc][Bc][0.5][0]{M}
\psfrag{1}[Bc][Bc][0.75][0]{Single}
\psfrag{2}[Bc][Bc][0.75][0]{Multiple}
\psfrag{D}[Bc][Bc][0.75][0]{{\color{myred} Data Streams}}
\psfrag{I}[Bc][Bc][0.75][0]{{\color{myred} Instruction}}
\psfrag{T}[Bc][Bc][0.75][0]{{\color{myred} Streams}}
\includegraphics[width=6 cm]{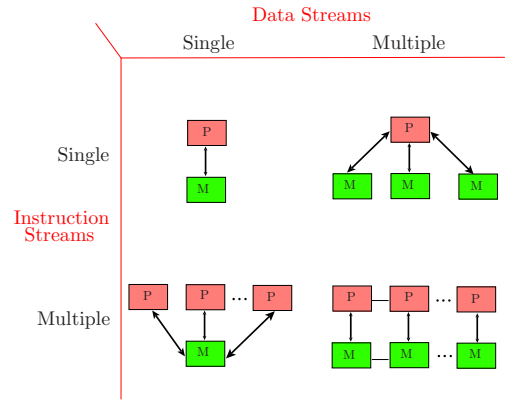}
\caption{Flynn's classification of parallel machines (P = processor, M
= memory).}
\label{flynn}
\end{figure}

Of these machines, those of type SIMD and MIMD are parallel machines,
the latter having a higher degree of parallelism.  In Fig.~\ref{flynn}
we show a schematic representation of Flynn's classification.  Only
processors and memory units are represented, without going into finer
details about the interconnection network, types of memories (shared,
distributed, cached, \ldots), pipelines,  etc.\footnote{Flynn's
classification is too coarse for classifying multiprocessor systems,
and there exist modifications to it (Hwang and Briggs, 1985) and new
ones as well like {\em H\"{a}ndler's classification} (1982) and
others.}

One may think at first glance that what counts in a PM is simply the
number of processors.  However, what really matters is the way the
many processors are organized and how the information is exchanged
among them.  The reason is because for two processors to
intercommunicate, it is necessary that they be synchronized and
consequently, they have to wait each other.  Thus, this slows the
functioning of a PM if only the number of processors is increased
without taking care of their organization.

Therefore, we arrive at the conclusion that to scale up a PM one has
to multiply the number of processors and to find out as well
interconnecting structures for them.  These structures or networks
need be regular, efficient and low cost.  The determination of the
best interconnecting network for the processors in a PM is specially
crucial when their number increases considerably.

For an interconnecting network (or lattice) to be good it has to
minimize at the same time the total number of physical connections (or
links) and the average distance between processors.  This average
distance is measured in terms of the number of connections to be
traversed.  Furthermore, the network has to be regular enough to allow
being scalable when more processors are added.

In order to understand these requirements let us enumerate and analyze
some archetypical networks.

\begin{figure}[ht]
\psfrag{a}{a)} \psfrag{b}{b)} 
\includegraphics[width=5cm]{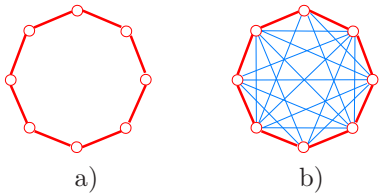}
\caption{Ring vs. fully connected processor lattices.}
\label{ring}
\end{figure}

{\em Fully connected lattice}: This is one extreme case which is made
up of, say, $N$ processors in such a way that all of them are
connected one another, as shown in Fig.~\ref{ring}.  The number of
connections is $\half N(N-1)$, and thus it is of order $O(N^2)$.  This
fact makes it non-practical because there are other more economical
alternatives for connections.

{\em Ring lattice}: The network of processors forms a ring (see
Fig.~\ref{ring}), which has the advantage of needing only two
connections per processors, no matter their number.  It this sense it
is opposite of the full lattice.  However, it has a very important
disadvantage, because in the worst case a message has to traverse
$N/2$ processors (half of the lattice) to reach its destiny.  This is
also non-practical when $N$ is large.

\begin{figure}[ht]
\psfrag{r}[Bc][Bc][0.8][0]{root}  
\psfrag{i}[Bc][Bc][0.8][0]{interior}
\psfrag{l}[Bc][Bc][0.8][0]{leaves} 
\includegraphics[width=3 cm]{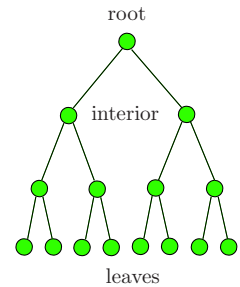}
\caption{Binary tree processor lattice.}
\label{bintree}
\end{figure}

{\em Binary Tree}: The processors are organized such that each node is
connected to three nodes, namely, one {\em parent} and two {\em
children} (Fig.~\ref{bintree}).  The problem with this type of lattice
is that the inner nodes deep inside the tree are very badly
communicated among themselves.

\begin{figure}[ht]
\psfrag{1}{1D} \psfrag{2}{2D} \psfrag{3}{3D} \psfrag{4}{4D}
\psfrag{a}{\tiny{0}} \psfrag{b}{\tiny{1}} \psfrag{x}{\tiny{00}}
\psfrag{y}{\tiny{01}} \psfrag{z}{\tiny{10}} \psfrag{t}{\tiny{11}}
\includegraphics[width=6 cm]{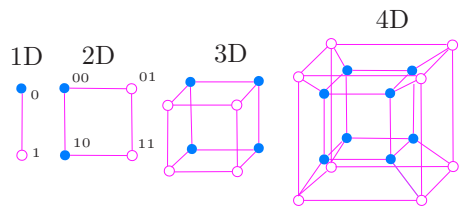}
\caption{Hypercube networks.}
\label{hypercube}
\end{figure}

{\em Hypercube}: This is the solution that has turned to be optimal in
meeting the desired requirements (Fig.~\ref{hypercube}).  In the
simplest possibility, one processor is installed at each vertex of the
cube, which can be of any dimension $D$.  In the familiar case of a
$D=3$ cube, each processor is connected to other $3$ and more
importantly, each one is at a maximum distance of $3$ connections from
any other.  For a $D$-dimensional hypercube the number of processors
is $2^D$, each one is connected to $D$ neighbor processors and is at
most a distance $D$ apart from any other.  The most famous PM based on
this hypercube architecture is the original Connection Machine and the
Crays.  It is not surprising that Feynman, who played a paramount role
in the beginning of quantum computers, worked in the design of this PM
and made some notorious contributions (Hillis, 1998).

The interconnecting networks of processors considered so far are
called {\em static} because the structure is fixed by construction.
There exists also the possibility of {\em dynamic} networks where its
configuration is changeable.  In this case the processors are
connected not directly but through commuters which can be switched in
different ways.

One of the fundamental problems posed by the parallel computers is its
control.  There are also several strategies to address this issue.
One possibility is to have a central processor working as a control
unit for the rest of processors, as in the SIMD. This is a model of
centralized control in which the control unit sends instructions to
the other processors which never interfere the central processor.  In
order to simplify their working, it is normal that the same
instruction is sent to all the processors which in turn operate on
different sets of data.  This mode of control has the same
disadvantages as the original VNM: it is slow.  The reason is because
the control unit has to send many electrical pulses to perform the
control task.

An alternative to centralized control consists in allowing each
processor to take its own decisions, usually consulting only its
nearest-neighbor processors.  This solution has also difficulties
because the programs must be written in a way very different from the
standard.  Moreover, such non-centralized control can become very
inefficient because the processors might spend most of their time
exchanging messages rather than making computations.

The problem of organizing and controlling the parallelism in a
classical computer resembles very much the organization problems in
the human societies, which is as open a problem there as for networks
of computers.  We shall see in Sec.~\ref{sec9:level1} that in a
quantum computer one also faces similar synchronization problems and
we shall discuss how they are solved in terms of physical principles.

\subsection{Classical Logic Gates and Circuits}
\label{sec8D:level2}

A Turing machine is by no means a practical computer, despite of being
a powerful theoretical machine.  In practice, computers are made of
electronic circuits, which in turn contain {\em logic gates}.  A logic
gate is a device that implements a classical {\em logic operator} like
the AND operator.  A logic operator or function $f$ is an application
$f: \{0,1\}^n \longmapsto \{0,1\}^m$, which maps an input of $n$
bit-valued operands into a $m$-bit-valued output.  When the target
space of $f$ is $\{0,1\}$, one usually says that $f$ is a {\em
Boolean} operator or function.  A {\em Boolean algebra} is a unital
algebra defined over the field $Z_2=\{0,1\}$.  Boolean algebras are
useful to elucidate situations which can be true or false, making
appropriate reasonings to draw conclusions correctly.  They are
therefore helpful in building practical computers and in programming.
Furthermore, it is possible to show that classical Turing machines are
equivalent to classical logic circuits.  This means that they both
have the same complexity classes.  This is a mathematical result that
legitimates the use of electronic circuits in the construction of real
computers.

Before stating this important result as a theorem, let us take a
closer look at some rudiments of Boolean logic that will also help in
understanding the peculiarities of quantum logic gates (see
Sec.~\ref{sec9:level1}).

An operator with one operand is called a {\em unary} operator, with
two operands is a {\em binary} operator.  There are three basic
Boolean or logic operators: 1/ The unary operator NOT: $x\mapsto{\rm
NOT}\ x:=\bar x:=1-x$, denoted also by overlining the argument
($\bar{\;\;}$).  2/ The binary operator AND: $(x,y)\mapsto x\ {\rm
AND}\ y:=x\wedge y:=xy$, also denoted by $\wedge$.  3/ And the binary
operator OR, $(x,y)\mapsto x\ {\rm OR}\ y:=x\vee y:=x+y-xy$, denoted
also by $\vee$.  As usual, Boolean arithmetics is done in the field
$\Z_2$: $1+1=0$.

\begin{table}
\begin{ruledtabular}
\begin{tabular}{c c | c c c c}
$x$ & $\bar{x}$ & $x$ & $y$ & $x\wedge y$ & $x\vee y$ \\ \colrule 0 &
1 & 0 & 0 & 0 & 0 \\ 1 & 0 & 0 & 1 & 0 & 1 \\ &   & 1 & 0 & 0 & 1 \\ &
& 1 & 1 & 1 & 1 \\
\end{tabular}
\end{ruledtabular}
\caption{Truth tables for the basic logic operators: NOT ($\bar{\;}$),
AND ($\wedge$), OR ($\vee $).}
\label{truth}
\end{table}

The action of a logic operator is represented by a {\em truth table}.
A truth table contains as many columns as input operands and ouput
bits, and $2^{\# {\rm operands}}$ rows.  The inputs are shown on the
left, and the output is shown on the right.  The truth tables for the
basic operators are shown in Table~\ref{truth}.  An important Boolean
expression involving 2 variables $x,y$ is $r=(\bar{x}\wedge y) \vee
(x\wedge\bar{y})$, i.e. $r(x,y)=x+y$.\footnote{This $r$ corresponds to
the XOR operation.} Expressions in the Boolean algebra can be
represented by {\em logic circuits}.  A logic circuit is a directed
acyclic graph with incoming lines carrying input Boolean variables
$x_1,x_2,\ldots,x_n$ and an outgoing line carrying the output variable
$y$ of the circuit.  Every node in the graph is a {\em logic gate}
which represents a logic operator of the Boolean algebra.  In real
computers, circuits consist of electronic devices such as switches and
wires.

\begin{figure}[h]
\psfrag{x}[Bc][Bc][0.75][0]{$x$} \psfrag{y}[Bc][Bc][0.75][0]{$y$}
\psfrag{a}[Bc][Bc][0.75][0]{AND} \psfrag{1}[Bc][Bc][0.75][0]{$x\wedge
y$} \psfrag{b}[Bc][Bc][0.75][0]{OR} \psfrag{4}[Bc][Bc][0.75][0]{$x\vee
y$} \psfrag{c}[Bc][Bc][0.75][0]{NOT}
\psfrag{2}[Bc][Bc][0.75][0]{$\bar{x}$}
\psfrag{d}[Bc][Bc][0.75][0]{NAND}
\psfrag{5}[Bc][Bc][0.75][0]{$\overline{x \wedge y}$}
\psfrag{e}[Bc][Bc][0.75][0]{NOR}
\psfrag{3}[Bc][Bc][0.75][0]{$\overline{x\vee y}$}
\psfrag{f}[Bc][Bc][0.75][0]{XOR} 
\psfrag{6}[Bc][Bc][0.75][0]{$x\oplus y$} 
\includegraphics[width=6 cm]{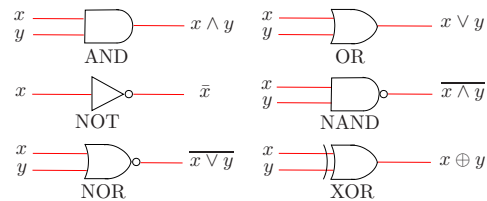}
\caption{Basic classical logic gates.}
\label{cgates}
\end{figure}

To each logic operator we can associate a logic gate with a specific
form.  That logic gate has a number of incoming lines, one per input
operand, and one outgoing line for the output result.  In
Fig.~\ref{cgates} we show the convention for the basic logic gates.
In the same way as the basic operators of the algebra make up more
complicated expressions, the basic gates are combined to construct
complex circuits.

\begin{figure}[t]
\psfrag{S}[Bc][Bc][0.75][0]{SUM} \psfrag{C}[Bc][Bc][0.75][0]{CARRY}
\psfrag{X}[Bc][Bc][0.75][0]{$y$} \psfrag{Y}[Bc][Bc][0.75][0]{$x$}
\psfrag{p}[Bc][Bc][0.75][0]{$xy$} \psfrag{a}[Bc][Bc][0.75][0]{$x+y$}
\psfrag{n}[Bc][Bc][0.75][0]{$1-y$} \psfrag{m}[Bc][Bc][0.75][0]{$1-x$}
\psfrag{o}[Bc][Bc][0.75][0]{$x+y-xy$}
\psfrag{r}[Bc][Bc][0.75][0]{$1-xy$} 
\includegraphics[height= 5 cm,width= 5 cm]{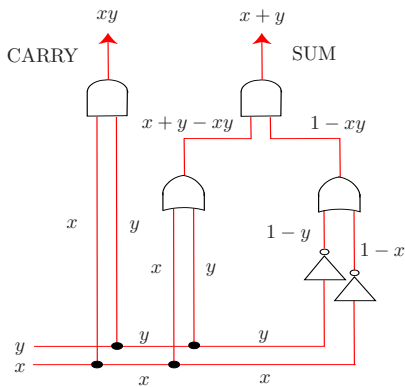}
\caption{A classical logic circuit: adder for two bits $x,y$.  The
bifurcating wires at the nodes are achieved with FANOUT gates.}
\label{ccircuit}
\end{figure}

Additional gates that duplicate the input values on wires are
frequently needed.  These are called FANOUT or COPY gates and they are
schematically represented by $-\!\bullet \kern -4.5pt <$ (see
Fig.~\ref{ccircuit}).  In classical computation, these are sort of
obvious gates for they simply correspond to splitting the wire into
two or more leads, which is an easy operation.  This is why they are
usually taken for granted throughout classical computing.
Nevertheless, these irreversible FANOUT gates are logically necessary
when discussing the important issue of universality of classical
gates.  On the contrary, these duplicating gates find no room in the
insides of a quantum circuit due to the linearity of quantum mechanics
(no-cloning theorem, Sec.~\ref{sec3:level1}).

A Boolean circuit computes a Boolean function in a natural way by
following its directed path (usually from left to right) upon
application of its constituent gates.  The {\em size} of a circuit $C$
is its number of gates, and the {\em depth} of $C$ is the length of
the longest directed path in it.  A typical circuit is depicted in
Fig.~\ref{ccircuit}.

Suppose that we are given a tractable decision problem, i.e.  a
problem in class {\bf P} (see Appendix).   This means that there
exists a Turing machine $M$ deciding it ($M(x_n)=0,1$) for initial
data $x_n$ of arbitrary length $n$, in polynomial time.  This problem
is said to have {\em polynomial circuits} when there is a family
$\{C_1,\ldots,C_n,\ldots\}$ of logic circuits, of polynomial size in
the input length $n$, such that $M(x_n)=0,1$ iff $C_n(x_n)=0,1$.

It can be shown that all problems in class {\bf P} have polynomial
circuits.  The converse, however, is not true: there exist undecidable
decision problems that have polynomial circuits (Papadimitriou, 1994).
This shortcoming is remedied by restricting the circuit family to be a
{\em uniform} circuit family: for each $n$, the description of each
$C_n$ is an output of an auxiliary Turing machine in polynomial time
when entered with an appropriate input of length
$n$.\footnote{Actually the auxiliary TM should be (log $n$)-space
bounded, what implies polynomial time boundedness.}

The equivalence between classical Turing machines and Boolean logic
circuits is stated in the following theorem (Savage, 1972; Schnorr,
1976; Pippenger and Fisher, 1979; Papadimitriou, 1994):

{\em  Turing machines and uniform circuit families}: A decision
problem is in class {\bf P}, i.e. it can be solved for inputs of
length $n$ by a Turing machine in polynomial time $p(n)$, iff it has a
uniform family of polynomial circuits.  Moreover, the minimum size of
$C_n$ is $O(p(n)\log p(n))$.

This theorem legitimates the simulation of Turing machines by logic
circuits.  Dealing with gates and circuits is simpler and more
practical than with Turing machines.  Actually, gates are packaged
into hardware chips.

So far we have introduced a set of three basic logic operators (NOT,
AND, OR).  It proves also convenient to introduce three additional new
gates: NAND, NOR and XOR. The gates NAND and NOR are the negation of
AND and OR, respectively.  The gate XOR is called {\em exclusive } OR,
and is also denoted by $\oplus$.  Their truth tables are shown in
Table~\ref{truth2}.

\begin{table}[ht]
\begin{ruledtabular}
\begin{tabular}{c c c c c}
$x$ & $y$ & $x \ {\rm NAND} \ y$ & $x \ {\rm NOR} \ y$ & $x \ {\rm
XOR} \ y$\\ \colrule 0 & 0 & 1 & 1 & 0 \\ 0 & 1 & 1 & 0 & 1\\ 1 & 0 &
1 & 0 & 1\\ 1 & 1 & 0 & 0 & 0\\
\end{tabular}
\end{ruledtabular}
\caption{Truth tables for the logic operators NAND, NOR, XOR.}
\label{truth2}
\end{table}

With the basic set \{NOT, AND, OR\} one can built any logic function
over the Boolean algebra, provided that FANOUT gates and ancilla or
work bits are freely used.  Because of this property, \{NOT, AND, OR\}
is called a {\em universal} set of logic gates.  However, this set is
not minimal.  To see this we use the so called de Morgan's laws, which
are the following Boolean identities:
\begin{equation}
\begin{split}
\overline{(x \vee y)} & = \bar{x} \wedge \bar{y}, \\ \overline{(x
 \wedge y)} & = \bar{x} \vee \bar{y}.
\label{qc11}
\end{split}
\end{equation}

\noindent These two algebraic equations are dual each other.  Negation
of the first produces $x \vee y = \overline{(\bar{x} \wedge
\bar{y})}$.  This is telling us is that OR gates are not essential:
the AND and NOT gates can by themselves reproduce the functionality of
the OR gate.  Similarly, the second relation in (\ref{qc11}) leads to
$(x \wedge y) = \overline{(\bar{x} \vee \bar{y})}$, that is, AND gates
can be implemented with OR and NOT gates.  Then the set $\{{\rm AND},
{\rm NOT}\}$ is universal, and so is the set $\{{\rm OR}, {\rm NOT}\}$.

Can we reduce further the number of elements in a universal set?  The
answer is yes.  The surprising result is that NAND gates alone (or,
similarly, NOR gates alone) are sufficient for constructing any
circuit (up to FANOUT and work bits).  We know this from the following
simple laws:
\begin{equation}
\begin{split}
 \bar{x} & = 1 {\rm \ NAND} \ x, \\ x \ \wedge \ y & = \overline{(x \
 {\rm NAND} \ y)} = 1 \ {\rm NAND} \ (x \ {\rm NAND} \
 y). \label{qc11b}
\end{split}
\end{equation}

\noindent Therefore we see that $\{ {\rm NAND} \}$ (or $\{ {\rm NOR}
\}$) can do everything that the set $\{ {\rm AND, NOT} \}$ does, and
hence $\{ {\rm NAND}\}$, $\{{\rm NOR}\}$ are also universal sets.

\section{Principles of Quantum Computation}
\label{sec9:level1}

\begin{figure}[ht]
\includegraphics[width=8 cm]{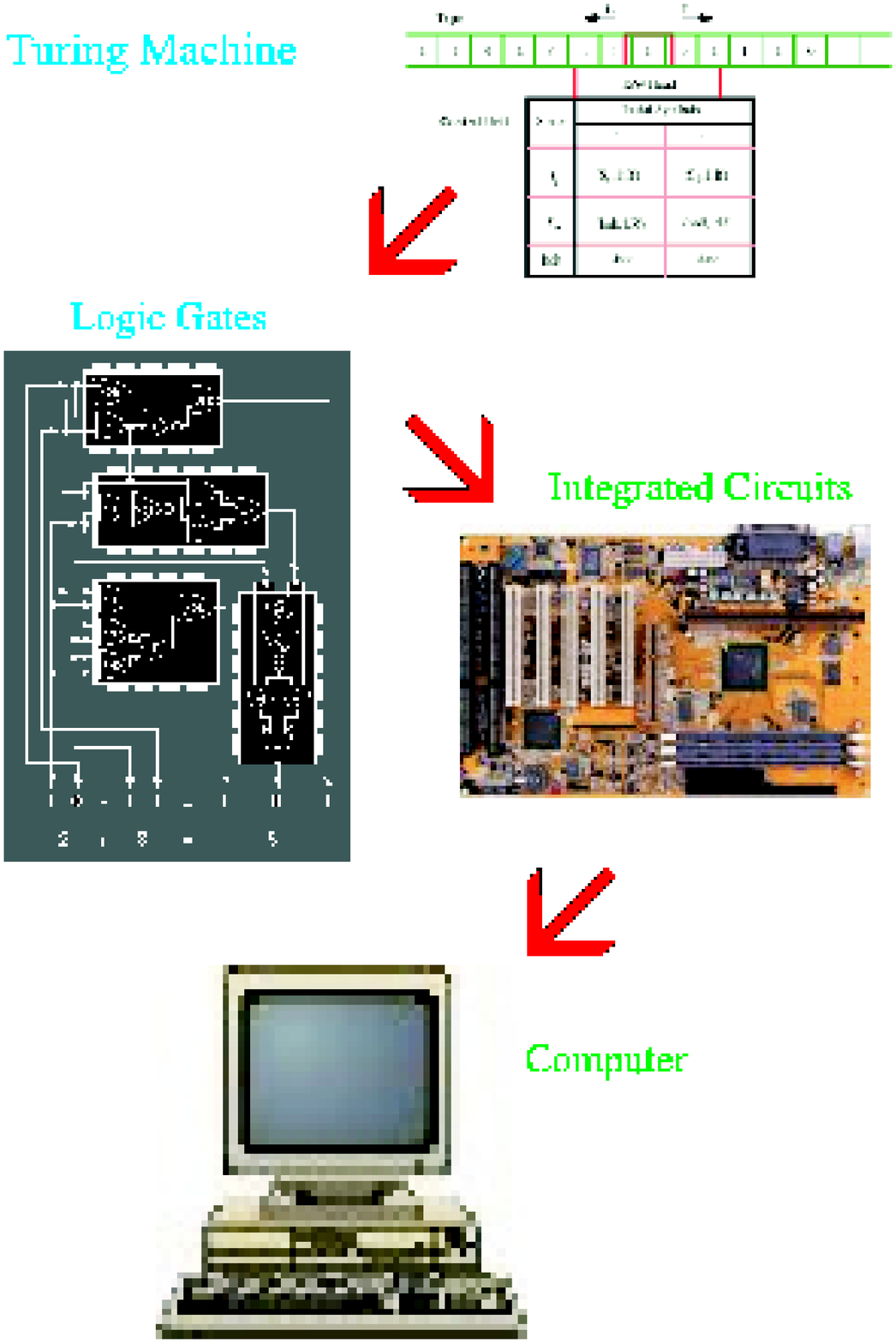}
\caption{From a Turing machine to a real computer.}
\label{history}
\end{figure}

In the previous section we have described some basic aspects of Turing
machines and their practical implementations by means of the Von
Neumann architecture.  Yet, there is a long way from there towards the
construction of a real computer as those we have on our desks.  In
Fig.~\ref{history} we provide a visualization of the route we have to
follow.  This long route starts with the abstract notion of a
classical computer embodied in a Turing machine.  No real computer has
a Turing machine inside.  Instead, the operations carried out by a
Turing machine can be substituted by {\em logic gates}.  These logic
gates can do sums, multiplications, logic operations, etc.  With just
a few logic gates we can do almost nothing of the daily tasks we are
used to nowadays.  To get the power and speed of an ordinary computer
we need millions of logic gates interconnected and integrated into
tiny circuits.  These are called {\em integrated circuits} or {\em
chips}.  Finally, these integrated circuits are arranged into the
computer motherboard with other components, and along with a screen,
keyboard, mouse, etc.  we have a universal machine capable of doing
many tasks, like writing this article.

All these four stages in Fig.~\ref{history} have been accomplished in
the case of the classical computers.  What is the current state of the
art in the case of quantum computers?  The first step in
Fig.~\ref{history} has also been achieved for quantum computers.  This
is the topic of Subsec.~\ref{sec9A:level2} where we discuss the notion
of quantum Turing machines, the quantum version of the classical
Turing machines introduced thus far.  Moreover, the second step
regarding the design of quantum logic gates has also been accomplished
as we shall explain in Subsec.~\ref{sec9B:level2}.  These quantum
gates are used as the basic components of a quantum computer to design
quantum algorithms that surpass certain very important classical
algorithms (see Sec.~\ref{sec10:level1}).  More important is the fact
that, in the recent years, an experimental realization of these
quantum gates have been made (see Sec.~\ref{sec11:level1}), which let
us cherish the possibility of building a real operative quantum
computer on equal footing as the current classical precursors.
However, to achieve this goal we need to move more steps farther like
finding the quantum equivalent of an integrated circuit (third step).
This step amounts to the problem of scalability in a quantum computer:
so far, the experimental realization mentioned previously are made of
a just a few gates and although a quantum gate is more powerful than a
classical one, we also need a large number of them to make non-trivial
tasks.  We need to scale up our current quantum technology.  Finally,
the last fourth step will be to have a real operative quantum computer
in our hands, with all the external devices to communicate with it.
Although there is still a long way ahead to achieve this goal, the
fact that the fundamental first and second steps have been already
done is very encouraging.  In the following we shall describe these
two steps for quantum computers.

From a fundamental point of view, a quantum computer (QC) is a quantum
Turing machine (QTM) and this is a concept that we shall next define.

\begin{figure}[ht]
\includegraphics[width=7 cm]{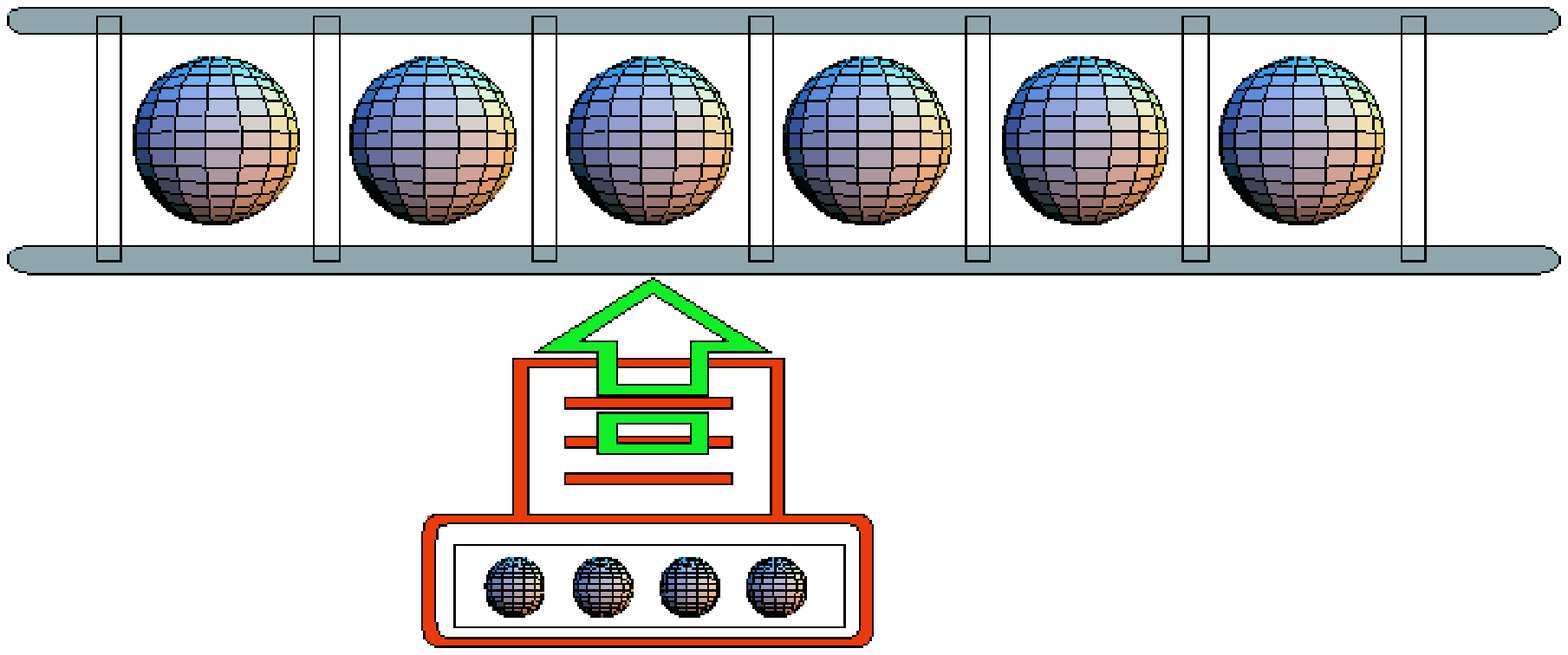}
\caption{Pictorical view of a quantum Turing machine: there are qubits
(Bloch's spheres, Fig.~\ref{sphere2}) in the tape and in the control
unit.}
\label{qturing}
\end{figure}

\subsection{The Quantum Turing Machine}
\label{sec9A:level2}

There have been several achievements before arriving at the concept of
a QTM and it is not our purpose to give a full account of all of them,
but instead we shall mention some of the most representative
constructions or {\em machines}.  We start mentioning the {\em
Benioff's machine}, which is a model for computation introduced by
P. Benioff (1980; 1981; 1982).  Benioff's goal was to use quantum
mechanical systems to construct reversible Turing Machines.  His
motivation was that the unitary evolution of an isolated quantum
system provides a way to implement reversible computations.  The issue
of reversibility had attracted much attention since Bennett (1973)
constructed a classical model of reversible computing machine
equivalent to a Turing machine.  Landauer (1961) had shown that
reversible operations dissipate no energy, while a Turing machine as
described in Sec.~\ref{sec8:level1} performs irreversible changes
during computations.  Although the Benioff's machine is a quantum
machine, it is not however a quantum computer for it is equivalent to
a reversible TM. Feynmann (1982) went one step further towards the
notion of quantum computer with his ``universal quantum simulator'' or
{\em Feynman's machine}.  He proposed to use quantum systems to
simulate quantum mechanics more efficiently than classical computers
do.\footnote{Manin (1980) had already envisaged that the complexity of
quantum systems surpassed the capabilities of classical computers.}
He showed (Feynmann, 1985) that classical TMs exponentially slow down
when simulating quantum phenomena while a universal quantum simulator
would do efficiently the job.  However, Feynman's machine is not fully
a quantum computer in the sense described below for it does not let
program an arbitrary task.

Deutsch (1985) gave the final step in the quest of a sensible
definition of a quantum computer.  His starting point is a critique of
the Church-Turing hypothesis (see Sec.~\ref{sec8A:level2}) which he
considers very vague as compared to physical principles such as the
gravitational equivalence principle.  Deutsch's proposes to make more
concrete the statement ``functions which would naturally be regarded
as computable'' in Church-Turing hypothesis.  He identifies such
functions as those which can be computed by a real physical system.
This is quite apparent, since it is hard to believe that something be
naturally computable if it cannot be computed in Nature.  Thus,
Deutsch goes on to promote the Church-Turing hypothesis into a
physical principle which he states as the {\em Church-Turing
Principle}: Every finitely realizable physical system can be perfectly
simulated by a universal model computing machine operating by finite
means.

The content of this principle is more physical than the corresponding
hypothesis since it appeals to objective concepts such as measurement,
physical system, etc. instead of the subjective notion of ``naturally
computable''.  The ``finite means machine'' in the Church-Turing
principle is more general and replaces the role of the Turing machines
in the corresponding hypothesis  (Sec.~\ref{sec8A:level2}).

Deutsch follows a natural way to introduce the definition of a {\em
Quantum Turing Machine} (QTM): starting from the knowledge we have of
its classical counterpart (see Sec.~\ref{sec8A:level2}) he replaces
some of the classical components of an ordinary TM, like bits,  by
quantum elements, like qubits.

A {\em Quantum Turing Machine} is a Finite State Machine which has
three components: a {\em finite processor}, an infinite {\em memory
unit} (of which only a finite portion is ever used) and a cursor.  The
description of these components is as follows:

i) {\em Finite Processor}: This is the control unit as in a TM but it
consists of a finite number $P$ of qubits.  Let us denote the Hilbert
space of these processor states as
\begin{equation}
{\cal H}_{\rm P}:={\rm span}\{\otimes_i| p_i\rangle:  p_i= 0,
1\}_{i=0}^{P-1}.
\label{qc2}
\end{equation}

ii) {\em Memory Tape}: This has a similar functionality as in a TM but
it consists of an infinite number of qubits.\footnote{Even if ideally
there is a qubit per cell, only a finite number of them are active
during each running of the QTM.} Let us denote the Hilbert space of
these memory states as
\begin{equation}
{\cal H}_{\rm M}:={\rm span}\{\otimes_i| m_i\rangle:  m_i= 0,
1\}_{i=-\infty}^{+\infty}.
\label{qc3}
\end{equation}

iii) {\em Cursor}: This is the interacting component between the
control unit and the memory tape.  Its position is scanned by a
variable $x\in {\cal H}_{\rm C}=\Z$, and the associated Hilbert space
is
\begin{equation}
{\cal H}_{\rm C}:={\rm span}\{|x\rangle: x\in\Z\}.
\label{qc3b}
\end{equation}

Therefore, there is a Hilbert space of states associated to a QTM
which altogether takes the form
\begin{equation}
{\cal H}_{\rm QC} := {\cal H}_{\rm C} \otimes {\cal H}_{\rm P} \otimes
{\cal H}_{\rm M}.  \label{qc4}
\end{equation}

\noindent
The basis vectors in the Hilbert space ${\cal H}_{\rm QC}$ of the QTM
are of the form
\begin{equation}
|x; {\bf p}; {\bf m}\rangle := |x; p_0, p_1,\ldots, p_{\rm P}; \ldots,
m_{-1}, m_{0},  m_{1},\ldots \rangle,
\label{qc5}
\end{equation}

\noindent and are called the {\em computational basis states}.

We may wonder about the relationship between the defining features of
a classical TM (see Sec.~\ref{sec8A:level2}) and those of a QTM. The
set of states ${\cal S}$ corresponds to the Hilbert space of states
${\cal H}_{\rm P}$ in a QTM.  The alphabet ${\cal A}$ is just the
qubit space $\C^2$.  As for the set of instructions ${\cal I}$ of a
TM, we need to specify the way a QTM works.

A QTM operates in steps of fixed duration $T$, and during each step
only the processor and a finite part of the memory unit interact via
the cursor.  We stress that a QTM, much like a TM, is a mathematical
construction; we shall present explicit experimental realizations in
Sec.~\ref{sec11:level1}.

The set of instructions ${\cal I}$ of a TM is replaced by the unitary
time evolution of the quantum states $|\Psi \rangle \in {\cal H}_{\rm
QC}$.  After a number $n\in \N$ of computational steps, the state of
the QTM will be transformed into
\begin{equation}
|\Psi (nT)\rangle = U^n |\Psi (0)\rangle, \label{qc6}
\end{equation}

\noindent with $U$ a unitary evolution operator, $UU^\dagger =
U^\dagger U = 1$.  A valid quantum program takes a finite number of
steps $n$.  To each QTM there is associated a unitary evolution
operator $U$ to make a certain job or program, much like a TM has a
unique set of instructions ${\cal I}$, and each TM makes a certain
task.  To specify the initial state $|\Psi(0)\rangle$, we set to zero
both the cursor position $x=0$ and the prepared processor states ${\bf
p}={\bf 0}$.  The memory states ${\bf  m}$ are prepared allocating the
{\em input data} and other program instructions, conveniently encoded
into a finite number of qubit strings, with the rest of the memory
qubits set to $\ket{0}$.   The initial state is then
\begin{equation}
|\Psi (0)\rangle = \sum_{{\bf  m}} c_{{\bf  m}}|0;{\bf 0 };{\bf
m}\rangle, \;\text{with} \; \sum_{{\bf  m}} |c_{{\bf  m}}|^2 = 1.
\label{qc7}
\end{equation}

The notion of a QTM operating ``by finite means" entering the
Church-Turing principle means that the machine cannot do infinitely
many operations at a given time nor at arbitrary positions along the
memory tape.  This notion suggests the following constraint on the
matrix elements of the evolution operator of a QTM:
\begin{equation}
\begin{split}
&\langle x^\prime;{\bf  p}^\prime;{\bf  m}^\prime|U| x;{\bf p};{\bf
 m} \rangle = [\delta_{x^\prime,x+1} U^{+}({\bf
 p}^\prime,m^\prime_{x^\prime}|{\bf  p},m_x) \\ & +
 \delta_{x^\prime,x-1} U^{-}({\bf  p}^\prime,m^\prime_{x^\prime}|{\bf
 p},m_x)] \prod_{x^\prime \neq x\pm 1}\delta_{m_{x^\prime},m_x}.
\label{qc8}
\end{split}
\end{equation}

\noindent In these matrix elements, the infinite product guarantees
that only a finite number of memory qubits participate  in a single
computational step. Once the qubit at the $x$th cursor position is
singled out, the two deltas appearing in the brackets guarantee that
the cursor position cannot change by more than one unit, either
backward, forward or both.  This operating mode amounts to locality in
the tape space.  We call the parts $U^{\pm}({\bf
p}^\prime,m^\prime_{x\pm 1}|{\bf  p},m_x)$ of $U$ forward and backward
matrices at $x$, respectively.  They represent the operators $P_{x\pm
1}U P_x$ in the computational basis, where $P_x$ is the  projection
onto the Hilbert subspace of ${\cal H}_{\rm QC}$ consisting of the
states with the cursor at the $x$th position.  Unitarity of $U$ is
equivalent to $U^{\pm\dagger}U^\mp=0$, $U^{+\dagger}U^+ +
U^{-\dagger}U^-=1$.  Each unitary operator $U\{U^-,U^+\}$ defines a
QTM.

As with any other computer, we need a mechanism to cause the QTM to
halt when the computation ends.  In a quantum machine there is a
severe constraint to do this because the principles of quantum
mechanics do not allow us to observe or measure the QTM until it
terminates.  To know when this happens, we may set aside one of the
qubits of the processor to signal the end.  Let us choose the first
qubit $| q_0\rangle$ to acquire the value $1$ when the computation is
over while it is $0$ during the operations.  The program does not
interact with $| q_0\rangle$ until when it has reached the end.  Thus,
the state $| q_0\rangle$ can be monitored periodically from the
outside without affecting the operation of a QTM.

So far we have set up several connections between the components of
quantum and classical Turing machines.  Moreover, to complete this
comparison, we can also think about the relationships concerning their
functioning.  Does a quantum TM extend somehow the notion of a
classical TM?  Yes, and this relation turns out to be very physical
and it will sound familiar to us.  Firstly, not all classical TMs are
closely related to a quantum TM, only those reversible classical TM
will be, as follows from the discussion above.  Then, it is possible
for a quantum TM to reproduce the functioning of a reversible
classical TM (Deutsch, 1985) if we choose its unitary evolution
operator to have the following form:
\begin{equation}
\begin{split}
& U^{\pm}({\bf  p}^\prime,m^\prime_{x\pm 1}|{\bf  p},m_x) = \\ &
\delta_{{\bf  p}^\prime,{\bf A}({\bf  p},m_x)} \delta_{m^\prime_{x\pm
1},B({\bf  p},m_x)} \half \left[ 1 \pm C({\bf  p},m_x) \right]
\label{qc9}
\end{split}
\end{equation}

\noindent where ${\bf A}, B, C$ are maps of $\Z_2^P\times
\prod_{-\infty}^{+\infty}\Z_2$ into $\Z_2^P,\Z_2$ and $\{-1,1\}$,
respectively.

\noindent This form of dynamics guarantees that this particular QTM
will remain in a computational basis state (\ref{qc5}) at the end of
each time step.  This is precisely the way a classical TM operates.
The requirement of reversibility is guaranteed by demanding that the
mapping $({\bf  p},m) \mapsto ({\bf A}({\bf  p },m),B({\bf
p},m),C({\bf  p},m))$ be bijective.

Therefore, there is a particular limiting case in which a quantum TM
becomes a reversible classical TM. This fact is somewhat reminiscent
of the familiar {\em correspondence principle} of quantum mechanics to
recover classical mechanics.  This principle played a fundamental role
in the development of the old quantum theory and the beginnings of the
modern quantum mechanics.  Here we are following a similar path by
starting with a revision of the classical fundamentals of information
and computation to thereby develop their quantum versions.

\subsubsection{Quantum Parallelism}
\label{sec9B:level3}

The capability of a quantum TM of being in several computational basis
states at the same time is called {\em quantum parallelism}, and is
one of the defining features of a QTM. The classical counterpart of
this is the notion of classical parallelism introduced in
Sec.~\ref{sec8C:level2}. The quantum version of doing ``many things
at a time'' in a classical parallel computer is the possibility of
being in {\em many states at a time} in a quantum computer.
Furthermore, in a classical computer it is not enough to have a large
number of processors connected in parallel in order to perform
computations efficiently.  It is also necessary to have all of them
appropriately synchronized to avoid message jams and disruptive
functioning of the several processors which would not operate
coherently.  Likewise, quantum parallelism is not enough to achieve a
successful quantum computation.  Recall that the result of a quantum
computation is probabilistic.  There is not a 100\% certainty that
after measuring the final output state it will contain the correct
result we are searching for.  We need to repeat the measurement
several times in order to retrieve the correct value of the function
or procedure for which the computer was devised.  If we program the
quantum computer carelessly, this number of measurements would be
exponentially large, and all the potential advantages of quantum
parallelism spoiled.  What do we need to make good quantum programs?
We need to reduce the number of trials to just a few.  This fact will
depend on how the evolution operator $U\{ U^+,U^- \}$ and the initial
memory states ${\bf | m\rangle }$ are prepared.  In order to become
good quantum programmers we must be smart enough so as to devise them
in such a way that the maxima of the probability distribution in the
output state correspond to the desired result, while the rest of
possible results, which are useless for the purpose of our
computation, must be somehow damped.  We recognize this pattern of
behaviour for the unitary operator $U\{ U^+,U^- \}$ as the phenomenon
of {\em constructive interference} of amplitudes in quantum mechanics.
The typical example is the two-slit experiment.

We shall present explicit examples of how quantum parallelism and
constructive interference work together when we deal with quantum
algorithms in Sec.~\ref{sec10:level1}. Now, we summarize these
correspondences between classical parallel and quantum computers as
follows:

\begin{center}
\underline{Classical Parallel Computers}

\medskip

i) many things at a time \\ ii) synchronization of many processors
$$
\updownarrow
$$
\underline{Quantum Computer}

\medskip

i) many states at a time \\ ii) constructive interference of many
states
\end{center}

The quantum version of parallelism exceeds the classical one, for
whereas in a quantum computer it is possible to have an exponentially
large number of available states within a reduced space, this capacity
seems unreachable in any known classical parallel computer.

In quantum mechanics there are some basic principles, like the
correspondence principle, Heisenberg's principle, Pauli's principle,
etc., which encode the fundamentals of that theory.  The knowledge of
those principles provide us with the essential understanding of
quantum mechanics at a glance, without going into the complete
formalism of that subject.  A similar thing happens with other areas
in physics.  In computer science there are also guiding rules to
devise the architecture of a computer (hardware) and the programs to
be run (software).  Likewise, in quantum computing we have seen that
there are basic principles that serve us as a guide to get the most
profit from a quantum computer.  These principles refer to the ideas
of quantum parallelism and quantum programming.  We know that
information and computation is physics.  Thus, there must be a
connection between the principles of quantum computation and the
principles of quantum physics.  It is useful to synthesize those
relationships between both fields in the form of basic principles, as
shown explicitly in Table~\ref{table3}.

\begin{table}[ht]
\caption{Principles of Quantum Computation.}
\begin{ruledtabular}
\begin{tabular}{ccc}
 \multicolumn{1}{c}{{\it Computer Science}}& & \multicolumn{1}{c}{{\it
 Quantum Physics}}\\ \colrule $1^{\rm st}$ Quantum Parallelism & = &
 Superposition Principle \\ \colrule $2^{\rm nd}$ Quantum Programming
 & = & Constructive Interference \\
\end{tabular}
\end{ruledtabular}
 \label{table3} \end{table}

By principles of quantum computation we mean those rules which are
specific to the act of computing according to the laws of quantum
mechanics.  In this table we indicate that the quantum version of
parallelism is realized through the superposition principle of quantum
mechanical amplitudes; likewise the act of quantum programming a
quantum computer should be closely related to constructive
interference of those amplitudes involved in the superposition of
quantum states in the registers of a quantum computer.  We shall see
these principles in action when studying quantum algorithms (see
Sec.~\ref{sec10:level1}) that supersede their classical counterparts.
This fact expresses that the capabilities of a quantum Turing machine
go well beyond those of a classical Turing machine.  The superposition
principle when applied to multipartite quantum systems like those of a
quantum register (see eq. (\ref{qc17}) below) yields the notion of
entanglement (Sec.~\ref{sec3O:level2}, Sec.~\ref{sec3D:level2}).

\subsubsection{Universal QTM}
\label{sec9C:level3}

The notion of universal Turing machine can also be extended to quantum
Turing machines.  A standard QTM is capable of performing only the job
for which it has been set up.  This is so because the unitary operator
$U\{ U^+,U^- \}$ and the memory quantum states ${\bf | m \rangle}$ are
chosen to do one specific task.  Deutsch (1985) has shown that the
elements $U\{ U^+,U^- \}$ and ${\bf | m \rangle}$ of a QTM can be
devised to simulate with arbitrary precision any other quantum
computer.  This is the concept of {\em universal quantum Turing
machine}.  A universal QTM is thus a programmable quantum computer.
We now give more explicit details about how a quantum TM is programmed.

Let $f$ be the any function that we want to compute with the universal
QTM, and let $\pi(f)$ be a quantum program to do the job.  The quantum
computer will take the program $\pi(f)$ and a given input value $i$
and then compute the desired value $f(i)$.  This process is
implemented in a QTM as follows.  There exists an integer $n_{\rm
fin}$ such that
\begin{equation}
U^{n_{\rm fin}} |0;{\bf 0}; \pi (f),i,{\bf 0}\rangle = |0;1,{\bf 0};
\pi (f),i,f(i),{\bf 0}\rangle,
\label{qc12}
\end{equation}

\noindent where the halting qubit is set to $\ket{1}$ after the
computation ends.  In this expression we assume that the initial
quantum memory states are
\begin{equation}
 {\bf | m}_{\rm in}\rangle = |\pi (f),i,{\bf 0}\rangle, \label{qc13}
\end{equation}

\noindent while the final memory states contain the answer $f(i)$:
\begin{equation}
 {\bf | m}_{\rm fin}\rangle = |\pi (f),i,f(i),{\bf 0}\rangle.
 \label{qc14}
\end{equation}

If in eq.(\ref{qc12}) we focus only on the memory states, then we can
use a short-hand notation for the unitary evolution,\footnote{See
Sec.~\ref{sec9C:level2} for more on quantum function evaluation.}
namely,
\begin{equation}
|\pi (f),i,j\rangle \mapsto |\pi (f),i,j \oplus f(i)\rangle.
\label{qc15}
\end{equation}

Although a QTM has an infinite-dimensional memory space, much like a
classical TM, we remark that only a finite-dimensional unitary
transformation needs be applied at every step of the computation to
simulate the associated QTM evolution.

The concept of a quantum Turing machine has many implications that we
shall continue to present.  Most of these implications amount to a
revision of the typical areas of classical computation in the light of
the new principles of computation.  For instance, now we can
immediately address how the theory of complexity gets affected in its
fundamentals.  In Sec.~\ref{sec8A:level2} we mentioned that this
theory deals with the issue of what a computer can do.  Namely, it
studies not only which function can be computed, but also how fast and
how much memory resources are needed.  This scheme must be modified to
convert it into a {\em quantum complexity theory}.  In this new theory
of complexity we must pose another question, ``with which
probability'' can a quantum computer achieve a certain task. See
Appendix for details.

\subsection{Quantum Logic Gates}
\label{sec9B:level2}

The quantum Turing machine is a basic model for quantum computation
that deals with the new characteristics posed by quantum principles at
a fundamental level, as compared with the classical functioning of a
classical Turing machine.  However, a quantum TM is not a practical
starting point for designing a quantum computer, much like the
classical Turing machine is not a handy computer.

The key idea is to decompose the functioning of a quantum computer
into the simplest possible primitive operations or gates.  The
identification of universal logic gates, such as NAND, in classical
computers (see Sec.~\ref{sec8D:level2}) was of great help in the
development of the field.  A universal gate such as NAND operates
locally on a very reduced number of bits, actually two.  However,
combining NAND gates in the appropriate number and sequence we can
carry out arbitrary computations on arbitrarily many bits.  This was
very useful in practice for it allowed device engineers to just focus
on creating only a few devices, leaving the rest to the circuit
designer.  The same rationale applies to a quantum computer and the
relation of a quantum Turing machine to quantum circuits.

When a quantum computer is working, it is an evolution unitary
operator that is effecting a predetermined action on a series of
qubits.  These qubits form the memory register of the machine or a
{\em quantum register}.  A quantum register is a string of qubits with
a predetermined finite length.  The space of all the possible register
states makes up the Hilbert space of states associated to the quantum
computer.  If ${\cal H}$ is the Hilbert space of a single qubit and
$|\Psi_i\rangle\in{\cal H}$, $i=1,2$, a given basis state, then a
basis vector $|\Phi\rangle$ for the states of the quantum register is
a tensor product of qubit states
\begin{equation}
|\Phi\rangle =|\Psi_1\rangle \otimes |\Psi_2\rangle \otimes \ldots
\otimes |\Psi_n\rangle \in {\cal H}^{\otimes n}.
\label{qc17}
\end{equation}

A quantum memory register can store multiple sequences of classical
bits in superposition.  This is a manifestation of the quantum
parallelism.

\begin{figure}[ht]
\psfrag{a}[Bc][Bc][0.75][0]{$|x_1\rangle$}
\psfrag{b}[Bc][Bc][0.75][0]{$|x_2\rangle$}
\psfrag{c}[Bc][Bc][0.75][0]{$|x_3\rangle$}
\psfrag{d}[Bc][Bc][0.75][0]{$|x_n\rangle$}
\psfrag{u}[Bc][Bc][0.75][0]{$|x^{\prime}_1\rangle$}
\psfrag{v}[Bc][Bc][0.75][0]{$|x^{\prime}_2\rangle$}
\psfrag{w}[Bc][Bc][0.75][0]{$|x^{\prime}_3\rangle$}
\psfrag{t}[Bc][Bc][0.75][0]{$|x^{\prime}_n\rangle$}
\includegraphics[width=6 cm]{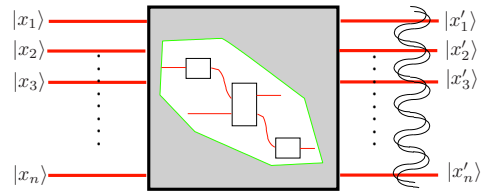}
\caption{A generic quantum logic gate. The wavy lines mean that the
output state is a generic  superposition of product quantum states.}
\label{qgengate}
\end{figure}

A {\em quantum logic gate} is a unitary operator acting on the states
of a certain set of qubits.  If the number of such qubits is $n$, the
quantum gate is represented by a $2^n\times 2^n$ matrix in the unitary
group ${\rm U}(2^n)$.  It is thus a {\em reversible} gate: we can
reverse backwards the action, thereby recovering the initial quantum
state from the final one.  Generically, a quantum logic gate can have
any finite number of input qubits, but in practice we shall be
interested in gates that are elementary for quantum computation, and
those have a small number of input qubits.  Diagrammatically, a
quantum gate is represented by a ``black box" wherein operation takes
place, and a number of input (output) lines, used to wire up a set of
gates, equal to the number of qubits involved in the computation (see
Fig.~\ref{qgengate}).  Let us see more explicitly how quantum gates
look like by giving some representative gates in increasing order of
complexity.

\vspace{10 pt}

\noindent {\em 1-Qubit Gates.} These are the simplest possible gates
for they take one input qubit and transform it into one output qubit.
The {\em quantum} NOT gate is a one-qubit gate.  Its unitary evolution
operator $U_{{\rm NOT}}$ is (\ref{xyz}):
\begin{equation}
U_{{\rm NOT}} =
\begin{pmatrix}
  0 & 1 \\ 1 & 0
\end{pmatrix}
\label{qc18}
\end{equation}
\begin{figure}[ht]
\psfrag{a}[Bc][Bc][1][0]{a)} \psfrag{b}[Bc][Bc][1][0]{b)}
\psfrag{c}[Bc][Bc][1][0]{c)} \psfrag{n}[Bc][Bc][1][0]{NOT}
\psfrag{x}[Bc][Bc][1][0]{$|x\rangle$}
\psfrag{y}[Bc][Bc][1][0]{$|1-x\rangle$}
\psfrag{s}[Bc][Bc][1][0]{$\sqrt{{\rm NOT}}$}
\psfrag{z}[Bc][Bc][1][0]{$|x\rangle$}
\psfrag{t}[Bc][Bc][1][0]{$U_{\sqrt{{\rm NOT}}} |x\rangle$}
\psfrag{h}[Bc][Bc][1][0]{H} \psfrag{u}[Bc][Bc][1][0]{$|x\rangle$}
\psfrag{v}[Bc][Bc][1][0]{$U_{{\rm H}} |x\rangle$}
\includegraphics[width=6 cm]{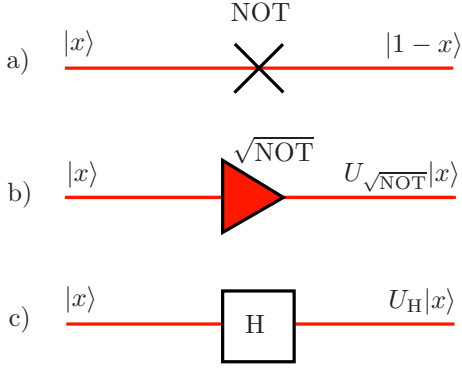}
\caption{Quantum unary gates: a) NOT gate, b) $\sqrt{{\rm NOT}}$ gate,
c) Hadamard gate.}
\label{unarygates}
\end{figure}

\noindent The truth table and the diagram representing this gate are
shown in Table~\ref{truth} and Fig.~\ref{unarygates}, respectively.
We see that this quantum NOT gate coincides with its classical
counterpart.  However, there is a basic underlying difference: the
quantum gate acts on qubits while the classical gate does it on bits.
This difference allows us to introduce a truly quantum one-qubit gate:
the $\sqrt{\rm NOT}$ gate.\footnote{{\em Square-root-of-}NOT gate.}
Its matrix representation is
\begin{equation}\label{qc19}
U_{\sqrt{{\rm NOT}}} := {1\over \sqrt{2}}
\ee^{\ii\pi/4}(1-\ii\sigma_x).
\end{equation}

\noindent This gate, when applied twice, gives NOT.  Explicitly
\begin{equation}
\begin{split}
\label{qc20}
U_{\sqrt{{\rm NOT}}}U_{\sqrt{{\rm NOT}}}& =
\begin{pmatrix}
 {1+\ii \over 2} & {1-\ii \over 2} \\ {1-\ii \over 2} & {1+\ii \over 2}
\end{pmatrix}
 \cdot
\begin{pmatrix}
 {1+\ii \over 2} & {1-\ii \over 2} \\ {1-\ii \over 2} & {1+\ii \over
 2} \end{pmatrix} \\ & =
\begin{pmatrix}
0 & 1 \\ 1 & 0 \end{pmatrix} = U_{\rm NOT}
\end{split}
\end{equation}

\noindent This gate has no counterpart in classical computers since it
implements nontrivial superpositions of basis states.

Another one-qubit gate without analogue in classical circuitry and
heavily used in quantum computations is the so called Hadamard gate H
(see Sec.~\ref{sec3:level1}).  It is defined as
\begin{equation}\label{qc21}
U_{{\rm H}} = {1\over \sqrt{2}}
\begin{pmatrix}
1 & 1\\ 1 & -1 \end{pmatrix}.
\end{equation}

\noindent {\em 2-Qubit Gates.} The XOR ({\em exclusive}-OR), or CNOT
({\em controlled}-NOT) gate, is an example of a quantum logic gate on
two qubits (\ref{qicnot}).  It is instructive to give the  unitary
action $U_{{\rm XOR,CNOT}}$ of this gate in several forms.  Its action
on the two-qubit  basis states is
\begin{equation}
\begin{split}
&U_{{\rm CNOT}} |00\rangle = |00\rangle,\quad  U_{{\rm CNOT}}
|10\rangle = |11\rangle,\\ &U_{{\rm CNOT}} |01\rangle = |01\rangle,
\quad U_{{\rm CNOT}} |11\rangle = |10\rangle.
\label{qc22}
\end{split}
\end{equation}

\noindent From this definition we see that the name of this gate is
quite apparent for it means that it executes a NOT operation on the
second qubit conditioned to have the first qubit in the state
$|1\rangle$.  Its matrix representation is
\begin{equation}
U_{\rm CNOT}=U_{\rm XOR}=
\begin{pmatrix} 1 & 0 & 0 & 0 \\
0 & 1 & 0 & 0 \\ 0 & 0 & 0 & 1 \\ 0 & 0 & 1 & 0 \\
\end{pmatrix}.
\label{qc23}
\end{equation}

\begin{figure}[ht]
\psfrag{a}{a)} \psfrag{x}{$|x_1\rangle$} \psfrag{y}{$|x_1\rangle$}
\psfrag{z}{$|x_2\rangle$} \psfrag{t}{$|x_2 \oplus x_1\rangle$}
\psfrag{b}{b)} \psfrag{p}{$\phi$} \psfrag{u}{$|x_1\rangle$}
\psfrag{v}{$|x_1\rangle$} \psfrag{w}{$|x_2\rangle$}
\psfrag{s}{$\ee^{\ii x_1 x_2 \phi}|x_2\rangle$} \psfrag{c}{c)}
\psfrag{d}{$|x_1\rangle$} \psfrag{g}{$|x_2\rangle$}
\psfrag{f}{$|x_2\rangle$} \psfrag{h}{$|x_1\rangle$}
\includegraphics[width=6cm]{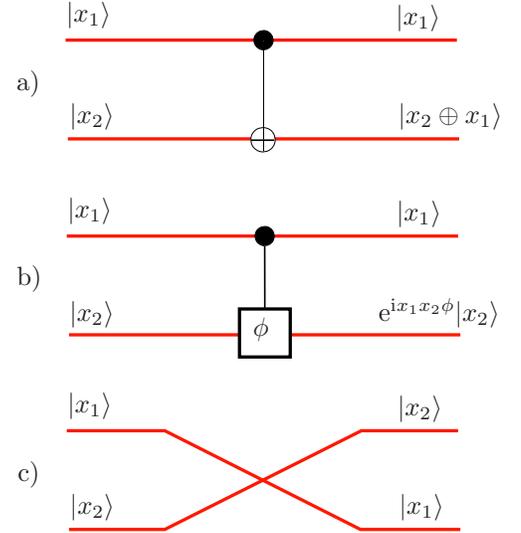}
\caption{Quantum binary gates: a) CNOT gate, b) CPHASE gate, c) SWAP
gate.}
\label{q2gates}
\end{figure}

The action of the CNOT operator (\ref{qc22}) immediately translates
into a corresponding truth table as in Table~\ref{truth3}.  The
diagrammatic representation of the CNOT gate is shown in
figure~\ref{q2gates}.

\begin{table}[ht]
\begin{ruledtabular}
\begin{tabular}{c c c c}
$x$ & $y$ & $x'$ & $y'$ \\  \hline 0 & 0 & 0 & 0  \\ 0 & 1 & 0 & 1 \\
1 & 0 & 1 & 1 \\ 1 & 1 & 1 & 0
\end{tabular}
\end{ruledtabular}
\caption{The truth table of the quantum CNOT gate.}
\label{truth3}
\end{table}

We shall see how this quantum CNOT gate plays a paramount role in both
the theory and experimental realization of quantum computers.  It
allows implementing conditional logic at a quantum level.

Unlike the CNOT gate, there are two-qubit gates with no analogue
classical gate.  One example is the {\em controlled-phase} gate or
CPHASE:
\begin{equation}
U_{{\rm CPh}(\phi)}:= \begin{pmatrix} 1 & 0 & 0 & 0 \\ 0 & 1 & 0 & 0
\\ 0 & 0 & 1 & 0 \\ 0 & 0 & 0 & \ee^{\ii \phi} \\
\end{pmatrix}
\label{qc24}
\end{equation}

\noindent It implements a conditional phase-shift $\ee^{\ii \phi}$ on
the second qubit.

\begin{figure}[h]
\psfrag{A}[Bc][Bc][0.75][0]{$U_H$} \psfrag{U}[Bc][Bc][1][0]{$\pi$}
\psfrag{B}[Bc][Bc][0.75][0]{$U_H$}
\psfrag{t}[Bc][Bc][1][0]{$\parallel$} 
\includegraphics[width=4 cm]{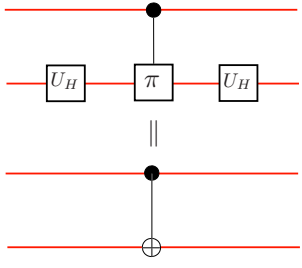}
\caption{Relation between CNOT gate and controlled-phase using
Hadamard gates.}
\label{q2pi}
\end{figure}

An important result is that we can reproduce the CNOT gate with a
controlled-phase gate of $\phi=\pi$ and two Hadamards transforms on
the target qubits as shown in Fig.~\ref{q2pi}.  This is a simply
consequence of the relation
\begin{equation}
U_H \sigma_z U_H = \sigma_x.
\label{qc24b}
\end{equation}

Other interesting two-qubit gates are the SWAP gate, which
interchanges the states of the two qubits, and the $\sqrt{{\rm SWAP}}$
gate,\footnote{Square-root-of-swap gate.} whose matrix representations
are
\begin{gather}
U_{{\rm SWAP}} :=
\begin{pmatrix}
1 & 0 & 0 & 0 \\  0 & 0 & 1 & 0 \\  0 & 1 & 0 & 0 \\  0 & 0 & 0 & 1 \\
\end{pmatrix}, \; U_{\sqrt{{\rm SWAP}}} :=
\begin{pmatrix}
1 & 0 & 0 & 0 \\  0 & {1+\ii \over 2} & {1-\ii \over 2}  & 0 \\  0
&{1-\ii \over 2}  & {1+\ii \over 2}  & 0 \\  0 & 0 & 0 & 1
\end{pmatrix}.
\label{qc25}
\end{gather}

\vspace{10 pt} \noindent {\em 3-Qubit Gates.} An immediate extension
of the CNOT construction to three-qubits yields the CCNOT gate (or
C$^2$NOT),\footnote{Controlled-controlled-not gate.} which is also
called Toffoli gate T (Toffoli, 1981).  The matrix representation is a
one-qubit extension of the CNOT gate, namely
\begin{equation}
U_{{\rm CCNOT}} =U_{{\rm T}} :=
\begin{pmatrix}
 1 & 0 & 0 & 0 & 0 & 0 & 0 & 0 \\  0 & 1 & 0 & 0 & 0 & 0 & 0 & 0\\  0
 & 0 & 1 & 0 & 0 & 0 & 0 & 0\\ 0 & 0 & 0 & 1 & 0 & 0 & 0 & 0\\  0 & 0
 & 0 & 0 & 1 & 0 & 0 & 0\\ 0 & 0 & 0 & 0 & 0 & 1 & 0 & 0\\  0 & 0 & 0
 & 0 & 0 & 0 & 0 & 1\\  0 & 0 & 0 & 0 & 0 & 0 & 1 & 0
\end{pmatrix}.
\label{qc26}
\end{equation}

\begin{figure}[ht]
\psfrag{a}{a)} \psfrag{x}[Bc][Bc][0.75][0]{$|x_1\rangle$}
\psfrag{y}[Bc][Bc][0.75][0]{$|x_1\rangle$}
\psfrag{z}[Bc][Bc][0.75][0]{$|x_2\rangle$}
\psfrag{t}[Bc][Bc][0.75][0]{$|x_2\rangle$}
\psfrag{l}[Bc][Bc][0.75][0]{$|x_3\rangle$}
\psfrag{m}[Bc][Bc][0.75][0]{$|x_3 \oplus x_1 x_2\rangle$}
\psfrag{b}{b)} \psfrag{R}[Bc][Bc][0.75][0]{$S(\theta)$}
\psfrag{u}[Bc][Bc][0.75][0]{$|x_1\rangle$}
\psfrag{v}[Bc][Bc][0.75][0]{$|x_1\rangle$}
\psfrag{w}[Bc][Bc][0.75][0]{$|x_2\rangle$}
\psfrag{s}[Bc][Bc][0.75][0]{$|x_2\rangle$}
\psfrag{n}[Bc][Bc][0.75][0]{$|x_3\rangle$} \psfrag{o}[Bc][Bc][0.65][0]
{$(\delta_{x_1x_2,0}I+\delta_{x_1x_2,1}U_{S(\theta)})|x_3\rangle$}
\psfrag{c}{c)} \psfrag{d}[Bc][Bc][0.75][0]{$|x_1\rangle$}
\psfrag{f}[Bc][Bc][0.75][0]{$|x_1\rangle$}
\psfrag{g}[Bc][Bc][0.75][0]{$|x_2\rangle$}
\psfrag{h}[Bc][Bc][0.75][0]{$|x_3\rangle$}
\psfrag{i}[Bc][Bc][0.75][0]{$|x_3\rangle$}
\psfrag{k}[Bc][Bc][0.75][0]{$|x_2\rangle$} 
\includegraphics[width=6 cm]{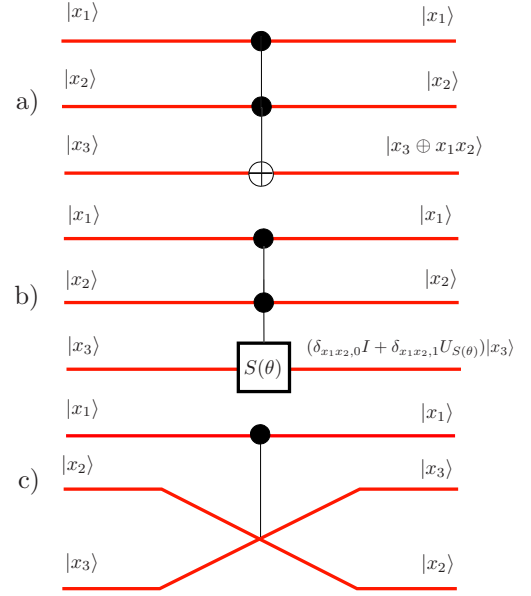}
\caption{A set of three-qubit gates: a) Toffoli gate, b) Deutsch gate,
c) Fredkin gate.}
\label{q3gates}
\end{figure}

\noindent The associated truth table is shown in Table~\ref{truth4}.
The first two input qubits $x,y$ are copied to the first two output
qubits $x',y'$ (see Fig.~\ref{q3gates}), while the third output qubit
$z'$ is the XOR of the third input $z$ and the AND of the first two
inputs $x,y$.

\begin{table}
\begin{ruledtabular}
\begin{tabular}{cccccc}
$x$ & $y$ & $z$ & $x'$ & $y'$ & $z'$ \\ \hline 0 & 0 & 0 & 0 & 0 & 0
  \\ 0 & 0 & 1 & 0 & 0 & 1 \\ 0 & 1 & 0 & 0 & 1 & 0 \\ 0 & 1 & 1 & 0 &
  1 & 1 \\ 1 & 0 & 0 & 1 & 0 & 0 \\ 1 & 0 & 1 & 1 & 0 & 1 \\ 1 & 1 & 0
  & 1 & 1 & 1 \\ 1 & 1 & 1 & 1 & 1 & 0
\end{tabular}
\end{ruledtabular}
\caption{Truth table for the Toffoli gate.}
\label{truth4}
\end{table}

The Deutsch gate D($\theta$)  (Deutsch, 1989) is also an important
three-qubit gate.  It is a {\em controlled-controlled}-$S$ or C$^2S$
operation (see Fig.~\ref{q3gates}), where

\begin{equation}\label{qc27}
U_{S(\theta)} := \ii\ee^{-\ii\half\theta\sigma_x} = \ii\cos\half\theta
+\sigma_x \sin\half\theta
\end{equation}

\noindent is a unitary operation that rotates a qubit about the $x$
axis by an angle $\theta$ and then multiplies it by a factor $\ii$.
We demand $\theta$ to be {\em incommensurate} to $\pi$, that is, not a
rational multiple of $\pi$.  Several properties follow: 1) Let
$|q\rangle$ be a given qubit, then for any fixed value of $\alpha \in
\R$ we can get arbitrarily close to $\ee^{\ii \alpha
\sigma_x}|q\rangle$ by successive application of $U_{S(\theta)}$ to
$|q\rangle$ a finite number of times.  2) The Deutsch gate generates
as closely as needed the Toffoli gate.  This is because the C$^2S^n$
gate is just the ${\rm D}^n$ gate.   And since we can make
$\fourth(n\theta/\pi-1)$, with $n=4k+1$, as near to a given arbitrary
integer as desired, ${\rm D}^n$ will thereby approach closely the
Toffoli gate.

Another instance of a three-qubit gate is the Fredkin gate F (Fredkin
and Toffoli, 1982).  It is a {\em controlled}-SWAP operation,
schematically shown in Fig.~\ref{q3gates} and represented by the matrix
\begin{equation}
U_{{\rm F}} =
\begin{pmatrix}
 1 & 0 & 0 & 0 & 0 & 0 & 0 & 0 \\  0 & 1 & 0 & 0 & 0 & 0 & 0 & 0\\  0
 & 0 & 1 & 0 & 0 & 0 & 0 & 0\\ 0 & 0 & 0 & 1 & 0 & 0 & 0 & 0\\  0 & 0
 & 0 & 0 & 1 & 0 & 0 & 0\\  0 & 0 & 0 & 0 & 0 & 0 & 1 & 0\\  0 & 0 & 0
 & 0 & 0 & 1 & 0 & 0\\  0 & 0 & 0 & 0 & 0 & 0 & 0 & 1\\
\end{pmatrix}
\label{qc28}
\end{equation}

Needless to say that these unitary linear gates not only act on the
basis states, but also on any linear combination of them.

We have enumerated a series of quantum logic gates whose use and
importance will be explained in the following sections.  We shall
address the experimental implementation of some of these quantum gates
in Sec.~\ref{sec11:level1}.

\subsection{Quantum Circuits}
\label{sec9C:level2}

The simple gates introduced in the previous section can be assembled
into a network-like arrangement that enable us to perform more
complicated quantum operations than those initially carried out by
those gates.  This is the basic idea of a {\em quantum circuit}.
Deutsch (1989) generalized the classical reversible circuit model to
produce the idea of quantum circuits.  A quantum circuit is a
computational network composed of interconnected elementary quantum
gates.

An example to illustrate a simple use of a quantum circuit is the
following.  Let us prepare initially a one-qubit state as an arbitrary
superposition of the logical states $|0\rangle, |1\rangle$, namely
\begin{equation}\label{qc29}
|\psi_0\rangle = a |0\rangle + b |1\rangle.
\end{equation}

\noindent We want to obtain a final state of GHZ type (\ref{ghz}):
\begin{equation}
|\psi_f\rangle = a |0 0 0\rangle + b |1 1 1\rangle.
\label{qc30}
\end{equation}

\begin{figure}[ht]
\psfrag{x}[Bc][Bc][1][0]{$a|0\rangle+b|1\rangle$}
\psfrag{y}[Bc][Bc][1][0]{$|0\rangle$}
\psfrag{z}[Bc][Bc][1][0]{$|0\rangle$} \psfrag{t}[Bc][Bc][1][0]
{$\begin{matrix} & \\ & a|000\rangle \\ & + \\ & b|111\rangle \\ &
\end{matrix}$}
\includegraphics[width=6 cm]{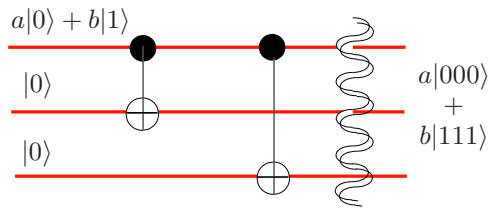}
\caption{An example of quantum circuit implementing a GHZ state.}
\label{qcircuit}
\end{figure}

\noindent To this purpose, instead of writing a pertinent sequence of
algebraic operations,  we can simply arrange the following quantum
circuit using the CNOT-gate as pictured in Fig.~\ref{qcircuit}.

Quantum circuits are widely used in quantum computation, where most of
the problems can be formulated in terms of them.  Moreover, it might
quite well be the case that standard quantum mechanics could be
flooded with quantum circuits in the future, something similar to what
happened with Feynman diagrams in quantum field theory.  The reason is
because quantum circuits are able to condensate graphically much more
information than the use of several formulas.  Besides, this form of
presenting and reasoning about is closer to what experimental
physicists really do with their devices.

In Sec.~\ref{sec8D:level2} we presented the basic result that a
classic Turing machine is equivalent to a classical logic circuit.  In
quantum computing there is a similar result due to Yao (1993) showing
that a quantum Turing machine is equivalent to a quantum circuit.
This theorem justifies replacing the more complicated study of quantum
Turing machines by that of quantum circuits, which are simpler to
analyze and design.  In fact, experimental approaches to quantum
computers are presented in terms of quantum circuits (see
Sec.~\ref{sec11:level1}).

Let $K$ be a quantum Boolean or logic circuit with $n$ input qubits.
Suppose that $\ket{\Psi_x}=\sum_{y\in \{ 0,1\}^n} c_{x}(y)\ket{y}$ is
the final quantum state of $K$ for an input $x \in \{ 0,1\}^n$.  The
{\em distribution generated} by $K$ for the input $x$ is defined as
the map $p_{x}:y \in \{ 0,1\}^n\mapsto |c_{x}(y)|^2$.  The quantum
circuit $K$ is said to ($n,t$)-{\em simulate} a quantum Turing machine
$Q$ if the family of probability distributions $p_{x}$, $x\in \{
0,1\}^n$, coincides with the probability distributions of the $Q$
configurations after $t$ steps with input $x$.\footnote{We assume that
a given configuration is encoded as a list of the tape symbols from
cell $-t$ to $t$, followed by the state and the position of the
cursor, all encoded as strings of qubits (see
Sec.~\ref{sec9A:level2}).}  Then Yao's theorem is the following
statement:

{\em Quantum Turing machines and quantum circuits}: Let $Q$ be a
quantum Turing machine and $n,t$ positive integers.  There exists a
quantum Boolean circuit $K$ of polynomial size in $n,t$, that
($n,t$)-simulates $Q$.

This result implies that quantum circuits can mimic quantum Turing
machines in polynomial time, and vice versa.  Thus, quantum circuits
provide a sufficient model for quantum computation that is easier to
implement and manipulate than QTMs.  This situation goes in parallel
with similar results about classical Boolean circuits and Turing
machines (Sec.~\ref{sec8D:level2}).  From now on when talking about a
quantum computer we shall usually refer to an underlying equivalent
quantum circuit.

\subsubsection{Universal quantum gates}

After the works of Deutsch (1989) and Yao (1993) the concept of a {\em
universal set} of quantum gates became central in the theory of
quantum computation.  A set ${\cal G}:=\{G_{1,n_1},\ldots,G_{r,n_r}\}$
of quantum gates $G_{j,n_j}$ acting on $n_j$ qubits, $j=1,\ldots,r$,
is called universal if {\em any} unitary action $U_N$ on $N$ input
quantum states can be decomposed into a product of succesive actions
of $G_{j,n_j}$ on different subsets of the input qubits.  More
explicitly, given any $U_N$ acting unitarily on $N$ qubits, there
exists a sequence $S_1,S_2,\ldots,S_s$ of subsets of
$\{1,2,\ldots,N\}$, with $n_{S_1},\ldots,n_{S_s}$ elements, and a map
$\pi: \{1,2,\ldots,s\}\to \{1,2,\ldots,r\}$ such that
$n_{\pi(j)}=n_{S_j}$, $\forall j$, and
\begin{equation}
U_N={U}_{N,G_{\pi(s)},S_s}\ldots{U}_{N,G_{\pi(1)},S_1}.
\label{qc31}
\end{equation}
Here
\begin{equation}
{U}_{N,G_{\pi(j)},S_j}:=I_{\{1,2,\ldots,N\}-S_j}\otimes
{U}_{G_{\pi(j)},S_j},
\label{qc32}
\end{equation}
where $I_{\{1,2,\ldots,N\}-S_j}$ is the identity on the qubits not in
$S_j$, and ${U}_{G_{\pi(j)},S_j}$ stands for the unitary action of the
gate $G_{\pi(j)}$ on the Hilbert space of the $n_{S_j}$ qubits in the
set $S_j$.

For instance, a generic unitary $k\times k$ matrix of dimension $k\geq
2$  can be represented as the product of $k(k-1)/2$ two-level  unitary
matrices (Reck et al., 1994).

This notion of universal set of gates is {\em exact} for the generic
transformation $U_N$ is reproduced exactly in terms of a finite number
of elements in ${\cal G}$. We denote this situation by writing the
universal set as ${\cal G}_{{\rm ex}}$.  However, this notion is too
strong.  Dealing with practical quantum devices, it is not conceivable
to work with a set of gates implementing any other gate with perfect
accuracy.  Thus, we are inevitably led to work with approximate
simulations of gates.  Underlying this idea there is the concept of
distance between two unitary gates.

A quantum gate $U_N$ is said to be approximated by another gate $U'_N$
with error $<\epsilon$, when the distance
$d(U_N,U'_N):=\inf_{\theta\in\R}||U_N - \ee^{\ii \theta}U'_N||$
between both matrices as projective operators is $ <
\epsilon$.\footnote{The norm $||A||$ of the (finite) matrix $A$ is
usually defined as $\sup_{x:||x||=1}||Ax||$.  Other norms are
topologically equivalent to it.}$^,$\footnote{ A compactness argument
shows that the $\inf$ in the definition of $d$ is attainable,
i.e. $\exists \theta_{0}$ such that $d(U_N,U'_N):=||U_N - \ee^{\ii
\theta_0}U'_N||$.  From now on, we will assume that the phase factor
is included in the approximating unitary operator $U'_N$.}  This means
that if the gate $U_N$ is replaced by gate $U'_N$ in a quantum circuit
$K$, then the unit rays of the associated output states will differ in
norm by at most $\epsilon$.\footnote{The unit ray of a state vector
$\ket{\phi}$ is the set $[\phi]:=\{\ee^{\ii \theta}
\ket{\phi}:\theta\in\R\}$.  A distance between unit rays can be
defined as $\dist([\phi_1],[\phi_2])=\inf_{\theta\in\R}||\phi_1-
\ee^{\ii \theta}\phi_2||$, what justifies the presence af a phase
factor in the notion of an appproximate gate.}

With this definition, we also introduce the notion of an {\em
approximate} set of universal quantum gates as before but with the
weaker requirement that it simulates any other quantum gate in an
approximate sense. We denote these sets as ${\cal G}_{{\rm ap}}$,  and
by universality we shall mean it in this sense henceforth, unless the
exact notion is explicitly indicated.

Some examples of universal sets of quantum gates, to be discussed
next, are the following (for a more mathematical and general approach,
see  Brilynski et al., 2001):

\begin{enumerate}

\item ${\cal G}_{{\rm ex}}^{{\rm I}}:=\{ U_2:U_2\in\rm U(2^2)\}$,
(DiVincenzo, 1995).

\item ${\cal G}_{{\rm ex}}^{{\rm II}}:=\{ U_1, {\rm CNOT}:U_1\in\rm
U(2)\}$,  (Barenco et al., 1995).

\item ${\cal G}_{{\rm ap}}^{{\rm III}}:=\{{\rm D}\}$, Deutsch gate
(\ref{qc27}), (Deutsch, 1989).

\item ${\cal G}_{{\rm ap}}^{{\rm IV}}:=\{\text{C$^2$-}U,
\text{C$^2$-}W\}$, with $U(\alpha):= R_{y}(4\pi \alpha)= \ee^{-\ii
2\pi \alpha \sigma_y}$, $W(\alpha):= {\rm diag}(1,\ee^{\ii 2\pi
\alpha})$, $\alpha $ an irrational root of a degree-2 polynomial
(Aharonov, 1998).

\item ${\cal G}_{{\rm ap}}^{{\rm V}}:=\{{\rm H}, {\rm CPh}({\pi \over
2}) \}$, (\ref{qc21}), (\ref{qc24}), (Solovay, 1995; Kitaev, 1997;
Cleve, 1999).

\item ${\cal G}_{{\rm ap}}^{{\rm VI}}:=\{{\rm H}, W, {\rm CNOT}\}$,
with $W:= {\rm diag}(1,\ee^{\ii \pi/4})$, (Cleve, 1999).

\end{enumerate}

Of these examples, 1/ and 2/ correspond to {\em infinite} sets of
universal gates. However, a practical quantum computer must have a set
with a {\em finite} number of universal gates.  Examples 3/ to 6/ are
finite suitable cases. Although with a finite set of gates we are
limited to simulate a countable subset of all possible quantum gates,
it is possible to reproduce an arbitrary gate within a given small
error $\epsilon$.  Moreover, a finite universal set ${\cal G}_{{\rm
ap}}$ is closer to the spirit of the Church-Turing principle stating
that a computing machine must operate by finite means
(Sec.~\ref{sec9A:level2}).

A first example of 3-qubit universal gate is the Deutsch gate
(Deutsch, 1989),\footnote{Previously Deutsch (1985) had already given
a universal set of eight 2$\times$2 matrices.}  which is an extension
of the Toffoli gate $U_{{\rm CCNOT}}$ (\ref{qc26}) (Toffoli, 1981) for
classical Boolean circuits. Toffoli gates are exactly universal for
reversible (classical) circuits.\footnote{To see that C$^2$-NOT is
classically universal, notice that: 1/
NOT$(x_3)=(\text{CCNOT}(1,1,x_3))_{3}$; 2/
AND$(x_1,x_2)=(\text{CCNOT}(x_1,x_2,0))_{3}$; and apply now the result
(Sec.~\ref{sec8D:level2}) that $\{\rm AND, NOT\}$ is a classical
universal set.  See in addition that the COPY operation is also
reproduced as COPY$(x_2)=(\text{CCNOT}(1,x_2,0))_{2,3}$.}  Deutsch
showed that his gate D($\theta_0$) with a fixed angle $\theta_0$ that
is an irrational multiple of $\pi$ is universal.

A further improvement in the analysis of quantum universal gates was
provided by DiVincenzo (1995) who showed that the set of two-qubit
gates is exactly universal for quantum computation.  This is a
remarkable result since it is known that its classical analogue is not
true: classical reversible two-bit gates are not sufficient for
classical computation.  The NAND gate, although binary, is not
reversible.

After DiVincenzo's result it was shown that a large subclass of
two-qubit gates are universal (Barenco, 1995) and moreover, that
almost any two-qubit gate is universal.

The reduction from three to two qubits amounts to a big simplification
in the analysis of quantum circuits and in their experimental
implementation.  It is much simpler to deal with two-body quantum
interactions than with a three-body problem.

The race towards bringing down the number of necessary qubits in the
elementary gates culminated with the joint work of Barenco et al.
(1995) in which it is shown that even one-qubit gates are enough for
quantum computation (in the exact sense) provided they are combined
with the CNOT gate. This result, another manifestation of the
superposition principle, is quite surprising since  in classical
computation the classical CNOT is not universal.

\begin{figure}[ht]
\begin{center}
\psfrag{a}[Bc][Bc][1][0]{a)} 
\psfrag{E}[Bc][Bc][1][0]{$E$} 
\psfrag{x}[Bc][Bc][1][0]{$|x_1\rangle$}
\psfrag{y}[Bc][Bc][1][0]{$|x_2\rangle$}
\psfrag{A}[Bc][Bc][1][0]{$U_3$} 
\psfrag{B}[Bc][Bc][1][0]{$U_2$}
\psfrag{C}[Bc][Bc][1][0]{$U_1$} 
\psfrag{t}[Bc][Bc][1][0]{$\parallel$}
\psfrag{b}[Bc][Bc][1][0]{b)} 
\psfrag{u}[Bc][Bc][1][0]{$|x_1\rangle$}
\psfrag{w}[Bc][Bc][1][0]{$|x_2\rangle$} 
\psfrag{U}[Bc][Bc][1][0]{$U$}
\includegraphics[width=5 cm]{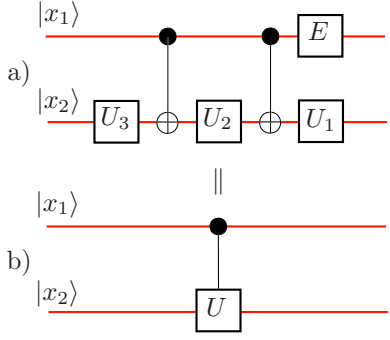}
\caption{Decomposition of an arbitrary  two-qubit C$U$ gate into
one-qubit gates and CNOTs. The symbol $E$ denotes the gate
$E:\ket{0}\mapsto \ket{0}, \ket{1}\mapsto\ee^{\ii\delta}\ket{1}$.}
\label{q2universal}
\end{center}
\end{figure}

\noindent We shall refer to this important result as the {\em
universality theorem of elementary quantum gates}.  The proof of this
result (Barenco et al., 1995) can be simply stated in terms of quantum
circuits and it has three parts.  Firstly, we need to prove that with
one-qubit gates plus CNOT it is possible to generate any
controlled-unitary two-qubit gate.  Secondly, this result is extended
to a controlled-unitary gate with an arbitrary number of qubits. And
thirdly, one applies these results to construct any unitary gate  with
one-qubit and CNOT gates.

1st {\em Part}. The proof of the first part is contained in the
identity between quantum circuits shown in Fig.~\ref{q2universal}.  In
the lower part we show a controlled-unitary C$U$ gate of two qubits
associated to a  unitary $2\times 2$ matrix $U$.  The upper part shows
its decomposition in terms of one-qubit gates $U_1,U_2,U_3,E$ and
CNOT's.  The rationale of this decomposition comes from group theory:
any unitary $2\times 2$ matrix $U$ can be decomposed as
\begin{equation}
U = {\rm Ph}(\delta)\bar U, \quad \bar U := R_z(\alpha) R_y(\beta)
R_z(\gamma) \in {\rm SU(2)}
\label{qc33}
\end{equation}

\noindent where  $\delta$ is the phase (mod $\pi$) of the U(1) factor
of U(2), and $\alpha, \beta, \gamma$ are the Euler angles
parameterizing the SU(2) matrix $\bar U$.  More explicitly,
\begin{equation}
\begin{aligned}
{\rm Ph}(\delta) & = \begin{pmatrix} \ee^{\ii \delta} & 0 \\ 0 &
\ee^{\ii \delta}
\end{pmatrix},
 & R_z(\alpha) & = \begin{pmatrix} \ee^{-\ii {\alpha \over 2}} & 0 \\
 0 & \ee^{\ii{\alpha \over 2}} \end{pmatrix}, \\ R_y(\beta) & =
 \begin{pmatrix} \cos {\beta \over 2} & -\sin {\beta \over 2} \\ \sin
 {\beta \over 2} & \cos {\beta \over 2}
\end{pmatrix},
 & R_z(\gamma) & = \begin{pmatrix} \ee^{-\ii {\gamma \over 2}} & 0 \\
 0 & \ee^{\ii {\gamma \over 2}}
\end{pmatrix}.
\label{qc34}
\end{aligned}
\end{equation}

With the help of this decomposition we can further show that for any
unitary matrix $\bar U$ in SU(2) there exist matrices $U_1,U_2,U_3$ in
SU(2) such that
\begin{equation}
\begin{aligned}
U_1 U_2 U_3 = 1, \\ U_1 \sigma_x U_2 \sigma_x U_3 = \bar U.
\label{qc35}
\end{aligned}
\end{equation}

\noindent The proof for this is by construction, namely,
\begin{equation}
\begin{aligned}
U_1 & = R_z(\alpha) R_y(\half\beta), \\ U_2 & = R_y(-\half\beta)
R_z(-\half(\alpha+\gamma)), \\ U_3 & = R_z(\half(-\alpha+\gamma)).
\label{qc36}
\end{aligned}
\end{equation}

Now, the equivalence between the quantum circuits of
Fig.~\ref{q2universal} proceeds by considering the two possibilities
for the first qubit.

i) $|x_1\rangle = |0\rangle$. In this case the CNOT gates are not
operative and using (\ref{qc35}) we find that the second qubit
$|x_2\rangle$ is not altered.

ii) $|x_1\rangle = |1\rangle$. In this case the CNOT gates do act on
the second qubit producing altogether the chain of operations ${\rm
Ph}(\delta) U_1 \sigma_x U_2 \sigma_x U_3 |x_2\rangle$, which using
(\ref{qc35}) turns out to be $U|x_2\rangle$. Recall that the
controlled-$\sigma_x$ gate is CNOT.

\begin{figure}[ht]
\begin{center}
\psfrag{a}[Bc][Bc][1][0]{a)} \psfrag{x}[Bc][Bc][1][0]{$|x_1\rangle$}
\psfrag{y}[Bc][Bc][1][0]{$|x_2\rangle$}
\psfrag{z}[Bc][Bc][1][0]{$|x_3\rangle$} \psfrag{A}[Bc][Bc][1][0]{$U$}
\psfrag{B}[Bc][Bc][1][0]{$U^{\dagger}$} \psfrag{C}[Bc][Bc][1][0]{$U$}
\psfrag{t}[Bc][Bc][1][0]{$\parallel$} \psfrag{b}[Bc][Bc][1][0]{b)}
\psfrag{u}[Bc][Bc][1][0]{$|x_1\rangle$}
\psfrag{w}[Bc][Bc][1][0]{$|x_2\rangle$}
\psfrag{U}[Bc][Bc][1][0]{$U^2$}
\includegraphics[width=5cm]{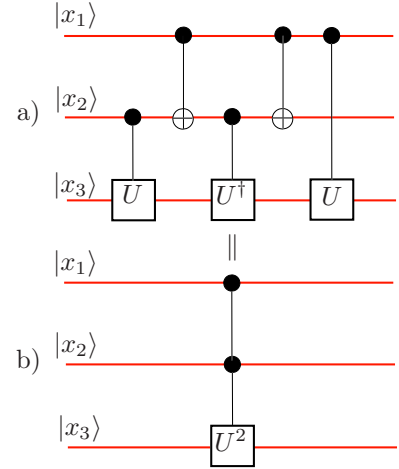}
\caption{Building-up a controlled-controlled-$U^2$ three-qubit gate
from elementary gates.}
\label{q3universal}
\end{center}
\end{figure}

2nd {\em Part}. The proof of the second part is represented in
Fig.~\ref{q3universal} by another identity between quantum circuits.
The proof is by induction on the number of qubits.  We illustrate the
simplest case.  In the lower part we show a
controlled-controlled-unitary C$^2U^2$ gate of three qubits associated
to the square of an arbitrary unitary $2\times 2$ matrix $U$.   The
upper part shows its decomposition in terms of controlled two-qubit
gates (which in turn were already decomposed into one-qubit gates and
CNOTs in the first part) and CNOTs.

The proof of this equivalence proceeds by considering the possible
actions on the third qubit depending on the state of the other two
qubits:

i) $|x_1\rangle=|0\rangle$.  In this case, the two CNOT gates become
inactive and so does the second controlled-$U$ gate.  We have two
possibilities: a) if $|x_2\rangle=|0\rangle$ then neither of the
remaining controlled gates operate and the net result is to leave
$|x_3\rangle$ unchanged; b) if $|x_2\rangle=|1\rangle$ then the effect
is now $U^{\dagger}U|x_3\rangle=|x_3\rangle$, as before.

ii) $|x_1 x_2\rangle=|10\rangle$.  Now the CNOT gates do operate on
the second qubit $|x_2\rangle$, and the second controlled-$U$ gate
acts on the third qubit.  However, the first $U$-gate is inactive.
Thus, the first CNOT gate changes the state of $|x_2\rangle$ to
$|1\rangle$ and this makes the $U^{\dagger}$-gate become operative.
Later, the action of the second CNOT brings the second qubit back to
$|0\rangle$.  Altogether, the final effect on $|x_3\rangle$ is to
yield $U U^{\dagger}|x_3\rangle=|x_3\rangle$, and remains unchanged
again.

iii) $|x_1 x_2\rangle=|11\rangle$.  In this case we need to produce
the action of $U^2$ on the third qubit.  Now, all the gates in
Fig.~\ref{q3universal} become operative and we make a sequential
counting of their effects.  As $|x_2\rangle=|1\rangle$, the first
$U$-gate does operate on the third qubit.  Next, the action of the
first CNOT gate sets $|x_2\rangle=|0\rangle$ so that the
$U^{\dagger}$-gate becomes inactive.  Then the second CNOT gate puts
the second qubit back to $|1\rangle$.  Altogether, the final effect on
$|x_3\rangle$ is to yield $U U|x_3\rangle=U^2|x_3\rangle$, as required.

Finally, we can always choose the initial matrix $U$ as the square
root of a unitary matrix, say $U^2=V$, such that the output in
Fig.~\ref{q3universal} is a C$^2V$-gate.  For instance, if we choose
$U=\ee^{\ii\pi/4} R_x(\half\theta)$  we reproduce the Deutsch gate
(\ref{qc27}).

Moreover, we can go on and provide a construction of an arbitrary
C$^nV$ transformation (useful in quantum algorithms) by extending the
construction in Fig.~\ref{q3universal} to an arbitrary number of
qubits.  For instance, for a controlled-$U^2$ gate of $4$ qubits we
would have another qubit line on Fig.~\ref{q3universal}b) and then the
construction holds by adding only a similar line to
Fig.~\ref{q3universal}a) so that the two CNOT gates become CCNOT
(C$^2$NOT) gates and the last C$^2U$ gate also picks up another
control qubit gate.  In general, for a $n$-qubit C$^{n-1}U^2$ gate
that has $n-1$ control qubits and one target qubit where $U^2$ acts,
the construction in Fig.~\ref{q3universal} is generalized by simply
using generalized C$^{n-2}$NOT gates with $n-2$ control qubits and a
last C$^{n-1}U$ gate with $n-1$ control qubits.  The proof of this
generalized construction follows straightforwardly.

3rd {\em Part}. Combining finally the results in Parts 1 and 2 with
the previuosly known construction of an arbitrary unitary matrix $U$
as  a product of two-level (not necessarily one-qubit) unitary
matrices of Reck et al. (1994), one can easily represent $U$ through
one-qubit  and CNOT gates, concluding this way  the proof that {\em
one-qubit gates plus CNOT is a set of elementary gates  for exact
universal computation} (Barenco et al., 1995).

So far we have only cared about the possibility of reconstructing a
generic quantum gate from a given set of gates.  The complexity of
these constructions, measured by the number of basic gates necessary
to achieve a certain gate simulation, is of great interest.

As an example of this issue, it is also interesting to count how many
elementary gates in ${\cal G}_{{\rm ex}}^{{\rm II}}$ are needed to
simulate a general C$^nU$ gate.  For instance, for a C$^2U$ gate the
first part of the proof yields 4 one-qubit gates and 2 CNOT's.  For a
generic controlled gate of $n$ control qubits C$^nU$, the second part
of the proof yields a quadratic dependence on $n$.  To see this, let
us denote by $C_{n}$ the cost of simulating a C$^nU$ gate.  From the
first part of the proof we know that the cost of simulating the $U$-
and $U^{\dagger}$-gates in Fig.~\ref{q3universal} is order
$\Theta(1)$;\footnote{One writes $y=\Theta(x)$ to denote that both
$y=O(x)$ and $x=O(y)$ hold simultaneously.}  on the other hand, it is
not difficult to show that the cost of the two C$^{n-1}$NOTs is
$\Theta(n+1)$ (Barenco et al., 1995).   The cost of the generalized
C$^{n-1}U$ gate is $C_{n-1}$, by recursive application of the
recursive construction.  Altogether, the cost of a gate satisfies a
recursion relation like this
\begin{equation}
C_{n} = C_{n-1} + \Theta(n+1),
\label{qc36b}
\end{equation}
whose solution yields $C_{n}=\Theta((n+1)^2)$.

What is the size (number of gates) for exactly simulating an arbitrary
gate of $n$ qubits in $U(2^n)$?  Barenco et al.  (1995) showed that
using the universal set ${\cal G}_{{\rm ex}}^{{\rm II}}$ this cost is
$O(n^3 4^n)$;\footnote{The factor $n^3$ arises from the cost  $O(n)$
to bring a generic two-level matrix to a C$^{n-1}$-unitary matrix
which in turn costs $O(n^2)$. The dominant factor $4^n$ just counts
asymptotically the maximum number of two-level unitary factors in the
Reck et al. decomposition.}  Knill (1995) reduced this bound to $O(n
4^n)$.

However, we are also interested in the efficiency of the approximate
simulation of a generic gate.  The universality property of a set of
gates ${\cal G}_{{\rm ap}}$ means that, given an arbitrary quantum
gate $U\in {\rm U}(2^n)$ and $\epsilon >0$, we can always devise an
approximate quantum gate $U'$ generated by ${\cal G}_{{\rm ap}}$ such
that $d(U,U')<\epsilon$. The errors scale up linearly with the number
of gates: given $N$ gates $U_i$ and their approximations $U_i^\prime$,
then the telescopic identity $U_1...U_N-U_1^\prime...U_N^\prime=
\sum_{1\leq k\leq N} U_1^\prime...U_{k-1}^\prime(U_k-U_k^\prime)
U_{k+1}...U_N$ yields immediately
$||U_1...U_N-U_1^\prime...U_N^\prime||<N\epsilon$.

This construction can be done efficiently using
$\text{poly}(1/\epsilon)$ gates from the universal set (Lloyd, 1995;
Preskill, 1998).  Although we will not prove it, the underlying reason
is simple: 1/ any universal set generates unitary matrices having
eigenvalues with phases incommensurate relative to $\pi$; 2/ if
$\theta/\pi\in\R$ is irrational, then the integral powers $\ee^{\ii
k\theta}, k\in\Z$ are dense in the unit circle $S_1$, and given
$\epsilon>0$, any $\ee^{\ii \alpha}\in S_1$ is within a distance
$\epsilon$ of some $\ee^{\ii n\theta}$ with $n=O(1/\epsilon)$.

As a matter of fact, we can do much better than approximating a given
$n$-qubit gate with circuits of size ${\rm poly}(1/\epsilon)$ in the
universal set ${\cal G}_{{\rm ap}}$.  A theorem of Solovay and Kitaev
shows that it is possible an exponentially improved approximation
(Solovay, 1995; Kitaev, 1997): Let ${\cal G}_{{\rm ap}}$ be an
arbitrary finite universal set of gates, i.e. ${\cal G}_{{\rm ap}}$
generates a dense subset in ${\rm U}(2^n)$.  Then, any matrix  $U\in
{\rm U}(2^n)$ can be approximated within an error $\epsilon$ by a
product of $O({\rm poly}({\rm log}(1/ \epsilon))$ gates in ${\cal
G}_{{\rm ap}}$ (more precisely, $O({\rm poly}({\rm
log}(1/\epsilon))=O({\rm log}^c(1/\epsilon))$, with $c\approx 2$).
The idea of the proof is to construct thinner and thinner nets of
points in ${\rm U}(2^n)$ by taking group commutators of unitaries in
previous nets.  It turns out that this way the width of the resulting
nets decreases exponentially.

Finally, when the above Solovay-Kitaev theorem is combined with the
complexity for exactly simulating gates with ${\cal G}_{{\rm
ex}}^{{\rm II}}$, and the linearity of the error propagation with the
number of gates, it immediately follows that any unitary gate
$U\in{\rm U}(2^n)$ can be approximated to within error $\epsilon$ with
$O(n 4^n \log^c(n 4^n/\epsilon))$ gates in any ${\cal G}_{{\rm
ap}}$. Note that this represents an exponential complexity in the
number of qubits, i.e. most gates will be hard to simulate.

\subsubsection{Arithmetics with QCs}

The universality theorem of elementary quantum gates is a central
result in the theory of quantum computation for it reduces the
implementation of conditional quantum logic to a small set of simple
operations.  However, with a computer we are typically interested in
doing arithmetic operations and thus we need to know how to perform
quantum arithmetics with universal quantum gates.  Vedral, Barenco and
Ekert (1995) provided efficient ways of doing arithmetic operations
such as addition, multiplication and modular exponentiation building
on the Toffoli gate.  The key point in their constructions is that we
have to preserve the coherence of quantum states and make those
operations reversible, unlike in a classical computer.  For instance,
the AND operation of Sec.~\ref{sec8D:level2} can be made reversible
by embedding it into a Toffoli gate (Ekert, Hayden and Inamori, 2000):
setting the third qubit to zero in (\ref{qc26}) we get
\begin{equation}
U_{{\rm CCNOT}} |x_1, x_2, x_3=0\rangle = |x_1, x_2, x_1 \wedge
x_2\rangle.
\label{qc37}
\end{equation}

\begin{figure}[ht]
\psfrag{x}[Bc][Bc][0.75][0]{$|x_1\rangle$}
\psfrag{y}[Bc][Bc][0.75][0]{$|x_1\rangle$}
\psfrag{z}[Bc][Bc][0.75][0]{$|x_2\rangle$}
\psfrag{t}[Bc][Bc][0.75][0]{$|x_1\oplus x_2\rangle$: Sum}
\psfrag{l}[Bc][Bc][0.75][0]{$|0\rangle$}
\psfrag{m}[Bc][Bc][0.75][0]{$|x_1 x_2\rangle$}
\includegraphics[width=4.5 cm]{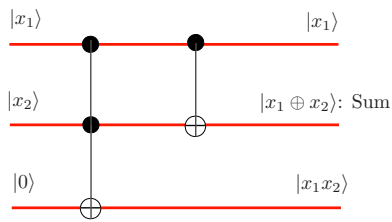}
\caption{The quantum addition from a Toffoli gate.}
\label{qadder}
\end{figure}

Similarly, the quantum addition can be embedded into a Toffoli gate as
shown in Fig.~\ref{qadder} with the help of a CNOT gate for the first
two qubits.  The result of the addition is stored in the second qubit.

A quantum multiplication can be implemented in a similar fashion and
also the exponentiation modulo $N$ (Vedral, Barenco and Ekert, 1995).
This latter operation is central in the Shor algorithm
(Sec.~\ref{sec10D:level2}).

\begin{figure}[ht]
\psfrag{u}[Bc][Bc][0.75][0]{$|x_1\rangle$}
\psfrag{v}[Bc][Bc][0.75][0]{$|x_1\rangle$}
\psfrag{m}[Bc][Bc][0.75][0]{$|x_m\rangle$}
\psfrag{w}[Bc][Bc][0.75][0]{$|x_{m+1}\rangle$}
\psfrag{s}[Bc][Bc][0.70][0]{$|x_{m+1} \oplus
f(x_1,\ldots,x_m)\rangle$} 
\psfrag{p}[Bc][Bc][1][0]{$U_f$}
\includegraphics[width=6cm]{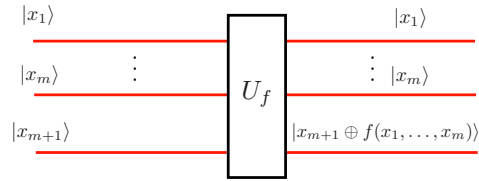}
\caption{A gate for function evaluation.}
\label{qfunction}
\end{figure}

Another important operation that must be implemented in a quantum
circuit is the evaluation of a function $f$.  This must again comply
with the requisite of reversibility, which is accomplished with a
$U_f$-gate as shown in Fig.~\ref{qfunction}, where $U_f$ is a unitary
transformation that implements the action of $f$ on certain qubits of
the circuit.  In this figure the box representing the evaluation of
the gate is a kind of {\em black box}, also called {\em quantum
oracle}, which represents the way in which we call or evaluate the
function $f$.  These evaluations are also called {\em queries}.

Reversible implementation of $f$ requires to split the quantum
register storing an initial state $|\Psi_0\rangle$ into two parts: the
{\em source register} and the {\em target register}, namely,
\begin{equation}
|\Psi_0\rangle = |\Psi_{\rm s}\rangle \otimes |\Psi_{\rm t}\rangle,
\label{qc38}
\end{equation}

\noindent where $|\Psi_{\rm s}\rangle$ stores the input data for the
computation and $|\Psi_{\rm t}\rangle$ stores the output data, that
is, the results of the quantum evolution or application of logic gates.

Thus, in order to implement a Boolean function $f: \{ 0,1\}^m \to \{
0,1\}$ in a quantum circuit we need the action of a unitary gate $U_f$
acting on the target register as follows
\begin{equation}
\begin{split}
&U_f|x_1x_2\ldots x_m\rangle_{\rm s} |x_{m+1}\rangle_{\rm t} = \\
&|x_1x_2\ldots x_m\rangle_{\rm s} |x_{m+1}\oplus
f(x_1,x_2,\ldots,x_m)\rangle_{\rm t}.
\label{qc39}
\end{split}
\end{equation}

\noindent Why is it not possible to evaluate directly the action of
$f$ by a unitary operation that evolves $|x\rangle$ into
$|f(x)\rangle$?  The answer lies in unitarity of computation: we know
that orthonormality is preserved under unitary transformations, thus
if $f$ is not a one-to-one mapping then two states $|x_1x_2\ldots
x_m\rangle$ and $|x_1^\prime x_2^\prime\ldots x_m^\prime\rangle$ that
are initially orthonormal could evolve into two non-orthonormal
states, say $|f(x_1,x_2,\ldots, x_m)\rangle = |f(x_1^\prime,
x_2^\prime,\ldots, x_m^\prime)\rangle$.

In the following we shall omit for simplicity the subscripts denoting
source and target registers.

\section{Quantum Algorithms}
\label{sec10:level1}

In this section we present a survey of the most representative quantum
algorithms to date, named after Deutsch-Jozsa, Simon, Grover and Shor,
without entering the many spinoffs and ramifications that they have
led to (Berstein and Vazirani, 1993; Hogg, 1998; Kitaev, 1995; etc.).
We also use these quantum algorithms to emphasize and see in action
the main ideas concerning the principles of quantum computation
introduced in Sec.~\ref{sec9:level1}.

Due to space constraints, we have left out some interesting
developments like {\em quantum clock synchronization}\footnote{A way
to make two atomic clocks start ticking at once.  This can also be
considered as an application of the quantum Fourier transform (see
Sec.~\ref{sec10D:level2} for quantum phase estimation (Cleve et al.,
1998)} (Chuang, 2000; Jozsa et al., 2000) and {\em quantum games}
(Meyer, 1999; Eisert, Wilkens and Lewenstein, 1999)\footnote{Quantum
games appear so far to be more related to quantum communication
protocols (Sec.~\ref{sec3:level1}) or to applications of the above
quantum algorithms.}.

The merging of Quantum Mechanics and Information Theory has proved to
be very fruitful.  One of the products of this is the discovery of
quantum algorithms that outperform classical ones.  It is appealing to
think that the outcome of this merging is the fact that we can take
classical algorithms and devise quantization processes in order to
discover new modified quantized versions of classical algorithms.  By
quantizing a classical algorithm it is simply meant the possibility of
using quantum bits in a quantum computer as oppossed to the classical
bits, and all the consequences thereof.  This way of thinking is
reminiscent of a well-known procedure of studying a quantum system by
starting with its classical analogue and making a quantization of it,
using for instance Dirac's prescription.  One instance of this
proposal is Shor's algorithm (Sec.~\ref{sec10D:level2}).  In fact,
Shor's algorithm relies on its ability to find the period of a simple
function in number theory.  The known classical algorithms for this
task are inefficient because, as mentioned in Sec.~\ref{sec6:level1},
they have subexponential complexity in the input length (unless hard
information is supplied aside).  However, when qubits are used to
implement the common algorithm (we quantize it in our language), then
the principles of quantum computation shorten the task to polynomial
time.  Of this drastic improvement are responsible the peculiar
properties of the discrete quantum Fourier transform
(Sec.~\ref{sec10D:level2}).

Shor's algorithm also illustrates another common feature of the
quantum algorithms known so far: they are best suited to study global
properties of a function or a sequence as a whole, like finding the
period of a function, the median of a sequence, etc., and not
individual details.  When the value of the function is needed for a
particular choice of the argument, no real advantage is gained: one
has to extract it from the quantum superposition and this may
generally require measuring many times on the output to compensate the
low probability, exponentially small in the register length, of
getting the desired result.

Let us point out that it is possible to give a unified picture of most
of the forthcoming algorithms in terms of  the {\em hidden subgroup
problem}: to find a generating set for a subgroup $K$ of a finitely
generated group $G$, given a function $f: G \rightarrow X$, where $X$
is a finite set and $f$ is constant and distinct on the
$K$-cosets. Some instances of this problem are the Deutsch-Jozsa,
Simon and Shor algorithms (Mosca and Ekert, 1999; Boneh and Lipton,
1995).  Likewise, one may profitably view the  quantum computation
process as a multiparticle quantum interference (Cleve et al., 1998).
However, we have adhered to a more traditional and historical pathway
of presenting these quantum algorithms.

\subsection{Deutsch-Jozsa Algorithm}
\label{sec10A:level2}

This is the quantum algorithm first introduced by Deutsch (1985),
providing an explicit and concrete example of how a quantum computer
can beat a classical computer.  Later, it was extended to more complex
situations by Deutsch and Jozsa (1992).  We shall present first an
improved version (Cleve et al., 1998) of this algorithm for the
simplest case of a Boolean function of a single qubit.

Suppose we are given an oracle which upon request computes a function
$f: \{ 0,1\}^n \to \{ 0,1\}$.  No other information on $f$ is
available, just the {\em promise} or assumption that $f$ is either
constant (i.e. $\forall x_1,x_2 \in \{ 0,1\}^n, f(x_1)=f(x_2)$) or
{\em balanced} (in the sense that $\#f^{-1}(0)=\# f^{-1}(1)$, i.e. the
numbers of arguments mapping to 0 and to 1 are equal).  The problem is
to ascertain whether $f$ is constant or balanced with as few queries
to the oracle as possible.

The result of the DJ algorithm is that we only need one query or
function evaluation to determine the nature of $f$, while classically
$2^{n-1}+1$ consultations would be necessary in the worst case.

\begin{figure}[ht]
\psfrag{u}[Bc][Bc][1][0]{$|0\rangle$}
\psfrag{v}[Bc][Bc][1][0]{measurement}
\psfrag{w}[Bc][Bc][1][0]{$|1\rangle$} \psfrag{h}[Bc][Bc][1][0]{$U_H$}
\psfrag{p}[Bc][Bc][1][0]{$U_f$} \psfrag{s}[Bc][Bc][1][0]{${1\over
\sqrt{2}} \sum_{x=0,1} (-1)^x|x\rangle$} \includegraphics[width=7
cm]{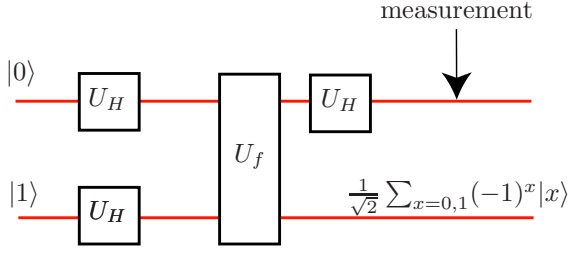}
\caption{Quantum circuit for the Deutsch-Jozsa algorithm.}
\label{DJalgo}
\end{figure}

Let us see this first when $n=1$.  Now $f$ is balanced iff $f(0)\neq
f(1)$, and thus the promise is worthless.  The quantum circuit in
Fig.~\ref{DJalgo} implements the DJ algorithm, and embodies the
following steps:

\smallskip
\noindent {\em Step 1.} An initial quantum register is prepared with
two qubits in the state $|\Psi_1\rangle:=|01\rangle$.

\noindent {\em Step 2.} The Hadamard gate (\ref{qc21}) is applied
bit-wise to this quantum register, producing the state
\begin{equation}
|\Psi_2\rangle:=U_H |0\rangle\otimes U_H |1\rangle =
\half(|0\rangle+|1\rangle)\otimes (|0\rangle-|1\rangle).
\label{qc40}
\end{equation}

\noindent {\em Step 3.} We query the $f$-oracle with the state
$|\Psi_2\rangle$, and get the answer $\ket{\Psi_3}:=U_f\ket{\Psi_2}$.
Using (\ref{qc39}) we readily find
\begin{equation}
\begin{split}
\ket{\Psi_3}&=U_f \half\sum_{x=0,1}|x\rangle (|0\rangle-|1\rangle) \\
&= \half\sum_{x=0,1} (-1)^{f(x)} |x\rangle (|0\rangle-|1\rangle).
\end{split}
\label{qc41}
\end{equation}

\noindent {\em Step 4.} The Hadamard gate is applied again to the
first qubit, what yields
\begin{equation}
\begin{split}
&\ket{\Psi_4}:={1\over 2}\sum_{x=0,1} (-1)^{f(x)} (U_H |x\rangle)
(|0\rangle-|1\rangle) \\ &={1\over 2^{3/2}} \sum_{x=0,1}
[(-1)^{f(x)}|0\rangle + (-1)^{x+f(x)}|1\rangle] \otimes
(|0\rangle-|1\rangle).
\end{split}
\label{qc42}
\end{equation}

\noindent {\em Step 5.} Finally, we measure (in the computational
basis) the first qubit (the second qubit plays no role anymore).
There are two possibilities: i) either $f$ is constant, and then the
first-qubit amplitude of $|1\rangle$ in (\ref{qc42}) vanishes and we
measure $|0\rangle$ with certainty; ii) or $f$ is not constant and
consequently it is balanced, in which case it is the amplitude of
$|0\rangle$ in (\ref{qc42}) which vanishes and we measure $|1\rangle$
with certainty.

Therefore, with this DJ algorithm we only need to call once the
function in order to determine whether it is constant or balanced.

Let us point out how the peculiarities of quantum mechanics enter in
the algorithm and provide its power.  In step 2 it is possible to
prepare a superposition of all the basis states using the Hadamard
gates which have no classical analogue.  In step 3 we evaluate the
function on all the basis states at one go.  However, this is not
enough and we need to use interference of the quantum amplitudes in
step 5 to discriminate between the two possibilities we were searching
for.  This is a simple manifestation of the idea of using constructive
interference to distill the desired results as was already advanced in
Sec.~\ref{sec9A:level2} (see Table~\ref{table3}).

\begin{figure}[ht]
\psfrag{u}[Bc][Bc][0.7][0]{$|0\rangle$}
\psfrag{v}[Bc][Bc][0.7][0]{measurement}
\psfrag{w}[Bc][Bc][0.7][0]{$|1\rangle$}
\psfrag{h}[Bc][Bc][0.7][0]{$U_H$} \psfrag{p}[Bc][Bc][0.7][0]{$U_f$}
\psfrag{s}[Bc][Bc][0.7][0]{${1\over \sqrt{2}} \sum_{x=0,1}
(-1)^x|x\rangle$} \includegraphics[width=6 cm]{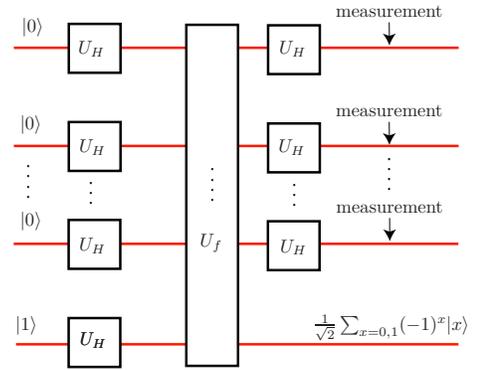}
\caption{Extended Deutsch-Jozsa algorithm.}
\label{DJalgo-gen}
\end{figure}

The extension of the DJ algorithm to a function of $n$ qubits $f: \{
0,1\}^n \to \{ 0,1\}$ constrained to be either constant or balanced
can be done with the help of the quantum circuit shown in
Fig.~\ref{DJalgo-gen}.  Following this circuit we can extend the
previous 5 steps immediately.  We prepare a source register with $n$
qubits initialized to $\ket{{0}}$ and a target register with one qubit
initialized to $|1\rangle$.  With ${x}$ we denote the integer ${
x}:=\sum_{i=0}^{n-1} x_i 2^i$ associated to the string of bits
$x_{n-1}\ldots x_1x_0$, and $\ket{{x}}:=\ket{x_{n-1}\ldots x_1x_0}$.

Let $\ket{\Phi_1}:=\ket{{0}}\ket{1}$. After the bit-wise application
of the Hadamard gate to $\ket{\Phi_1}$ we find
\begin{equation}
\begin{split}
&\ket{\Phi_2}:=U_H^{\otimes (n+1)}\ket{\Phi_1}=(U_H |0\rangle) (U_H
|0\rangle) \ldots (U_H |0\rangle) (U_H |1\rangle) \\ & = {1\over
2^{n/2}} \sum_{{x=0}}^{2^n-1}|{x}\rangle {1\over \sqrt{2}}
\sum_{y=0,1} (-1)^y |y\rangle.
\end{split}
\label{qc43}
\end{equation}

Using (\ref{qc39}), the function evaluation on $\ket{\Phi_2}$ yields
the following state
\begin{equation}
\ket{\Phi_3}:={1\over 2^{n/2}} \sum_{{x=0}}^{2^n-1} (-1)^{f({x})}
|{x}\rangle \ {1\over \sqrt{2}} \sum_{y=0,1} (-1)^y |y\rangle.
\label{qc44}
\end{equation}

In the next step we apply again the Hadamard gates but only on the $n$
source qubits.  After some algebra we arrive at the final state
$\ket{\Phi_4}$ given by
\begin{equation}
\begin{split}
&\ket{\Phi_4}:=(U_H^{\otimes n}\otimes 1)\ket{\Phi_3}\\ &={1\over
2^{n}} \sum_{{x=0}}^{2^n-1} \sum_{{ x}'={0}}^{2^n-1} (-1)^{{x}\cdot
{x}' + f({ x})}|{x}'\rangle \ {1\over \sqrt{2}} \sum_{y=0,1} (-1)^y
|y\rangle,
\end{split}
\label{qc45}
\end{equation}

\noindent where ${x}\cdot {x}':=\sum_{i=0}^{n-1} x_i x'_i\in \Z_2$.

If $f$ is constant, then it produces an overall sign factor in
(\ref{qc45}), and after the double summation only the state
$\ket{{x}'} = \ket{{0}}$ survives.  On the contrary, if $f$ is
balanced, then the same reasoning shows that such state has zero
amplitude in $\ket{\Phi_4}$.  In summary, only when all the final
source qubits are $|0\rangle$ the function is constant; otherwise, it
is balanced.

Thus, measuring the state of the source qubits we can determine the
nature of $f$ with certainty.

This final measurement step allow us to take advantage of the
interference among amplitudes obtained in previous stages.

A single query to the function black box has proved sufficient.
However, with the classical algorithms known so far we would require a
number of $2^{n-1}+1$ function evaluations (in the worst case) to
determine with certainty which type of function $f$ is.  This
represents an exponential speed-up for this quantum algorithm.

Let us point out that classically, given any $1>\epsilon >0$, it is
also possible to devise an efficient probabilistic algorithm such that
running it a large enough number of times $M$ (independent of the
input length $n$) will determine whether any given function $f$ is
constant or balanced, with error probability $<\epsilon$.  This is
the procedure: the function $f$ is evaluated for $M$ random choices of
the argument.  When any two of the values differ, then we know that
$f$ is balanced.  However, when all values are equal then  the error
probability in claiming that $f$ is constant will be less than
$2^{-M}$. Thus, it suffices to choose $M$ such that $2^{-M}
<\epsilon$. In this sense, the quantum DJ algorithm is not such an
impressive improvement over classical algorithms.

\subsection{Simon Algorithm}
\label{sec10B:level2}

Simon's algorithm (1994) uses several tools of the DJ algorithm.  It
deals with a vector-valued Boolean function $f: \{ 0,1\}^n \to \{
0,1\}^n$ which is constrained by the following condition or promise:
There exists a non-null vector ${p}\in \{ 0,1\}^n$, called the {\em
period} of $f$, such that $f({x})=f({y})$ if and only if either
${x}={y}$ or ${x}={y}\oplus{p}$.  Note that such an $f$ is forcefully
a 2-to-1 function.

This algorithm finds the period ${p}$ after a number $O(n)$ of
function evaluations, while the known classical algorithms would
require an exponential number of queries.

\begin{figure}[ht]
\psfrag{u}[Bc][Bc][0.7][0]{$|0\rangle$}
\psfrag{v}[Bc][Bc][0.7][0]{measure}
\psfrag{w}[Bc][Bc][0.7][0]{$|0\rangle$}
\psfrag{h}[Bc][Bc][0.7][0]{$U_H$} 
\psfrag{p}[Bc][Bc][0.7][0]{$U_f$}
\includegraphics[width=6 cm]{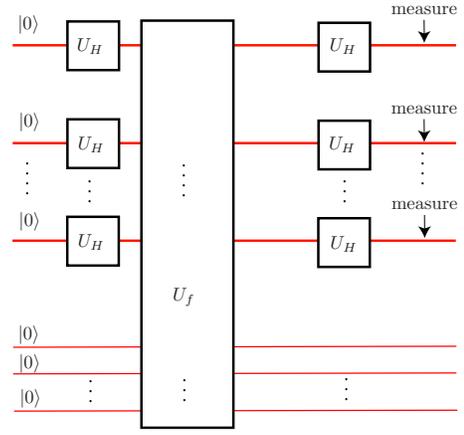}
\caption{Quantum circuit for Simon's algorithm.}
\label{simon}
\end{figure}

The steps in Simon's algorithm  can be seen in Fig.~\ref{simon}.  Both
the source and target registers have $n$ qubits each.  The algorithm
proceeds as  follows:\footnote{Sometimes one introduces, for
didactical purposes, a further step  in which the target qubits are
measured (Jozsa, 1997).}

\noindent {\em Step 1.} The quantum registers are initialized to the
state $\ket{\Psi_1}:=\ket{{0}}\ket{{0}}= |00 \ldots 0\rangle|00 \ldots
0\rangle$.

\noindent {\em Step 2.} The Hadamard gate (\ref{qc21}) is applied
bit-wise to the source register, producing the state
\begin{equation}
\ket{\Psi_2}:=(U_H |0\rangle) \ldots (U_H |0\rangle) |{ 0}\rangle =
{1\over 2^{n/2}} \sum_{{x=0}}^{2^n-1}|{ x}\rangle |{0}\rangle.
\label{qc47}
\end{equation}

\noindent {\em Step 3.} The vector-valued function $f$ is evaluated on
the target qubits by applying the gate $U_{f}$.  Using (\ref{qc39}) we
readily find the entangled state (Sec.~\ref{sec3:level1})
\begin{equation}
\ket{\Psi_3}:= U_{f}\ket{\Psi_2}={1\over 2^{n/2}}
\sum_{{x=0}}^{2^n-1}|{x}\rangle |f({x})\rangle.
\label{qc48}
\end{equation}

\noindent {\em Step 4.} A further application of the Hadamard gates to
the source qubits results in the state
\begin{equation}
\begin{split}
&\ket{\Psi_4}:={1\over 2^n}\sum_{x=0}^{2^n-1}\sum_{y=0}^{2^n-1}
(-1)^{x\cdot y}\ket{y}\ket{f(x)}\\ &={1\over
2^{n+1}}\sum_{x,y=0}^{2^n-1} [(-1)^{x\cdot y}+(-1)^{(x\oplus p)\cdot
y}] \ket{y}\ket{f(x)}.
\label{qc50}
\end{split}
\end{equation}

\noindent Note that only those qubit states $\ket{y}$ such that
${p}\cdot y=0$ enter with non-vanishing amplitudes in $\ket{\Psi_4}$.
The remaining ones are washed out by destructive interference.

\noindent {\em Step 5.} An ideal measurement of the source qubits (in
the computational basis) will necessarily yield a state $\ket{y}$ such
that $p\cdot y=0$ with probability $2^{-(n-1)}$.

\noindent {\em Step 6.} Repeating the previous steps $M$ times we will
get $M$ vectors $y_{(i)}$ such that
\begin{equation}
{p}\cdot y_{(i)} = 0, \ i=1,\ldots, M.
\label{qc52}
\end{equation}

\noindent Solving this linear system with the Gaussian elimination
algorithm will yield the period $p$ with probability large enough
provided $M=O(n)$.

The cost of Simon's algorithm is $O(n^2+nC_f(n))$, where $C_f(n)$ is
the cost of evaluating the function $f$ on inputs of length $n$. The
term $n^2$ is just the cost of the Gaussian elimination over $\Z_2$.

However, a classical blind search would require $2^{n-1}+1$ calls to
the oracle in the worst case, and on the average a number $O(2^{n/2})$
of function evaluations (Shor, 2000).  Thus, Simon's algorithm
represents an exponential speed-up.

We note in passing that Simon's algorithm resorts to a classical
algorithm (Gaussian elimination) to finish off the job.  We shall find
another interesting collaboration between quantum and classical
procedures in Shor's algorithm.

\subsection{Grover Algorithm}
\label{sec10C:level2}

The previous quantum algorithms show explicitly some instances where a
quantum computer beats a classical computer, as was advanced in
Sec.~\ref{sec8A:level2} devoted to quantum Turing machines.  However,
they also present several drawbacks:

i) {\em utility}: it is not clear what they are useful for in
practical applications.

ii) {\em structure}: the searched functions are constrained to comply
with certain promises.  These are called {\em structured problems}.
Thus, we may feel as if those constraints quantumly conspire in favor
of the DJ and Simon algorithms.

Grover's algorithm (1996, 1997) represents an example of {\em
unstructured problem}: one in which no assumptions are made about the
function $f$ under scrutiny.  Thus, we can contrast classical and
quantum algorithms on equal footing.  Although it came after Shor's
algorithm (1994), we present it first for it is quite related to the
previous algorithms.

The algorithm by Grover solves the problem of searching an element in
a list of $N$ unsorted elements.  For instance, searching a database
like a telephone book when we know the number but not the person's
name.  When the size of the database becomes very large it is known to
be one of the basic problems in computational science (Knuth, 1975).
The utility of one such algorithm is guaranteed.  Classically, one may
devise many strategies to perform that search, but if the elements in
the list are randomly distributed, then we shall need to make $O(N)$
trials in order to have a high confidence of finding the desired
element.  Grover's quantum searching algorithm takes advantage of the
quantum mechanical properties to perform the searching problem with an
efficiency of order $O(\sqrt{N})$ (Grover, 1996; 1997).

Let us state the searching problem in terms of a list ${\cal
L}[0,1,\ldots, N-1]$ with a number $N$ of unsorted elements.  We shall
denote by $x_0$ the marked element in ${\cal L}$ that we are looking
for.  The quantum mechanical solution of this searching problem goes
through the preparation of a quantum register in a quantum computer to
store the $N$ items of our list.  This will allow exploiting quantum
parallelism.  Thus, let us assume that our quantum registers are made
of $n$ source qubits so that $N=2^n$.  We shall also need a target
qubit to store the output of function evaluations or calls.

To implement the quantum search we need to construct a unitary
operation that discriminates between the marked item $x_0$ and the
rest.  The following function
\begin{equation}
f_{x_0}(x) := \begin{cases} 0 & {\rm if} \ x \neq x_0, \\ 1 & {\rm if}
\ x = x_0,
\end{cases}
\label{qc53}
\end{equation}

\noindent and its corresponding unitary operation (\ref{qc39})
\begin{equation}
U_{f_{x_0}}|x\rangle |y\rangle = |x\rangle |y \oplus f_{x_0}(x)\rangle
\label{qc54}
\end{equation}

\noindent will do the job.  We shall need to count how many
applications of this operation or oracle calls are needed to find the
item.  The rationale behind the Grover algorithm is: 1/ to start with
a quantum register in a state where all the computational basis states
are equally present; 2/ to apply several unitary transformations to
produce an output state in which the probability of catching the
marked state $|x_0\rangle$ is large enough.

\begin{figure}[ht]
\psfrag{u}[Bc][Bc][0.7][0]{$|0\rangle$}
\psfrag{v}[Bc][Bc][0.7][0]{measure}
\psfrag{w}[Bc][Bc][0.7][0]{$|1\rangle$}
\psfrag{h}[Bc][Bc][0.7][0]{$U_H$}
\psfrag{p}[Bc][Bc][0.7][0]{$U_{f_{x_0}}$}
\psfrag{q}[Bc][Bc][0.7][0]{$U_{f_{0}}$}
\psfrag{x}[Bc][Bc][0.7][0]{$U_{x_{0}}$}
\psfrag{0}[Bc][Bc][0.7][0]{$-D$}
\includegraphics[width=6cm]{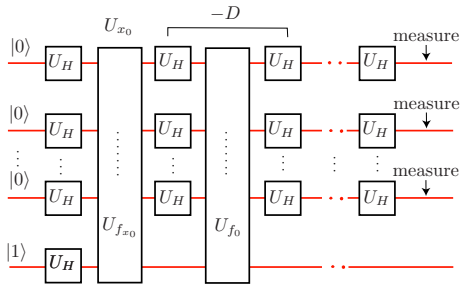}
\caption{The quantum circuit (up to an irrelevant global sign factor) 
for Grover's  algorithm.}
\label{qgrover}
\end{figure}

We present now the steps in Grover's algorithm, with the quantum
circuit shown in Fig.~\ref{qgrover}.

\noindent {\em Step 1.} Initialize the quantum registers to the state
$\ket{\Psi_1}:=|00 \ldots 0\rangle|1\rangle$.

\noindent {\em Step 2.} Apply bit-wise the Hadamard one-qubit gate
(\ref{qc21}) to the source register, so as to produce a uniform
superposition of basis states in the source register, and also to the
target register:
\begin{equation}
\ket{\Psi_2}:=U_{\rm H}^{\otimes (n+1)}\ket{\Psi_1} = {1\over
2^{(n+1)/2}} \sum_{{x}={0}}^{2^n-1}|{x}\rangle \sum_{y=0,1} (-1)^y
|y\rangle.
\label{qc55}
\end{equation}

\noindent {\em Step 3.} Apply now the operator $U_{f_{{x}_0}}$:
\begin{equation}
\begin{split}
\ket{\Psi_3} &:= U_{f_{{x}_0}}\ket{\Psi_2} \\ & =
2^{-(n+1)/2}\sum_{{x}={0}}^{2^n-1} (-1)^{f_{{x}_0}({x})} |{x}\rangle
\sum_{y=0,1} (-1)^y |y\rangle.
\label{qc57}
\end{split}
\end{equation}

\begin{figure}[ht]
\psfrag{u}[Bc][Bc][0.7][0]{uniform} \psfrag{x}[Bc][Bc][0.7][0]{$x_0$}
\psfrag{0}[Bc][Bc][0.7][0]{$0$} \psfrag{1}[Bc][Bc][0.7][0]{$1$}
\psfrag{2}[Bc][Bc][0.7][0]{$\ldots$}
\psfrag{3}[Bc][Bc][0.7][0]{$\ldots$}
\psfrag{N}[Bc][Bc][0.65][0]{$N-1$} 
\includegraphics[width=6cm]{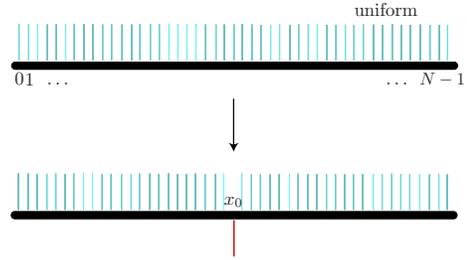}
\caption{Schematic representation of Grover's operator $U_{{x}_0}$ in
(\ref{qc56}).}
\label{qcomb}
\end{figure}

Let $U_{{x}_0}$ be the operator defined by
\begin{equation}
U_{{x}_0} |{x}\rangle := (1 - 2 |{x}_0\rangle \langle
{x}_0|)|{x}\rangle=
\begin{cases}
-|{x}_0\rangle & {\rm if} \ {x}={x}_0, \\ |{x}\rangle & {\rm if} \ {x}
\neq {x}_0,
\end{cases}
\label{qc56}
\end{equation}

\noindent that is, it flips the amplitude of the marked state leaving
the remaining source basis states unchanged.  Grover presents this
operator graphically as in Fig.~\ref{qcomb}, with a sort of ``quantum
comb'' where the spikes denote the uniform amplitudes of state
(\ref{qc55}) and the action of $U_{{x}_0}$ is to flip over the spike
corresponding to the marked item.

\noindent We realize that the state in the source register of
(\ref{qc57}) equals precisely the result of the action of $U_{{x}_0}$,
i.e.
\begin{equation}
\begin{split}
\ket{\Psi_3}=([1 - 2 |{x}_0\rangle \langle {x}_0|]\otimes
1)\ket{\Psi_2}.
\label{qc57a}
\end{split}
\end{equation}

\begin{figure}[ht]
\psfrag{u}[Bc][Bc][0.7][0]{average} 
\psfrag{x}[Bc][Bc][0.7][0]{$x_0$}
\psfrag{0}[Bc][Bc][0.7][0]{$0$} 
\psfrag{1}[Bc][Bc][0.7][0]{$1$}
\psfrag{2}[Bc][Bc][0.7][0]{$\ldots$}
\psfrag{3}[Bc][Bc][0.7][0]{$\ldots$}
\psfrag{N}[Bc][Bc][0.65][0]{$N-1$}
\includegraphics[width=6cm]{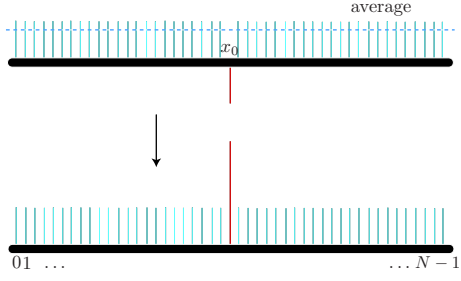}
\caption{Schematic representation of Grover's operator $D$ in
(\ref{qc58}). The dashed line represents the mean amplitude.}
\label{qaverage}
\end{figure}

\noindent {\em Step 4.} Apply next the operation $D$ known as {\em
inversion about the average} (Grover, 1996; 1997).  This operator is
defined as follows
\begin{equation}
D := -(U_H^{\otimes n}\otimes I) U_{f_0} (U_H^{\otimes n}\otimes I),
\label{qc58}
\end{equation}

\noindent where $U_{f_0}$ is the operator in (\ref{qc57}) for ${x}_0 =
{0}$.  The effect of this operator on the source qubits is to
transform $\sum_x \alpha_x\ket{x}\mapsto \sum_x(-\alpha_x+
2\ve{\alpha})\ket{x}$, where $\ve{\alpha}:=2^{-n}\sum_x\alpha_x$ is
the mean of the amplitudes, so its net effect is to amplify the
amplitude of $\ket{x_0}$ over the rest.  This is graphically
represented in Fig.~\ref{qaverage} (Grover, 1996; 1997).

\noindent {\em Step 5.} Iterate steps 3 and 4 a number of times $m$.

\noindent {\em Step 6.} Measure the source qubits (in the
computational basis).  The number $m$ is determined such that the
probability of finding the searched item $x_0$ is maximal.

The basic component of the algorithm is the quantum operation encoded
in steps 3 and 4 which is repeatedly applied to the uniform state
$\ket{\Psi_2}$ in order to find the marked element.

Although this procedure resembles the classical strategy, Grover's
neatly designed operation enhances by constructive interference of
quantum amplitudes (see Table~\ref{table3}) the presence of the marked
state one looks for.

It is possible to give a more general formulation to the operators
entering steps 3 and 4 of the algorithm (Galindo and Martin-Delgado,
2000).  To this end it is sufficient to focus on the source qubits and
introduce the following definitions:

{\em i)} A {\em Grover operator} $G$ is any unitary operator with at
most two different eigenvalues; i.e., $G$ a linear superposition of
two orthogonal projectors $P$ and $Q$:
\begin{equation}
G = \alpha P + \beta Q, \; \; P^2 = P, \; Q^2 = Q, \; \; P + Q = 1,
\label{qc59}
\end{equation}

\noindent where $\alpha, \beta \in \C$ are complex numbers of unit
norm.

{\em ii)} A {\em Grover kernel} $K$ is the product of two Grover
operators:
\begin{equation}
K = G_2 G_1.
\label{qc60}
\end{equation}

Some elementary properties follow immediately from these definitions:

a) Any Grover kernel $K$ is a unitary operator.

b) Let the Grover operators $G_1, G_2$ be chosen such that
\begin{equation}
\begin{split}
& G_1 = \alpha P_{x_0} + \beta Q_{x_0},  P_{x_0} + Q_{x_0} = 1,\\ &G_2
= \gamma \bar{P} + \delta \bar{Q}, \quad \bar{P} + \bar{Q} = 1,
\label{qc62}
\end{split}
\end{equation}

\noindent with $P_{x_0} = |x_0\rangle \langle x_0|,$ and $\bar{P}$
given by the rank 1 matrix
\begin{equation}
\bar{P} := {1\over N} \left(
\begin{array}{ccc}
1 & \ldots & 1 \\ \vdots & & \vdots \\ 1 & \ldots & 1
\end{array}
\right).
\label{qc63}
\end{equation}

\noindent This is clearly a projector $\bar{P} = |k_0\rangle \langle
k_0|$ on the subspace spanned by the state $|k_0\rangle = {1\over
\sqrt{N}} (1,\ldots ,1)^{\rm t}$, where the superscript denotes the
transpose.  Then, if we take the following set of parameters,
\begin{equation}
\alpha = -1, \; \beta = 1, \; \gamma = -1, \; \delta = 1,
\label{qc64}
\end{equation}

\noindent the Grover kernel (\ref{qc60}) reproduces the original
Grover's choice (1996; 1997).  This property follows immediately by
construction.  In fact, we have in this case $G_1 = 1 - 2
P_{x_0}=:G_{x_0}$ whilst the operator $G_2 = 1 - 2 \bar{P}$ coincides
(up to a sign) with the diffusion operator $D$  (\ref{qc58})
introduced by Grover to implement the inversion about the average of
step 4.

The iterative part of the algorithm in step 5 corresponds to applying
$m$ times the Grover kernel $K$ to the initial state $|x_{\rm
in}\rangle :=2^{-n/2}\sum_x \ket{x}$, which describes the source
qubits after step 2, searching for a final state $|x_{\rm f}\rangle$
of the form
\begin{equation}
|x_{\rm f}\rangle:= K^m |x_{\rm in}\rangle,
\label{qc65-9}
\end{equation}

\noindent such that the probability $p(x_0)$ of finding the marked
state is above a given threshold value.  We may take this value to be
$1/2$, meaning that we choose a probability of success of $50 \%$ or
larger.  Thus, we are seeking under which circumstances the following
condition
\begin{equation}
p(x_0) = |\langle x_0 | K^m |x_{\rm in}\rangle|^2 \geq 1/2
\label{qc66-10}
\end{equation}

\noindent holds true.

The analysis of this probability gets simplified if we realize that
the evolution associated to the searching problem can be mapped onto a
reduced 2D-space spanned by the vectors
\begin{equation}
\{|x_0\rangle, |x_{\perp}\rangle := {1\over {\sqrt{N-1}}} \sum_{x\neq
x_0} |x\rangle\}.
\label{qc67-11}
\end{equation}

\noindent Then we can easily compute the projections of the Grover
operators $G_1,G_2$ in the reduced basis with the result
\begin{equation}
G_1 = \left(
\begin{array}{cc}
\alpha & 0 \\ 0 & \beta
\end{array}
\right),
\label{qc68-12}
\end{equation}
\begin{equation}
G_2 = \left(
\begin{array}{cc}
\delta & 0 \\ 0 & \gamma
\end{array}
\right) + (\gamma - \delta) \left(
\begin{array}{cc}
{1\over N} & {\sqrt{N-1}\over N} \\ {\sqrt{N-1}\over N} & {-1\over N}
\end{array}
\right).
\label{qc69-13}
\end{equation}

\noindent From now on, we shall fix two of the phase parameters using
the freedom we have to define each Grover factor in (\ref{qc60}) up to
an overall phase.  Then we decide to fix them as follows:
\begin{equation}
\alpha = \gamma = -1.
\label{qc70-14}
\end{equation}

\noindent With this choice, the Grover kernel (\ref{qc58}) takes the
following form in this basis:
\begin{equation}
K = {1\over N} \left(
\begin{array}{cc}
1 + \delta (1-N) & -\beta (1+\delta) \sqrt{N-1} \\ (1+\delta)
\sqrt{N-1} & \beta (1+\delta-N)
\end{array}
\right).
\label{qc71-15}
\end{equation}

The source state $|x_{\rm in}\rangle$ has the following components in
the reduced basis
\begin{equation}
|x_{\rm in}\rangle = {1\over {\sqrt{N}}} |x_0\rangle + \sqrt{{N -
1\over N}} |x_{\perp}\rangle.
\label{qc72-16}
\end{equation}

In order to compute the probability amplitude in (\ref{qc66-10}), we
introduce the spectral decomposition of the Grover kernel $K$ in terms
of its eigenvectors $\{ |\kappa_1\rangle, |\kappa_2\rangle \}$, with
eigenvalues ${\rm e}^{{\rm i}\omega_1}, {\rm e}^{{\rm i}\omega_2}$.
Thus we have
\begin{equation}
    \begin{split} &a(x_0) := \langle x_0 | K^m |x_{\rm in}\rangle = \\
&{1\over \sqrt{N}} \sum_{j=1}^2 \left\{|\langle x_0|\kappa_j\rangle|^2
+ {\sqrt{N-1}} \langle x_0|\kappa_j\rangle \langle
\kappa_j|x_{\perp}\rangle\right\} {\rm e}^{{\rm i}m\omega_j}.
\label{qc73-17}
    \end{split}
\end{equation}

\noindent This in turn can be cast into the following closed form:
\begin{equation}
    \begin{split} &\langle x_0 | K^m |x_{\rm in}\rangle = \\ &{\rm
e}^{{\rm i}m\omega_1} \left( {1\over \sqrt{N}} + ({\rm e}^{{\rm i}m
\Delta \omega} - 1) \langle x_0|\kappa_2\rangle \langle
\kappa_2|x_{\rm in}\rangle \right),
\label{qc74-18}
    \end{split}
\end{equation}

\noindent with $\Delta \omega := \omega_2 - \omega_1$.

In terms of the matrix invariants
\begin{equation}
{\rm Det} K = \beta \delta, \quad {\rm Tr} K = - (\beta + \delta) +
(1+\beta)(1+\delta) {1\over N},
\label{19}
\end{equation}
the eigenvalues $\zeta_{1,2} := {\rm e}^{{\rm i}\omega_{1,2}}$ are
given by
\begin{equation}
\zeta_{1,2} =\half{\rm Tr} K \mp \sqrt{-{\rm Det} K + \fourth({\rm Tr}
K)^2}.
\label{20}
\end{equation}

The corresponding unnormalized eigenvectors are
\begin{equation}
|\kappa_{1,2}\rangle \propto \left(
\begin{array}{c}
{A \mp \sqrt{-4({\rm Det} K) N^2 + A^2}\over 2(1+\delta)\sqrt{N-1}} \\
1
\end{array}
\right),
\label{qc75-21}
\end{equation}

\noindent with
\begin{equation}
A := (\beta - \delta) N + (1-\beta)(1+\delta).
\label{qc76-22}
\end{equation}

\begin{figure}[ht]
\begin{center}
\includegraphics[width = 7 cm]{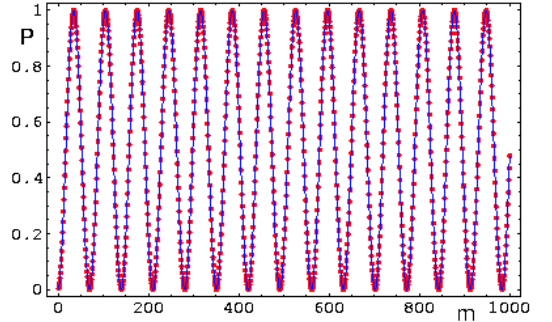}
\vskip 0.25cm
\caption[]{Probability of success $p$ as a function of the time step
for $N=1000$ and $\beta=\delta={\rm e}^{{\rm i}\pi/2}$.}
\label{fig2}
\end{center}
\end{figure}

\noindent Although we could work out all the expressions for a generic
value $N$ of elements in the list, we shall restrict our analysis to
the case of a large number of elements, $N \rightarrow \infty$ (see
Fig.~\ref{fig2}).  Thus, in this asymptotic limit we need to know the
behaviour for $N\gg 1$ of the eigenvector $|\kappa_2\rangle$, which
turns out to be
\begin{equation}
|\kappa_{2}\rangle \propto \left(
\begin{array}{c}
{\beta - \delta \over 1 + \delta} \sqrt{N} + O({1\over\sqrt{N}}) \\ 1
\end{array}
\right).
\label{qc77-23}
\end{equation}

\noindent Thus, for generic values of $\beta, \delta$ we observe that
the first component of the eigenvector dominates over the second one,
meaning that asymptotically $|\kappa_2\rangle \sim |x_0\rangle$ and
then $\langle x_0|\kappa_2\rangle \langle \kappa_2|x_{\rm in}\rangle =
O({1\over \sqrt{N}})$.  This implies that the probability of success
in (\ref{qc74-18}) will never reach the threshold value
(\ref{qc66-10}).  Then we are forced to tune the values of the two
parameters in order to have a well-defined and nontrivial algorithm,
and we demand
\begin{equation}
\beta = \delta\neq -1.
\label{qc78-24}
\end{equation}

Now the asymptotic behaviour of the eigenvector changes and is given
by a balanced superposition of marked and unmarked states, as follows
\begin{equation}
|\kappa_{2}\rangle \sim {1\over\sqrt{2}} \left(
\begin{array}{c}
{\rm i}\delta^{1/2} \\ 1
\end{array}
\right).
\label{qc79-25}
\end{equation}

\noindent This is normalized and we see that none of the components
dominates.  When we insert this expression into (\ref{qc74-18}) we find
\begin{equation}
|\langle x_0 | K^m |x_{\rm in}\rangle| \sim \half |\delta| |{\rm
e}^{{\rm i}m \Delta \omega} - 1| \sim   \left|\sin (\half
m\Delta\omega)\right|.
\label{qc80-26}
\end{equation}

\noindent This result means that we have succeeded in finding a class
of algorithms which are appropriate for solving the quantum searching
problem.  Now we need to find out how efficient they are.  To do this
let us denote by $M$ the smallest value of the time step $m$ at which
the probability becomes maximum; then,  asymptotically,\footnote{The
symbol $[x]$ stands for the  closest integer to $x$.}
\begin{equation}
M \sim [\left|\pi /\Delta \omega\right|].
\label{qc81-27}
\end{equation}

\noindent As it happens, we are interested in the asymptotic behaviour
of this optimal period of time $M$.  From the equation (\ref{20}) we
find the following behaviour as $N \to \infty$:
\begin{equation}
\Delta \omega \sim {4 \over \sqrt{N}}{\rm Re}\sqrt{\delta}.
\label{qc82-28}
\end{equation}

Thus, if we parameterize $\delta = {\rm e}^{{\rm i}\phi}$, then we
finally obtain the expression
\begin{equation}
M \sim \left[{\pi \over 4 \cos {\phi \over 2}}\sqrt{N}\right].
\label{qc83-29}
\end{equation}

Therefore, we conclude that the Grover algorithm of the class
parameterized by $\phi$ is a well-defined quantum searching algorithm
with an efficiency of order $O(\sqrt{N})$.

There have been many applications of Grover's work to quantum
searching: finding the mean and median of a given set of values
(Grover, 1996), searching the maximum/minimum (Durr and Hoyer, 1996),
searching more than one marked item (Boyer et al., 1998), quantum
counting, i.e., finding the number of marked items without caring
about their location (Brassard, Hoyer and Tapp, 1998), etc.  There is
also a nice geometrical interpretation of the Grover kernel
$K=-G_2G_1$ in terms of two reflections $G_1$ and $-G_2$, one about
$\ket{x_{\perp}}$ and the other about $\ket{x_{\rm in}}$, producing a
simple rotation of the initial state (Jozsa, 1999) by an angle $\theta
=2 \arcsin\frac{1}{\sqrt{N}}$ in the plane spanned by $\ket{x_0}$ and
$\ket{x_{\perp}}$. With this construction it is straightforward to
arrive at the following exact condition for the optimal value $m$ of
iterations:
\begin{equation}
m =\left[\half\left( \displaystyle\frac{\pi}{2
\arcsin\frac{1}{\sqrt{N}}}-1\right) \right].
\label{qcgeometrical}
\end{equation}

Finally, it has been shown that Grover's algorithm is optimal (Bennett
et al., 1997; Zalka, 1999), that is, its quadratic speed-up cannot be
improved for unstructured lists.

\subsection{Shor Algorithm}
\label{sec10D:level2}

Shor's algorithm (1994) came as a wake-up call for cryptographers
working with codes based on the difficulty of factoring large integer
numbers\footnote{``The problem of distinguishing prime numbers from
composite numbers and of resolving the latter into their prime factors
is known to be one of the most important and useful in arithmetic''
(Gauss, 1801).}  (see Sec.~\ref{sec6A:level2}), and now it represents
a Damocles' sword hanging over this type of cryptosystems.

The algorithm of Shor has several parts that make it somewhat
involved.  It may be useful to keep in mind the main ingredients
entering this algorithm:

i) A periodic function.

ii) Quantum parallelism.

iii) Quantum Fourier transform.

iv) Quantum measurement.

v) Euclid's classical algorithm for finding the greatest common
divisor gcd($n_1,n_2$) of two integers $n_1,n_2$.

Quantum computation opens the door to a new factorization method in
polynomial time (Shor, 1994).  This is why, although the technological
difficulties to succeed in their construction are
enormous,\footnote{As Preskill (1997) recalls, it is quite risky to
make guesses in this field; fifty years ago it was foreseen that {\em
``Where a calculator on the ENIAC is equipped with 18,000 vacuum tubes
and weighs 30 tons, computers in the future may have only 1,000 tubes
and perhaps only weigh 1 1/2 tons"} (Popular Mechanics, March 1949),
and the ``future'' has surpassed these expectations amply.} it is
highly interesting to find systems for key distribution whose security
(see Sec.~\ref{sec6B:level2}) does not rely upon the practical
difficulty of factoring large integers.  Quite ironically, quantum
physics provides both a fast factorization method and a secure key
distribution (Sec.~\ref{sec6B:level2}).

Let $N\geq 3$ be an odd integer  to factorize.  Let $a$ be an integer
in $(1,N)$.  Let us assume that gcd$(N,a)=1$, that is, $N$ and $a$ are
coprimes; otherwise gcd$(N,a)$ would be a nontrivial factor $f$ of
$N$, and we would restart with $N/f$.  The integral powers $a^x$ of
$a$ form a cyclic group in ${\Z}_N:={\Z}/N{\Z}$, and there exists a
smallest integer $r\in(1,N)$, called the {\em order} of $a$ mod $N$,
such that $a^r=1$ in ${\Z}_N$.  Several cases may arise:

1) $r$ is odd;

2) $r$ is even and $a^{r/2}=-1$ in ${\Z}_N$;

3) $r$ is even and $a^{r/2}\neq -1$ in ${\Z}_N$.

\noindent Only the case 3) is of interest for then gcd$(N,a^{r/2}\pm
1)$ are nontrivial factors of $N$.

It can be shown that, for any given odd $N$, the probability of
picking up at random an integer $a\in[1,N]$ coprime to $N$ and
fulfilling 3) is $\geq 1/(2\log N)$, provided that $N$ is not a pure
prime power (Ekert and Jozsa, 1996).\footnote{There are fast power
tests to detect whether $N$ is a prime power, say $N=p^s$, and to find
$p$ in that case (Cohen, 1993).  A rudimentary transcendental and not
very efficient procedure consists in trying with the integers $\lfloor
N^{1/k}\rfloor, \lceil N^{1/k}\rceil, k=2,3,\ldots,\lceil \log_2
N\rceil$, until hopefully finding one being a divisor of $N$.}
Therefore it will be enough to analyze $O(\log(1/\epsilon)\log N)$
randomly chosen values of $a$ to succeed in obtaining a nontrivial
factor of $N$ with a probability larger than $1-\epsilon$.  For
example, if $N=21823$, and $a=12083$, the order of $a$ mod $N$ is
$r=3588$, and $12083^{1794}\equiv 4866\;\mod\;21823$, thereby
gcd$(12083^{1794}\mp 1,21823)=\{139,157\}$ are factors of 21823.  On
the contrary, although the order  of $a=14335$ mod $N$ is also even,
namely $r=1794$, however $14335^{897}\equiv -1\;\mod\;21823$, and
gcd$(14335^{897}\mp 1,21823)=\{1,21823\}$, so that no nontrivial
factor of $N$ is now obtained.

The big problem lies in computing the order $r$ of $a$ mod $N$ for
large $N$.  And here is where the Shor algorithm comes in to quantumly
search for the order $r$ of an integer $x$ in the multiplicative group
${\Z}_{N}^\ast$ of integers modulo $N$, by producing a state with
periodicity $r$.

As usual, we need two quantum registers: a source register with $K$
qubits such that $Q:=2^K\in(N^2,2N^2)$, and a target register with at
least $N$ basis states (i.e. with $\lceil\log_2 N\rceil$ qubits).

\begin{figure}[ht]
\psfrag{u}[Bc][Bc][1][0]{$|0\rangle$}
\psfrag{w}[Bc][Bc][1][0]{$|0\rangle$} 
\psfrag{h}[Bc][Bc][1][0]{$U_H$}
\psfrag{p}[Bc][Bc][1][0]{$U_f$} 
\psfrag{v}[Bc][Bc][0.75][0]{measure} 
\psfrag{q}[Bc][Bc][1][0]{QFT}
\includegraphics[width=8 cm]{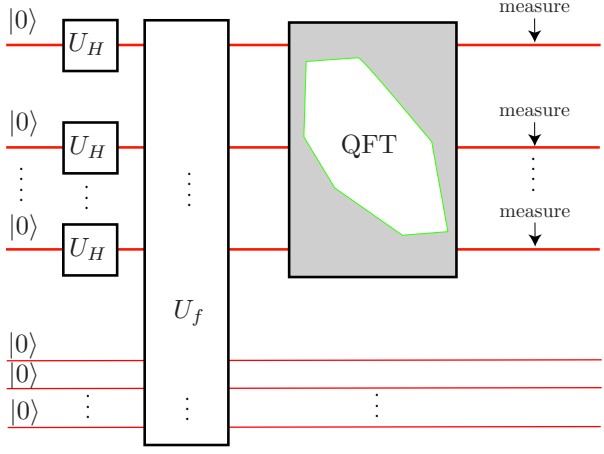}
\caption{A quantum circuit representing the Shor algorithm.}
\label{qcshor}
\end{figure}

These are the main steps of Shor's algorithm (see Fig.~\ref{qcshor}):

\medskip

\noindent {\em Step 1.} Initialize the source and target qubits to the
state $\ket{\Psi_1}:=\ket{0}\otimes\ket{0}$.

\noindent {\em Step 2.} Apply on the source register the {\em quantum
Fourier transform} (which is just the discrete Fourier transform $F_Q$
in ${\Z}_{Q}$):\footnote{This is specially fast when $Q=2^{K}$.}
\begin{equation}
U_{{\rm F}_Q}: \ket{q}\mapsto \frac{1}{\sqrt Q}
\sum_{q^\prime=0}^{Q-1}{\ee}^{2\pi\ii qq^\prime/Q} \ket{q^\prime}.
\label{qc84}
\end{equation}
Here, as usual, $q:=\sum_{j=0}^{Q-1}q_j2^j$, $q_j=0,1$, and
$\ket{q}:=\ket{q_{Q-1} \ldots q_1q_0}$.  The following output state is
produced:
\begin{equation}
\ket{\Psi_2}:=(U_{{\rm F}_Q}\otimes
1)\ket{\Psi_1}=Q^{-1/2}\sum_{q=0}^{Q-1}\ket{q}\otimes\ket{0}.
\label{qc85}
\end{equation}

\noindent This particular case of the quantum Fourier transform
corresponds to the Hadamard gate acting bit-wise on the source qubits.

\noindent {\em Step 3.} Next apply the gate $U_{a}$ implementing the
modular exponentiation function $q\mapsto a^q \;\mod\;N$:
\begin{equation}
\ket{\Psi_3}:=U_{a}\ket{\Psi_2}=Q^{-1/2}\sum_{q=0}^{Q-1}
\ket{q}\otimes\ket{a^q \;\mod\;N}.
\label{qc86}
\end{equation}

\noindent This operation computes at one go $a^q \;\mod\;N$ for all
$q$ as a manifestation of the quantum parallelism (see
Sec.~\ref{sec9A:level2}).

\noindent {\em Step 4.} Apply again the Fourier transform $U_{{\rm
F}_Q}$ on the source register.  Then the state becomes
\begin{equation}
\begin{split}
\ket{\Psi_4}&:=(U_{{\rm F}_Q}\otimes 1)\ket{\Psi_3}\\
&=\frac{1}{Q}\sum_{q=0}^{Q-1}\sum_{q'=0}^{Q-1} {\ee}^{2\pi\ii qq'/Q}
\ket{q}\otimes\ket{a^{q'}\;\mod\;N}.
\label{qc88}
\end{split}
\end{equation}

\noindent {\em Step 5.} Measure the source qubits in the computational
basis.  The probability of finding them in the state $\ket{q}$ is
$\prob(q)=\sum_{j=0}^{r-1}\prob_j(q)$, where
\begin{equation}
\prob_j(q):=\frac{1}{Q^2}\left|\sum_{k=0}^{B_j-1}\left(\ee^{2\pi\ii
qr/Q}\right)^k\right|^2,
\label{qc89}
\end{equation}
with $B_j:=1+\lfloor(Q-1-j)/r\rfloor$.

To simplify the algebra, an intermediate step is introduced  in most
discussions of Shor's algorithm in which the target qubits are
measured prior to the second application of the QFT (Shor, 1995; Ekert
and Jozsa, 1996).  If $\ket{b}$ is the result, the source register
will be projected onto a state
$B^{-1/2}\sum_{k=0}^{B-1}\ket{d_b+kr}$, superposition of basis states
with the periodicity $r$ of $a^q$. Here $d_{b}$ is the minimum
non-negative integer such that $a^{d_{b}} \;\mod\;N=b$, and
$B:=1+\lfloor (Q-1-d_b)/r\rfloor$ is the length of the series. After
applying the QFT and measuring the source qubits, the probability to
obtain now $\ket{q}$ is just $(Q/B_{d_b})\prob_{d_b}(q)$.

\begin{figure}[ht]
\psfrag{x}[Bc][Bc][1][0]{\color{blue}$q$}
\psfrag{y}[Bc][Bc][1][0]{\color{blue}$\prob(q)$}
\psfrag{l}[Bc][Bc][1][0]{\color{red} $Q=2^8$}
\psfrag{r}[Bc][Bc][1][0]{\color{red} $r=10$}  \centering
\hspace{-1cm}  \includegraphics[width=7 cm]{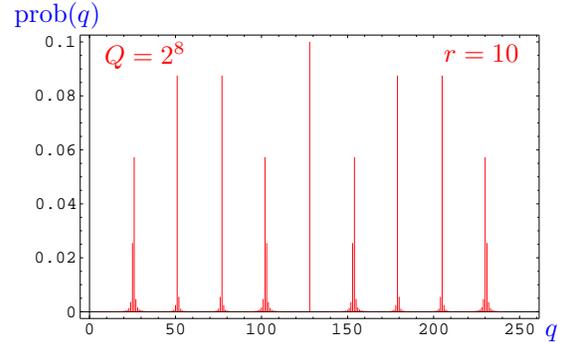}
\caption{The probability $\prob(q)$ for the case $Q=2^8, r=10$.  It
gets concentrated around the integers $\left\lfloor sQ/r \right\rfloor
$, with $s$ integer.}
\label{shorB}
\end{figure}

Let us see how to pull out the order $r$ of $a$ from the study of the
above probability $\prob(q)$.  The analysis of the geometrical series
in (\ref{qc89}) shows that $\prob(q)$ peaks around those $q$s for
which all the complex numbers in the sum fall in a same half-plane of
${\C}$, and thus they enhance each other constructively.  It can be
shown that such $q$s are characterized by $|(qr\;\mod\;Q)|\leq \half
r$, they number $r$, and satisfy $\prob(q)\geq (2/\pi)^2 r^{-1}$;
therefore the probability of hitting upon anyone of them is $\geq
(2/\pi)^2=0.405...$.  In Fig.~\ref{shorB} the form of $\prob(q)$ is
shown.

The condition of constructive interference (see Table~\ref{table3})
for each $q>0$ amounts to the existence of an integer
$q^\prime\in(0,r)$ such that $|(q/Q)-(q^\prime/r)|\leq\half Q^{-1}$.
As we have chosen $Q>N^2$, and $r<N$, there exists a unique $q^\prime$
such that the fraction $q^\prime/r$ satisfies that inequality.  This
rational number $q^\prime/r$ can be easily found as a convergent to
the (finite simple) continued fraction expansion of $q/Q$.  If this
convergent is the irreducible fraction $q_1/r_1$, it may happen that
$a^{r_1}\equiv 1 \;\mod\;N$, which implies $r=r_1$, and we are over.
Otherwise, we would only know that $r_1$ is a divisor of $r$, and we
would have to carry on, choosing another $q$ with constructive
interference, to see if this time we are luckier.  It can be shown
that the probability of finding an appropriate $q$ is order ${\cal
O}(1/\log\log r)$, and therefore with a number ${\cal O}(\log\log N)$
of trials it is highly probable to obtain $r$.

For example, let be $N=15$ (this is a sort of ``toy model''), and
$a=7$.  We can effortlessly see by brute force that $r=4$.  Suppose,
however, that we insist in following the Shor way (quite a luxury in
this case, but a necessity if $N$ had half a thousand digits).  We
would take $Q=2^8$ to comply with $N^2<Q<2N^2$.  After step 5 we would
obtain the state $\ket{q}$ of the source qubits, where, for instance,
$q=0,64,128,192$ with probabilities $0.25,0.25,0.25,0.25$.  The first
value is useless, for $q/Q$ does not allow us to determine $r$ if
$q=0$.  From the continued fraction series expansion $\{a_0, a_1, a_2,
...\} := a_0+1/(a_1+1/(a_2+...))$ of $q/Q$ ($64/256 = \{0, 4\}$,
$128/256 = \{0, 2\}$, $192/256 = \{0, 1, 3\}$) we see that for $q =
64$ (resp.  $128, 192$), the fraction $1/4$ (resp.  $1/2, 3/4$)
approximates $q/Q$ with an error less than $1/2Q$.  Thus, 4 is a
divisor of $r$, i.e. $r = 4, 8, 12$, etc.  A direct check selects $r =
4$ as the order of 7 mod 15. And since $7^{4/2}\not\equiv -1$ mod 15,
then gcd$(49\pm 1,15)=\{5,3\}$ are factors of 15.

As a little more complicated example, take $N = 25397, a = 71$.  Then
$Q = 2^{30} = 1073741824$.  There are many values of $q$ for which the
probability is appreciable and similar.  One of those is $q =
6170930$, for which $\prob(q)$ is about $2\times 10^{-3}$.  The
approximation 1/174 to $q/Q$ is the only convergent with denominator
$< N$ provided us by the continued fraction expansion $\{0, 174,
1542732, 2\}$ of $q/Q$.  Therefore, the order $r$ of 71 mod 25397 is a
multiple of 174, say $r = 174, 348, 522$, etc.  A direct check shows
that $r = 522$. Also in this case $a^{r/2}\not\equiv -1$ mod $N$, and
gcd$(71^{261}\pm 1,25397)=\{109,233\}$ are divisors of 25397.

\begin{figure}[ht]
\psfrag{y}[Bc][Bc][0.8][1]{\color{red}$t(n)$}
\psfrag{x}[Bc][Bc][0.8][1]{\color{red}number of bits}
\psfrag{q}[Bc][Bc][0.8][1] {\color{red}$\#_{\_}{q}-{\rm
logic}_{\_}{\rm operations}({n })\sim 25 \; n^3$}
\psfrag{l}[Bc][Bc][0.8][1]{\color{red} $t(136000)=20$ years} \centering
\hspace{-1cm} 
\includegraphics[width=7 cm]{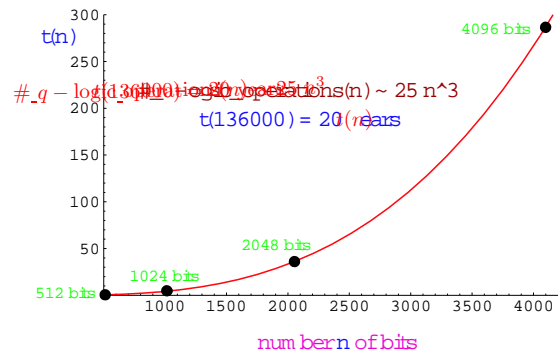}
\caption{Factorization times with a hypothetical QC at a nominal clock
frequency of 100 MHz.  The time $t(n)$, in minutes, is shown as a
function of the number of bits.}
\label{fac2}
\end{figure}

In Fig.~\ref{fac2} the factorization time with an hypothetical quantum
computer at 100 MHz is represented as a function of binary length of
the integer to be factorized.  The spectacular efficiency of the Shor
algorithm stands out, with a time of 20 years for an integer of about
40\,000 digits (Hughes, 1997).

Shor's algorithm may seem a bit miraculous after those several
``manipulations'' or steps.  The rationale is the same as described in
Sec.~\ref{sec9:level1}: to drive the system into an appropriate
outcome state that upon measurement yields the desired result with
high probability. Where does the constructive interference ingredient
(Table~\ref{table3}) come into the algorithm? It is by means of the
second QFT operation. This is designed to produce the interference
among qubit amplitudes in such a way as to enhance those aspects of
the output that favors the determination of the order $r$.

\subsubsection{The Quantum Fourier Transform}

\begin{figure*}[ht]
\psfrag{6}[Bc][Bc][0.75][0]{$|q_0\rangle$}
\psfrag{5}[Bc][Bc][0.75][0]{$|q_1\rangle$}
\psfrag{2}[Bc][Bc][0.75][0]{$|q_{K-3}\rangle$}
\psfrag{1}[Bc][Bc][0.75][0]{$|q_{K-2}\rangle$}
\psfrag{0}[Bc][Bc][0.75][0]{$|q_{K-1}\rangle$}
\psfrag{h}[Bc][Bc][0.75][0]{$U_{\rm H}$}
\psfrag{o}[Bc][Bc][0.75][0]{$U_{2}$}
\psfrag{r}[Bc][Bc][0.75][0]{$U_{K}$}
\psfrag{l}[Bc][Bc][0.75][0]{$U_{2}$}
\psfrag{m}[Bc][Bc][0.70][0]{$U_{K-1}$}
\psfrag{i}[Bc][Bc][0.75][0]{$U_{2}$}
\psfrag{f}[Bc][Bc][0.75][0]{$\ket{0}+\ee^{2\pi \ii 0.q_0} \ket{1}$}
\psfrag{e}[Bc][Bc][0.75][0]{$\ket{0}+\ee^{2\pi \ii 0.q_1q_0} \ket{1}$} 
\psfrag{c}[Bc][Bc][0.75][0]{$\ket{0}+\ee^{2\pi \ii q_{K-3} \ldots 
q_1q_0} \ket{1} $} 
\psfrag{b}[Bc][Bc][0.75][0]{$\ket{0}+\ee^{2\pi \ii
0.q_{K-2} \ldots q_0} \ket{1} $}
\psfrag{a}[Bc][Bc][0.75][0]{$\ket{0}+\ee^{2\pi \ii 0.q_{K-1} \ldots
q_0} \ket{1} $} 
\includegraphics[width=14 cm]{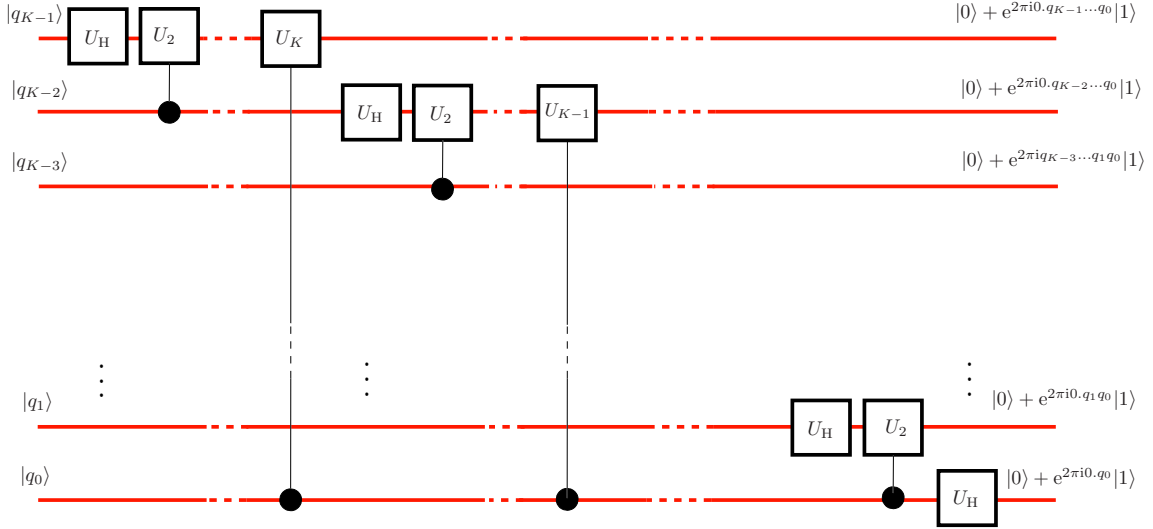}
\caption{Implementation of the quantum Fourier transform with Hadamard
and controlled-phase gates (up to a reversion of output qubits).  By
$U_j$ we denote the unary gate $U_j:=\ket{0}\bra{0}+
\ee^{2\pi\ii/2^j}\ket{1}\bra{1}$. For typographical reasons a factor
$2^{-1/2}$ has been omitted in each output qubit.}
\label{qft}
\end{figure*}

Let us take a closer look at the discrete Fourier transform $U_{{\rm
F}_Q}$ when $Q=2^K$.  It is at the core of Shor's algorithm and is
responsible for its exponential speed-up.  To analyze the efficiency
of the Shor algorithm it proves convenient to implement the QFT by
means of one- and two-qubit gates.  The result, shown in
Fig.~\ref{qft}, will follow from the expression (\ref{qc84}), duly
worked out.

The phase factor ${\ee}^{2\pi\ii qq^\prime/2^K}$ in (\ref{qc84}) is a
periodic function of $q$, and of $q^\prime$ as well, with period
$2^K$.  The numbers $q$ and $q^\prime$ have the following binary
decompositions: $q=\sum_{j=0}^{K-1} q_j 2^j, q_j=0,1$ and
$q^\prime=\sum_{l=0}^{K-1} q^\prime_l 2^l, q^\prime_l=0,1$.  Then
their product can be written as
\begin{equation}
qq^\prime=\sum_{j,l=0}^{K-1} q_jq^\prime_l 2^{j+l}=\sum_{0\leq
j+l<K}q_jq^\prime_l 2^{j+l}\;\mod\;\Z_Q.
\end{equation}
By entering this expression into (\ref{qc84}), and defining $\bar
q^\prime_l:=q^\prime_{K-1-l}, l=0,\ldots,K-1$,
$0.abc\ldots:=2^{-1}a+2^{-2}b+2^{-3}c+\ldots$, we find
\begin{equation}
\begin{split}
& U_{{\rm F}_Q}\ket{q}=\frac{1}{\sqrt Q} \sum_{q^\prime=0}^{Q-1}
\exp(2\pi\ii qq^\prime/2^K)\ket{q^\prime} \\ & = \frac{1}{\sqrt
Q}\sum_{q^\prime=0}^{Q-1} \exp(2\pi\ii\sum_{0\leq j+l<K}q_jq^\prime_l
2^{j+l-K})\ket{q^\prime} \\ & = \frac{1}{\sqrt Q}\sum_{\bar
q^\prime=0}^{Q-1} \exp(2\pi\ii\sum_{0\leq j\leq l<K}q_j\bar q^\prime_l
2^{j-l-1})\ket{\bar q^\prime},
\end{split}
\label{qc90}
\end{equation}
and hence

\begin{equation}
\begin{split}
& U_{{\rm F}_Q}\ket{q}= \frac{1}{\sqrt Q}\sum_{\bar q^\prime=0}^{Q-1}
\bigotimes_{l=0}^{K-1} \exp(2\pi\ii\sum_{0\leq j\leq l} q_j
2^{j-l-1}\bar q^\prime_l)\ket{\bar q^\prime_l} \\ & = \frac{1}{\sqrt
Q} \bigotimes_{l=0}^{K-1}\sum_{\bar q^\prime_l=0}^{1}
\exp(2\pi\ii\sum_{0\leq j\leq l} q_j 2^{j-l-1}\bar
q^\prime_l)\ket{\bar q^\prime_l} \\ & = \frac{1}{\sqrt Q}
\bigotimes_{l=0}^{K-1}(\ket{0}+ \exp(2\pi\ii 0.q_lq_{l-1}\ldots
q_0)\ket{1}).
\end{split}
\label{qc90a}
\end{equation}
In particular, the transformed state $U_{{\rm F}_Q}\ket{q}$ is
separable. The QFT gate $U_{{\rm F}_Q}$ can be explictly written as a
product of Hadamard, controlled-phase and SWAP gates:
\begin{equation}
\begin{split}
U_{{\rm F}_Q}=&\left(\prod_{i=0}^{{\lfloor K/2\rfloor-1}} U_{{\rm
SWAP},i,K-1-i}\right)\times\\ &\prod_{l=
K-1,...,1,0}\left(\left(\prod_{0\leq j\leq l-1}
U_{j,l}(\theta_{l-j})\right)U_{{\rm H},l}\right),
\label{qc90b}
\end{split}
\end{equation}
where $\theta_j:=\pi/2^j$, $U_{{\rm SWAP},i,j}$ exchanges the qubit
states labelled by $i,j$, and
\begin{equation}
\begin{split}
&U_{{\rm H},l}\ket{...q_l...}:=2^{-1/2} \sum_{\bar
q^\prime_l=0,1}\ee^{\ii\pi q_l\bar q^\prime_l} \ket{...\bar
q^\prime_l...},  \\ & U_{j,l}(\theta) \ket{...q_l...q_j...}  :=
\ee^{\ii q_lq_j\theta}\ket{...q_l...q_j...}
\end{split}
\label{qc90c}
\end{equation}
are the Hadamard gate action of the one-qubit $\ket{q_l}$, and the
controlled-phase gate action on the two-qubit state $\ket{q_lq_j}$,
respectively.  From the factorization (\ref{qc90b}) we can read off
the quantum circuit (see Fig.~\ref{qft}) implementing the QFT (up to a
reversion of the output qubits).

The number of Hadamard gates in this implementation of the QFT is $K$,
and that of the controlled-gates is $\half K(K-1)$.  Altogether this
implies that the size of quantum circuit for Shor's algorithm is order
$O(K^2)$  regardless of the SWAP gates for the final reversion
(Coppersmith, 1994).\footnote{In contrast, the classical fast Fourier
transform requires order $O(K2^K)$ elementary operations to transform
a $K$-bit vector (Press et al., 1992).}

The quantum Fourier transform can be extended to deal with qubits with
a number of states $d$ not necessarily equal to $2$ (see
Sec.~\ref{sec3:level1}).  In this case the dimension of the Hilbert
space of $K$ source qubits is $Q=d^K$, and equations
(\ref{qc84},\ref{qc90c}) for the QFT, the Hadamard and the
controlled-phase gates hold true provided the phase angle is taken to
be
\begin{equation}
\theta_j = {2\pi \over d^{j+1} }
\label{qc94}
\end{equation}

For instance, for qubits with $d=3$ state or qutrits, the Hadamard
gate takes the following explicit form
\begin{equation}
\begin{aligned}
U_{\rm H}^{(3)} |0\rangle & = {1\over\sqrt{3}} [|0\rangle + |1\rangle
+ |2\rangle] \\ U_{\rm H}^{(3)} |1\rangle & = {1\over\sqrt{3}}
[|0\rangle + \omega |1\rangle + \omega^2 |2\rangle] \\ U_{\rm H}^{(3)}
|2\rangle & = {1\over\sqrt{3}} [|0\rangle + \omega^2 |1\rangle +
\omega^3 |2\rangle]
\end{aligned}
\label{qc95}
\end{equation}
with $\omega:={\ee}^{2\pi\ii/3}$.

In this general case, the sequence of one- and two-qubit gates for the
decomposition of the QFT remains valid, as well as their counting.
This implies that using qudits for QFT does not spoil its superb
performance, while retaining the advantage of reducing by a factor of
$\lfloor\log_{2}d \rfloor$ the length of the quantum registers (see
Sec.~\ref{sec3:level1}).

\subsubsection{Cost of Shor's Algorithm}

We finally evaluate the complexity of Shor's algorithm.  The first QFT
transform (step 2) is just a Hadamard operation applied bit-wise and
its cost is $O(\log_2 N)$.  The modular exponentiation in step 3
consumes $O(\log_2^2 N \log_2\log_2 N \log_2 \log_2 \log_2 N)$ time
(Shor, 1994).  The second QFT gate (step 4) is, according to the
results just mentioned, $O(\log_2^2 N)$. Therefore the total cost to
determine the order $r$ of $a$ mod $N$, with a probability of success
$O(1)$, is $O(\log_2^{2+\epsilon} N)$, any $\epsilon>0$.

Once $r$ is determined, there remains to calculate gcd($a^{r/2}\pm
1,N$) in order to find a factor of $N$.  This arithmetical operation
is more resource demanding, since it takes $O(\log_2^3 N)$ time steps
when Euclid's celebrated algorithm is applied.\footnote{Actually, a
more refined implementation of the gcd algorithm (Knuth, 1981) reduces
its cost to $O(\log N(\log\log N)^2\log\log\log N)$.}

Altogether we end up with a total cost $O(\log_2^3 N)$ for the
complete factorization algorithm with high probability,\footnote{Or
better $O(\log_2^{2+\epsilon} N)$, if the previous footnote is
considered.}  what represents in practice a subexponential gain over
the classical best algorithms (QS, GNFS) known nowadays.

\subsection{On the Classification of Algorithms}
\label{sec10E:level2}

One of the most important issues in quantum computing is the design of
quantum algorithms.  There are known very few of them.  Apparently, we
are lacking the basic principles underlying the quantum version of
{\em algorithm problem solving}.  We want in part to address this
question and we believe that one attempt to understand the basic
principles of quantum algorithm design may proceed with the comparison
with the known strategies of designing classical algorithms in
Computational Science.  This is suggested by the studies about the
relationships between fundamentals of classical and quantum
computations presented in Sec.~\ref{sec8:level1} and
Sec.~\ref{sec9A:level2}.  In this regard, we need to distinguish
between fundamentals of quantum computation and strategies for
designing algorithms.  Although the latter are still unknown, the
former have been described in Table~\ref{table3}.  The fact that we
can understand the fundamentals of quantum computation does not mean
in principle that we know the keys to set up quantum algorithms,
although it can be of great help.

Now let us come to the point of analysing the classical strategies of
algorithm design from the point of view of quantum computation.  To
this end, we shall consider the classification introduced by Levitin
(1999) who has done a reformulation which includes and categorizes in
a nice fashion other classifications schemes (Brassad and Bratley,
1996).  Following Levitin, there are four classical general design
techniques which we shall describe briefly by its definition and with
a simple example to illustrate them.  This example is the problem of
computing $a^n$ mod $p$, which is of great importance in public-key
encryption algorithms (Sec.~\ref{sec6:level1},
Sec.~\ref{sec10D:level2}).  Then we have the following generic types:

1) {\em Brute Force Algorithms}

It amounts to solving a problem by directly applying its crude
formulation.  Example: $a^n = a \cdot a \cdots a$, $n$ times.

2) {\em Divide-and-Conquer Algorithms}

The original problem is partitioned into a number of smaller
subproblems, usually of the same kind.  These in turn are then solved
and their solutions combined to get a solution of the bigger problem.
This strategy usually employs recursivity in order to obtain a greater
profit.  Example: $a^n = a^{\lfloor n/2\rfloor} \cdot a^{\lfloor
n/2\rfloor} \cdot a^{n-2\lfloor n/2\rfloor}$.

3) {\em Decrease-and-Conquer Algorithms}

The original problem is reduced to a smaller one, which is usually
solved by recursion and the solution so obtained is applied to find a
solution of the original problem.  Examples: a) $a^n = a^{n-1}\cdot a$
(decrease-by-one variety); b) $a^n = (a^{\lfloor n/2\rfloor})^2$ if
$n$ even, $a^n = (a^{\lfloor n/2\rfloor})^2\cdot a$ if $n$ odd
(decrease-by-half variety).

4) {\em Transform-and-Conquer Algorithms}

The original problem is transformed into another equivalent problem
which is more amenable to solution with simpler techniques.  Example:
$a^n$ is computed by exploiting the binary representation of $n$.

These four types of strategies have in turn several subtypes we shall
not dwell upon.

Table~\ref{tableclalgo} contains these classical strategies with some
well-known and less trivial examples of representative algorithms.
There are important algorithms built upon a mixture of these basic
techniques; for example, the Fast Fourier Transform employs both
divide-and-conquer and transform-and-conquer techniques.

\begin{table}[t]
\begin{ruledtabular}
\begin{tabular}{c|c}
\hline Classical Technique  &  Algorithm Example  \\ \colrule \hline
Brute Force & Searching the Largest\\ \hline Divide-and-Conquer &
Quicksort \\ \hline Decrease-and-Conquer &  Euclid's Algorithm \\
\hline Transform-and-Conquer & Gaussian Elimination \\
\end{tabular}
\end{ruledtabular}
\caption{Classification of Classical Algorithms.}
\label{tableclalgo}
\end{table}

Now, it can be quite revealing to set up the quantum version of
Table~\ref{tableclalgo} by classifying the most useful of the so-far
known quantum algorithms.  This we do in Table~\ref{tableqalgo}.

\begin{table}[h]
\begin{ruledtabular}
\begin{tabular}{c|c}
\hline Quantum Technique  &  Algorithm Example  \\ \colrule \hline &
Grover's Algorithm\\ Brute Force & Deutsch-Jozsa' Algorithm\\ & Simon's
Algorithm\\ \hline Divide-and-Conquer &  $\emptyset$ \\ \hline
Decrease-and-Conquer &   $\emptyset$ \\ \hline Transform-and-Conquer &
Shor's Algorithm \\
\end{tabular}
\end{ruledtabular}
\caption{Classification of quantum algorithms.}
\label{tableqalgo}
\end{table}

Several remarks are in order.

Firstly, we have placed Grover's algorithm in the category of Brute
Force algorithms.  The strategy is similar to its classical
counterpart, which is of Brute Force type.  The difference lies in the
fact that the quantum operation is realized through a unitary operator
which implements the reversible quantum computation.\footnote{By a
similar rationale, we have placed Deutsch-Jozsa and Simon algorithms
in the same class} Although the Brute Force technique gives usually
low efficient algorithms, it is very important for several reasons.
One is that there are important cases, like the searching problem,
where the Brute Force method outperforms more sophisticated strategies
like divide-and-conquer.  We find Grover's algorithm as a realization
of the Brute Force technique at the quantum level and this is why it
is so simple and of general purpose  at the same time.

Secondly, we have included Shor's algorithm in the category of
transform-and-conquer algorithms.  As we have explained in
Sec.~\ref{sec10D:level2}, Shor solves the factorization problem by
reducing it to the problem of finding the period of a certain function
in number theory, which in turn is solved with the aid of the
fundamentals of quantum computation.  Having realized this, we point
out that the classical version of transform-and-conquer algorithms are
very rare (Anany, 1999).  This may explain why Shor's algorithm,
although more powerful than Grover's, it has a more reduced range of
applications.

Thirdly, the most notorious aspect of Table~\ref{tableqalgo} is the
absence of quantum algorithms based on the divide-and-conquer
technique, which is by far the most general and used strategy in
classical computation.  This may partly account for the list of
quantum algorithms being so short.  Moreover, if we resort to the
basic features of quantum computation (Table~\ref{table3}) we may
explain somehow why this entry is empty in Table~\ref{tableqalgo}.  We
know that a quantum register supports the superposition of many states
at the same time.  This implies that the qubits of the quantum
registers are strongly correlated (entangled) and their joint state is
not separable into a product of states of smaller subregisters.  Thus
quantum parallelism and entanglement render unnatural any try to
implement the strategy of divide-and-conquer in a quantum register at
least in a straightforward and naive fashion.\footnote{A blend of
classical and quantum algorithms might make room for a
divide-and-conquer strategy.}

\section{Experimental Proposals of Quantum Computers}
\label{sec11:level1}

The great challenge of quantum computation is to build real quantum
computers capable of implementing the quantum logic operations of
Sec.~\ref{sec9:level1} and of performing the quantum algorithms of
Sec.~\ref{sec10:level1}.  In this section we present some of the
experimental proposals to this end.  Some of these proposals have been
actually carried out, and this is already a significant advance for it means
that the theoretical constructs can be checked experimentally.
However, these devices are very modest in size and the real
breakthrough will be to scale them up to sizes capable of doing tasks
not yet done with classical computers, like code-breaking with Shor's
algorithm or database searching with Grover's algorithm.

Before giving an overview of a few experimental proposals, it is
convenient to summarize what they all have in common. There is a
generic setting to build a quantum computer.\footnote{At least with
our present knowledge.}  We basically need:

{i)} any two-level quantum system,

{ii)} interaction between qubits,

{iii)} external manipulation of qubits.

\noindent The two-level system is used as a qubit and the interaction
between qubits is used to implement the conditional logic of the
quantum logic gates (Sec.~\ref{sec9:level1}).  The system of qubits
must be accessible for external manipulations: to read in the input
state and read out the output, as well as during the computation if
the quantum algorithm requires it.

Interestingly enough, some of the possible qubits and quantum logic
gates have been with us since the early times of Bohr.  For example,
the quantum NOT-gate is obtained, at least in principle, either by
exciting an atomic ground state to an upper level with a photon of
apppropriate frecuency and time length, or by induced emission.  If
the length of light pulses is halved, a Hadamard-like gate will 
result.\footnote{Strictly speaking, this halved-pulse produces the action of
the so-called  {\em pseudo-Hadamard} gate.}  Quantum computation has
provided us with a new insight on these operations.

There are several settings in which one can illustrate the very basics
of realizing experimental quantum computers and seeing the above three
requirements in action. We shall choose as our qubit system a spin
$\half$ massive particle with magnetic moment, whose translational
motion will be ignored.\footnote{Other simple choices are the
polarization of a photon, an atomic system with just two relevant
levels, etc.} Placing this qubit in a suitably oscillating external
magnetic field will allow us to theoretically implement the unary
quantum gates.

We shall not dwell upon all the practical technicalities of the
experimental proposals below but instead present the basic physical
foundations underlying some of the quantum computers.

\subsubsection{One- and Two-Qubit Logic Gates with Spin Qubits}

This is one of the few examples where one can follow exactly the
evolution of the quantum system, and it is versatile enough to let
building some of the basic logic gates.  We present it as a
preparation for more complex setups.

Suppose that our qubit, a spin $\half$ particle, has a magnetic moment
$\bfmu=\gamma \vec{S}$, where $\vec{S}=\half \hbar\bfsigma$ is the
spin operator.  In the presence of a uniform but time-dependent
magnetic field $\vec{B}(t)$ the qubit state $\ket{\psi(t)}$ will
evolve with the Hamiltonian $H(t)=-\gamma \vec{S}\cdot\vec{B}(t)$
(Rabi, 1937):
\begin{equation}
\ii\hbar{\dd\over\dd t}\ket{\psi(t)} =
-\gamma\mathbf{S}\cdot\mathbf{B}(t)\ket{\psi(t)}.
\label{qc97}
\end{equation}

When the magnetic field rotates uniformly around a fixed axis (say
$Oz$), namely
\begin{equation}
\mathbf{B}(t) = (B_1 \cos\omega t, B_1\sin\omega t, B_0),
\label{qc98}
\end{equation}
then Eq. (\ref{qc97}) can be solved explicitly, with the result
(Galindo and Pascual, 1990b):
\begin{equation}
\begin{split}
&|\psi(t)\rangle= U(t)|\psi(0)\rangle, \\ &U(t):=\ee^{-\ii \omega t
\sigma_z/2} \ee^{-\ii [(\omega_0-\omega)
\sigma_z+\omega_1\sigma_x]t/2}= \\ &(\cos\half\omega
t-\ii(\sin\half\omega t)\sigma_z) (\cos\half\Omega
t-\ii(\sin\half\Omega t)\sigma^\prime),
\label{qc103}
\end{split}
\end{equation}
where $\omega_0:=-\gamma B_0, \omega_1 = -\gamma B_1$,
$\Omega:=((\omega_0-\omega)^2+\omega_1^2)^{1/2}$ is the so-called Rabi
frequency, and $\sigma^\prime:=\Omega^{-1}[(\omega_0-\omega)
\sigma_z+\omega_1\sigma_x]$.

As the computational basis (Sec.~\ref{sec9A:level2}) we will take the
eigenvectors of $\sigma_z$: $|0\rangle:=\ket{\!\uparrow}$ (spin-up
state), $|1\rangle:=\ket{\!\downarrow}$ (spin-down state).\footnote{
With this choice, $\ket{0}$ will be the ground state of the magnetic
Hamiltonian provided that the spin corresponds to a positively charged
particle ($\gamma > 0$).}

The probability of spin flip $\uparrow \leftrightarrow \downarrow$ is
one if and only if $\omega=\omega_0$ (resonance condition), hence
$\Omega=|\omega_1|$, and $t\Omega\in 2\pi(\Z+\half)$.  When the
oscillating part of the magnetic field (\ref{qc98}) is resonant, i.e.
it satisfies $\omega=\omega_0$, then such field is known as a {\em Rabi pulse}.

Let us see how to induce one-qubit operations using Rabi pulses of
appropriate durations.  In view of (\ref{qc33}), and up to the global
phase factor represented by ${\rm Ph}(\delta)$ in (\ref{qc34}), it suffices
to do it for the rotations $R_z(\alpha)$, $R_y(\beta)$:

{a)} The rotation $R_z(\alpha)$ is emulated by taking a constant field
along the $z$-axis and setting to zero the oscillating part ($B_1=0$,
i.e. $\Omega=0$).  The angle is simply $\alpha =\half\omega_0  T$, $T$
being the pulse length. The rotation $R_z(\gamma)$ is obtained
similarly.

{b)} To reproduce the rotation $R_y(\beta)$ in the decomposition
(\ref{qc33}), note that $R_y(\beta)= R_z
(\half\pi)R_x(\beta)R_z(-\half\pi)$, and that $U(t)=R_z(\omega
t)R_x(\Omega t)$.  Therefore, to build $R_y(\beta)$ it suffices to
compose with suitable rotations around $Oz$, implemented as above, the
action of a Rabi pulse with $\Omega T=\beta$.

For instance, a {\em $\pi$-pulse}, i.e. a pulse with duration
$T=\pi/\Omega$, reproduces in the interaction picture a quantum
NOT-gate (up to a global factor -$\ii$).\footnote{At resonance, the
time evolution operator $U(t)$ factorizes as $U(t)=\ee^{-\ii \omega_0
t \sigma_z/2} \ee^{-\ii \Omega t\sigma_x/2}$. The first factor
represents the evolution operator $U_0(t)$ under the static magnetic
field, whereas the second factor is just the total unitary propagator
$U_{\rm I}(t):=U_0^{-1}(t)U(t)$ in the interaction picture.}
Similarly, a ${\pi\over 2}$-pulse produces essentially a Hadamard gate.

So far we have manipulated externally the spins $\half$ to produce
one-qubit gates.  To generate two-qubit gates we need a pair of interacting
qubits at sites 1, 2.  For simplicity's sake, let us assume the
simplest possible type of interaction between them, namely, an Ising
interaction:
\begin{equation}
H_{12} = -(\gamma_1 S_1^z + \gamma_2 S_2^z) B^z + 2(J/\hbar) S_1^z
S_2^z.
\label{qc109}
\end{equation}

\begin{figure}[ht]
\psfrag{0}[Bc][Bc][1][0]{$|00\rangle$}
\psfrag{1}[Bc][Bc][1][0]{$|01\rangle$}
\psfrag{2}[Bc][Bc][1][0]{$|10\rangle$}
\psfrag{3}[Bc][Bc][1][0]{$|11\rangle$}
\psfrag{e}[Bc][Bc][1][0]{energy}
\psfrag{x}[Bc][Bc][0.75][0]{$|\omega_2|$}
\psfrag{y}[Bc][Bc][0.75][0]{$|\omega_1|$}
\psfrag{r}[Bc][Bc][0.75][0]{$|\omega_2|-J$}
\psfrag{s}[Bc][Bc][0.75][0]{$|\omega_1|-J$}
\psfrag{t}[Bc][Bc][0.75][0]{$|\omega_1|+J$}
\psfrag{u}[Bc][Bc][0.75][0]{$|\omega_2|+J$} 
\includegraphics[width=6cm]{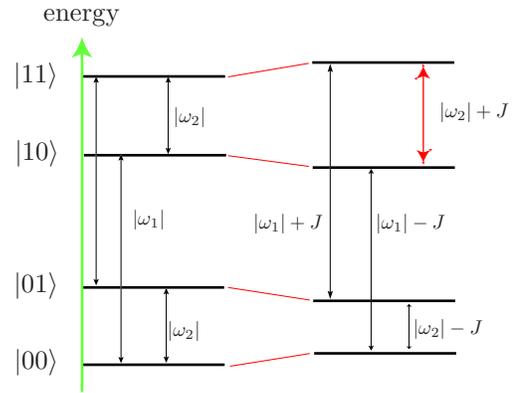}
\caption{Energy levels of a two-qubit spin system with Ising
interaction (units $\hbar=1$).  On the left, the non-interacting
Zeeman levels, and on the right the levels perturbed by the Ising term
(when $\omega_1<\omega_2<-J<0$).}
\label{niveles}
\end{figure}

\noindent The origin of the single spin terms may be the presence of an
external magnetic field.  In
case (\ref{qc109}), this field is constant and directed along $Oz$,
and the two spins may have different magnetic moments. 
The coupling constant $J$
measures the spin-spin interaction.  Defining the frequencies
$\omega_i:=-\gamma_i B^z, i=1,2$, the eigenvalues of this Hamiltonian
are
\begin{equation}
E_{x_1 x_2} =\half\hbar[(-1)^{x_1}\omega_1 + (-1)^{x_2}\omega_2 +
(-1)^{x_1+x_2} J],
\label{qc110}
\end{equation}
where $x_i=0,1, i=1,2$.

These energy levels are represented in Fig.~\ref{niveles} for
$\omega_1<\omega_2<-J<0$.  We clearly see that if we apply a
$\pi$-pulse with frequency $\omega = |\omega_2|+J$, the states
$|11\rangle$ and $|10\rangle$ get swapped while the rest are not
excited.  This is precisely what does a CNOT-gate with the first spin
acting as control qubit and the second spin as a target qubit (Berman
et al., 1997).

Other useful two-qubit gates such as the controlled-phase gate
(\ref{qc24}), that enters Shor's algorithm, can be built-up similarly
using the Ising interaction.  An explicit construction of this gate is
the following (Jones, Hansen and Mosca, 1998)
\begin{equation}
U_{{\rm CPh}}(\phi) = \exp\left(-\ii\half\phi [-\half + \bar S_1^z +
\bar S_2^z -2 \bar S_1^z\bar S_2^z]\right),
\label{qc111}
\end{equation}

\noindent where $\bar S_k^z:=S_k^z/\hbar=\half \sigma_k^z$.  Of
particular interest is the case $\phi = \pi$ for, as remarked in
Sec.~\ref{sec9B:level2}, with this controlled gate plus two Hadamard
gates (on the target qubit) we can reconstruct the important CNOT gate
(\ref{qc24b}).

\subsection{The Ion-Trap QC}
\label{sec11A:level2}

The ion-trap quantum computer was introduced by Cirac and Zoller
(1995) and since then many other potential and actual realizations of
quantum computers have been pursued by many groups.  The quantum
hardware is the following: a qubit is a single ion held in a trap by
laser cooling and the application of appropriate electromagnetic
fields; a quantum register is a linear array of ions; operations are
effected by applying laser Rabi pulses; information transmission is
achieved as a result of the Coulomb interaction between ions and the
exchange of phonons from collective oscillations.  We see again, at a
very fundamental level, that information is physical.  Using the
Cirac-Zoller (CZ) technique it was possible to construct soon
afterward a single quantum gate by Monroe et al.  (1995).

The ion-trap proposal has several advantages: it needs manipulation of
quantum states that were already known from precision spectroscopy
techniques; it has low decoherence rates due to decay of excited
states and the heating of the ionic motion; there exist very efficient
experimental methods to retrieve the information from the quantum
computer like the mechanism of quantum jumps.
\begin{figure}[h]
\includegraphics[width=5 cm]{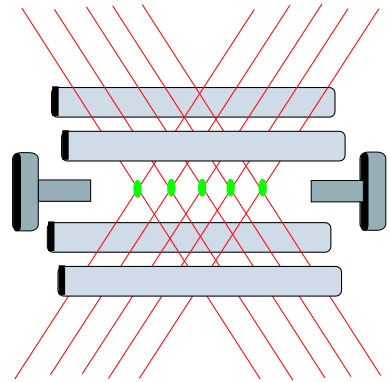}
\caption{Schematic geometry of a radio-frequency quadrupole linear
ion-trap.  Laser beams address a string of ions in the middle of the
setup with 4 linear rods and 2 end-caps.}
\label{geomion}
\end{figure}

\subsubsection{Experimental setup}

The geometry of a radio frequency (RF) ion-trap or Paul trap is
schematically shown in Fig.~\ref{geomion}.  A RF Paul trap uses static
and oscillating electric potentials to confine particles within small
($\sim 1$ $\mu$m) regions.  To obtain a string of ions forming the
quantum register we need a quadrupole ion trap with a cylindrical
geometry.  The confining mechanism of ions is twofold:

i) A strong {\em radial confinement}, achieved by RF potentials
generally produced with four rod electrodes.

ii) An {\em axial confinement} achieved by applying a harmolic-like
electrostatic potential through two end caps.

The ions lie along the trap axis and their oscillations are controlled
by the axial potential.  The collective oscillations of the string
center of mass (CM) are used as a sort of computational bus,
transferring information from one ion to another by phonon exchange.
The dimensions of the ion-traps used by Los Alamos group are typically
1 cm long and 1-2 mm wide (Hughes et al., 1998).

Before any computation takes place, the CM of the ion string must be
set to its ground state.  This is accomplished by a laser cooling
process that cools down the ions to the ground state of their
vibrational motion.  The result of this cooling is an ion string
configuration as shown in Fig.~\ref{geomion}, crystallizing into a
linear array which makes possible to address each ion individually by
lasers.  The inter-ion spacing can be controlled as a balance of the
ion Coulomb repulsion and the axially confining potential (Wineland et
al., 1997).
\begin{figure}[h]
\psfrag{1}[Bc][Bc][1][0]{$|1\rangle$}
\psfrag{0}[Bc][Bc][1][0]{$|0\rangle$}
\psfrag{2}[Bc][Bc][1][0]{$|2\rangle$}
\psfrag{e}[Bc][Bc][1][0]{energy} 
\psfrag{u}{$4\,^2S_{1/2}$}
\psfrag{x}{$3\,^2D_{3/2}$} 
\psfrag{y}{$3\,^2D_{5/2}$}
\psfrag{r}{$4\,^2P_{1/2}$} 
\psfrag{s}{$4\,^2P_{3/2}$} 
\psfrag{3}{732 nm}
\psfrag{4}{729 nm} 
\psfrag{6}{397 nm} 
\includegraphics[width=7 cm]{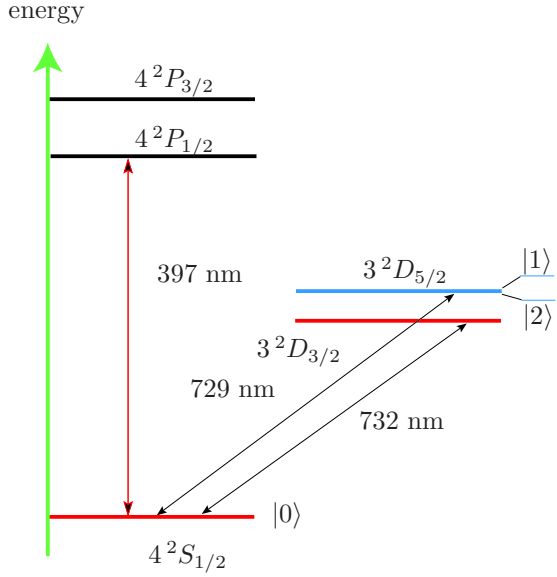}
\caption{Relevant energy levels in Ca$^+$ ions.}
\label{caions}
\end{figure}

Several kinds of ions (Be$^+$, Ca$^+$, Ba$^+$, Mg$^+$, Hg$^+$,
Sr$^+$, etc.) and qubit schemes have been proposed.  The CZ qubit
$\{|0\rangle, |1\rangle\}$ is built using some appropriate electronic
ion states.  For instance, Los Alamos group (Hughes et al., 1998) have
chosen Ca$^+$ ions, whose more relevant levels are shown in
Fig.~\ref{caions}.  The state qubits $\{|0\rangle, |1\rangle \}$ and
one extra auxiliary level $|2\rangle$ (to be described below) are
identified as follows (see Fig.~\ref{caions}):
\begin{equation}
\begin{split}
|0\rangle & = |4\,^2S_{1/2},M_J=\half\rangle, \\ |1\rangle & =
|3\,^2D_{5/2},M_J=\textstyle\frac{3}{2}\rangle, \\ |2\rangle & =
|3\,^2D_{5/2},M_J=-\half\rangle.
\end{split}
\label{qc111b}
\end{equation}

The level ($4\,^2S_{1/2},M_J=\half$) is the ground state while
$(3\,^2D_{5/2},M_J=\frac{3}{2})$ is a metastable level with a long
lifetime (1.06 s).  Both the electric-dipole transition $4\,^2S_{1/2}
\rightarrow 4\,^2P_{1/2}$ at 397 nm wavelength and the electric
quadrupole transition $4\,^2S_{1/2} \rightarrow 3\,^2D_{3/2}$ at 732 nm
are suitable for Doppler and sideband laser cooling, respectively.  In
Doppler cooling the laser radiation pressure slows down the axial
motion of the ions until temperatures $T\sim$ a few mK. To further
reduce the temperature ($T\sim$ a few $\mu$K) until no phonons are
present, one resorts to sideband cooling (Hughes et al.  1997).

\begin{figure}[h]
\psfrag{c}[Bc][Bc][0.75][0]{$|1\rangle$}
\psfrag{d}[Bc][Bc][0.75][0]{$|0\rangle$}
\psfrag{f}[Bc][Bc][0.75][0]{$|2\rangle$}
\psfrag{a}[Bc][Bc][0.75][0]{auxiliar}
\psfrag{g}[Bc][Bc][0.75][0]{$|g\rangle$}
\psfrag{e}[Bc][Bc][0.75][0]{$|e\rangle$}
\psfrag{p}[Bc][Bc][0.75][0]{$\otimes$} 
\psfrag{x}[Bc][Bc][0.75][0]{0
phonons} 
\psfrag{y}[Bc][Bc][0.75][0]{1 phonons}
\psfrag{v}[Bc][Bc][0.75][0]{$V(\pi,\phi)$}
\psfrag{1}[Bc][Bc][0.75][0]{$U_1(\pi,\phi)$}
\psfrag{2}[Bc][Bc][0.75][0]{$U_2(2\pi,\phi)$} 
\includegraphics[width=6cm]{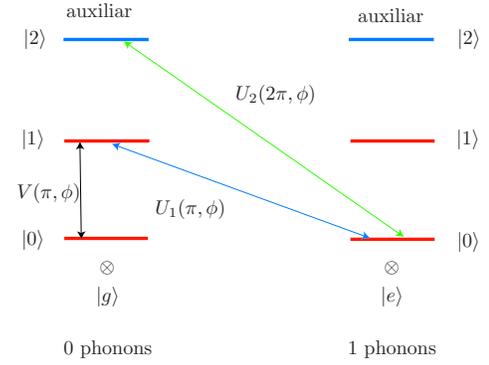}
\caption{Schematic representation of the transitions generated by the
$V$- and $U$-pulses.}
\label{ionlevels}
\end{figure}

The interaction between CZ qubits is achieved using two types of
degrees of freedom: internal (the electronic states of the ions), and
external (the vibrational states of their collective excitations).
Thus, an active state for information processing is the tensor product
of an electronic state and a quantum oscillator state of the axial
potential, namely,
\begin{equation}
|\Psi \rangle = |x\rangle |\alpha\rangle, x=0,1; \ \alpha={\rm g,e},
\label{qc112}
\end{equation}

\noindent where $|x\rangle$ refer to the electronic levels and $|{\rm
g}\rangle$, $|{\rm e}\rangle$ denote the ground state and first
excited state of the vibrational motion, respectively.  In $|{\rm
g}\rangle$ there are no phonons present in the system while there is
one phonon in $|{\rm e}\rangle$ (see Fig.~\ref{ionlevels}).

\subsubsection{Laser pulses}

With this structure of states one can apply two types of laser Rabi
pulses to the ions in order to achieve quantum logic operations.
These are called $V$- and $U$-pulses:

{\em V-pulse}. This pulse implements one-qubit operations.  Its
frequency is tuned to resonate with the optical transition between the
qubit states.  It swaps the electronic states $|0\rangle
\leftrightarrow |1\rangle$ and leaves the vibrational mode in the
ground state $|{\rm g}\rangle$.  The unitary evolution operator
induced by this pulse is
\begin{equation}
\begin{split}
& V(\theta,\phi) := \ee^{-\ii t H_V/\hbar}, \\ & H_V := \half\hbar\Omega 
[\ee^{-\ii \phi}|1\rangle \langle 0| + \ee^{\ii
\phi}|0\rangle \langle 1|],
\end{split}
\label{qc113}
\end{equation}

\noindent where $\theta := \Omega t$, $H_V$ is the $V$-pulse
Hamiltonian, $\Omega$ is the Rabi frequency (proportional to the
square root of the laser intensity), and $\phi $ is the laser phase.
Then, this pulse produces the following action on the electronic
states:
\begin{equation}
V(\theta,\phi):
\begin{cases}
|0\rangle \mapsto \cos {\theta \over 2} |0\rangle - \ii \ee^{-\ii
\phi} \sin {\theta \over 2} |1\rangle, & \\ |1\rangle \mapsto \cos
{\theta \over 2} |1\rangle - \ii \ee^{\ii \phi} \sin {\theta \over 2}
|0\rangle. &
\end{cases}
\label{qc114}
\end{equation}

{\em U-pulse}. This pulse is used to implement two-qubit operations.
The laser frequency is now adjusted to induce simultaneously both an
electronic and a vibrational transition.  To help performing the
desired logic gates, an auxiliary electronic state $|2\rangle$ (see
Fig.~\ref{ionlevels}) is available.  The time evolution operator led
by this pulse is
\begin{equation}
\begin{split}
& U_{\hat{x}}(\kappa,\phi) := \ee^{-\ii t H_U(\hat{x})/\hbar}, \quad
\hat{x} =1,2, \\ & H_U(\hat{x}) := \half\hbar \eta \Omega [\ee^{-\ii
\phi}|\hat{x}\rangle \langle 0| a + \ee^{\ii \phi}|0\rangle \langle
\hat{x}| a^{\dagger}],
\end{split}
\label{qc115}
\end{equation}
\noindent where: $H_U$ is the $U$-pulse Hamiltonian, $\kappa :=
\eta\Omega t$, $\eta$ is the Lamb-Dicke parameter\footnote{This
quantity is the ratio between the width of the ion oscillation in the
vibrational ground state of the register and the (reduced) laser
wavelength $\lambda_L/2\pi$: $\eta:= (\hbar/2NM_{\rm
ion}\omega_z)^{1/2}(2\pi/\lambda_L)$, where $N$ is the number of cold
ions, and $\omega_z$ is the vibrational frequency of the register CM
along the trap axis.  The Lamb-Dicke criterion $\eta\ll 1$ is demanded
for Eq.  (\ref{qc115}) to be a good approximation (Cirac and Zoller,
1995).  For the Ca$^+$ trap, with $N\sim 10$, $\omega_z\sim 100$ kHz,
then $\eta\sim 0.2$.} and $a^{\dagger},a$ are creation and
annihilation phonon operators satisfying
\begin{equation}
a^{\dagger} |{\rm g}\rangle = |{\rm e}\rangle, \ a |{\rm e}\rangle =
|{\rm g}\rangle, \ [a,a^{\dagger}] = 1.
\label{qc116}
\end{equation}
Several physical constraints on these parameters in a linear ion-trap
are to be fulfilled for it to function stably and as required (Cirac
and Zoller, 1995).

The $U$-pulse acts as follows:
\begin{equation}
U_{\hat{x}}(\kappa,\phi):
\begin{cases}
|0\rangle|{\rm g}\rangle \mapsto |0\rangle|{\rm g}\rangle, & \\
|0\rangle|{\rm e}\rangle \mapsto \cos {\kappa \over 2} |0\rangle|{\rm
|e}\rangle - \ii \ee^{-\ii \phi} \sin {\kappa \over 2} |\hat{x}\rangle
|{\rm g}\rangle, & \\ |\hat{x}\rangle|{\rm g}\rangle\mapsto \cos
|{\kappa \over 2} |\hat{x}\rangle|{\rm g}\rangle - \ii \ee^{\ii \phi}
|\sin {\kappa \over 2} |0\rangle|{\rm e}\rangle.  &
\end{cases}
\label{qc117}
\end{equation}

\begin{figure}
\psfrag{a}[Bc][Bc][0.75][0]{a)}
\psfrag{1}[Bc][Bc][0.75][0]{$|x_1\rangle$}
\psfrag{i}[Bc][Bc][0.75][0]{ion $i$}
\psfrag{2}[Bc][Bc][0.75][0]{$|x_2\rangle$}
\psfrag{j}[Bc][Bc][0.75][0]{ion $j$}
\psfrag{3}[Bc][Bc][0.75][0]{$|g\rangle$}
\psfrag{p}[Bc][Bc][0.75][0]{phonon}
\psfrag{R}[Bc][Bc][0.75][0]{$U_1(\pi,0)$}
\psfrag{S}[Bc][Bc][0.70][0]{$U_2(2\pi,0)$}
\psfrag{u}[Bc][Bc][0.75][0]{$(-1)^{x_1}|x_1\rangle$}
\psfrag{v}[Bc][Bc][0.75][0]{$|g\rangle$}
\psfrag{w}[Bc][Bc][0.75][0]{$(-1)^{x_1x_2+x_1}|x_2\rangle$}
\psfrag{o}[Bc][Bc][0.75][0]{$(-\ii)^{x_1}|0\rangle$}
\psfrag{n}[Bc][Bc][0.75][0]{$|{\rm p}(x_1)\rangle$}
\psfrag{l}[Bc][Bc][0.75][0]{$|x_2\rangle$}
\psfrag{m}[Bc][Bc][0.75][0]{$(-1)^{x_1x_2+x_1}|x_2\rangle$}
\psfrag{b}[Bc][Bc][0.75][0]{b)} \psfrag{x}[Bc][Bc][0.75][0]{$U$-pulse
1} \psfrag{y}[Bc][Bc][0.75][0]{$U$-pulse 2}
\psfrag{z}[Bc][Bc][0.75][0]{$U$-pulse 3}
\psfrag{c}[Bc][Bc][0.75][0]{$|1\rangle_i$}
\psfrag{d}[Bc][Bc][0.75][0]{$|0\rangle_i$}
\psfrag{f}[Bc][Bc][0.75][0]{$|1\rangle_j$}
\psfrag{h}[Bc][Bc][0.75][0]{$|0\rangle_j$}
\psfrag{g}[Bc][Bc][0.75][0]{$|g\rangle$}
\psfrag{e}[Bc][Bc][0.75][0]{$|e\rangle$}
\psfrag{k}[Bc][Bc][0.75][0]{$\pi$} \psfrag{t}[Bc][Bc][0.75][0]{$2\pi$}
\includegraphics[width=9 cm]{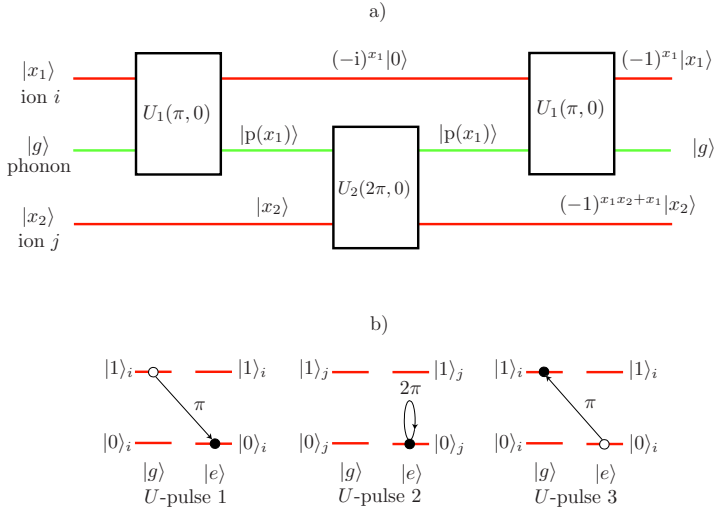}
\caption{a) Quantum circuit for the controlled-phase gate in an
ion-trap QC. We denote by $|{\rm p}(x_1)\rangle$ the phonon states
${\rm p}(0):={\rm g}$, ${\rm p}(1):={\rm e}$.  Note also that the
overall final phase is $(-1)^{x_1x_2}$, as it corresponds to a
controlled phase $\phi=\pi$.  b) Evolution of a state under the
sequence of $U$-pulses in (\ref{qc118}).}
\label{upulses}
\end{figure}

\subsubsection{Building logic gates}

By controlling the duration of the laser pulses in (\ref{qc114}) and
(\ref{qc117}) we can perform logic operations in a fashion akin to
those for spin qubits with Rabi pulses.  The nice thing abouth the
ion-trap QC is that the same Rabi pulses can drive conditional logic
when phonons are suitably put to work.

For instance, a CNOT gate can be constructed using a series of $V$-
and $U$-pulses.  To this end, we first reproduce a $\pi$
controlled-phase (\ref{qc24}) gate between qubits at sites $i,j$ as
follows:
\begin{equation}
U^{(i,j)}_{{\rm CPh}}(\pi) = U_1^{(i)}(\pi, 0) U_2^{(j)}(2\pi,0)
U_1^{(i)}(\pi, 0)
\label{qc118}
\end{equation}
\noindent The explicit action of this squence of operations is shown
in Fig.~\ref{upulses}.  This two-bit gate is constructed only out of
$U$-pulses.

In order to construct CNOT from this gate (see (\ref{qc24b}),
Fig.~\ref{q2pi}) we need to resort to $V$-pulses, namely
\begin{equation}
U^{(i,j)}_{{\rm CNOT}} = V^{(j)}(\half\pi, \half\pi) U^{(i,j)}_{{\rm
CPh}}(\pi) V^{(j)}(\half\pi, \half\pi)
\label{qc119}
\end{equation}

\noindent where these $V$-pulses correspond to Hadamard gates.  Other
logic gates involving a larger number of qubits can be constructed
similarly using theses basic pulse operations (Cirac and Zoller, 1995).

Let us note that the $2\pi$ auxiliary rotations in (\ref{qc118}) do
not produce any population of the auxiliary atomic levels nor the CM
levels.  Thus, a variation of the population of these levels by the
gate operation would indicate a faulty experimental realization.

Upon completion of the quantum operations in the ion-trap QC, we need
to readout the outcome result (see Sec.~\ref{sec9:level1}).  This is
done by measuring the state of each qubit in the quantum register
using the quantum jump technique (Nagourney et al, 1986; Bergquist et
al., 1986; Sauter et al., 1986).  For instance, for the Ca$^+$ qubits
(\ref{qc111b}), the laser is tuned to the dipole transition
$4\,^2S_{1/2} \rightarrow 4\,^2P_{1/2}$ at 397 nm (see Fig.~\ref{caions}).
Now, there are two possibilities for the ion being addressed with the
laser: i) if the ion radiates (fluoresce), this means that its state
is $|0\rangle$; ii) if the ion does not radiate (remains dark), then
it was in the $|1\rangle$ state.  Therefore, just by observing which
ions fluoresce and which remain dark we can retrieve the bit values of
the register.  Actually, there is a third possibility in which
$4\,^2P_{1/2} \rightarrow 3\,^2D_{3/2}$.  In order to prevent this
metastable level from being populated, a pump-out laser is
also required.

\subsubsection{Further applications}

\begin{figure}[h]
\psfrag{s}[Bc][Bc][1][0]{$200\ \mu$m}
\includegraphics[width=7 cm]{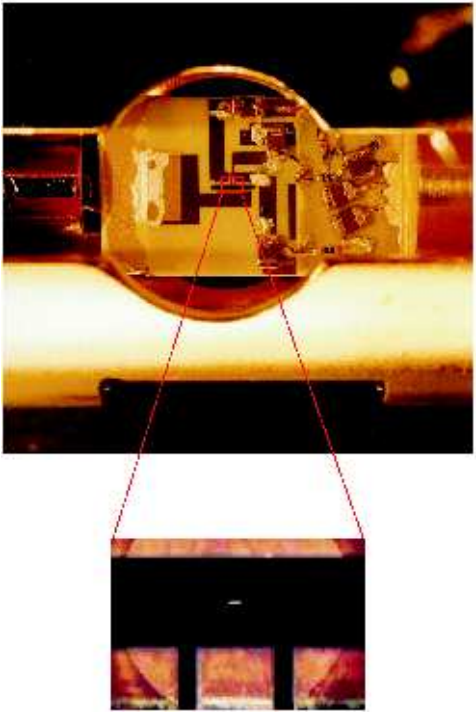}
\caption{Micromachined ion trap showing a four-qubit register in the
inset (Sackett et al., 2000).}
\label{iontrap}
\end{figure}

The ion-trap technique has also found applications in the preparation
of entangled states (Molmer and Sorensen, 1999).  This has been
experimentally realized by the NIST group (Sackett et al., 2000) with
the generation of entangled states of two and four trapped ions.  In
Fig.~\ref{iontrap} a 4-qubit quantum register used in these
experiments is shown.

Unavoidable errors put computational limits in ion-trap quantum
computers.  Sources of these constraints are the spontaneous decay of
the metastable state, laser phase decoherence, ion heating and other
kinds of errors.  Using simple physical arguments it is possible to
place upper bounds on the number of laser pulses $N_U$ sustained by
the ion trap before entering a decoherence regime (Hughes et al.,
1996), namely,
\begin{equation}
N_U L^{1.84} < {2 Z (\tau/1\,{\rm s}) \over  A^{1/2}
F^{3/2}(\lambda/1\,{\rm m})^{3/2}}
\label{qc119b}
\end{equation}

\noindent where $Z$ is the ion degree of ionization, $\tau$ is the
lifetime of the metastable state, $L$ is the number of ions and $A$ their
atomic mass, $F$ parameterizes the focusing capability of the laser
and $\lambda$ is the laser wavelength.  This bound depends on the ion
parameters $A$ and $\tau$, making some ion
species more suitable than others.\footnote{The number $N_U$ refers
only to the number $U$-pulses for they last much longer than the
$V$-pulses, which are thus neglected.} With this bound it is possible
to estimate the number of ions needed to factorize a 438-bit number
using Ytterbium (with the transition 4f$^{14}$6s $^2$S$_{1/2}
\leftrightarrow$ 4f$^{13}$6s$^2$ $^2$F$_{1/2}$, which has a very long
lifetime (1533 days) and a wavelength of 467 nm).  Around 2200 trapped
ions and $4.5 \times 10^{10}$ pulses would be required to perform the
sought factorization, in about 100 hours of computation time (Hughes
et al., 1996).

Scalability of the ion-trap QC is a central issue if we want to have a
real useful machine for number factoring and the like.  With current
techniques, it is believed that prospects for reaching a few tens of
qubits are good (Hughes et al., 1998).  Cirac and Zoller (2000) have
proposed an ion-trap based quantum computer with a two-dimensional
array of independent ion traps and a different ion (head) that moves
above this plane.  This setup is still conceptually simple and it is
believed to be within reach of present experimental technologies.

\subsection{NMR Liquids: Quantum Ensemble Computation}
\label{sec11B:level2}

We have seen that using spin qubits and spin resonance is a natural
choice for doing quantum computations.  Nuclear spins are good
candidates for spin qubits but they pose both theorical and
experimental challenges.  There have been independent proposals to
overcome these difficulties: the {\em logical labelling formalism} by
Gershenfeld, Chuang and Lloyd (1996), Gershenfeld and Chuang (1997),
and the {\em spatial averaging formalism} by Cory, Fhamy and Havel
(1997).  They have been addressed experimentally by several groups.
Later, a {\em time averaging formalism} was introduced by Knill,
Chuang and Laflamme (1997).

The quantum hardware in this case consists of a liquid containing a
large number of molecules of a certain type.  A qubit is the spin of a
nucleus in a molecule, and a quantum register is a molecule as a
whole, i.e., each molecule is an independent quantum computer;
operations are effected using nuclear magnetic resonance techniques
(Rabi oscillations) and information transmission between nuclei is based
on the spin interactions within each molecule.

\subsubsection{Spins at thermal equilibrium}

The choice of nuclear spins as qubits has several pros and cons.  On
one hand, nuclear spins in a molecule of a liquid are very robust
quantum systems, for they are well screened from other sources of
magnetic fields by the electron cloud that surrounds them.  This
results in decoherence times of the order of seconds, long enough to
let quantum computations going on.  On the contrary, in a liquid at
finite temperature the nuclear spins form a highly mixed state, not a
pure state as we have been assuming in the formalism for quantum
computation introduced so far.  Such formalism needs be modified
accordingly, by describing with {\em density matrices} the mixed
states of spins and their evolution.

A consequence of the finite temperature is that the precise initial
conditions of a particular nuclear spin are not known as required for
standard quantum computation.  Instead, we can only know the
probability of finding the spin in one of the two states
$|0\rangle=\ket{\!\uparrow}$ or $|1\rangle=\ket{\!\downarrow}$.  In the
following, we shall assume that the molecules in the solution are in
thermal equilibrium at some temperature $T$.  
Hence the density matrix describing the quantum
state of the relevant nuclear spins in each single molecule is
\begin{equation}
\rho := {\ee^{-\beta H}\over {\rm Tr}[\ee^{-\beta H}]},
\label{qc120}
\end{equation}

\noindent where $H$ is the Hamiltonian of the system, $\beta=1/k_{\rm
B}T$ the inverse temperature, and the trace is over any orthonormal
basis of the Hilbert space.  Let us take the simplest case of a single
spin qubit with a Zeeman splitting Hamiltonian $H=\omega S^z$,
$\omega=-\gamma B_0$.  Then, (\ref{qc120}) becomes
\begin{equation}
\begin{aligned}
\rho_{00} & = {\ee^{-\beta \hbar \omega/2} \over \ee^{\beta \hbar
\omega/2}+\ee^{-\beta \hbar \omega/2}}, 
\\ \rho_{11} & = {\ee^{\beta
\hbar \omega/2} \over \ee^{\beta \hbar \omega/2}+\ee^{-\beta \hbar
\omega/2}}, 
\\ \rho_{01} & = 0 = \rho_{10}.
\end{aligned}
\label{qc121}
\end{equation}

\noindent The diagonal terms of $\rho$ represent the probability of
finding the spin in the state $|0\rangle$ or $|1\rangle$.  In contrast,
the density matrix of a pure
state $|\psi(t)\rangle:= \alpha_0(t) |0\rangle + \alpha_1(t) |1\rangle
$ is
\begin{equation}
\rho_{\psi} := |\psi \rangle \langle \psi | =
\begin{pmatrix}
|\alpha_0|^2 & \alpha_0 \alpha^{\ast}_1 \\ \alpha^{\ast}_0 \alpha_1 &
|\alpha_1|^2
\end{pmatrix}.
\label{qc122}
\end{equation}

\noindent Therefore we see that at finite temperature and thermal equilibrium,
the off-diagonal elements of the density matrix average to zero while
they are non-vanishing for a generic pure quantum state.

\subsubsection{Liquid state NMR spectroscopy}

To overcome these difficulties, the proposal for a NMR quantum
computer takes advantage of the highly developed techniques in liquid
state NMR spectroscopy accumulated for fifty years (Ernst et al.,
1987).

\begin{figure}[h]
\psfrag{a}[Bc][Bc][0.75][0]{a)} \psfrag{b}[Bc][Bc][0.75][0]{b)}
\psfrag{c}[Bc][Bc][0.75][0]{c)} 
\includegraphics[width=7 cm]{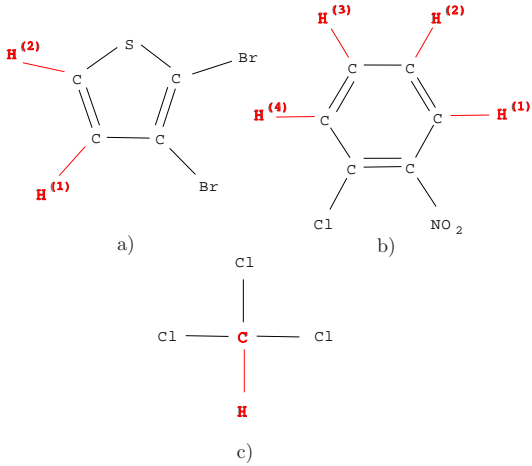}
\caption{Some examples of molecules used in NMR liquid quantum
computation: a) 2,3-dibromo-thiophene (homonuclear), b)
1-chloro-2-nitro-benzene (homonuclear), c) chloroform (heteronuclear)}
\label{molecules}
\end{figure}

In a NMR liquid the molecules are in solution.  In each molecule only
some of its nuclei are active for doing quantum computation.  When the
qubits consist of atomic nuclei of the same chemical element the
molecules are called {\em homonuclear}, and {\em heteronuclear}
otherwise.  Examples of homonuclear
molecules are shown in Fig.~\ref{molecules}, like the
2,3-dibromo-thiophene where the active nuclear spins are those of the
two Hydrogen atoms, or the 1-chloro-2-nitro-benzene with four active
Hydrogen atoms.  An example of heteronuclear molecule is the
$^{13}$C-labelled chloroform\footnote{The nucleus of the most common 
isotope $^{12}$C is spinless.
Adding one extra neutron endows it with an overall operative spin 
$\half$.}  in which the two active qubits come from the atoms of Hydrogen
and Carbon.  The number of qubits in the working register narrows the
choice of the molecule structure.

\begin{figure}[h]
\includegraphics[width=5 cm]{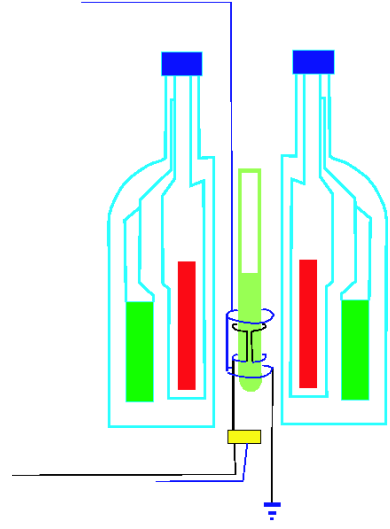}
\caption{Schematic setup of a NMR experiment}
\label{spec}
\end{figure}

An appropriate experimental setup for NMR computation is much like any
other instrumentation used in NMR spectroscopy.  In Fig.~\ref{spec}
the basic structure of a NMR spectrometer is shown.  The liquid sample
is held in a probe inside a radio-frecuency cavity subjected to a
strong homogeneous magnetic field of around 10 T, usually produced by
a superconducting magnet.  The RF cavity is tuned to the resonance
frequencies of the active nuclear spins.

In a typical sample there are $N\sim 10^{18}$ molecules in solution.
The dipole-dipole interactions between the spins in different
molecules as well as other intermolecular interactions average to zero
due to the random rotational motion of the molecules in the usual time
scale for controlling the spin dynamics and the measurement (Slichter,
1990).  Hence, only interactions within each molecule are observable
and the sample can be regarded as an ensemble of independent and
mutually incoherent quantum computers.  This reasonable approximation
yields a huge reduction in the large density matrix of dimension $\sim
2^{O(N)}$ describing the whole ensemble of active nuclear spins, which
may be replaced by a much smaller density matrix of dimension $2^n$,
where $n$ is the number of active nuclei in a single molecule.

Within each molecule, the total Hamiltonian $H(t)$ of the active spins
has two parts (Cory et al., 2000), one internal and another external:
\begin{equation}
H(t) := H_{\rm int} + H_{\rm ext}(t).
\label{qc123}
\end{equation}

\noindent The internal Hamiltonian describes the interactions among
spins within the molecule, while the external Hamiltonian controls the
spin dynamics under Rabi pulses.  The operator $H_{\rm int}$ embodies:
a) the molecule interaction energy with a strong homogeneous magnetic
field that causes a Zeeman splitting of the nuclear spin levels; b)
the spin-spin interactions between active nuclei, modelled by a
magnetic exchange interaction $2(J_{ij}/\hbar){\bf S}_i\cdot {\bf
S}_j$ mediated by electrons in molecular orbitals that overlap both
nuclear spins $i,j$.  In most cases this interaction can be further
simplified using the weak coupling aproximation $|J_{ij}|
\ll|\omega_i-\omega_j|$, which assumes that the spin-spin coupling is
much smaller than the Zeeman splitting.  This simplification produces
a {\em scalar coupling} of Ising type between the spins, and yields
the following good approximation to the internal Hamiltonian:
\begin{equation}
H_{{\rm int}} \approx \sum_{i=1}^n \omega_i S_i^z + 2 \sum_{i\neq
j=1}^n (J_{ij}/\hbar) S_i^z S_j^z,
\label{qc124}
\end{equation}
\noindent where $J_{ij}$ measures the coupling between the active
spins at sites $i,j$,\footnote{In NMR spectroscopy $J_{ij}$ are
typically $\sim 100$ Hz.} and $\omega_i$ are the resonance frequencies
for each spin.  They are different even for homonuclear molecules due
to the unlike screening of each nuclear spin from the surrounding
electrons.  This effect is called {\em chemical shift}.  Thus, in
(\ref{qc124}) the one-body terms may be used to distinguish qubits,
while the two-body terms serve to implement the conditional logic of
two-qubit gates.  The values of the parameters $\omega_i$ and $J_{ij}$
are determined by standard NMR spectroscopy techniques prior to the
computation.  Standard NMR spectroscopy and NMR quantum computation
share the means but differ in goals: in the former we aim to determine
the parameters of the Hamiltonian (\ref{qc124}) to study the chemistry
and dynamics of the molecules in solution, while in the latter the
form of (\ref{qc124}) is already known and we set out to use it to
perform controlled logic operations.

The external time dependent Hamiltonian $H_{{\rm ext}}(t)$ helps to
control the evolution of the spins.  These form an ensemble of
systems, initially described by the thermal density matrix $\rho$
(\ref{qc121}) and its time evolution is
\begin{equation}
\rho (t) = U(t) \rho (0) U^{\dagger}(t),
\label{qc125}
\end{equation}
\noindent where $U(t)$ is the unitary propagator generated by the
total Hamiltonian in (\ref{qc123}) and $\rho(0)$ is the thermal
density matrix (\ref{qc121}).

\subsubsection{High temperature regime: pseudo-pure states}

The evolution of the density matrix (\ref{qc120}) is simplified in the
high temperature limit $k_{{\rm B}}T \gg \hbar \omega_i$, where the
Zeeman splittings are much smaller than the Bolzmann energy.  Then, we
can approximate (\ref{qc120}) as follows
\begin{equation}
\rho \simeq {1-\beta H \over {\rm Tr}(1-\beta H)} \simeq \rho_n :={1
\over 2^n} - {\beta H \over 2^n}.
\label{qc126}
\end{equation}

\noindent Thus, in NMR quantum computing there is no need for cooling
down the system until reaching its ground state as in other types of
QCs.

Let us analyze step by step the approximation (\ref{qc126}) for
quantum computing.  First, let us consider the case of a single spin.
Then, the density matrix is simply given by
\begin{equation}
\begin{aligned}
\rho_1 & := \half - \epsilon_1 \delta_{1}, \\ \delta_{1} & := \bar
S_1^z, \ \epsilon_1 := \half \hbar \omega_1/k_{\rm B}T,
\end{aligned}
\label{qc127}
\end{equation}
\noindent where $\delta_{1}$ is called the {\em deviation density
matrix}\footnote{Sometimes it is also called {\em reduced density
matrix}.} and $|\epsilon_1|\sim 10^{-5}$ at room temperature for
conventional NMR liquids.  Thus, the factor $\epsilon_1$ gives the
strength of the NMR signal relative to background noise.  This expression can
be further simplified by dropping out the unit term, which does not
change under time evolution (\ref{qc125}):  in a NMR
experiment the expectation value of an observable ${O}$
is given by
\begin{equation}
\langle {O} \rangle = \tr(O\rho),
\label{qc128}
\end{equation}
\noindent and, as it happens, all NMR observables are traceless.
Thus, all the information is in $\epsilon_1\delta_{1}$.  As $\epsilon_1$ enters
only as an overall scale factor, we can also drop it out in all this
description and write the effective thermal density matrix simply as
\begin{equation}
\rho_1 \sim \bar S_1^z.
\label{qc129}
\end{equation}

Now let us recall that for a qubit in the ground state or excited
state the density matrices are
\begin{equation}
\begin{aligned}
\rho_{|0\rangle} = |0\rangle \langle 0| = \half + \bar S^z, \\
\rho_{|1\rangle} = |1\rangle \langle 1| = \half - \bar S^z,
\end{aligned}
\label{qc130}
\end{equation}

\noindent and discarding the unit terms, we see that for NMR purposes
the one-qubit states $|0\rangle$, $|1\rangle$, are equivalent to $\bar
S^z$, $-\bar S^z$, respectively.  The spin operators 
representing one-qubit states in this correspondence are 
called {\em pseudo-pure} or {\em
effective pure states}. It also works for a superposition state; for
instance, the pure state $|\Psi\rangle=2^{-1/2} (|0\rangle
+|1\rangle)$ has a density matrix
\begin{equation}
\rho_{|\Psi\rangle} = \half + \bar S^x,
\label{qc131}
\end{equation}
\noindent equivalent to $\bar S^x$.  Actually, the correspondence is
one-to-one in the case of one-qubit states, for the density matrix of
a single pure state (\ref{qc122}) is a Hermitean operator that can be
expanded as a real linear combination of the Pauli matrices $\{1,
\sigma^x, \sigma^y, \sigma^z \}$.

Then, the time evolution of a NMR density matrix is that of the spin
$\half$ operators.  When the external Hamiltonian corresponds to a
Rabi pulse, the transformation laws are simple.  The evolution
operator for a single spin with Zeeman Hamiltonian $H_1:=\hbar
\omega_1 \bar S_1^z$ is
\begin{equation}
U_{\rm Z}(t) := \ee^{-\ii t \omega_1 \bar S_1^z} = \cos (\half\omega_1
t) 1 - 2\ii \sin (\half\omega_1 t) \bar S_1^z,
\label{qc132}
\end{equation}
\noindent whence the evolution of the one-qubit effective pure states:
\begin{equation}
\begin{split}
U_{{\rm Z}}(t) \bar S_1^x U^{\dagger}_{{\rm Z}}(t) & = \cos (\omega_1
t) \bar S_1^x + \sin (\omega_1 t) \bar S_1^y, \\ U_{{\rm Z}}(t) \bar
S_1^y U^{\dagger}_{{\rm Z}}(t) & = -\sin (\omega_1 t) \bar S_1^x +
\cos (\omega_1 t) \bar S_1^y, \\ U_{{\rm Z}}(t) \bar S_1^z
U^{\dagger}_{{\rm Z}}(t) & = \bar S_1^z.
\end{split}
\label{qc133}
\end{equation}

The Zeeman propagator $U_{\rm Z}(t)$ rotates the spin around the
$z$-axis an angle $\varphi:= \omega_1 t$.  It is customary to use the
spectroscopist notation to denote the unitary action of the RF pulses
in the rotating frame or interaction picture:
\begin{equation}
[\varphi]^{\alpha}_i:=\ee^{-\ii \varphi \bar S_i^{\alpha}}, \
\alpha=x,y, \quad i=1,2,\ldots,n,
\label{qc134}
\end{equation}
\noindent where $\varphi$ is the rotation angle, $\alpha$ is the
rotation axis, and $i$ the index  labelling the rotating qubit.  Thus,
the effect of a $[\pi ]_1^x$ pulse
\begin{equation}
[\pi]^{x}_1  = \ee^{-\ii \pi \bar S_1^x} =
\begin{pmatrix}
0 & -\ii \\ -\ii & 0
\end{pmatrix}
\label{qc136a}
\end{equation}
\noindent is,
\begin{equation}
\bar S_1^z \overset{[\pi]^{x}_1}{\longrightarrow} -\bar S_1^z \quad
{\rm i.e.} \;  |0\rangle\langle 0| \leftrightarrow |1\rangle\langle 1|
.
\label{qc136b}
\end{equation}
\noindent Therefore, with a $[\pi]_1^x$ pulse effected on a
non-interacting ensemble of single spins in thermal equilibrium, we
can effectively simulate the quantum transition between the qubit
states $|0\rangle$ and $|1\rangle$.  In the thermal equilibrium
ensemble, there is an excess of populated ground states with respect
to the populations of excited states.  After applying the pulse, the
populations are reversed.  Likewise, a $[\half\pi]_1^x$ pulse produces
off-diagonal terms in the density matrix at finite temperature that
simulates quantum superpositions of pure states.

For multiqubit states, the correspondence between pure states and spin
density matrices is not so simple.  Let us consider the case of
two-qubit states.  It is possible to extend the description of
multi-spin density matrix using the so-called {\em product
operator formalism} by the NMR spectroscopists.  Thus, the density
matrix for the pure ground state $|\Psi\rangle=|00\rangle$ is
\begin{equation}
\rho_{|\Psi\rangle} := |00\rangle \langle 00| = \half(\half + \bar
S_1^z + \bar S_2^z + 2 \bar S_1^z\bar S_2^z).
\label{qc137}
\end{equation}

In general, any density matrix can be expanded in a tensor product
basis of one-spin operators $\{\bar S_i^x, \bar S_i^y, \bar
S_i^z\}_{i=1,\ldots,n}$.  For $n$  qubits,
\begin{equation}
\begin{split}
&\rho = \sum_{\alpha_1,...,\alpha_n}
c_{\alpha_1,...,\alpha_n}\sigma_1^{\alpha_1} ...  \sigma_n^{\alpha_n},
\\ & c_{\alpha_1,...,\alpha_n}:= 2^{-n}\tr(\rho \,\sigma_1^{\alpha_1}
...  \sigma_n^{\alpha_n}),
\end{split}
\end{equation}
where $\alpha_i=0,x,y,z$, and $\sigma_i^0:=1$.

This has the advantage that the evolution of the ensemble density
matrix is then simply determined through the evolution rules for
single spin operators.  The problem that we face now is that the
thermal equilibrium matrix in the high-temperature limit $k_{\rm
B}T\gg \hbar \omega_i$ for the Hamiltonian (\ref{qc124}) is
\begin{equation}
\begin{split}
\rho_2 = \textstyle{1 \over 4} - {1 \over 8} {\hbar \beta} \,& {\rm
diag}(\omega_1+\omega_2+J_{12},\omega_1-\omega_2-J_{12}, \\ &
-\omega_1+\omega_2-J_{12},-\omega_1-\omega_2+J_{12}),
\end{split}
\label{qc138}
\end{equation}
\noindent which is further approximated assuming a weak coupling
regime $|\omega_1-\omega_2|, |J_{1,2}| \ll|\omega_1+\omega_2|/ 2$ to
\begin{equation}
\rho_2  \simeq \fourth-\epsilon_2 (\bar S_1^z+\bar S_2^z), \quad
\epsilon_2  := \textstyle\frac{1}{8}\hbar(\omega_1+\omega_2)/k_{\rm B}T,
\label{qc139}
\end{equation}
\noindent and the corresponding deviation matrix $\delta_2:=\bar
S_1^z+\bar S_2^z$ is not equivalent to the initial quantum ground
state (\ref{qc137}) we want to simulate.  This is the {\em
initialization problem} in NMR computing.

\subsubsection{Logic gates with NMR}

To prepare the ensemble of spins in the referencial state
(\ref{qc137}) as well as to implement the logical operations for
quantum processing, we need to resort to a series of well-known
techniques in NMR liquid spectroscopy to carry out controlled time
evolution of spins:

i) {\em Rabi pulses.}  The associated external Hamiltonian
(\ref{qc123}) corresponds to a harmonically oscillating magnetic field
perpendicular to the Zeeman axis.  It is applied at resonance and its
effect on a single spin in the $z$-direction is the following
\begin{equation}
\begin{split}
& [\varphi ]_1^x : \;  S_1^z \mapsto \cos(\varphi) S_1^z -
\sin(\varphi) S_1^y, \\ & [\varphi ]_1^y :\;  S_1^z \mapsto
\cos(\varphi) S_1^z + \sin(\varphi) S_1^x,
\end{split}
\label{qc140}
\end{equation}

\noindent where $\varphi:=\Omega t$, $t$ being the time duration and
$\Omega$ the Rabi frequency.

ii) {\em  Chemical-shift pulses.}  They act as the propagator
generated by the Zeeman part of the internal Hamiltonian
(\ref{qc123}).  Their effect on the spin operators is given by
(\ref{qc133}).

iii) {\em Scalar pulses.} These induce the time evolution under the
scalar coupling (two-spin) part of the internal Hamiltonian
(\ref{qc123}).  For two qubits labelled 1,2, this scalar coupling
propagator is also diagonal in the computational basis:
\begin{equation}
U_J(t)=\ee^{-\ii 2J_{12}t \bar S_1^z \bar S_2^z } = \cos(\half J_{12}
t) - 4\ii \sin(\half J_{12} t) \bar S_1^z \bar S_2^z,
\label{qc141}
\end{equation}
\noindent and its effect on single spin operators is
\begin{equation}
\begin{split}
U_J(t)\bar S_1^x U^{\dagger}_J(t) & = \cos(J_{12} t)\bar S_1^x + 2
\sin(J_{12} t)\bar S_1^y\bar S_2^z, \\ U_J(t)\bar S_1^y
U^{\dagger}_J(t) & = \cos(J_{12} t)\bar S_1^y - 2 \sin(J_{12} t)\bar
S_1^x\bar S_2^z, \\ U_J(t)\bar S_1^z U^{\dagger}_J(t) & =\bar S_1^z.
\end{split}
\label{qc142}
\end{equation}

The NMR spectroscopist notation for these pulses is
\begin{equation}
[\varphi]_{12}^J:=\ee^{-\ii 2J_{12}t \bar S_1^z \bar S_2^z },
\label{qc143a}
\end{equation}
\noindent where the rotation angle is $\varphi = J_{12}t$ and the
subscript denotes the spins involved in the scalar pulse.

iv) {\em Gradient pulses.} This is the technique used in the spatial
averaging formalism of Cory et al.  (1996; 1997).  It consists in
applying an external Hamiltonian (\ref{qc123}) in the form of a field
gradient along the liquid sample:
\begin{equation}
H_{\rm grad} = -\sum_{i=1}^n \gamma_i (z \partial_z B^z)_{z=z_i} S_i^z,
\label{qc143b}
\end{equation}
\noindent where $z_i$ is the coordinate of the $i$-th spin in the
sample along the direction of the applied field gradient.  This
produces a spatially varying distribution of states throughout the
sample.  Its effect is to create a position-dependent phase shift with
zero average, causing the vanishing of non-diagonal elements of the
density matrix.  The notation for these pulses is $[{\rm grad}]^z$.

This gradient method is used to selectively turn off the tranverse
($x,y$) spin factors in the product operator expansion of the density
matrix, while leaving untouched the rest.  For example, it is possible
to induce the following transformation
\begin{equation}
[{\rm grad}]^z: \bar S_1^z+\bar S_2^x \mapsto \bar S_1^z.
\label{qc145}
\end{equation}

Now, the combined effect of the following series of pulses (Jones,
2000) produces the reference state (\ref{qc137}) starting from the
thermal ensemble of spins (\ref{qc139}):\footnote{This sequence is not
necessarily unique.}
\begin{equation}
\begin{split}
& \bar S_1^z +  \bar S_2^z \\ & \overset{[\pi/3]_2^x}{\mapsto} \bar
S_1^z + {1\over 2}\bar S_2^z - {\sqrt{3}\over 2} \bar S_2^y \\ &
\overset{[{\rm grad}]^z}{\mapsto} \bar S_1^z + {1\over 2}\bar S_2^z \\
& \overset{[\pi/4]_1^x}{\mapsto} {1\over \sqrt{2}} \bar S_1^z -
{1\over \sqrt{2}}\bar S_1^y + {1\over 2}\bar S_2^z \\ &
\overset{[\pi/2]_{12}^J}{\mapsto} {1\over \sqrt{2}}\bar  S_1^z +
{1\over \sqrt{2}}2 \bar S_1^x  \bar S_2^z  + {1\over 2}\bar S_2^z \\ &
\overset{[-\pi/4]_1^y}{\mapsto} {1\over 2} \bar S_1^z - {1\over 2}
\bar S_1^x + {1\over 2}2 \bar S_1^x  \bar S_2^z  + {1\over 2}\bar
S_2^z + {1\over 2}2 \bar S_1^z  \bar S_2^z \\ & \overset{[{\rm
grad}]^z}{\mapsto} {1\over 2} \bar S_1^z + {1\over 2}\bar S_2^z +
{1\over 2}2 \bar S_1^z  \bar S_2^z.
\end{split}
\label{qc146}
\end{equation}

Once we have the reference state available, we can proceed to
effectively simulate other quantum states applying series of pulses to
produce the desired ensemble of spin states.  For instance, the
density matrix of the Bell
state $|\Psi\rangle=(\ket{00}+\ket{11})/\sqrt{2}$ 
in the product operator formalism is
\begin{equation}
\rho_{|\Psi\rangle} = {1\over 2}\left({1\over 2} + 2 \bar S_1^z \bar
S_2^z + 2 \bar S_1^x \bar S_2^x - 2 \bar S_1^y \bar S_2^y \right),
\label{qc147}
\end{equation}
\noindent which can be reached from the ground state $\ket{00}$ with
the unitary operator
\begin{equation}
U = \ee^{-\ii\pi\bar S_1^x\bar S_2^y}.
\label{qc148}
\end{equation}

This propagator, in turn, can be simulated with the following series
of NMR pulses (from right to left):
\begin{equation}
[\half\pi]_2^x[-\half\pi]_1^y[\half\pi]_{12}^J[\half\pi]_1^y
[-\half\pi]_2^x: \rho_{|00\rangle}\mapsto\rho_{|\Psi\rangle}.
\label{qc149}
\end{equation}

Likewise, the controlled-NOT gate is simulated by the following
sequence:
\begin{equation}
[-\half\pi]_2^y [-\half\pi]_{2}^z [\half\pi]_{1}^z
[\half\pi]_{12}^J[\half\pi]_2^y.
\label{qc151}
\end{equation}

In a similar fashion, one can implement other quantum states and
logic gates.  Actually, this NMR pulse technique has been so
highly developed that it is possible to simulate the propagator of a
set of interacting spins with any desired couplings, even turning on
and off certain spin couplings at will.  For this reason, this
capability for controlling the NMR dynamics is referred to as {\em
spin choreography} (Freeman, 1998).

The logical labelling formalism of Gershenfeld and Chuang (1997) uses
a different strategy to prepare pseudo-pure states.  It is based in
the appropriate embedding of a set of spin states into a larger
system.  It does not resort to field gradients but instead these
auxiliary spin states are used to implement the quantum computation
with several qubits.  There are also experimental realizations of
this scheme (Vandersypen et al., 1999).

\subsubsection{Measurements}

Once the NMR computation is over, we have to read out the result from
the spectrometer.  This is done by measuring the macroscopic
magnetization of the liquid sample with a detection coil (see
Fig.~\ref{spec}).  This bulk magnetization induces currents in the
transverse RF coil which is tuned to the resonance frequency.  The RF
coil generates a dipole field and only the dipolar components of
the density matrix oriented along the transversal magnetic field will
couple to the measurement device.

\begin{figure}[h]
\psfrag{D}[Bc][Bc][0.75][0]{$\Delta \omega$}
\psfrag{j}[Bc][Bc][0.75][0]{$2J_{12}$}
\psfrag{1}[Bc][Bc][0.75][0]{$\omega_2-J_{12}$}
\psfrag{2}[Bc][Bc][0.75][0]{$\omega_2+J_{12}$}
\psfrag{3}[Bc][Bc][0.75][0]{$\omega_1-J_{12}$}
\psfrag{4}[Bc][Bc][0.75][0]{$\omega_1+J_{12}$}
\includegraphics[width=6 cm]{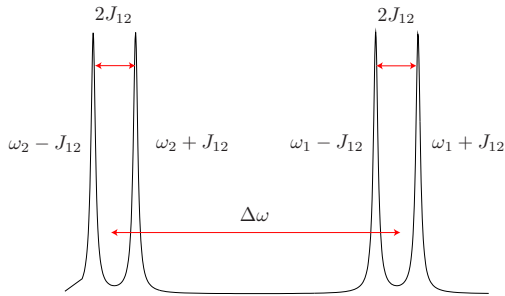}
\caption{Schematic signal from a NMR liquid spectrometer corresponding
to an in-phase doublet for a two-spin system with energy levels as in
Fig.~\ref{niveles}. Notice that here the frequencies are positive.}
\label{2pulses}
\end{figure}

In computing with NMR ensembles, measuring an observable (\ref{qc128})
entails a perturbation softer than for pure states, where measurement
is a strong projective process.  The measured currents are
proportional to the following trace (Cory et al., 2000)
\begin{equation}
{\rm Tr}\left( \sum_{i=1}^n\bar S_i^+ \rho \right),
\label{qc152}
\end{equation}
\noindent with $\bar S_i^+:=\bar S_i^x+\ii \bar S_i^y$.  For instance,
the signal (\ref{qc152}) due to the precession induced on $S_i^x,
i=1,2$, by the chemical-shifts and scalar-coupling pulses acting on a
two-qubit molecule such as the 2,3-dibromo-thiophene
(Fig.~\ref{molecules}~a)), is shown in Fig.~\ref{2pulses}.  This is
the Fourier-transformed real part of the signal (Cory, Price and
Havel, 1997) and clearly shows the populations peaks corresponding to
the 4 states of a two-spin system depicted in Fig.~\ref{niveles}.
This is called an {\em in-phase doublet} for both peaks have the same
sign.  For different series of pulses the pattern of the signal
changes accordingly and this allows to retrieve the information
contained in the ensemble of states.  When implementing simple quantum
algorithms with NMR liquid spectroscopy, the output retrieval is
performed by analysing a subset of  resonances, but in more general
situations a technique called {\em quantum state tomography}  is used
to systematically obtain the final quantum state (Knill, Chuang and
Laflamme, 1997).

\subsubsection{Achievements and limitations}

There is an extensive list of experimental achievements in NMR quantum
computing (Cory et al., 2000).  Just to quote a few of them, two-qubit
gates have been constructed by several groups (Cory, Fahmy and Havel,
1996; Chuang et al., 1997; Collins et al., 1999), the Toffoli gate has
also been implemented (Price et al., 1999), as well as the quantum
Fourier transform (Weinstein, Lloyd, and Cory, 1999), quantum
teleportation (Nielsen, Knill and Laflamme, 1998), etc.,  and there
are NMR experiments involving 7-qubits (Knill et al., 2000).  An
alternative approach to implement NMR quantum computation uses
geometric phase-shift gates (Jones et al., 2000) where the controlled
phases are Berry phases.

Despite the list of successes in NMR quantum computing, there are
currently strong limitations in the scalability of the
pseudo-pure state preparation: it is clear from (\ref{qc126}) that the
deviation density matrix used in high-temperature NMR scales
exponentially down with the factor $2^{-n}$.  This is a severe
limitation that reduces the ratio of the observable signal to the
background noise.  To overcome this inefficiency we would need an
exponentially large system.\footnote{ Something that happens in
classical DNA computing (Adleman, 1994), where there is a trade-off
between exponential computing time for solving a problem and
exponential space for molecular states.} It is currently estimated
that it is not possible to go well beyond 10 qubits using NMR liquid
state methods.  This and other shorthcomings has led to pursue other
NMR-like proposals, but this time based on solid state samples (Cory
et al., 2000), with the aim at using true pure states.  The goals set
for these proposals are to reach 10-30 qubits, still not far enough
for competitive purposes.

The use of mixed states in NMR computing and the fact that they are
exponentially inefficient have raised doubts about the truly quantum
nature of the computations carried out by NMR liquid spectroscopy.
The main objection comes from the result by Braunstein et al.  (1999)
showing that all the pseudo-pure states used so far in NMR are
separable, with no entanglement.  This does not invalidate the
exponential speed-up obtained with the implementation of quantum
algorithms.\footnote{Whether working with separable states in NMR
spectroscopy is a truly quantum computation or not is still a
controversial issue (Jones, 2000).}

\subsection{Solid-State Quantum Computers}
\label{sec11C:level2}

There are several proposals for building a quantum computer with some
sort of solid-state device.  We have just mentioned that a possible
cure for the shorthcomings of bulk NMR liquid computation is precisely
resorting to solid NMR techniques.  One type of proposals uses
macroscopic superconducting devices with a radio frequency SQUID as
the qubit (Averin, 1998).  The presence of 0 or 1 quanta of flux is
the two-state system.  Several ways to couple the SQUIDs to make logic
circuits exist, like using Josephson tunnel junctions (Makhlin,
Sch\"on and Shnirman, 2001).  Other type of designs rely on quantum
dot nanotechnology: Barenco et al.  (1995) proposed using both charge
and spin degrees of freedom for qubits in quantum dots, addressed
respectively with electric and magnetic fields.  Loss and DiVincenzo
(1998) also propose using spin states of electrons in quantum dots as
qubits.

The list of experimental proposals is too large by now to be covered
in detail.  Instead, we shall focus on one of the most original
proposals for doing solid-state quantum computation: this is Kane's
idea (1998) for building a silicon-based quantum computer.  This is an
appealing program for Kane envisages the possibility of using the
semiconductors used in most conventional computer electronics for
building also a quantum computer, although the challenges to achieve
this goal are still enormous.  The belief though is that the silicon
technology is a very rapidly developing field and there are chances to
overcome those challenges.

The quantum hardware in Kane's proposal is an array of nuclear spins
located on donors in silicon.  Then, a qubit is the individual nuclear
spin of Phosphor $^{31}$P atoms; a quantum register is the whole array
of $^{31}$P dopants in Silicon $^{28}$Si; operations are effected
using a combination of magnetic resonance techniques (Rabi pulses)
with static electric fields; information is exchanged between nearby
$^{31}$P nuclear spins by means of the surrounding electrons.
\begin{figure}[h]
\includegraphics[width=7 cm]{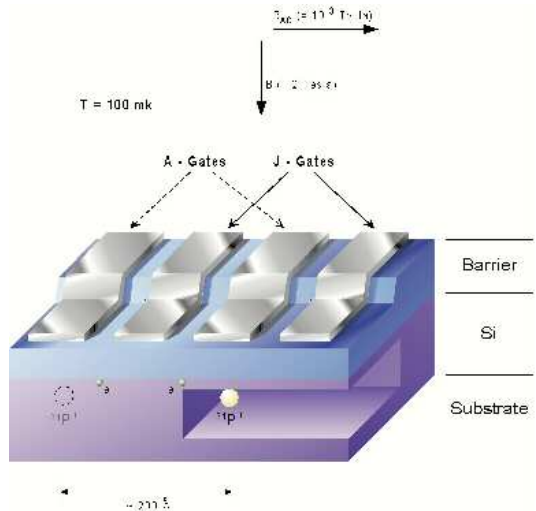}
\caption{Schematic design of a silicon-based quantum computer pursued
by the group of South Wales university.}
\label{kaneqc}
\end{figure}

\subsubsection{Semiconductors for quantum computation}

The choice of nuclear spins in this case is again motivated by their
extremely well isolation from the environment, like in the NMR
proposal.  A further requirement now is that the dopant spins must not
interact appreciably with the spins of the host semiconductor.  To
guarantee this we demand that the chemical elements of the host have
zero nuclear spin $S=0$, to avoid undesired spin couplings.  This
singles out the semiconductor group V as a host candidate and removes
other groups like III (with Ga) and IV (with As).  Silicon $^{28}$Si
is an example of stable isotope in group V.

Unlike the NMR liquid spectroscopy, Kane's QC is neither a bulk spin
quantum computation nor resorts to macroscopic magnetization
measurements.  Instead, it truly needs addressing spins individually
for initialization and readout, and this is precisely one of the open
challenges.

The basic ingredient in Kane's proposal is to trade direct nuclear
spin interactions by electronic detections, which are likely to be
easier to handle.  Thus, the spin state of an individual nucleus
dopant on a semiconductor will not be detected directly, but through
its hyperfine interaction with the surrounding electrons.  The
hyperfine interaction is proportional to the probability density of
the electrons at the nucleus.  The electronic cloud is sensitive to
electric voltages and can in principle be externally manipulated.
Moreover, in certain cases the electronic wave functions extend far
enough so as to overlap with a neighbouring atom, thereby producing an
indirect coupling between nuclear spins mediated by the atomic
electrons.  This indirect electron coupling can also be enhanced by
applying external electric fields.

These conditions are met by shallow level donors like $^{31}$P, for
which the range of the electron wave function is of order 10-100
\AA. In addition, within the group V, the only shallow donor in Si
with nuclear spin $S=\half$ is precisely $^{31}$P. Therefore, the
$^{31}$P:Si system is a good candidate for a silicon based quantum
computer.  For instance, at low $^{31}$P concentrations and low
temperature $T=1.5$ K, the electron spin relaxation time is order
$10^3$ s, and the nuclear spin relaxation time is over 10 hours.  If
the temperature is further reduced to $T \sim$ mK, the phonon limited
$^{31}$P relaxation time is likely of the order of $10^{18}$ s (Kane,
1998).

\subsubsection{External control fields}

We see that in Kane's idea the electrons play a role similar to
phonons in the Cirac-Zoller gate: they both mediate the conditional
interactions between the real qubits.  Likewise, we also need external
electric fields to bring dopant nuclei close enough to interact.  In
all, we need to control three types of external fields:

1) Electric gates above the donors to control individual electronic
states (see Fig.~\ref{kaneqc}).

2) Electric gates between the donors to control interactions between
qubits.

3) Constant $B$ and oscillating $B_{{\rm ac}}$ magnetic fields to
execute operations on the individual spins much akin to those we have
described for nuclear spin resonance.

The scenario for replacing a Si vacancy by a P dopant atom is possible
because both elements have similar sizes.  Of the five outer (3p)
electrons in a $^{31}$P atom (one more than in Si), four of them will
form covalent bonds with neighouring Si atoms, while the remaining
fifth electron is loosely bound to the $^{31}$P atom.  This outer
electron and the rest of the dopant atom behave in first approximation
as a Hydrogen-like atom embedded into a Si environment.  At low
temperatures, the electron state is 1s and this yields a large
hyperfine interaction.  The effective Bohr radius is estimated at 30
\AA. To proceed with the quantum computation we need this electron to
remain in its ground state, and to apply an external constant magnetic
field to break the spin degeneracy.  These conditions are met if
$2\mu_{\rm B} B \gg k_{\rm B} T$, as for the typical values
$B\geq 2$ T and $T\leq 100$ mK.

\subsubsection{Logic gates}

The description of the basic gate operations is the following:

i) {\em One-qubit $A$-gate.} The terminology is due to the $A$
coupling constant of the hyperfine interaction between nuclear and
electron spins.  Single spin control is achieved by externally
changing the voltage on a gate electrode ($A$-gate) located on top of
each nucleus (see Fig.~\ref{kaneqc}); spin-flips are then driven by a
Rabi pulse tuned to the resonance frequency for the particular spin.

The one-qubit Hamiltonian $H_1$ modelling the interaction between the
nuclear spin (denoted by n) and the electronic spin (denoted by e) in
the presence of a constant magnetic field $B$ is
\begin{equation}
\begin{split}
H_1 & := H_{1,{\rm Z}} + (A/\hbar^2) {\bf S}_{{\rm n},1}\cdot {\bf
S}_{{\rm e},1}, \\ H_{1,{\rm Z}} & := -\gamma_{\rm n} S_{{\rm n},1}^z
B -\gamma_{\rm e} S_{{\rm e},1}^z B,
\end{split}
\label{qc153}
\end{equation}
\noindent where ${\bf S}_{{\rm n},1}$, ${\bf S}_{{\rm e},1}$ are the
nuclear and electron spins, $\gamma_{\rm n}{\bf S}_{{\rm n},1}$,
$\gamma_{\rm e}{\bf S}_{{\rm e},1}$ their corresponding magnetic
moments, and
\begin{equation}
A:= -{8\pi\over 3}\bar\gamma_{\rm n}\bar\gamma_{\rm e}|\Psi(0)|^2,\;
{\rm with}\;\bar\gamma_{\rm n}:= \hbar\gamma_{\rm n}, \bar\gamma_{\rm
e}:= \hbar\gamma_{\rm e},
\label{qc154}
\end{equation}
\noindent is the contact hyperfine interaction energy, with
$|\Psi(0)|^2$ the probability density of the electron wave function at
the nucleus position.  Note that $\bar\gamma_{\rm e}=-g_{\rm
e}\mu_{\rm B}, \bar\gamma_{\rm n}=g_{\rm n}\mu_{\rm N}$, where $g_{\rm
e}=2$, $g_{\rm n}\approx 2\times 1.13$  are, respectively, the
relevant electron Land\'e $g$-factor and the nuclear gyromagnetic
factor in $^{31}$P:Si.  Under operating conditions the electron
remains in its ground state, and the separation of the nuclear spin
levels is, to second order in the hyperfine coupling
$A\ll\bar\gamma_{\rm n} B$:\footnote{We have also approximated
$-\gamma_{\rm e} B + \gamma_{\rm n} B$ by $-\gamma_{\rm e} B$ in the
denominator of (\ref{qc155}).}
\begin{equation}
\hbar \omega_A = \bar\gamma_{\rm n} B + {A\over 2} - {A^2\over
4\bar\gamma_{\rm e} B}.
\label{qc155}
\end{equation}

\begin{figure}[h]
\psfrag{A}[Bc][Bc][0.80][0]{A-gate}
\psfrag{B}[Bc][Bc][0.80][0]{Barrier} \psfrag{S}[Bc][Bc][1][0]{Si}
\psfrag{P}[Bc][Bc][1][0]{$^{31}$P} \psfrag{e}[Bc][Bc][1][0]{${\rm
e}^-$} \psfrag{V}[Bc][Bc][1][0]{$V=0$} \psfrag{W}[Bc][Bc][1][0]{$V>0$}
\includegraphics[width=7 cm]{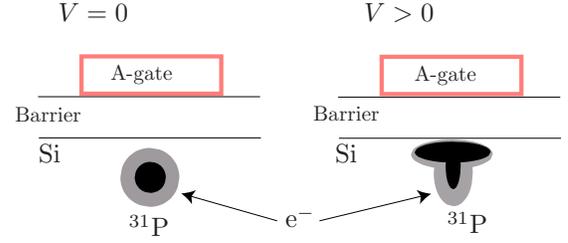}
\caption{Pictorical representation of an $A$-gate that controls the
nucleus-electron system (\ref{qc153}).  An externally applied electric
field shifts the electron wavefunction from the donor $^{31}$P,
reducing the contact hyperfine interaction (\ref{qc154}).}
\label{kaneAgate}
\end{figure}

\begin{figure}[h]
\psfrag{A}[Bc][Bc][0.75][0]{$A$-gate}
\psfrag{J}[Bc][Bc][0.75][0]{$J$-gate} \psfrag{a}[Bc][Bc][1][0]{a)}
\psfrag{b}[Bc][Bc][1][0]{b)} \psfrag{g}[Bc][Bc][1][0]{$J=0$}
\psfrag{h}[Bc][Bc][1][0]{$J>0$} \psfrag{V}[Bc][Bc][1][0]{$V>0$}
\psfrag{W}[Bc][Bc][1][0]{$V>0$} 
\includegraphics[width=7 cm]{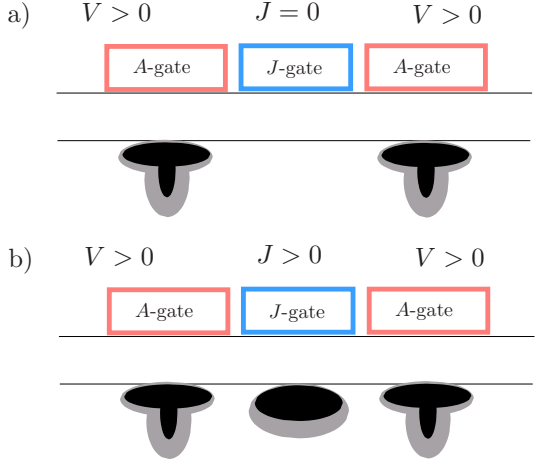}
\caption{Pictorical representation of a $J$-gate that controls the
nucleus-electron-nucleus system (\ref{qc156}).  When the electrostatic
potentitial of the $J$-gate is off (a)) or on (b)), the $J$-exchange
coupling in (\ref{qc156}) gets reduced or enhanced, respectively.}
\label{kaneJgate}
\end{figure}

In $^{31}$P:Si, $A/2h=58$ MHz and therefore $A>\bar\gamma_{{\rm n}} B$
for $B<3.5$ T.  We can have control over this energy gap with the
static electric field applied with the $A$-gate (see
Fig.~\ref{kaneqc}).  This shifts the electron wave function away from
the nucleus (see Fig.~\ref{kaneAgate}) and reduces the hyperfine
interaction $A$ in (\ref{qc154}).  Thus, the frequency (\ref{qc155})
of the nuclear spins is controlled externally and this allows us to
bring them into resonance with the oscillating pulse $B_{{\rm ac}}$ in
order to effect arbitrary one-spin rotations.

ii) {\em Two-qubit $J$-gate}. The name is suggested by the $J$
spin-exchange coupling between electron spins.  Conditional logic
operations are possible because of electron-mediated interactions
between the nuclear spins of two Kane's qubits when brought
sufficiently close by an externally applied voltage ($J$) gate (see
Fig.~\ref{kaneqc}).  The two-qubit Hamiltonian is then
\begin{equation}
H_{12} = \sum_{i=1}^2 (H_{i,{\rm Z}} + A_i\bar{\bf S}_i^{{\rm n}}\cdot
\bar{\bf S}_i^{{\rm e}}) + J \bar{\bf S}_1^{{\rm e}}\cdot \bar{\bf
S}_2^{{\rm e}},
\label{qc156}
\end{equation}
\noindent where $H_{i,{\rm Z}}$ are the Zeeman Hamiltonians for each
qubit (\ref{qc153}), $A_i$ are the hyperfine couplings for each
nucleus-electron system and $J$ is the exchange coupling interaction
between electron spins.  This exchange energy depends on the overlap
of the electron wave functions.  Treating the $^{31}$P dopants as
Hydrogen-like atoms in first approximation, the $J$ coupling can be
estimated for well separated donors as (Herring and Flicker, 1964)
\begin{equation}
J(r) \simeq 1.6 {e^2\over \epsilon a_{{\rm B}}} \left({r\over a_{{\rm
B}}}\right)^{5/2} \ee^{-2r/a_{{\rm B}}}
\label{qc157}
\end{equation}
\noindent with $r$ the inter-donor distance, $\epsilon=11.7$ the Si
dielectric constant and $a_{{\rm B}}$ the Bohr radius of the atom.  As
the $J$ coupling depends on the electron overlapping, 
we can use again a voltage gate between the donors to
distort the electron clouds in order to control their coupling
strength (see Fig.~\ref{kaneJgate}).  This coupling will be
significant when $J\simeq |\bar\gamma_{\rm e}|B/2$ and this
corresponds to a donor separation of order 100-200 {\AA} (Kane, 1998),
which is not far from the current limits of atom-scale lithography.

The relevant energy levels for doing quantum computation with a
two-qubit Hamiltonian (\ref{qc156}) are easily found (Berman et al.,
1999).  This Hamiltonian is a $16\times 16$ matrix.  We shall label
the basis states with the components of the nuclear and electron spins
at each donor site, with $\ket{0}_{\rm n}, \ket{1}_{\rm n}$ denoting
nuclear spins (up and down) and $\ket{\!\uparrow}_{\rm e},
\ket{\!\downarrow}_{\rm e}$ for the electron spins; for instance,
\begin{equation}
\ket{11}_{\rm n}\ket{\!\downarrow\downarrow}_{\rm e}
\label{qc158}
\end{equation}
\noindent represents a state with both nuclear and electron spins down.

In the presence of a static magnetic field and for low temperatures
($k_{\rm B}T \ll|\bar\gamma_{\rm e}|B$), the electrons remain
with the spins down polarized $\ket{\!\downarrow \downarrow}_e$.  For
example, $B=2$ T, $T=100$ mK meet this requirement.  However, we shall
see that switching the $J$-gate on may change such state, which will
be the basis for doing spin measurements.

The essence of the functioning of the $J$-gate is to enhance the
overlap between the electron wave functions of two nearest $^{31}$P
donors. In this way, the $^{31}$P nuclear spins (Kane qubits) can be
indirectly coupled one another through the electron mediated
interaction $J$. To perform two-qubit quantum logic gates, we need to
address individually the 4 nuclear spin  states $\{\ket{00}_{\rm
n},\ket{01}_{\rm n},\ket{10}_{\rm n},\ket{11}\}_{\rm n}$.  For
simplicity, we assume $A_1=A_2=A$.  In the absence of $J$-coupling the
states $\ket{01}_{\rm n}\ket{\!\downarrow\downarrow}_{\rm e},
\ket{10}_{\rm n}\ket{\!\downarrow\downarrow}_{\rm e}$ are
degenerate. These states belong to the sector of total $z$-component
of spin $\bar S^z_{\rm tot}:=(\bar S_{1,{\rm n}}^z+\bar S_{2,{\rm
n}}^z)+(\bar S_{1,{\rm e}}^z+\bar S_{2,{\rm e}}^z)=-1$.  The role of
the $J$-gate is precisely to control this energy splitting, which we
now try to estimate.

Let us consider the Kane  implementation of the CNOT-gate (Goan and
Milburn, 2000).  There are four steps involved:

1/ We start with $J=A_2-A_1=0$, so that the states
$\{\ket{00}_{\rm n}\ket{\!\downarrow\downarrow}_{\rm e}, \ket{01}_{\rm
n}\ket{\!\downarrow\downarrow}_{\rm e}, \ket{10}_{\rm
n}\ket{\!\downarrow\downarrow}_{\rm e}, \ket{11}_{\rm
n}\ket{\!\downarrow\downarrow}_{\rm e}\}$ have energies
\begin{equation}
\begin{split}
& E_{\ket{00}_{\rm n}\ket{\!\downarrow\downarrow}_{\rm e}} = -
\sqrt{(-\bar \gamma_{\rm e}+\bar \gamma_{\rm n})^2B^2+A^2} -\half A, 
\\ 
& E_{\ket{01}_{\rm n}\ket{\!\downarrow\downarrow}_{\rm e}} = E_{\ket{10}_{\rm
n}\ket{\!\downarrow\downarrow}_{\rm e}} = 
\\
&\half
((\bar \gamma_{\rm e}+\bar \gamma_{\rm n})B - \sqrt{(-\bar \gamma_{\rm
e}+\bar \gamma_{\rm n})^2B^2+A^2}),
\\ & E_{\ket{11}_{\rm
n}\ket{\!\downarrow\downarrow}_{\rm e}} = (\bar \gamma_{\rm e}-\bar
\gamma_{\rm n})B + \half A.
\end{split}
\label{qc158b}
\end{equation}

2/ Next one introduces a bias between the two $A$-gates by
adiabatically switching on a difference $\triangle A:=A_1-A_2$ in
their couplings, while keeping still $J=0$. This splits the degeneracy
of the $\ket{01}_{\rm n}\ket{\!\downarrow\downarrow}_{\rm e}$,
$\ket{10}_{\rm n}\ket{\!\downarrow\downarrow}_{\rm e}$ states,
allowing us to choose one as a control qubit and the other as a target
qubit. The energies in (\ref{qc158b}) become
\begin{equation}
\begin{split}
& E_{\ket{00}_{\rm n}\ket{\!\downarrow\downarrow}_{\rm e}} = - \half
(\sqrt{(-\bar \gamma_{\rm e}+\bar \gamma_{\rm n})^2B^2+A_1^2} 
\\ & \quad +\sqrt{(-\bar \gamma_{\rm e}+\bar \gamma_{\rm n})^2B^2+A_2^2})
-\fourth(A_1+A_2),
\\ & E_{\ket{01}_{\rm
n}\ket{\!\downarrow\downarrow}_{\rm e}} = -\fourth \triangle A
\\ &\quad +\half ((\bar \gamma_{\rm
e}+\bar \gamma_{\rm n})B - \sqrt{(-\bar \gamma_{\rm e}+\bar
\gamma_{\rm n})^2B^2+A_1^2}),
\\ & E_{\ket{10}_{\rm n}\ket{\!\downarrow\downarrow}_{\rm e}} = 
\fourth \triangle A
\\ &\quad +\half
((\bar \gamma_{\rm e}+\bar \gamma_{\rm n})B - \sqrt{(-\bar \gamma_{\rm
e}+\bar \gamma_{\rm n})^2B^2+A_2^2}),
\\
& E_{\ket{11}_{\rm n}\ket{\!\downarrow\downarrow}_{\rm e}} = (\bar
\gamma_{\rm e}-\bar \gamma_{\rm n})B + \fourth(A_1+A_2),
\end{split}
\label{qc158c}
\end{equation}
and the corresponding eigenstates are still $\{\ket{00}_{\rm
n}\ket{\!\downarrow\downarrow}_{\rm e}$, $\ket{01}_{\rm
n}\ket{\!\downarrow\downarrow}_{\rm e}$, $\ket{10}_{\rm
n}\ket{\!\downarrow\downarrow}_{\rm e}, \ket{11}_{\rm
n}\ket{\!\downarrow\downarrow}_{\rm n}\}$, predominantly.

3/ Once the two qubits are distinguished energetically it is time to
introduce, again adiabatically, the $J$-coupling to bring the states
$\ket{10}_{\rm n}$ and $\ket{01}_{\rm n}$ to the symmetric and
antisymmetric combinations, namely
\begin{equation}
\begin{split}
& \ket{10}_{\rm n}\mapsto \ket{\rm s}_{\rm n}:=2^{-1/2} (\ket{01}_{\rm
n}+\ket{10}_{\rm n}),
\\ & \ket{01}_{\rm n}\mapsto \ket{\rm a}_{\rm
n}:=2^{-1/2} (\ket{01}_{\rm n}-\ket{10}_{\rm n}).
\end{split}
\label{qc158d}
\end{equation}
For this purpose it is necessary to keep $J$ at full strength before
switching off adiabatically $\triangle A$.

The energies of the new eigenstates both in presence of $A$- and
$J$-couplings, with $\triangle A=0$, can be computed exactly by
diagonalizing $H_{12}$ in the sectors of fixed total 3th component
$S_{{\rm tot}}^z$  of the spin, since this is a conserved quantity.
Only the values $S_{{\rm tot}}^z=-2,-1,0$ are relevant for our
discussion, since our initial states lie there. First we need to know
the energy splitting $\hbar \omega_J$ between the symmetric and
antisymmetric qubit states in the sector $S_{{\rm tot}}^z=-1$. Second,
to control the Rabi pulse in the coming step, the gap energy $\hbar
\omega_{\rm ac}$ between $\ket{\rm s}_{\rm n}
\ket{\!\downarrow\downarrow}_{\rm e}$ and $\ket{11}_{\rm
n}\ket{\!\downarrow\downarrow}_{\rm e}$ must also be known.

To calculate $\hbar \omega_J$ we use the reduced basis
\begin{equation}
\{ \ket{01}_{\rm n}\ket{\!\downarrow \downarrow}_{\rm e},
\ket{10}_{\rm n}\ket{\!\downarrow \downarrow}_{\rm e}, \ket{11}_{\rm
n}\ket{\!\downarrow \uparrow}_{\rm e}, \ket{11}_{\rm n}\ket{\!\uparrow
\downarrow}_{\rm e} \}
\label{qc159}
\end{equation}
\noindent to express the Hamiltonian $H_{12}$ in the sector $S_{{\rm
tot}}^z=-1$ as the following matrix
\begin{equation}
\begin{split}
& H_{(-1)}=\\ & \begin{pmatrix} \fourth J +\bar\gamma_{\rm e} B & 0 &
0 & \half A \\ 0 & \fourth J +\bar\gamma_{\rm e} B &  \half A & 0  \\
0 & \half A  &  -\fourth J+\bar\gamma_{\rm n} B & {\half J} \\ \half A
& 0  & {\half J} & -\fourth J+\bar\gamma_{\rm n} B
\end{pmatrix}.
\end{split}
\label{qc160}
\end{equation}

As $A_1=A_2=A$, the two-qubit Hamiltonian is symmetric under the site
labels and its eigenvectors can either be symmetric or antisymmetric
under this exchange. The two symmetric (unnormalized) eigenstates  are
given by
\begin{equation}
\begin{split}
& \ket{{\rm s},\pm}:=  \\ & (\bar\gamma_{\rm n} B + \fourth J -E_{{\rm
s},\pm}) \ket{{\rm s}}_{\rm n}\ket{\!\downarrow\downarrow}_{\rm e}
+\half A \ket{00}_{\rm n}\ket{{\rm s}}_{\rm e},
\end{split}
\label{qc161}
\end{equation}
where
\begin{equation}
\begin{split}
\ket{{\rm s}}_{\rm e} &:= {1\over \sqrt{2}}(\ket{\!\downarrow
\uparrow}_{\rm e} + \ket{\!\uparrow \downarrow}_{\rm e}),\\ E_{{\rm
s},\pm} &:= \half(\bar\gamma_{\rm e}+\bar\gamma_{\rm n})B + \fourth J
\pm \half\sqrt{(-\bar\gamma_{\rm e}+\bar\gamma_{\rm n})^2B^2 + A^2}.
\end{split}
\label{qc161b}
\end{equation}

Similarly the two antisymmetric (unnormalized) eigenstates are
\begin{equation}
\begin{split}
& \ket{{\rm a},\pm}:=  \\ & -(-\bar\gamma_{\rm e} B - \fourth J
+E_{{\rm a},\pm}) \ket{00}_{\rm n}\ket{{\rm a}}_{\rm e} -\half A
\ket{{\rm a}}_{\rm n}\ket{\!\downarrow\downarrow}_{\rm e},
\end{split}
\label{qc162}
\end{equation}
with
\begin{equation}
\begin{split}
\ket{{\rm a}}_{\rm e} := & {1\over
\sqrt{2}}(\ket{\!\downarrow\uparrow}_{\rm e} - \ket{\!\uparrow
\downarrow}_{\rm e}),\\ E_{{\rm a},\pm} := & \half(\bar\gamma_{\rm
e}+\bar\gamma_{\rm n})B - \fourth J\\ &\pm
\half\sqrt{((-\bar\gamma_{\rm e}+\bar\gamma_{\rm n})B-J)^2 + A^2}.
\end{split}
\label{qc162b}
\end{equation}

\begin{figure}[h]
\psfrag{x}[Bc][Bc][1][0]{$J/2|\bar\gamma_{\rm e}|B$}
\psfrag{y}[Bc][Bc][1][0]{$E/2|\bar\gamma_{\rm e}|B$}
\psfrag{s}[Bc][Bc][0.8][0]{$\ket{{\rm s},+}$}
\psfrag{t}[Bc][Bc][0.8][0]{$\ket{{\rm s},-}$}
\psfrag{a}[Bc][Bc][0.8][0]{$\ket{{\rm a},+}$}
\psfrag{b}[Bc][Bc][0.8][0]{$\ket{{\rm a},-}$} 
\includegraphics[width=7 cm]{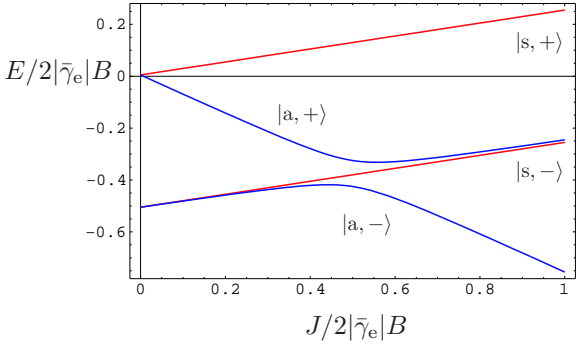}
\caption{Energy levels for a two-donor interacting system as a
function of the exchange coupling $J$, for $A=0.2|\bar\gamma|_{\rm
e}B$.}
\label{kanelevels}
\end{figure}

In Fig.~\ref{kanelevels} the energies $E_{{\rm s},\pm},E_{{\rm
a},\pm}$ are plotted against the exchange coupling constant $J$. For a
two-electron spin system with antiferromagnetic coupling ($J > 0$),
the exchange interaction lowers the energy of the spin singlet with
respect to the triplets.  When the static magnetic field is applied,
the electron ground state is $\ket{\!\downarrow \downarrow}_{\rm e}$
for $J < |\bar\gamma_{\rm e}|B$.  The exchange coupling can be
increased adiabatically by external manipulation of the $J$ voltage
gate.  For $J > |\bar\gamma_{\rm e}|B$, the electron ground state is
the singlet.  The value $J = |\bar\gamma_{\rm e}|B$ corresponds to the
case where levels $E_{a,+}$ and $E_{s,-}$ avoid their crossing
(Fig.~\ref{kanelevels}).  The energy splitting to be controlled with
the $J$-gate is $\hbar \omega_J:= E_{s,-}-E_{a,-}$, which can be
estimated using the exact formulas (\ref{qc161b}), (\ref{qc162b}) and
treating the hyperfine interaction as a small perturbation (assuming
$J < |\bar\gamma_{\rm e}|B$):
\begin{equation}
\hbar \omega_J \simeq {A^2\over 4}\left( {1\over |\bar\gamma_{\rm e}|B
-J} - {1\over |\bar\gamma_{\rm e}|B}\right)
\label{qc163}
\end{equation}
\noindent For the $^{31}$P:Si system at $B=2$ T and $J/h = 30$ GHz,
(\ref{qc163}) gives $\nu_J=75$ kHz as the nuclear spin exchange
frequency.  This is roughly the rate at which binary operations can be
performed in the purported quantum computer. Recall that the speed for
individual spin operations is determined by the oscillating field
$B_{{\rm ac}}$, and this speed is comparable to $75$ kHz when $B_{{\rm
ac}}\sim 10^{-3}$ T.

To calculate finally the gap $\hbar \omega_{\rm ac}$, we just need the
energy of the state $\ket{11}_{\rm n}\ket{\!\downarrow\downarrow}_{\rm
e}$ which lies in the trivial sector $S_{{\rm tot}}^z=-2$:
\begin{equation}
E_{\ket{11}_{\rm n}\ket{\!\downarrow\downarrow}_{\rm
e}}=(\bar\gamma_{\rm e}+ \bar\gamma_{\rm n})B+\fourth J+\half A.
\label{qc163b}
\end{equation}

\begin{figure}[t]
\psfrag{a}[Bc][Bc][1][0]{a)} \psfrag{b}[Bc][Bc][1][0]{b)}
\psfrag{t}[Bc][Bc][0.75][0]{time $t$}
\psfrag{M}[Bc][Bc][0.75][0]{$B_{{\rm ac}}/1$ T}
\psfrag{D}[Bc][Bc][0.65][0]{$\Delta A/\bar\gamma_{{\rm n}}B$}
\psfrag{J}[Bc][Bc][0.65][0]{$J/\bar\gamma_{{\rm n}}B$}
\psfrag{E}[Bc][Bc][0.65][0]{$E/\bar\gamma_{{\rm n}}B$}
\psfrag{u}[Bc][Bc][0.55][0]{$\ket{00}_{\rm n}$}
\psfrag{v}[Bc][Bc][0.55][0]{$\ket{01}_{{\rm n}}$}
\psfrag{w}[Bc][Bc][0.55][0]{$\ket{10}_{{\rm n}}$}
\psfrag{x}[Bc][Bc][0.55][0]{$\ket{11}_{{\rm n}}$}
\psfrag{s}[Bc][Bc][0.65][0]{$\ket{s}_{{\rm n}}$}
\psfrag{n}[Bc][Bc][0.65][0]{$\ket{a}_{{\rm n}}$}
\includegraphics[width=7 cm]{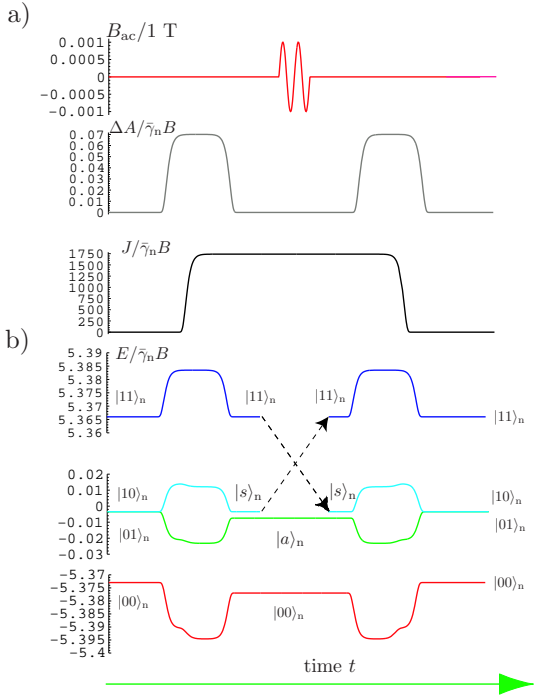}
\caption{Implementation of the CNOT-gate in a Kane quantum computer as
described in steps 1/-4/ in text (time $t$ runs along the horizontal
axis). In a) the externally driven couplings are shown, and in b) the
qubits energies are plotted, conveniently shifted by $E\mapsto
E-\bar\gamma_{\rm e}B-\fourth J$.}
\label{kanecnot}
\end{figure}

4/ The moment is right to  enforce the CNOT operation. This amounts to
swap the states $\ket{{\rm s}}_{\rm n}$ and $\ket{11}_{\rm n}$, which
are well separated in energies by previous steps, while leaving the
two other states untouched. To this aim, it suffices now to apply a
Rabi pulse $H_{\rm ac}(t):=-\gamma_{\rm n}(S_{\rm n,1}^x+S_{\rm
n,2}^x)B_{{\rm ac}} \sin\omega_{\rm ac} t$ resonant with the
separation energy between the states to be exchanged. Although the
gaps $E_{\ket{11}_{\rm n}\ket{\!\downarrow\downarrow}_{\rm e}}-
E_{\ket{\rm s}_{\rm n}\ket{\!\downarrow\downarrow}_{\rm e}}$ and
$E_{\ket{\rm a}_{\rm n}\ket{\!\downarrow\downarrow}_{\rm e}}-
E_{\ket{00}_{\rm n}\ket{\!\downarrow\downarrow}_{\rm e}}$ are very
close one each other, however the spin part of the magnetic
interaction $H_{\rm ac}(t)$ only couples in first order the states
$\ket{{\rm s}}_{\rm n}$ and $\ket{11}_{\rm n}$ and thus it does not
affect essentially the states $\ket{{\rm a}}_{\rm n}$ and
$\ket{00}_{\rm n}$.  To complete the CNOT-gate one applies backwards
the steps 3/, 2/ and 1/ (see Fig.~\ref{kanecnot}).

Other computer operations such as spin measurements and initialization
of the quantum register are also based on the adiabatic manipulation
of the $A$- and $J$-voltages.  The underlying idea has been to
correlate nuclear spin states adiabatically with states of electron
spins, which in turn are affect the symmetry of the electron orbital
wave function (Kane, 2000).

Unlike the QC proposals based on ion-traps or NMR spectroscopy, the
silicon-based QC has not been yet implemented
experimentally.\footnote{There is a funded project in the
Semiconductor Nanofabrication Facility of the South Wales University
(Australia) for building a Kane's quantum computer.}  This will
require nanofabrication at the atomic scale involving at least
specialized techniques such as quantum electronic measurements with
Single Electron Transistors (SET) for addressing individual qubits,
atom-scale lithography to place Phosphorus donors in a Silicon crystal
with near-atomic precission, combined with electron beam lithography
for building the quantum array of qubits, etc.  (Kane, 2000).  It
remains an open issue whether the current developments in these
technologies will be enough to build a Kane quantum computer.

\section{Conclusions}
\label{sec12:level1}

Although this may look an extensive review, the field has grown at
such a pace that it is not possible to cover in detail all the
interesting developments going on, and many have been left out.  Just
to mention a few of them: universal sets of fault-tolerant quantum
gates, a thorough study of decoherence problems, quantum erasure,
further experimental proposals for quantum computers, etc.

We share the belief in the mutual benefit of the symbiosis between
quanta and information. The  very knowledge of the foundations of
physics can benefit from the theory of information and computation
(Landauer, 1991; 1996).  We have reviewed some of the aspects coming
out from the fruitful idea that information is physics. We could
further speculate all the way around: physics is also information. It
might quite well be the case that a fundamental theory of physics
could be based on the notion of qubit from which all the rest would be
derived (Wheeler, 1990; Zeilinger, 1999).

We have made an effort to present both classical and quantum aspects
of information and computation.  Classical aspects have been
traditionally linked to computer science, of interest both to computer
and electronic engineers, and to mathematicians addressing its
theoretical and abstract foundations.  Quantum aspects, on the
contrary, have been almost uniquely associated to quantum physicists.
Thus, each community finds its own barrier in order to jump over and
to enter the field of quantum computation: an engineer lacks
frequently the necessary training in quantum theory while most
physicists are not used to deal with elementary aspects of information
and the insides of a real computer.  These shorthcomings make
traditionally difficult to bring together both type of researchers.
Our work is aimed in part at setting up a bridge between both
communities in the belief that it will be rewarding for both of them.
We are confident that after this quantum information revolution time
will be ripe for quantum mechanics to be taught regularly at engineer
schools, and for information theory to figure among background courses
in physics.  Moreover, by presenting a brief account of the
experimental realization of quantum computers we also stress the close
relationship with other particular fields like condensed matter and
its many branches, specially with the area of strongly correlated
systems.

There is currently a big interest in building real quantum computers,
capable of doing non-trivial tasks.  Also, a bunch of new proposals
have been presented and this trend is likely to continue.  Each
physical system or interaction in nature is scrutinized as a possible
realization of a quantum computer. Marvelous machines,
like aircrafts, were envisaged in the past by
Leonardo da Vinci; he described them on a piece of paper and were not
actually built up until hundreds of years later.  Likewise, nowadays
we find theoretical designs of prospective quantum computers.  We hope
that in the case of quantum computers this process will not take that
long.  At least for the current modest realizations the elapsed time
has been short.  Even these modest realizations are remarkable since
they allow for testing some of the theoretical principles.

Now we come necessarily to an end.  And we close with a grand query.
We have talked about a large variety of computer machines: classical
-- both sequential and parallel machines of many types -- and quantum
mechanical -- both theoretical and experimental.  Yet, there is a
marvellous machine which plays a paramount role in all those
constructions, because after all, it is the one that has devised them
all.  And thus, it is also natural to ask: what type of computer
machine is the human brain?

\begin{acknowledgments}

We would like to thank I. Cirac and P. Zoller for their enthusiasm in
embracing this project and for pushing us to carry through this long
process. We have benefited from discussions and correspondence with
I. Cirac, H-S. Goan, L. Grover, P. Hoyer, B. King,  A.K. Lenstra,
A. Levitin, H.te.Riele, A. Trill and P. Zoller.

We are partially supported by the 
CICYT project AEN97-1693 (A.G.) and by the 
DGES Spanish grant PB98-0685 (M.A.M.-D.).
\end{acknowledgments}

\section*{List of Symbols and Acronyms}

\noindent BB84: Bennett-Brassad 1984

\noindent B92: Bennett 1992

\noindent BBPSSW96: Bennett-Brassard-Popescu-Schumacher-Smolin-Wooters
1996

\noindent CPU: Central Processing Unit

\noindent E91: Ekert 1991

\noindent ECCC: Error-Correcting Classical Code

\noindent EDP: Entanglement Distillation Protocol

\noindent EPR: Einstein-Podolsky-Rosen

\noindent GNFS: General Number Field Sieve

\noindent LOCC: Local Operations Classical Communications

\noindent MIPS: Million Instructions Per Second

\noindent NDTM: Nondeterministic Turing Machine

\noindent NMR: Nuclear Magnetic Resonance

\noindent NP: Class of nondeterministic polynomial-time problems

\noindent P: Class of deterministic polynomial-time problems

\noindent PKC: Public Key Cryptography

\noindent PTM: Probabilistic Turing Machine

\noindent QC: Quantum Computer

\noindent QECC: Quantum Error Correction Code

\noindent QFT: Quantum Fourier Transform

\noindent QKD: Quantum Key Distribution

\noindent QTM: Quantum Turing Machine

\noindent RF:  Radio Frequency

\noindent RSA: Rivest-Shamir-Adleman

\noindent TM: Turing Machine

\noindent VNM: Von Neumann Machine

\section*{Appendix: Computational Complexity}
\label{appx:level1}

There are non-solvable problems like the halting problem of TM
(Sec.~\ref{sec8A:level2}). In fact, their number is uncountable. On
the other hand, solvable  problems can be classified according to
their difficulty.  There are easy problems (computationally {\em
tractable}), like computing the determinant of any $n\times n$ matrix,
and there are difficult problems (computationally {\em hard} or {\em
untractable}), like computing the permanent of the same
matrix.\footnote{The definition of the permanent is similar to the
determinant.  In fact the only difference is the missing sign of the
permutations.}

The {\em complexity classes} have been devised  to group solvable
problems according to their  degree of difficulty. Three aspects are
addressed (Nielsen and Chuang, 2000) : 1/ time or space resources
required by its solution, 2/ the machine used in its solution (DTM,
NDTM, PTM, or QTM), and 3/ the type of problem (decision, number of
solutions, optimization, etc.).

\subsection{Classical Complexity Classes}

When the computation is done with DTMs or NDTMs, the relevant classes
are the following (Papadimitriou, 1994; Welsh, 1995; Yan, 2000;
Salomaa 1989; Li and Vit\'anyi, 1997):\footnote{Although the
complexity classes  {\bf P}, {\bf NP}, etc., that we shall consider
here usually contain only decision problems (problems
whose solution is either YES (1) or NO (0)), we shall implicitly
enlarge them  by including other computational  problems, searching,
etc., which are defined in a similar fashion to decision problems by
means of the costs in time or space invested in its solution.}

i/  Class {\bf P} (Polynomial), containing  those problems that 
a DTM solves in
{\em polynomial time}, i.e., the time taken
for the DTM to find the solution increases at most polynomially with
the length $n$ (in bits) of the initial data.

Examples: 1/ arithmetic operations such as the addition and
multiplication of integers, 2/ Euclid's
algorithm, 3/ modular exponentiation, 4/ computation of determinants,
5/ sorting a list (SORT), and 6/ multiplication of 
of points on elliptic curves by large integers. 

ii/ Class {\bf NP} (Nondeterministic Polynomial), 
containing  those problems that a NDTM solves in
polynomical time.\footnote{As there may be several computational
pathways leading to the solution, the one of shortest duration 
marks the cost (Salomaa, 1989).}

As there are not NDTMs in practice, it is convenient to know this
other equivalent characterization of the {\bf NP} class in which only
DTMs are involved: a problem is {\bf NP} if, given an arbitrary
initial data $x$ of binary length $n$, it admits any {\em succint
certificate} or {\em polynomial witness} $y$ (i.e., of polynomial
length in $n$), such that there exists a DTM which, with those data $x,y$,
can solve the given problem in polynomial time in $n$.

Clearly, $\text{\bf P}\subseteq\text{\bf NP}$. A central conjecture
in computation theory is $\text{\bf P}\varsubsetneqq\text{\bf NP}$.

Examples: 1/ the DISCRETE LOGARITHM problem (computation in $\Z_N$ of
the solution $x$ to $a^x = b \ \mod \ N$), 2/ 
the PRIMALITY problem (given $N$, is it prime?), 3/ COMPOSITENESS, 
complement to PRIMALITY
(given $N$, is it composite?), 4/ the FACTORIZATION problem (find the
decomposition of $N$ into prime factors), 5/ the satisfiability
problem SAT (check whether a given
Boolean expression $\phi$ in normal conjunctive form
$\phi=\bigwedge_1^n C_i$, $C_i:=z_{i1}\vee z_{i2}\vee \ldots \vee
z_{ir_i}$, with $z_{ij}\in (x_{ij},\neg x_{ij})$ Boolean variables or
their negations, is satisfiable, that is, there exists a choice of
variables that make $\phi$ true), and 6/ the traveling salesman problem
TSD(D) (given $n$ cities, their mutual distances $d_{ij}\geq 0$ and a
cost or ``travel budget'',   find whether  there exists a cyclic
permutation $\pi$ such that $\sum_{i=1}^{n}d_{i,\pi(i)}\leq C$).

FACTORIZATION is {\bf NP}  since it is apparent that given $N$, and
the succint certificate consisting of its prime divisors, the
decomposition of $N$ into primes is trivial and of polynomial cost.

iii/ Class {\bf PSPACE} (Polynomial Space) ({\bf NSPACE}, Nondeterministic
polynomial Space), containing those problems
that some DTM (NDTM) solves in polynomial space, i.e., 
using a number of cells that grows at most
polynomially with the length (in bits) of the initial data.

It is known that $\text{\bf NP}\subseteq\text{\bf PSPACE}=
\text{\bf NSPACE}$.

Examples: 1/ In the two-players game GEOGRAPHY, player
$A$ chooses the name of a city, say MADRID, and $B$ has to name another
city, like DUBLIN, starting with the last letter D of the previous city; 
then the turn is on $A$ for naming another city starting with N, like NEWYORK,
$B$ says next KYOTO, and  so on and so forth. The cities' names must not
be repeated.  The loser is the  player who cannot name another city
because there are not more names left.  The GEOGRAPHY problem
is: given an arbitrary set of cities (strings, all different, of
alphabet symbols), and $A$'s initial choice of one of them, can $A$
win?. It can be shown that  GEOGRAPHY is 
{\bf PSPACE}-complete.\footnote{Given a complexity class {\bf X}, 
a decision problem $P\in\text{\bf X}$ is called {\bf X}-complete when
any $Q\in\text{\bf X}$ is polinomially reducible to $P$, i.e.,
$\exists$ a polynomial-time map $f:x\mapsto f(x)$ from the inputs
of $Q$ to the inputs of $P$ such that $Q(x)=0,1$ iff $P(f(x))=0,1$.} 2/ Also
the game GO suggests a GO problem on $n\times n$ boards and
the associated question of whether there exists some winning strategy
for the starting player.  This GO Problem is likewise {\bf
PSPACE}-complete.

iv/ Class {\bf EXP} (Exponential) ({\bf NEXP}, (Nondeterministic 
Exponential)), containing those problems that
some DTM (NDTM) solves in exponential time, i.e.,
a time that grows at most exponentially with
the length (in bits) of the initial data.

Examples: Consider the problems related to the games
GO, CHECKERS and CHESS on $n\times n$ fields: are always there winning
strategies for the first player? Since the number of movements to 
analyse grows exponentially
with the board size, such problems are in {\bf EXP}. Furthermore, 
it is believed that they are not in class {\bf NP}.

The following inclusions among the previous classes hold:
$$
\text{\bf P}\subseteq\text{\bf NP}\subseteq\text{\bf PSPACE}
\subseteq\text{\bf EXP}\subseteq\text{\bf NEXP}.
$$
Moreover, it is also known that $\text{\bf P} \varsubsetneqq\text{\bf
EXP}$. Thus, at least one of the three firts inclusions 
in the long previous chain
must be proper. But it is ignored which one.

The classification does not end here. There are even more
``monstrous'' problems, as far as complexity is concerned.  For
instance, pertaining to the Presburger arithmetic there exists a 
problem doubly exponential at least (time complexity $O(2^{2^n})$ in the
size $n$ of the initial data).

Let us now assume that our computers are PTMs.  The corresponding
classes are called {\em random}, and some of them stand out:

i/ Class {\bf RP} (Randomized Polynomial),  consisting of those decision
problems that a PTM $T$, always working in polynomial time (for every
initial data), decides with error $\leq\half$. These problems are
called polynomial Monte Carlo. In other words, if $L$ denotes the
set of input data having answer YES, i.e., 1, then
\begin{equation*}
\begin{split}
&x\in L\implies \prob(T(x)=1)\geq \half, \\ &x\notin L\implies
\prob(T(x)=1)=0.
\end{split}
\end{equation*}
This means that all computational pathways that a PTM $T$ can take
from a data $x\notin L$ end up with rejection ($T(x)=0$, i.e., NO), while if
$x\in L$, then at least a fraction  $\half$ of the possible paths
end up with acceptance ($T(x)=1$).  Therefore, there cannot be false
positives, and at most a fraction $\half$ of false negatives can
happen (cases in which $x\in L$ and however the followed path ends
with rejection). Repeating the computation with the same $x\in L$ a
number of times $n\gtrsim\lceil\log_2\delta^{-1}\rceil$, where $0<\delta<1$,
we will be able to get that the probability of $n$ consecutive
false negatives be $\leq \delta$ and thus as small as desired by
appropriately choosing  $\delta$, or equivalently, that the
probability to obtain in that series of  $n$ trials some acceptance of
$x$ be  $\geq (1-\delta)$ and thus as close to 1 as we wish.  In cases
of real ``bad luck'' it might happen that very long series would not
contain any acceptance of $x$; that is why it is often said that such $T$
decides the problem in average case polynomial time.

ii/ Class $\text{\bf
ZPP}:=\text{\bf RP}\cap\text{\bf coRP}$ 
(Zero-error Probabilistc Polynomial), where the class {\bf coRP} is the
complement of {\bf RP}, that is, it contains those decision
problems that answer (YES, NO) to an input iff there exists a
problem in {\bf RP} which answers (NO, YES) to the same input.

The class {\bf ZPP} thus contains those decision problems for
which there exist two  PTM $T_{\rm RP}$ and $T_{\rm coRP}$, 
always working in polynomial time and
satisfying
\begin{equation*}
\begin{split}
&x\in L\Rightarrow \prob(T_{\rm RP}(x)=1)\geq \half,\prob(T_{\rm coRP}(x)=1)=0,
\\ 
&x\notin L\Rightarrow \prob(T_{\rm RP}(x)=1)=0,\prob(T_{\rm coRP}(x)=1)
\geq \half.
\end{split}
\end{equation*}

These problems are called polynomial Las Vegas: they are Monte Carlo,
and so are their complements.  In other words, they have two Monte
Carlo algorithms, one without false positives, and another one without
false negatives. Most likely any input data will be decidable with
certainty: it is enough that the algorithm without false positives
says YES, or the one without false negatives says NO. In case of
real bad luck, we shall have to repeat both until one of them yields a
conclusive answer.

Example: PRIMALITY is in {\bf ZPP}. The Miller-Selfridge-Rabin algorithm
(pseudo-primality strong test, 1974) is of coMonteCarlo type, that is,
PRIMALITY is in {\bf coRP} (in fact, the probability of false positives,
i.e., that one probable prime  be composite, is $\leq 1/4$).  That PRIMALITY
in also in {\bf RP} is a harder issue, and was proved by Adleman and Huang
(1987), with the theory of Abelian varieties (generalization of
elliptic curves to higher dimensions).\footnote{Given an integer $N$,
there exists a deterministic primality-testing algorithm, due to
Adleman-Pomerance-Rumely-Cohen-Lenstra (1980-81), with complexity
$O((\log_2 N)^{c \log_2 \log_2 \log_2 N})$, where $c$ is a constant. A
current typical computer takes about 30 s for $N$ with 100 decimal
digits, about 8 min if  $N$ has 200 digits, and a reasonable time for
1000 digits.}

iii/ Class {\bf BPP} (Bounded-error Probabilistic Polynomial). 
It contains those decision problems for which there exists a PTM
$T$ always working in  polynomial time and satisfying
\begin{equation*}
\begin{split}
&x\in L\implies \prob(T(x)=1)\geq \textstyle\frac{3}{4}, \\ &x\notin L\implies
\prob(T(x)=1)\leq\textstyle \frac{1}{4}.
\end{split}
\end{equation*}

{\bf BPP} problems are perhaps those representing best the notion of
realistic computations. They are accepted or rejected by a
PTM with the possibility to err.  But the error probability 
is $\leq\fourth$
both on the acceptance as well as on the rejection. Repetition of the
algorithm with the same input allows to amplify the probability of
success, and, using  the majority rule, to decide within polynomial
time (average case time, except in bad luck instances) and with an error as
small as required. It is not known whether ${\text{\bf BPP}}\subseteq
\text{{\bf NP}}$, 
although it is believed that ${\text{\bf NP}}\not\subseteq
\text{{\bf BPP}}$. It is clear that  
$\text{\bf RP}\subseteq\text{\bf BPP}$,  and
likewise $\text{\bf BPP}=\text{\bf coBPP}$.  Generically:
\begin{equation*}
\begin{split}
&\text{\bf P}\subseteq\text{\bf ZPP}\subseteq\text{\bf
RP}\subseteq(\text{\bf BPP},\text{\bf NP})\subseteq\\
&\subseteq\text{\bf PSPACE} \subseteq\text{\bf EXP}\subseteq\text{\bf
NEXP}.
\end{split}
\end{equation*}

\begin{figure}[tbp]
\begin{center}
\includegraphics[width=0.48\linewidth]{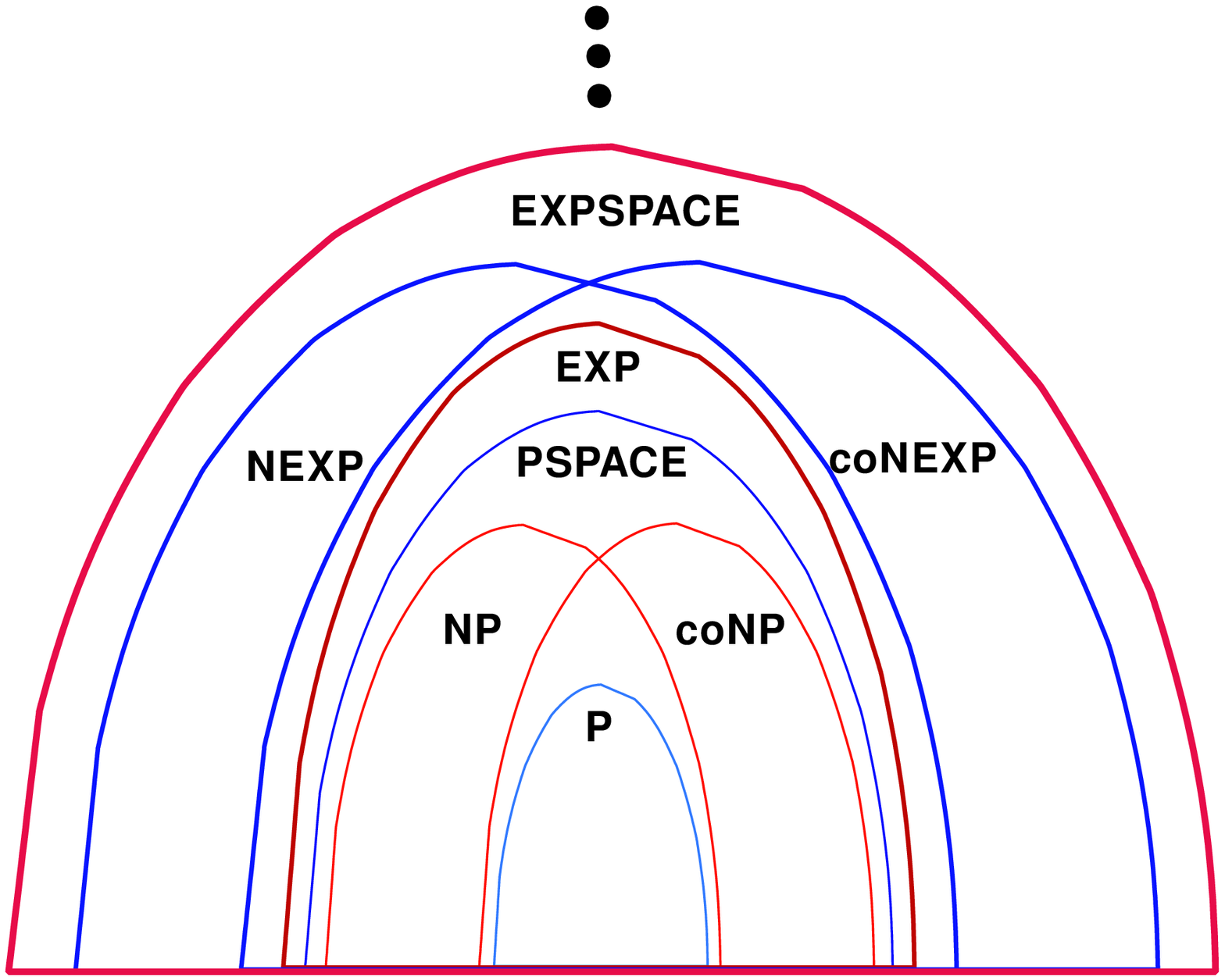} \hfill
\includegraphics[width=0.48\linewidth]{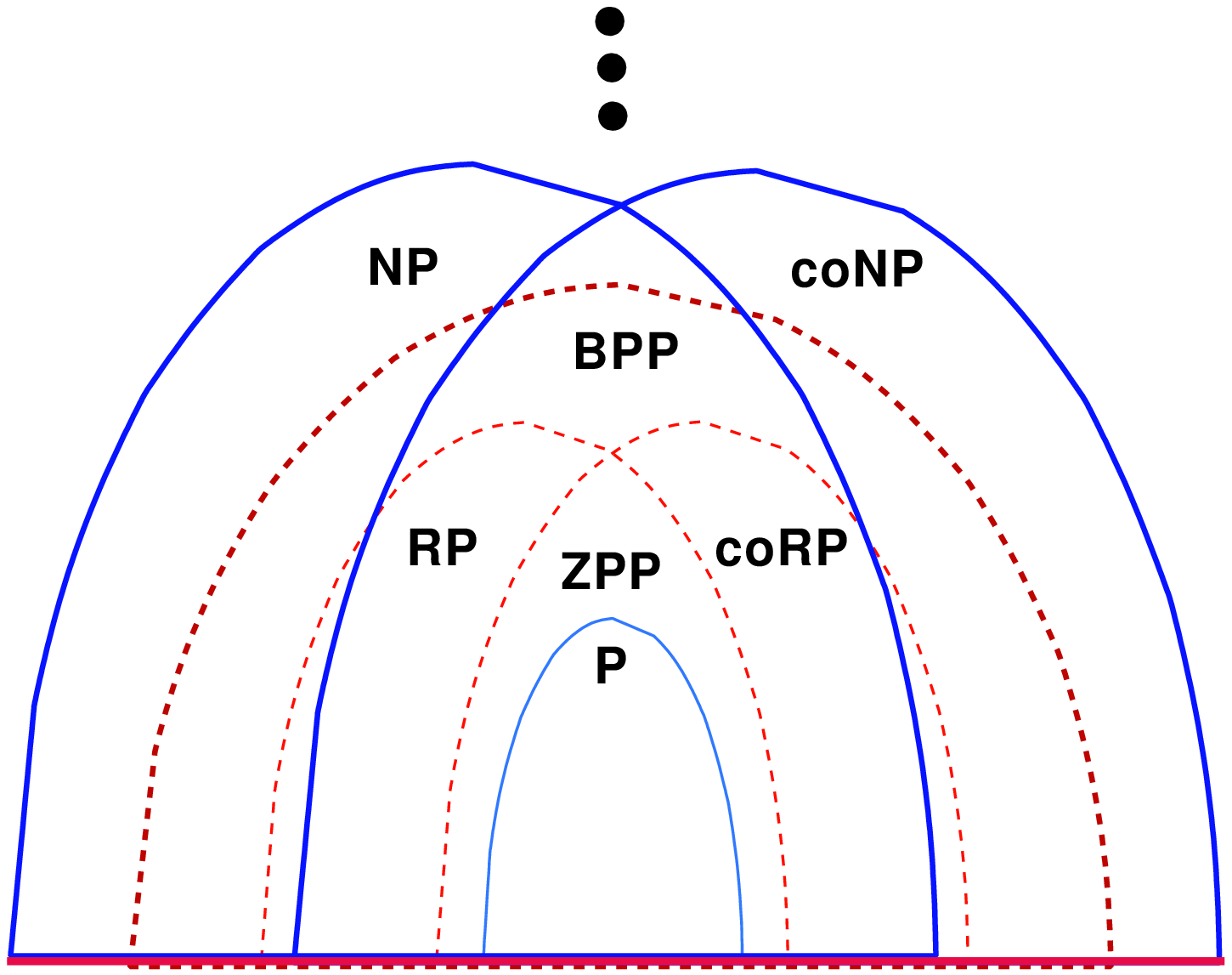}
\caption[]{Different classical complexity classes. On the right, we
provisionally accept that {\bf BPP} class is not a subset of {\bf NP}.}
\label{fig:clasescomplejidad}
\end{center}
\end{figure}

Fig.~\ref{fig:clasescomplejidad} shows the inclusions among the
classical complexity classes (Papadimitriou, 1995).

\subsection{Quantum Complexity Classes}

When the computers employed in the computations are QTMs, the
associated complexity classes are called quantum. We shall quote some
of the most relevant:

i/ Class {\bf QP} (Quantum Polynomial), containing those (decision)
problems solvable in polynomial time with a QTM.

ii/ Class {\bf
BQP} (Bounded-error Quantum Polynomial). It contains those problems 
solvable with error $\leq 1/4$ in
polynomial time with a QTM.

iii/ Class $\text{\bf ZQP}$ (Zero-error probability Quantum Polynomial). 
Set of problems solvable with zero error probability
in expected polynomial time with a QTM.

The following relations with the classical complexity classes hold:
\begin{equation*}
\text{\bf P} \varsubsetneqq\text{\bf QP}, \quad \text{\bf BPP}
\subseteq\text{\bf BQP}\subseteq\text{\bf PSPACE}.
\end{equation*}

The proper inclusion of {\bf P} in {\bf QP}, shown by  Berthiaume and
Brassard (1992), is very remarkable. It means that quantum computers
can solve efficiently more problems than their classical kin. 
It amounts to the first clear victory in the strict
separation of classical and quantum complexities.

The second chain of inclusions is due to Bernstein and Vazirani
(1993). It remains open the crucial question of whether $\text{\bf
BPP} \varsubsetneqq\text{\bf BQP}$ or not. That is, are there
quantum ``tractable'' problems which are classically hard?  Simon's
algorithm (Subsec.~\ref{sec10B:level2})  is a first positive
indication in the presence of a quantum oracle.  
Another fact supporting this point comes from Shor's
algorithm (Subsec.~\ref{sec10D:level2}), showing that FACTORIZATION
and DISCRETE LOGARITHM
are in {\bf BQP}, whereas the current state of the art does not allow us to
assert that they are in {\bf BPP}.  The inclusion of {\bf BQP} in {\bf
PSPACE} implies that it is possible to classically simulate, and with as
good aproximation as desired, quantum problems with reasonable memory
resources, although the simulation would be exponentially slow
in time.  That is why there are not solvable problems
with QTMs escaping the domain of DTMs. Stated in a different way,
quantum computation does not contradict the Church-Turing hypothesis
(Subsec.~\ref{sec8A:level2}). Only invoking efficiency might classical TMs
yield to QTMs.

Even though we do not know whether {\bf BPP} is a proper subset of
{\bf BQP}, we do know classical particular cases of  algorithms (not
complexity classes as a whole) that can be speeded-up quantumly with
respect to their classical running.  Simon's algorithm 
shows an exponential gain
$O(2^n)\to O(n)$ (Subsec.~\ref{sec10B:level2}), and Grover's shows a
quadratic improvement $O(N)\to O(N^{1/2})$
(Subsec.~\ref{sec10C:level2}). But is not always possible to speed-up
the algorithm substancially.  There are oracle problems which do not
admit an essential quantum speed-up; at the most it is possible to
go from $N$ classical queries down to $N/2$ quantum queries. An example is
the PARITY problem (to find the parity of the number of non-zero bits
of a string in $\{0,1\}^n$, (Farhi et al., 1998)).

\end{document}